\documentclass[reprint,longbibliography,twocolumn,superscriptaddress,
showpacs,preprintnumbers,
notitlepage,amsmath,amssymb,
aps,
prx,
]{revtex4-1}

\usepackage{graphicx}
\usepackage{dcolumn}
\usepackage{bm}
\usepackage{diagbox}

\usepackage{graphicx}
\usepackage{enumerate}
\usepackage{subfigure}
\usepackage{epstopdf}
\usepackage[colorlinks,
linkcolor=red,
anchorcolor=black,
citecolor=blue]{hyperref}
\usepackage{amssymb}
\usepackage{framed}
\usepackage{tabularx}
\usepackage{booktabs}
\usepackage{float}
\usepackage[table]{xcolor}
\usepackage{array}
\usepackage{tikz}
\hypersetup{colorlinks=true}

\begin{document}
\title{Real-space construction of crystalline topological superconductors and insulators in 2D interacting fermionic systems}
\author{Jian-Hao Zhang}
\affiliation{Department of Physics, The Chinese University of Hong Kong, Shatin, New Territories, Hong Kong, China}
\author{Shuo Yang}
\affiliation{State Key Laboratory of Low Dimensional Quantum Physics and Department of Physics, Tsinghua University, Beijing 100084, China}
\affiliation{Frontier Science Center for Quantum Information, Beijing 100084, China}
\author{Yang Qi}
\email{qiyang@fudan.edu.cn}
\affiliation{Center for Field Theory and Particle Physics, Department of Physics, Fudan University, Shanghai 200433, China}
\affiliation{State Key Laboratory of Surface Physics, Fudan University, Shanghai 200433, China}
\affiliation{Collaborative Innovation Center of Advanced Microstructures, Nanjing 210093, China}
\author{Zheng-Cheng Gu}
\email{zcgu@phy.cuhk.edu.hk}
\affiliation{Department of Physics, The Chinese University of Hong Kong, Shatin, New Territories, Hong Kong, China}

\begin{abstract}
The construction and classification of crystalline symmetry protected topological (SPT) phases in interacting bosonic and fermionic systems have been intensively studied in the past few years. Crystalline SPT phases are not only of conceptual importance, but also provide us great opportunities towards experimental realization since space group symmetries naturally exist for any realistic material. In this paper, we systematically classify the crystalline topological superconductors (TSC) and topological insulators (TI) in 2D interacting fermionic systems by using an explicit real-space construction. In particular, we discover an intriguing fermionic crystalline topological superconductor that can only be realized in interacting fermionic systems (i.e., not in free-fermion or interacting bosonic systems). Moreover, we also verify the recently conjectured crystalline equivalence principle for generic 2D interacting fermionic systems. 
\end{abstract}

\newcommand{\lra}{\longrightarrow}
\newcommand{\xra}{\xrightarrow}
\newcommand{\ra}{\rightarrow}
\newcommand{\bs}{\boldsymbol}
\newcommand{\ul}{\underline}
\newcommand{\1}{\text{\uppercase\expandafter{\romannumeral1}}}
\newcommand{\2}{\text{\uppercase\expandafter{\romannumeral2}}}
\newcommand{\3}{\text{\uppercase\expandafter{\romannumeral3}}}
\newcommand{\4}{\text{\uppercase\expandafter{\romannumeral4}}}
\newcommand{\5}{\text{\uppercase\expandafter{\romannumeral5}}}
\newcommand{\6}{\text{\uppercase\expandafter{\romannumeral6}}}

\maketitle

\tableofcontents

\section{Introduction}
\subsection{The goal of this paper}
In the past decade, a lot of efforts have been made on the theoretical prediction and experimental searching for topological superconductors (TSC) and topological insulators (TI) in non-interacting or weakly-interacting systems \cite{KaneRMP,ZhangRMP}. However, in realistic materials, strong electronic interactions typically play a very important role and can not be neglected or treated as perturbations, especially in low dimensional systems. Therefore, a complete construction and classification of TSC/TI in interacting fermionic systems become a very important but challenging problem. It turns out that a large class of TSC/TI require certain symmetry protection and they can be connected to a trivial disorder phase (e.g. an s-wave BCS-superconductor or an atomic insulator) in the absence of global symmetry. Such kind of ``integer" TSC/TI are short-range entangled quantum states and they are actually the simplest examples of symmetry-protected topological (SPT) phases \cite{ZCGu2009}.   

Thanks to the cutting-edge breakthrough in the classification and construction of SPT phases for interacting bosonic and fermionic systems recently \cite{pollmann10,chen11a,chen11b,schuch11,XieChenScience,cohomology,invertible1,invertible2,invertible3,Kapustin2014,wen15,ZCGu2012,Kapustin2015,Kapustin2017,general1,general2}, a complete understanding of ``integer" TSC/TI protected by internal symmetry(e.g., time reversal symmetry or spin rotational symmetry) for interacting electronic systems has been achieved \cite{fidkowski10, fidkowski11,wangc-science,invertible2,ChongWang2014,Witten}. In general, by ``gauging'' the internal (unitary) symmetry~\cite{LevinGu,Gu-Levin,gauging1,threeloop,ran14,wangj15,wangcj15,lin15,gauging3,dimensionalreduction,gauging2,2DFSPT,braiding} and investigating the braiding statistics of the corresponding gauge fluxes/flux lines, different SPT phases can be uniquely identified.
Moreover, gapless edge states or anomalous surface topological orders have also been proposed as another very powerful way to characterize different SPT phases in interacting systems~\cite{Ashvin2013,ChongWang2013,XieChen2015,ChenjieWang2016,XLQi2013,Senthil2013,Lukasz2013,XieChen2014,ChongWang2014}.

In recent years, the notion of SPT phases was further extended to systems with crystalline symmetry protection and the so-called crystalline SPT phases have been intensively studied \cite{TCI,Fu2012,ITCI,reduction,building,correspondence,SET,230,BCSPT,Jiang2017,Kane2017,Shiozaki2018,ZDSong2018,defect,realspace,KenX,rotation,LuX,YMLu2018,Cheng2018,Hermele2018,Po2020,Huang2020PRR,Huang2021PRR}. 
Crystalline SPT phases are not only of conceptual importance, but also provide us great opportunities towards experimental realization since space group symmetries naturally exist for any realistic material.
The crystalline TI first proposed in free fermion systems is the simplest example of crystalline SPT phases, and it has already been realized in many different materials \cite{TCIrealization1,TCIrealization2,TCIrealization3,TCIrealization4}.  For free fermion systems, there are two systematic methods for classifying and characterizing the crystalline TI: one is the so-called ``\textit{symmetry indicators}'' \cite{230,indicator3,indicator5,indicator1,indicator2,Po2020}, which classifies and characterizes the crystalline TI by symmetry representations of band structures at high-symmetry momenta; another is a real-space construction based on the concept of topolgical crystal \cite{reduction,realspace}. Very recently,  
boundary modes \cite{surfaceTCI,d-2,rotationsurface} of the so-called higher-order TSC/TI \cite{higher4,higher2,higher3,higher1,higher5,RXZhang2020} protected by crystalline symmetry(with additional time reversal symmetry in certain cases) 
also attract a lot of interest in both 2D and 3D.
In general, an $nth$-order TSC/TI protects gapless modes at the
system boundary of codimension $n$. For example, a second-order 3D TI has gapless states on
its hinges, while its surfaces are gapped, and a third-order 3D TI has gapless states on its corners,
while both its surfaces and hinges are gapped. Nevertheless, most of these studies are still focusing on free fermion systems and it is not quite clear whether the corresponding gapless boundary modes are stable or not against interactions.
On the other hand, for interacting bosonic systems, it was pointed out that the classification of crystalline SPT phases is closely related to the SPT phases with internal symmetry. In Ref.~\cite{correspondence}, 
a ``\textit{crystalline equivalence principle}'' was proposed with a rigorous mathematical proof: i.e., crystalline topological phases with space group symmetry $G$ are in one-to-one correspondence with topological phases protected by the same internal symmetry $G$, but acting in a twisted way, where if an element of $G$ is a mirror reflection (orientation-reversing symmetry), it should be regarded as a time-reversal symmetry (antiunitary symmetry). This principle indicates the profound relationship between crystalline SPT phases and SPT phases protected by internal symmetry. Thus, the classification of crystalline SPT phases for free-fermion and interacting bosonic systems can be computed systematically.

Despite the huge success in understanding crystalline SPT phases for free-fermion and interacting bosonic systems, a systematical understanding of crystalline SPT phases for interacting fermionic systems is still lacking. Although it has been believed that the strategy of classification schemes \cite{correspondence,defect,realspace,KenX} should still work and some simple examples have been studied~\cite{rotation,YMLu2018,dihedral}, most studies focus on the systems with point group symmetry only and the generic cases are unclear.  Recent study on generalizing ``\textit{crystalline equivalence principle}'' into interacting fermionic systems shed new light towards a complete understanding of crystalline SPT phases for interacting fermion.
In Ref. \onlinecite{dihedral}, by some explicit calculations for both crystalline SPT phases and SPT phases protected by internal symmetry, it has been demonstrated that the \textit{crystalline equivalence principle} is still valid for 2D crystalline SPT phases protected by point group symmetry, but in a twisted way, where spinless (spin-1/2) fermions should be mapped into spin-1/2 (spinless) fermions.

In this paper, we aim at systematically constructing and classifying crystalline TSC/TI for 2D interacting fermionic systems and establishing a general paradigm of real-space construction for interacting fermionic crystalline SPT phases. We will consider both spinless and spin-1/2 fermionic systems. In particular, we obtain an intriguing fermionic TSC that cannot be realized in either free-fermion or interacting bosonic systems: a $p4m$ (\#11 wallpaper group) symmetric 2D system with spinless fermions. These TSC can be realized in systems with co-planar spin order and might have very interesting experimental implementations. Furthermore, we compare all our results with the classifications of 2D fermionic SPT (FSPT) phases protected by corresponding internal symmetries. We confirm the crystalline equivalence principle for generic 2D interacting fermionic systems, where a mirror reflection symmetry action should be mapped onto a time-reversal symmetry action, and that spinless (spin-1/2) fermionic systems should be mapped into spin-1/2 (spinless) fermionic systems. 

Our general real-space construction scheme includes following three major steps:
\begin{description}
\item[Cell decomposition] For a specific wallpaper group, firstly we can divide it into an assembly of unit cells; then we divide each unit cell into an assembly of lower-dimensional blocks.
\item[Block-state decoration] For a specific wallpaper group with cell decomposition, we can decorate lower-dimensional block-state on different blocks. A gapped assembly of block-states is called \textit{obstruction free} decoration.
\item[Bubble equivalence] For a specific obstruction-free decoration, we need to further examine 
whether such a decoration can be trivialized or not. Finally, the obstruction and trivialization free block state decoration corresponds to a 2D fermionic crystalline SPT phase.
\end{description}
In addition, we also need to examine the possible non-trivial stacking relation between block-states with different dimensions to determine the actual group structure of 2D fermionic crystalline SPT phases. 

\subsection{Space group symmetry for spinless and spin-1/2 systems\label{spinSec}}
Here we would like to clarify the precise meaning of ``spinless'' and ``spin-1/2'' fermions for systems with and without $U^f(1)$ charge conservation symmetry. 

For a fermionic system with total symmetry group $G_f$, there is always a subgroup $\mathbb{Z}_2^f=\{1,P_f=(-1)^{F}\}$, where $F$ is the total number of fermions. $\mathbb{Z}_2^f$ is the center of $G_f$ because all physical symmetries commute with $P_f$, i.e., cannot change fermion parity of the system. In particular, for systems without $U^f(1)$ charge conservation symmetry, we can define the bosonic (physical) symmetry group by a quotient group $G_b=G_f/\mathbb{Z}_2^f$. In reverse, for a given physical symmetry group $G_b$, there are many different fermionic symmetry groups $G_f$ which are the central extension of $G_b$ by $\mathbb{Z}_2^f$. It can be expressed by the following short exact sequence:
\begin{align}
0\rightarrow\mathbb{Z}_2^f\rightarrow G_f\rightarrow G_b\rightarrow0
\label{SES1}
\end{align}
and different extensions $G_f$ are characterized by different factor systems of Eq. (\ref{SES1}) that are 2-cocycles $\omega_2\in\mathcal{H}^2(G_b,\mathbb{Z}_2)$. Consequently, we denote $G_f$ as $\mathbb{Z}_2^f\times_{\omega_2}G_b$. 

For systems with additional $U^f(1)$ charge conservation, the group element is $U_\theta=e^{i\theta F}$. Aforementioned fermion parity operator $P_f=U_\pi$ is the order 2 element of $U^f(1)$, hence we denote this charge conservation symmetry by $U^f(1)$ with a superscript $f$. It is easy to notice that $U^f(1)$ charge conservation is a normal subgroup of the total symmetry group $G_f$, which can be expressed by the following short exact sequence:
\begin{align}
0\rightarrow U^f(1)\rightarrow G_f\rightarrow G\rightarrow0
\label{SES2}
\end{align}
where $G:=G_f/U^f(1)$. In reverse, for a given physical symmetry group $G$, we can define $G_f=U^f(1)\rtimes_{\omega_2}G$. 
Here $\omega_2$ is related to the extension of the physical symmetry group $G$. 
The multiplication of the total symmetry group $G_f$ is defined as:
\begin{align}
(1,g)\times(1,h)=\left(e^{2\pi i\omega_2(g,h)F},gh\right)\in G_f
\end{align}
with $\omega_2\in\mathbb{R}/\mathbb{Z}=[0,1)$ as a $U(1)$ phase, associated with $g,h\in G$. Therefore $\omega_2$ is a 2-cocycle in $\mathcal{H}^2(G,\mathbb{R}/\mathbb{Z})$. 

The spin of fermions (spinless or spin-1/2) is characterized by different choices of 2-cocycles $\omega_2$, i.e. the spinless corresponds to a trivial $\omega_2$ while spin-1/2 fermion corresponds to specific choice of nontrivial $\omega_2$. 

For example, consider even-fold dihedral group $D_{2n}$ symmetry with two generators $\bs{R}$ and $\bs{M}$ satisfying $\bs{R}^{2n}=\bs{M}^2=I$ ($n\in\mathbb{Z}$ and $I$ is identity) for systems without $U^f(1)$ symmetry.
Different extensions of fermion parity are characterized by different 2-cocycles $\omega_2$:
\begin{align}
\omega_2\in\mathcal{H}^2(D_{2n},\mathbb{Z}_2)=\mathbb{Z}_2^3
\end{align}
In particular, the spinless fermions corresponding to the trivial 2-cocycle $\omega_2$ satisfy:
\begin{align}
\left\{
\begin{aligned}
&\bs{R}^{2n}=1\\
&\bs{M}^2=1\\
\end{aligned}
\right.,
\end{align}
while the spin-1/2 fermions corresponding to the 2-cocycle $\omega_2$ satisfy:
\begin{align}
\left\{
\begin{aligned}
&\bs{R}^{2n}=P_f\\
&\bs{M}^2=P_f\\
&\bs{M}\bs{R}\bs{M}^{-1}\bs{R}=1
\end{aligned}
\right.
\end{align}

To satisfy these conditions, we consider the 2-cocycle $\omega_2$ as following. For $\forall a_g,b_h\in D_{2n}$ defined as:
\begin{align}
D_{2n}=\left\{(a,g)=a_g\Big|0\leq a\leq(2n-1),0\leq g\leq1\right\},
\end{align}
we choose:
\begin{align}
\omega_2(a_g,b_h)=&\left\lfloor\frac{\left[(-1)^{g+h}a\right]_{2n}+\left[(-1)^{h}b\right]_{2n}}{2n}\right\rfloor\nonumber\\
&+(1-\delta_a)(a+1)h+g\cdot h
\end{align}
where we define $[x]_n\equiv x(\mathrm{mod}~n)$, $\lfloor x\rfloor$ as the greatest integer less than or equal to $x$, and
\begin{align}
\delta_a=
\left\{
\begin{aligned}
&1~~\mathrm{if}~a=0\\
&0~~\mathrm{otherwise}
\end{aligned}
\right.
\end{align}

For systems with $U^f(1)$ symmetry, spinless and spin-1/2 fermions are characterized by different 2-cocycle $\omega_2$:
\begin{align}
\omega_2\in\mathcal{H}^2(D_{2n},\mathbb{R}/\mathbb{Z})=\mathbb{Z}_2
\end{align}
In particular, the spinless fermions corresponding to the trivial 2-cocycle $\omega_2$ satisfy:
\begin{align}
\bs({MR}^n)^2=1
\end{align}
while the spin-1/2 fermions corresponding to the 2-cocycle $\omega_2$ satisfy:
\begin{align}
\bs({MR}^n)^2=P_f
\end{align}

\subsection{Summary of main results}


\begin{table}[t]
\renewcommand\arraystretch{1.2}
\begin{tabular}{|c|c|c|c|c|c|c|}
\hline
$~~G_b~~$&$~~~E_{0}^{\mathrm{1D}}~~~$&$~~~E_{0}^{\mathrm{0D}}~~~$&~~~$\mathcal{G}_0$~~~\\
\hline
$p1$&${\color{red}\mathbb{Z}_2^2}$&${\color{red}\mathbb{Z}_2}$&${\color{red}\mathbb{Z}_2^3}$\\
\hline
$p2$&$\mathbb{Z}_1$&${\color{red}\mathbb{Z}_2^3}\times{\color{blue}\mathbb{Z}_2}$&${\color{red}\mathbb{Z}_2^3}\times{\color{blue}\mathbb{Z}_2}$\\
\hline
$pm$&${\color{red}\mathbb{Z}_2^3}$&$~{\color{red}\mathbb{Z}_2^2}\times{\color{blue}\mathbb{Z}_2}~$&${\color{red}\mathbb{Z}_2^5}\times{\color{blue}\mathbb{Z}_2}$\\
\hline
$pg$&${\color{red}\mathbb{Z}_2^2}$&${\color{red}\mathbb{Z}_2}$&${\color{red}\mathbb{Z}_2^3}$\\
\hline
$cm$&${\color{red}\mathbb{Z}_2^2}$&${\color{red}\mathbb{Z}_2}\times{\color{blue}\mathbb{Z}_2}$&${\color{red}\mathbb{Z}_2^3}\times{\color{blue}\mathbb{Z}_2}$\\
\hline
$pmm$&$\mathbb{Z}_1$&${\color{red}\mathbb{Z}_2^4}\times{\color{blue}\mathbb{Z}_2^4}$&${\color{red}\mathbb{Z}_2^4}\times{\color{blue}\mathbb{Z}_2^4}$\\
\hline
$pmg$&${\color{red}\mathbb{Z}_2}$&${\color{red}\mathbb{Z}_2^2}\times{\color{blue}\mathbb{Z}_2^2}$&${\color{red}\mathbb{Z}_2^3}\times{\color{blue}\mathbb{Z}_2^2}$\\
\hline
$pgg$&${\color{red}\mathbb{Z}_2}$&${\color{red}\mathbb{Z}_2}\times{\color{blue}\mathbb{Z}_2}$&${\color{red}\mathbb{Z}_2^2}\times{\color{blue}\mathbb{Z}_2}$\\
\hline
$cmm$&$\mathbb{Z}_1$&${\color{red}\mathbb{Z}_2^3}\times{\color{blue}\mathbb{Z}_2^2}$&${\color{red}\mathbb{Z}_2^3}\times{\color{blue}\mathbb{Z}_2^2}$\\
\hline
$p4$&$\mathbb{Z}_1$&${\color{red}\mathbb{Z}_2^2}\times{\color{blue}\mathbb{Z}_4\times\mathbb{Z}_2}$&${\color{red}\mathbb{Z}_2^2}\times{\color{blue}\mathbb{Z}_4\times\mathbb{Z}_2}$\\
\hline
$p4m$&${\color{red}\mathbb{Z}_2}$&${\color{red}\mathbb{Z}_2^3}\times{\color{blue}\mathbb{Z}_2^3}$&${\color{red}\mathbb{Z}_2^4}\times{\color{blue}\mathbb{Z}_2^3}$\\
\hline
$p4g$&$\mathbb{Z}_1$&${\color{red}\mathbb{Z}_2^2}\times{\color{blue}\mathbb{Z}_2^2}$&${\color{red}\mathbb{Z}_2^2}\times{\color{blue}\mathbb{Z}_2^2}$\\
\hline
$p3$&$\mathbb{Z}_1$&${\color{red}\mathbb{Z}_2}\times{\color{blue}\mathbb{Z}_3^3}$&$~{\color{red}\mathbb{Z}_2}\times{\color{blue}\mathbb{Z}_3^3}~$\\
\hline
$~~~p3m1~~~$&${\color{red}\mathbb{Z}_2}$&${\color{red}\mathbb{Z}_2}\times{\color{blue}\mathbb{Z}_2}$&${\color{red}\mathbb{Z}_2^2}\times{\color{blue}\mathbb{Z}_2}$\\
\hline
$p31m$&${\color{red}\mathbb{Z}_2}$&${\color{red}\mathbb{Z}_2}\times{\color{blue}\mathbb{Z}_2\times\mathbb{Z}_3}$&${\color{red}\mathbb{Z}_2^2}\times{\color{blue}\mathbb{Z}_2\times\mathbb{Z}_3}$\\
\hline
$p6$&$\mathbb{Z}_1$&${\color{red}\mathbb{Z}_2^2}\times{\color{blue}\mathbb{Z}_3^2}$&${\color{red}\mathbb{Z}
_2^2}\times{\color{blue}\mathbb{Z}_3^2}$\\
\hline
$p6m$&$\mathbb{Z}_1$&${\color{red}\mathbb{Z}_2^2}\times{\color{blue}\mathbb{Z}_2^2}$&${\color{red}\mathbb{Z}_2^2}\times{\color{blue}\mathbb{Z}_2^2}$\\
\hline
\end{tabular}
\caption{Interacting classification of 2D crystalline TSC for spinless fermionic systems. The results are listed layer by layer, together with their group structures (represented by $\mathcal{G}_0$). We label the classification indices with fermionic/bosonic root phases with red/blue. The fermionic $\mathbb{Z}_4$ indices are obtained from nontrivial extensions between 1D and 0D block-states, thus stacking two root phases will become another fermionic crystalline TSC. In particular, 1D block-state of the $p4m$ case is an intriguing fermionic SPT phase that cannot be realized by free-fermion and interacting bosonic systems.}
\label{spinless}
\end{table}

\begin{table}[t]
\renewcommand\arraystretch{1.2}
\begin{tabular}{|c|c|c|c|c|}
\hline
$~~G_b~~$&$~~~E_{1/2}^{\mathrm{1D}}~~~$&$~~~E_{1/2}^{\mathrm{0D}}~~~$&$~~~\mathcal{G}_{1/2}~~~$\\
\hline
$p1$&${\color{red}\mathbb{Z}_2^2}$&${\color{red}\mathbb{Z}_2}$&${\color{red}\mathbb{Z}_2^3}$\\
\hline
$p2$&${\color{red}\mathbb{Z}_2^3}$&${\color{red}\mathbb{Z}_4^4}$&${\color{red}\mathbb{Z}_4\times\mathbb{Z}_8^3}$\\
\hline
$pm$&${\color{red}\mathbb{Z}_2}$&${\color{red}\mathbb{Z}_4^2}$&${\color{red}\mathbb{Z}_4\times\mathbb{Z}_8}$\\
\hline
$pg$&${\color{red}\mathbb{Z}_2^2}$&$~{\color{red}\mathbb{Z}_2}~$&${\color{red}\mathbb{Z}_2^3}$\\
\hline
$cm$&${\color{red}\mathbb{Z}_2}$&${\color{red}\mathbb{Z}_4}$&${\color{red}\mathbb{Z}_2\times\mathbb{Z}_4}$\\
\hline
$pmm$&$~\mathbb{Z}_1~$&${\color{blue}\mathbb{Z}_2^8}$&${\color{blue}\mathbb{Z}_2^8}$\\
\hline
$pmg$&${\color{red}\mathbb{Z}_2^2}$&${\color{red}\mathbb{Z}_4^3}$&${\color{red}\mathbb{Z}_4\times\mathbb{Z}_8^2}$\\
\hline
$pgg$&${\color{red}\mathbb{Z}_2^2}$&${\color{red}\mathbb{Z}_4^2}$&${\color{red}\mathbb{Z}_2\times\mathbb{Z}_4\times\mathbb{Z}_8}$\\
\hline
$cmm$&${\color{red}\mathbb{Z}_2}$&${\color{red}\mathbb{Z}_4}\times{\color{blue}\mathbb{Z}_2^4}$&${\color{red}\mathbb{Z}_8}\times{\color{blue}\mathbb{Z}_2^4}$\\
\hline
$p4$&${\color{red}\mathbb{Z}_2^2}$&${\color{red}\mathbb{Z}_8^2\times\mathbb{Z}_4}$&${\color{red}\mathbb{Z}_2\times\mathbb{Z}_8^3}$\\
\hline
$p4m$&$\mathbb{Z}_1$&${\color{blue}\mathbb{Z}_2^6}$&${\color{blue}\mathbb{Z}_2^6}$\\
\hline
$p4g$&${\color{red}\mathbb{Z}_2}$&${\color{red}\mathbb{Z}_8}\times{\color{blue}\mathbb{Z}_2^2}$&${\color{red}\mathbb{Z}_2\times\mathbb{Z}_8}\times{\color{blue}\mathbb{Z}_2^2}$\\
\hline
$p3$&$\mathbb{Z}_1$&${\color{red}\mathbb{Z}_2}\times{\color{blue}\mathbb{Z}_3^3}$&${\color{red}\mathbb{Z}_2}\times{\color{blue}\mathbb{Z}_3^3}$\\
\hline
$~~~p3m1~~~$&$\mathbb{Z}_1$&${\color{red}\mathbb{Z}_4}$&${\color{red}\mathbb{Z}_4}$\\
\hline
$p31m$&$\mathbb{Z}_1$&${\color{red}\mathbb{Z}_4}\times{\color{blue}\mathbb{Z}_3}$&${\color{red}\mathbb{Z}_4}\times{\color{blue}\mathbb{Z}_3}$\\
\hline
$p6$&${\color{red}\mathbb{Z}_2}$&${\color{red}\mathbb{Z}_{12}\times\mathbb{Z}_4}\times{\color{blue}\mathbb{Z}_3}$&${\color{red}\mathbb{Z}_{12}\times\mathbb{Z}_8}\times{\color{blue}\mathbb{Z}_3}$\\
\hline
$p6m$&$\mathbb{Z}_1$&${\color{blue}\mathbb{Z}_2^4}$&${\color{blue}\mathbb{Z}_2^4}$\\
\hline
\end{tabular}
\caption{Interacting classification of 2D crystalline TSC for spin-1/2 fermionic systems. The results are listed layer by layer, together with their group structure (represented by $\mathcal{G}_{1/2}$). We label the classification indices with fermionic/bosonic root phases with red/blue. We note that except for $p1$ and $pg$ cases(spinless fermion and spin-1/2 fermion are the same for these two cases), all $\mathbb{Z}_4$ indices are from 2-fold rotation(the on-site symmetry group of arbitrary 0D block is $\mathbb{Z}_4^f$ which is the nontrivial $\mathbb{Z}_2^f$ extension of $\mathbb{Z}_2$.) and stacking two root phases will become a bosonic crystalline SPT phase.
Similarly, for $p4$ case, two of three $\mathbb{Z}_8$ fermionic indices are from 4-fold rotation and stacking two root phases will also become a bosonic SPT phase. The fermionic $\mathbb{Z}_8$ index of the $p4g$ case can be understood in the same way. All other fermionic $\mathbb{Z}_8$ indices are obtained from nontrivial extension between 1D and 0D block-states. For these cases, stacking two fermionic root phases will become another fermionic crystalline TSC. In addition, the $\mathbb{Z}_{12}$ index of $p6$ case is obtained from 6-fold rotation and stacking two fermionic root phases will also lead to a bosonic phase.}
\label{spin-1/2}
\end{table}

Here we first summarize all classification results of 2D crystalline TSC for both spinless and spin-1/2 fermionic systems. We label the classification attributed to $p$-dimensional block-state decorations by $E^{d\mathrm{D}}$. For the systems with spinless fermions, the classification results are summarized in Table. \ref{spinless}, and the classification data are listed layer by layer, i.e., classification contributed by 0D/1D block-state decorations, respectively. For the systems with spin-1/2 fermions, the classification results are summarized in Table. \ref{spin-1/2} layer by layer. Furthermore, we also study the group structure of the classifications by explicitly investigating the possible nontrivial stacking relation between 1D and 0D block-states: For certain cases, stacking of several 1D block-states can be deformed into a 0D block-state, hence the total group could be a nontrivial extension between 1D and 0D block-states. 
In particular, we label the classification indices with fermionic root phase by red, and the classification indices with bosonic root phase by blue. 

\begin{table}[t]
\renewcommand\arraystretch{1.2}
\begin{tabular}{|c|c|c|c|c|}
\hline
$~~G_b~~$&~spinless~~&~spin-1/2~~\\
\hline
$p1$&${\color{red}\mathbb{Z}}$&${\color{red}\mathbb{Z}}$\\
\hline
$p2$&${\color{red}\mathbb{Z}\times\mathbb{Z}_4^3}\times{\color{blue}\mathbb{Z}_2}$&${\color{red}\mathbb{Z}\times\mathbb{Z}_4^3}\times{\color{blue}\mathbb{Z}_2}$\\
\hline
$pm$&${\color{red}\mathbb{Z}\times\mathbb{Z}_4}\times{\color{blue}\mathbb{Z}_2}$&${\color{red}\mathbb{Z}\times\mathbb{Z}_4}\times{\color{blue}\mathbb{Z}_2}$\\
\hline
$pg$&${\color{red}\mathbb{Z}}$&${\color{red}\mathbb{Z}}$\\
\hline
$cm$&${\color{red}\mathbb{Z}}\times{\color{blue}\mathbb{Z}_2}$&${\color{red}\mathbb{Z}}\times{\color{blue}\mathbb{Z}_2}$\\
\hline
$pmm$&${\color{red}\mathbb{Z}\times\mathbb{Z}_4^{3}}\times{\color{blue}\mathbb{Z}_2^4}$&${\color{red}2\mathbb{Z}}\times{\color{blue}\mathbb{Z}_2^{8}}$\\
\hline
$pmg$&${\color{red}\mathbb{Z}\times\mathbb{Z}_4^2}\times{\color{blue}\mathbb{Z}_2}$&${\color{red}\mathbb{Z}\times\mathbb{Z}_4^2}\times{\color{blue}\mathbb{Z}_2}$\\
\hline
$pgg$&${\color{red}\mathbb{Z}\times\mathbb{Z}_4}\times{\color{blue}\mathbb{Z}_2}$&${\color{red}\mathbb{Z}\times\mathbb{Z}_4}\times{\color{blue}\mathbb{Z}_2}$\\
\hline
$cmm$&${\color{red}\mathbb{Z}\times\mathbb{Z}_4^2}\times{\color{blue}\mathbb{Z}_2^2}$&${\color{red}2\mathbb{Z}\times\mathbb{Z}_4}\times{\color{blue}\mathbb{Z}_2^4}$\\
\hline
$p4$&${\color{red}\mathbb{Z}\times\mathbb{Z}_8\times\mathbb{Z}_4}\times{\color{blue}\mathbb{Z}_4\times\mathbb{Z}_2}$&${\color{red}\mathbb{Z}\times\mathbb{Z}_8\times\mathbb{Z}_4}\times{\color{blue}\mathbb{Z}_4\times\mathbb{Z}_2}$\\
\hline
$p4m$&${\color{red}\mathbb{Z}\times\mathbb{Z}_8\times\mathbb{Z}_4}\times{\color{blue}\mathbb{Z}_2^3}$&${\color{red}2\mathbb{Z}\times\mathbb{Z}_2}\times{\color{blue}\mathbb{Z}_2^6}$\\
\hline
$p4g$&$~{\color{red}\mathbb{Z}\times\mathbb{Z}_8}\times{\color{blue}\mathbb{Z}_2^2}~$&$~{\color{red}\mathbb{Z}\times\mathbb{Z}_2}\times{\color{blue}\mathbb{Z}_4\times\mathbb{Z}_2^2}~$\\
\hline
$p3$&${\color{red}\mathbb{Z}\times\mathbb{Z}_3^2}\times{\color{blue}\mathbb{Z}_3^3}$&${\color{red}\mathbb{Z}\times\mathbb{Z}_3^2}\times{\color{blue}\mathbb{Z}_3^3}$\\
\hline
$~~p3m1~~$&$~{\color{red}\mathbb{Z}\times\mathbb{Z}_3^2}\times{\color{blue}\mathbb{Z}_2}~$&$~{\color{red}\mathbb{Z}\times\mathbb{Z}_3^2}\times{\color{blue}\mathbb{Z}_2}~$\\
\hline
$p31m$&${\color{red}\mathbb{Z}\times\mathbb{Z}_3}\times{\color{blue}\mathbb{Z}_6}$&${\color{red}\mathbb{Z}\times\mathbb{Z}_3}\times{\color{blue}\mathbb{Z}_6}$\\
\hline
$p6$&$~{\color{red}\mathbb{Z}\times\mathbb{Z}_{12}}\times{\color{blue}\mathbb{Z}_6\times\mathbb{Z}_3}~$&$~{\color{red}\mathbb{Z}\times\mathbb{Z}_{12}}\times{\color{blue}\mathbb{Z}_6\times\mathbb{Z}_3}~$\\
\hline
$p6m$&${\color{red}\mathbb{Z}\times\mathbb{Z}_{12}}\times{\color{blue}\mathbb{Z}_2^2}$&${\color{red}2\mathbb{Z}\times\mathbb{Z}_6}\times{\color{blue}\mathbb{Z}_2^3}$\\
\hline
\end{tabular}
\caption{The interacting classification of crystalline TI for 2D interacting fermionic systems. The results for both spinless and spin-1/2 fermions are summarized together. We note that the classifications are the same for those wall paper groups with only one reflection axis. We label the classification indices with fermionic/bosonic root phases with red/blue. }
\label{insulator U(1)}
\end{table}

For 2D crystalline TI protected by both wall paper group and $U^f(1)$ charge conservation symmetry, 
we generalize the procedures of real-space construction highlighted in Sec. \ref{general} to include the internal $U^f(1)$ symmetry.
It turns out that 1D block-state decoration does not contribute any nontrivial crystalline topological phase because of the absence of nontrivial 1D root phase in the presence of $U^f(1)$ symmetry. 
All classification results are summarized in Table \ref{insulator U(1)}. Again we label the classification indices with fermionic root phase by red, and the classification indices with bosonic root phase by blue.

The rest of the paper is organized as follows: In Sec. \ref{general}, we introduce the general paradigm of the real-space construction of crystalline SPT phases protected by wallpaper group in 2D interacting fermionic systems. In Sec. \ref{example}, we explicitly show how to construct and classify the crystalline TSC in 2D interacting fermionic systems for five different crystallographic systems  by using real-space construction, for both spinless and spin-1/2 fermions. All classification results are summarized in Tables \ref{spinless} and \ref{spin-1/2}.  Furthermore, we also classify the crystalline TI in 2D interacting fermionic systems with additional $U^f(1)$ charge conservation by using similar real-space construction scheme in Sec. \ref{insulator}, and the results are summarized in Table \ref{insulator U(1)}. 
In Sec. \ref{principle}, by comparing these results with the classification results of 2D FSPT phases protected by the corresponding on-site symmetry groups, we verify the \textit{crystalline equivalence principle} for generic 2D interacting fermionic systems.
Finally, conclusions and discussions about further applications of real-space construction and experimental implications are presented in Sec. \ref{conclusion}. 
In Supplementary Materials, we first discuss the 2D crystalline TI protected by point group symmetry and compare the results with the classifications of 2D FSPT phases protected by the corresponding internal symmetry, then we discuss the real space construction of TSC and TI for all remaining cases of wallpaper groups \cite{supplementary}.

\section{General paradigm of real-space construction\label{general}}
In this section, we highlight the general paradigm of real-space construction of crystalline SPT phases   for 2D interacting fermionic systems. There are three major steps: Firstly, we decompose the whole system into an assembly of unit cells, each of which is composed by several lower-dimensional blocks; secondly we decorate some proper lower-dimensional block-states on them and check their validity (for SPT phases, we require a fully gapped bulk ground state without ground-state degeneracy), that is, if the bulk state of a block-state construction cannot be fully gapped, we call such a decoration \textit{obstructed}; finally we consider the so-called bubble equivalence to investigate all possible \textit{trivializations} (We note that certain block-states decorations actually lead to a trivial crystalline SPT phase). An obstruction-free and trivialization-free decoration corresponds to a nontrivial crystalline SPT phase. Below we demonstrate these procedures in full details by using the \#14 wallpaper group $p3m1$ as an example.

\subsection{Cell decomposition}
\begin{figure*}
\centering
\includegraphics[width=\textwidth]{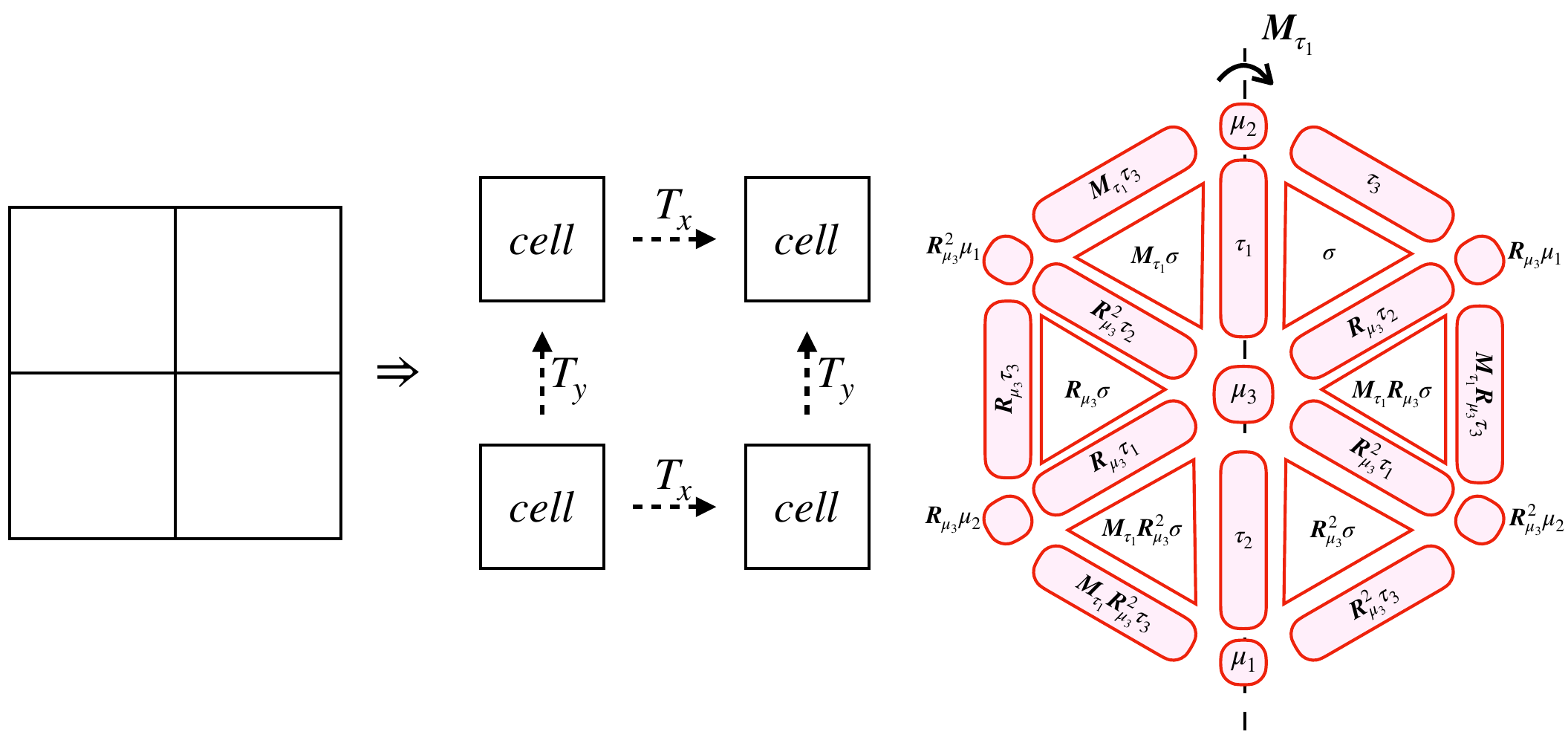}
\caption{Cell decomposition of \#14 wallpaper group $p3m1$. Left panel illustrates the inter-cell decomposition that decomposes a lattice to an assembly of unit cells; right panel illustrates the intra-cell decomposition that decompose a unit cell to an assembly of lower-dimensional blocks.}
\label{cell}
\end{figure*}

For a 2D system with an arbitrary wallpaper group symmetry, we can divide the whole system into an assembly of unit cells, where different unit cells are identical and related by translation symmetries, as illustrated in the left panel of Fig. \ref{cell}. Therefore, we should only specify the physics in each unit cell. 

Then we decompose a specific unit cell of the wallpaper group $p3m1$ into an assembly of lower-dimensional blocks (see the right panel of Fig. \ref{cell}). Here $\bs{R}_{\mu_3}$ represents 3-fold rotational symmetry operation centred at the 0D block labeled by $\mu_3$, and $\bs{M}_{\tau_1}$ represents the reflection symmetry operation with the axis (indicated by the vertical dashed line in right panel of Fig. \ref{cell}) coincided with the 1D block labeled by $\tau_1$.

The physical background of the ``intra-cell'' decomposition is the ``extended trivialization'' in each cell \cite{reduction}. Suppose $|\psi\rangle$ is an SPT state that cannot be trivialized by a symmetric finite-depth local unitary transformation. Due to the translational symmetry, $|\psi\rangle$ can be expressed in terms of a direct product of the wavefunctions of all cells:
\begin{align}
|\psi\rangle=\bigotimes\limits_{c}|\psi_c\rangle
\end{align}
Because of the translational symmetry, investigation of a specific $|\psi_c\rangle$ in a cell is enough for understanding the SPT state $|\psi\rangle$. As a consequence, $|\psi_c\rangle$ will inherit the property that cannot be trivialized by a symmetric finite-depth local unitary transformation $O^{\mathrm{loc}}$. Nevertheless, we can still define an alternative local unitary to \textit{extensively} trivialize $|\psi_c\rangle$. First we can trivialize the region $\sigma$ (see the right panel of Fig. \ref{cell}): restrict $O^{\mathrm{loc}}$ to $\sigma$ as $O_{\sigma}^{\mathrm{loc}}$ and act it on $|\psi_c\rangle$:
\begin{align}
O_\sigma^{\mathrm{loc}}|\psi_c\rangle=|T_\sigma\rangle\otimes|\psi_c^{\bar\sigma}\rangle
\end{align}
where the system is in the product state $|T_\sigma\rangle$ in region $\sigma$ and the remainder of the system $\bar\sigma$ is in the state $|\psi_c^{\bar\sigma}\rangle$. To trivialize the system symmetrically, we denote that $V_gO^{\mathrm{loc}}_\sigma V_g^{-1}$ trivializes the region $g\sigma$, where $g\in D_3$. Therefore, we act on $|\psi_c\rangle$ with:
\begin{align}
O^{\mathrm{loc}}=\bigotimes_{g\in D_3}V_gO^{\mathrm{loc}}_\sigma V_g^{-1}
\end{align}
which results in an \textit{extensively trivialized} wavefunction:
\begin{align}
&|\psi_c'\rangle=O_R^{\mathrm{loc}}|\psi_c\rangle\nonumber\\
&=\bigotimes_{g\in D_3}|T_{g\sigma}\rangle\otimes\bigotimes_{j=1,h\in D_3}^3|\psi_{h\tau_j}\rangle\otimes\bigotimes_{k=1,p\in D_3}^3|\psi_{p\mu_k}\rangle
\end{align}
where $\tau_j,j=1,2,3$ and $\mu_k,k=1,2,3$ label the 1D and 0D blocks as illustrated in the right panel of Fig. \ref{cell}. Now all nontrivial topological properties of $|\psi_c\rangle$ are encoded in lower-dimensional block-states $|\psi_{h\tau_j}\rangle$ and $|\psi_{p\mu_k}\rangle$, hence all nontrivial properties of $|\psi\rangle$ are encoded in lower-dimensional blocks in different unit cells.

\subsection{Block-state decoration}
Subsequently, with cell decompositions, we can decorate some proper lower-dimensional block-states on the corresponding lower-dimensional blocks. Some symmetry operations act internally on some lower-dimensional blocks, hence the lower-dimensional block-states should respect the corresponding on-site symmetry on which they decorate. As an example, we still consider the \#14 wallpaper group $p3m1$ with the cell decomposition as illustrated in Fig. \ref{cell}, the 3-fold rotational symmetry operations act on $g\mu_j$ ($g\in D_3$ and $j=1,2,3$) internally, and reflection symmetry operations act on $h\tau_k$ ($h\in D_3$ and $k=1,2,3$) internally, hence the root phases decorated on 0D and 1D blocks are 0D FSPT phases protected by $\mathbb{Z}_3\rtimes\mathbb{Z}_2$ on-site symmetry and 1D FSPT phases protected by $\mathbb{Z}_2$ on-site symmetry, respectively. All $d$D block-states form the group $\{\mathrm{BS}\}^{d\mathrm{D}}$, and all block-states form the following group:
\begin{align}
\{\mathrm{BS}\}=\bigotimes_{d=0}^1\{\mathrm{BS}\}^{d\mathrm{D}}
\end{align}
Here ``BS'' is the abbreviation of ``block-states''.

Furthermore, the decorated states should respect the \textit{no-open-edge condition}. Once we decorate some lower-dimensional block-states on the corresponding blocks, they might leave several gapless modes on the edge of the corresponding blocks, and there are several gapless edge modes coinciding near the blocks with lower dimension. Repeatedly consider the wallpaper group $p3m1$ as an example, if we decorate a Majorana chain on the 1D block labeled by $\tau_1$ (because of the rotational symmetry, there are also two Majorana chains decorated at the 1D blocks labeled by $\bs{R}_{\mu_3}\tau_1$ and $\bs{R}_{\mu_3}^2\tau_1$, respectively), leaving 3 dangling Majorana modes near the 0D block labeled by $\mu_3$.
In order to contribute an SPT state, the bulk of the system should be fully gapped, hence the aforementioned gapless modes should be gapped out (by some proper interactions, mass terms, entanglement pairs, etc.) in a symmetric way. If the bulk of the system cannot be fully gapped (i.e., several aforementioned 0D modes cannot be gapped in a symmetric way), we call the corresponding decoration \textit{obstructed}. Equivalently, an obstruction-free decoration should satisfy the no-open-edge condition. All obstruction-free $d$D block-states form the group $\{\mathrm{OFBS}\}^{d\mathrm{D}}\subset \{\mathrm{BS}\}^{d\mathrm{D}}$ as a subgroup of $\{\mathrm{BS}\}^{d\mathrm{D}}$, and all obstruction-free block-states form the following group:
\begin{align}
\{\mathrm{OFBS}\}=\bigotimes_{d=0}^1\{\mathrm{OFBS}\}^{d\mathrm{D}}\subset\{\mathrm{BS}\}
\end{align}
Here ``OFBS'' is the abbreviation of ``obstruction-free block-states'', and $\{\mathrm{OFBS}\}$ is a subgroup of $\{\mathrm{BS}\}$.

\begin{figure}
\centering
\includegraphics[width=0.48\textwidth]{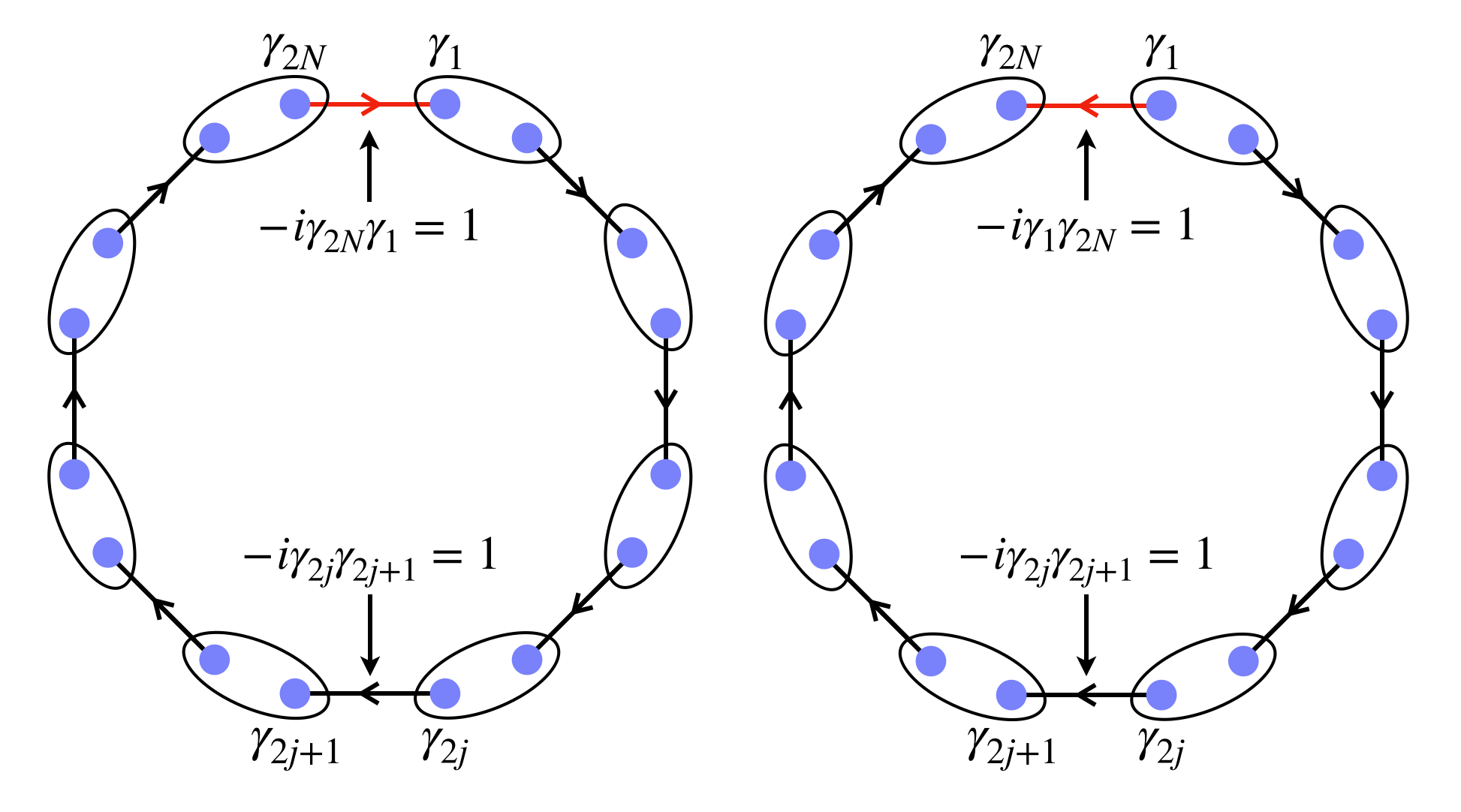}
\caption{Majorana chain with periodic boundary condition (PBC, left panel) and anti-periodic boundary condition (anti-PBC, right panel). The boundary conditions are indicated by the red arrows in both panels. Here ellipses represent the physical sites, and the solid oriented line from $j$ to $k$ indicates the paring direction, corresponding to the $i\gamma_j\gamma_k$ term in parent Hamiltonian. For the Majorana chain with PBC, the graph is not Kasteleyn oriented, and 
the ground state has odd fermion parity; for the Majorana chain with anti-PBC, the graph is Kasteleyn oriented, thus 
the ground state has even fermion parity.}
\label{anti-PBC}
\end{figure}

\subsection{Bubble equivalence}
In order to obtain a nontrivial SPT state from obstruction-free block state decorations, we should further consider possible trivializations. For blocks with dimension larger than 0, we can further decorate some codimension 1 degree of freedom that could be trivialized when they shrink to a point.
This construction is called \textit{bubble equivalence}, and we demonstrate it for different dimensions:

\paragraph{2D bubble equivalence}For 2D blocks, we can consider a 1D chain which can be shrunk to a point inside each 2D block, and there is no on-site symmetry on them for all possible cases. In fermionic systems, the only possible state we can decorate is Majorana chain. There are two distinct boundary conditions: periodic boundary condition (PBC) with odd fermion parity and anti-periodic boundary condition (anti-PBC) with even fermion parity, as seen Fig. \ref{anti-PBC}. According to the definition of bubble equivalence, we only choose the ``Majorana bubbles'' with anti-PBC because it can be trivialized if we shrink it into a point: if we decorate a Majorana chain with anti-PBC on a 2D block, we can shrink it to a smaller one by a 2D local unitary (LU) transformation without breaking any symmetry. Repeatedly apply this LU transformation on ``Majorana'' bubble, we can shrink it to a point and eliminate it (because a Majorana chain with anti-PBC has even fermion parity) by a symmetric finite-depth circuit.

Techniquely, it is well known that for two Majorana modes $\gamma_j$ and $\gamma_k$, their entanglement pair $i\gamma_j\gamma_k$ can be created by the following projection operator \cite{general1,general2}:
\begin{align}
P_{j,k}=\frac{1}{2}\left(1-i\gamma_j\gamma_k\right)
\end{align}
and the direction is from $\gamma_j$ to $\gamma_k$. Consequently the creation operator of a Majorana chain containing $2N$ Majorana modes with anti-PBC on the 2D block $\sigma$ can be generated by an assembly of these projection operators:
\begin{align}
A_\sigma=\prod\limits_{i=1}^{N-1}P_{2i,2i+1}\times\frac{1}{2}\left(1+i\gamma_{2N}\gamma_1\right)
\end{align}
Here the last bracket shows the direction of the Majorana entanglement pair $\langle\gamma_1,\gamma_{2N}\rangle$ is from $\gamma_1$ to $\gamma_{2N}$, as an explicit indication of the anti-PBC of the Majorana chain we have created. Finally the operator of creating a 2D ``Majorana'' bubble in the entire lattice is:
\begin{align}
A=\bigotimes_{\sigma}A_\sigma
\end{align}
In particular, 2D ``Majorana'' bubble cannot change the parity of Majorana chains on 1D blocks: each 1D block is the shared border of two nearby 2D blocks, hence the number of Majorana chains on this 1D block can only be changed by 0 or 2 by 2D ``Majorana'' bubble.

\paragraph{1D bubble equivalence}For 1D blocks, we can consider two 1D irreducible representation of the corresponding total on-site symmetry of the 1D blocks that should be trivialized if they shrunk to a point. There are two possibilities: 

The first one is fermionic 1D bubble: consider two complex fermions with the following geometry:
\begin{align}
\centering
\includegraphics[width=0.15\textwidth]{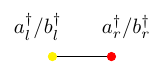}
\label{1D bubble}
\end{align}
Where yellow and red dots represent two complex fermions $a_l^\dag$ and $a_r^\dag$ who are trivialized when they are fused, i.e., $a_l^\dag a_r^\dag|0\rangle$ is a trivial atomic insulating state with even fermion parity. We demonstrate that this 1D bubble can be shrunk to a point and trivialized by a finite-depth circuit: if we decorate a 1D bubble, we can enclose $a_l^\dag$ and $a_r^\dag$ by an LU transformation. Repeatedly apply this LU transformation, we can shrink these two modes to a point. Therefore, the creation operator of fermionic 1D bubbles in the entire lattice is:
\begin{align}
B_j^f=\bigotimes_{\tau}(a_l^\tau)^\dag(a_r^\tau)^\dag
\end{align}

Another one is bosonic 1D bubble: consider two complex fermions with the geometry indicated in Eq. (\ref{1D bubble}), where yellow and red dots represent two bosons $b_l^\dag$ and $b_r^\dag$ that carry 1D irreducible representations of the physical symmetry group (total symmetry group quotient by fermion parity $\mathbb{Z}_2^f$) of corresponding 1D blocks. They should be trivialized by shrinking them to a point: $b_l^\dag b_r^\dag|0\rangle$ carries trivial 1D irreducible representation of the physical symmetry group. The creation operator of bosonic 1D bubbles in the entire lattice is:
\begin{align}
B_j^b=\bigotimes_{\tau}(b_l^\tau)^\dag(b_r^\tau)^\dag
\end{align}
And the creation operator of general 1D bubbles is:
\begin{align}
B_j=B_j^f\otimes B_j^b
\end{align}

Enlarge these bubbles and approximate to the nearby lower-dimensional blocks, the FSPT phases decorated on the bubble can be fused with the original states on the nearby lower-dimensional blocks, which leads to some possible \textit{trivializations} of lower-dimensional block-state decorations. 




Suppose there are $m$ different kinds of 1D bubble constructions, labeled by $B_j,j=1,...,m$. With this notation we can label an arbitrary bubble construction by an operator:
\[
A^{l_0}\prod\limits_{j=1}^{\beta}B_j^{l_j},~~l_0,l_j\in\mathbb{Z},
\]
where $l_0/l_j$ means that we take 2D/1D bubble construction $A/B_j$ by $l_0/l_j$ times. According to the definition of the bubble construction, taking an arbitrary bubble construction on the trivial state will lead to another trivial state, and all these trivial states form another group as following:
\begin{align}
\{\mathrm{TBS}\}=\left\{A^{l_0}\prod\limits_{j=1}^{\beta}B_j^{l_j}|0\rangle\Big|l_0,l_j\in\mathbb{Z}\right\}
\end{align}
Here ``TBS'' is the abbreviation of ``trivial block-states'', and $\{\mathrm{TBS}\}\subset\{\mathrm{OFBS}\}$ because all trivial block-states are obstruction-free block-states. $\{\mathrm{TBS}\}$ includes trivial block-states with different dimensions: $\{\mathrm{TBS}\}^{d\mathrm{D}}$ ($d=0,1$). Therefore, an obstruction and trivialization free block-state can be labeled by a group element of the following quotient group:
\begin{align}
\mathcal{G}=\{\mathrm{OFBS}\}/\{\mathrm{TBS}\}
\end{align}
and all group elements in $G$ are not equivalent because we have already divided all trivial states connected by bubble constructions. Equivalently, group $G$ gives the classification of the corresponding crystalline topological phases. 

In particular, we note that the block-states are constructed layer-by-layer. Therefore, we should specify the $d$-dimensional obstruction-free and trivialization-free block-states:
\begin{align}
E^{d\mathrm{D}}=\{\mathrm{OFBS}\}^{d\mathrm{D}}/\{\mathrm{TBS}\}^{d\mathrm{D}}
\end{align}
We should note that $E^{d\mathrm{D}}$ is not a group in the sense of SPT classification, because in order to obtain the ultimate classification of SPT phases, we should further consider the possible stacking between block-states with different dimensions. $E^{d\mathrm{D}}$ can only be treated as a group only in the sense of $d$D block-states. 

With all obstruction and trivialization free block-states with different dimensions, the ultimate classification with accurate group structure of 2D crystalline fSPT phases is extension between $E^{1\mathrm{D}}$ and $E^{0\mathrm{D}}$:
\begin{align}
\mathcal{G}=E^{1\mathrm{D}}\times_{\omega_2}E^{0\mathrm{D}}
\end{align}
here the symbol $\times_{\omega_2}$ depicts the possible extensions of $E^{1\mathrm{D}}$ and $E^{0\mathrm{D}}$ that is characterized by following short exact sequence:
\begin{align}
0\rightarrow E^{1\mathrm{D}}\rightarrow\mathcal{G}\rightarrow E^{0\mathrm{D}}\rightarrow0
\end{align}

In the following, we explicitly apply these procedures to calculate the classification of crystalline TSC and TI by several representative examples for each crystallographic systems.

\section{Construction and classification of crystalline topological superconductor\label{example}}
In this section, we describe the details of real-space construction for crystalline TSC in 2D interacting fermionic systems by analyzing several typical examples. It is well known that all 17 wallpaper groups can be divided into five different crystallographic systems: 
\begin{description}
\item[Square lattice]with rotational symmetry of order 4, including $p4$, $p4m$, $p4g$.
\item[Parallelogrammatic lattice]with only rotational symmetry of order 2, and no other symmetry than translational, including $p1$, $p2$. 
\item[Rhombic lattice]with reflection combined with glide reflection, including $cm$, $cmm$
\item[Rectangle lattice]with reflection or glide reflection, but not both, including $pm$, $pg$, $pmm$, $pmg$, $pgg$.
\item[Hexagonal lattice]with rotational symmetry of order 3 or 6, including $p3$, $p3m1$, $p31m$, $p6$, $p6m$.
\end{description}
The key distinction between different crystallographic systems is 0D blocks as centers of different point group.

In particular, we apply the general paradigm of real-space construction highlighted in Sec. \ref{general} to investigate five representative cases that belong to different crystallographic systems:
\begin{enumerate}
\item square lattice: $p4m$;
\item parallelogrammatic lattice: $p2$.
\item rhombic lattice: $cmm$;
\item rectangle lattice: $pgg$;
\item hexagonal lattice: $p6m$;
\end{enumerate}
All other cases are assigned in Supplementary Materials \cite{supplementary}. The classification results are summarized in Table \ref{spinless} and \ref{spin-1/2}, for spinless and spin-1/2 fermions, respectively. Furthermore, for a 2D spinless system with $p4m$ wallpaper group symmetry, there is an intrinsic interacting fermionic crystalline TSC  that cannot be realized in free fermion systems or interacting bosonic systems.

\subsection{Square lattice: $p4m$}
\begin{figure}
\centering
\includegraphics[width=0.46\textwidth]{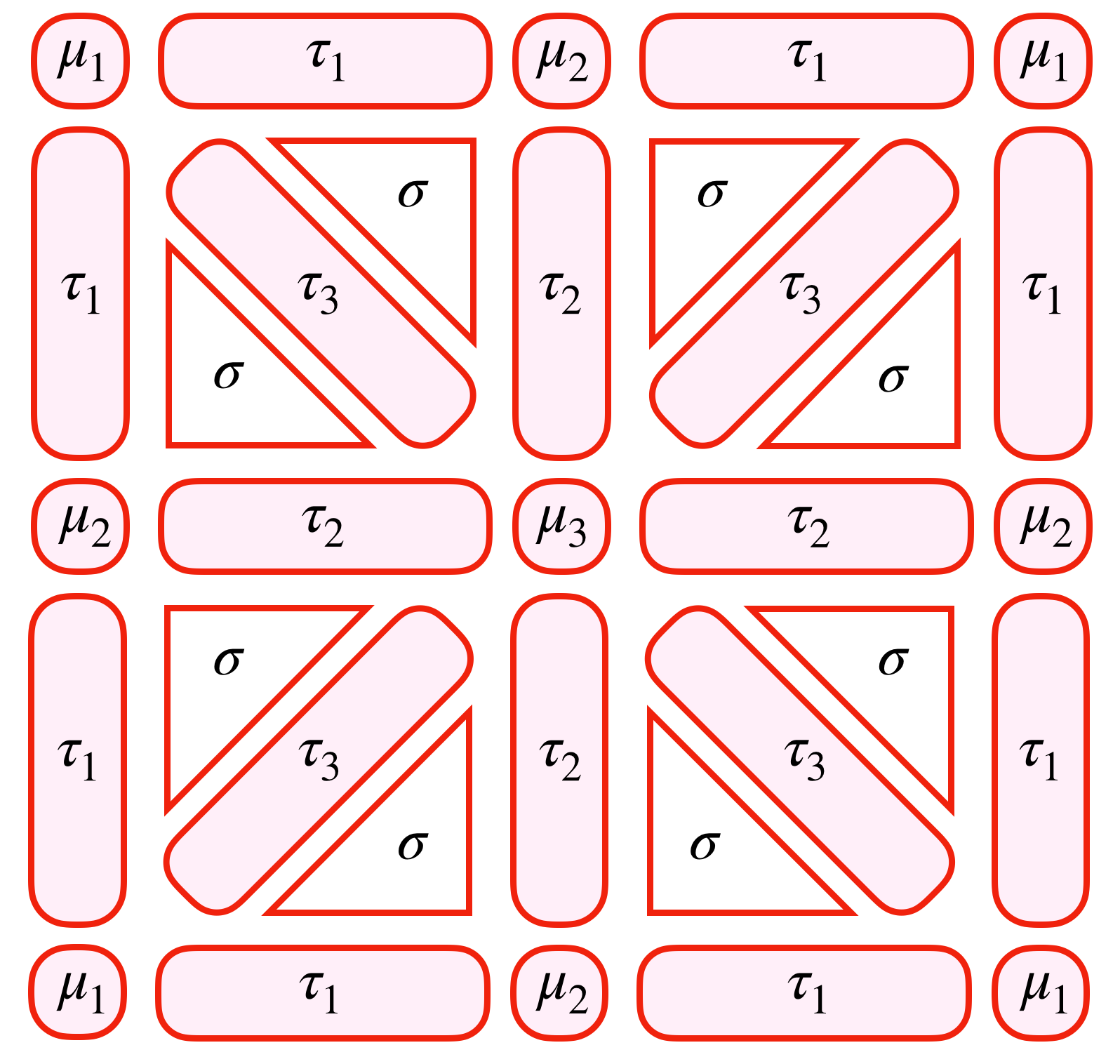}
\caption{\#11 wallpaper group $p4m$ and its cell decomposition.}
\label{p4m}
\end{figure}

For square lattice, we demonstrate the TSC protected by $p4m$ symmetry as an example. In the remainder of this paper, we use the same label of $p$-dimensional blocks that can be related by symmetry actions for abbreviation. The corresponding point group is dihedral group $D_4$, and for 2D blocks $\sigma$, there is no on-site symmetry group; for 1D blocks $\tau_1,\tau_2,\tau_3$, the on-site symmetry group is $\mathbb{Z}_2$ via the reflection symmetry acting internally; for 0D blocks $\mu_1$ and $\mu_3$, the on-site symmetry group is $\mathbb{Z}_4\rtimes\mathbb{Z}_2$ via the $D_4$ symmetry acting internally; for 0D blocks $\mu_2$, the on-site symmetry group is $\mathbb{Z}_2\rtimes\mathbb{Z}_2$ via the $D_2 \subset D_4$ symmetry acting internally, as seen in Fig. \ref{p4m}.

We discuss systems with spinless and spin-1/2 fermions separately. The ``spinless''/``spin-1/2'' fermion means that the point subgroup is extended trivially/nontrivially by fermion parity $\mathbb{Z}_2^f$ \cite{dihedral}. 

\subsubsection{Spinless fermions}
For spinless systems, we first consider the 0D block-state decoration, for 0D blocks $\mu_j$ ($j=1,2,3$), the classification data of the corresponding 0D block-states can be characterized by different 1D irreducible representations of the full symmetry group ($n=2,4$):
\begin{align}
\mathcal{H}^1\left[\mathbb{Z}_2^f\times\left(\mathbb{Z}_n\rtimes\mathbb{Z}_2\right),U(1)\right]=\mathbb{Z}_2^3
\label{p4m classification data}
\end{align}

For arbitrary 0D block [whose classification data are calculated in Eq. (\ref{p4m classification data})], three $\mathbb{Z}_2$ have different physical meanings: the first $\mathbb{Z}_2$ represents the parity of complex fermion (even or odd), the second $\mathbb{Z}_2$ represents the rotation eigenvalue $-1$, and the third $\mathbb{Z}_2$ represents the reflection eigenvalue $-1$. So at each 0D block, the block-state can be labeled by $(\pm,\pm,\pm)$, where these three $\pm$ represent the fermion parity and eigenvalues of two independent reflection generators, respectively. We should note that even-fold dihedral group can also be generated by two independent reflection operations: for 0D blocks $\mu_1/\mu_3$, $D_4$ symmetry can be generated by reflection operations $\bs{M}_{\tau_1}/\bs{M}_{\tau_2}$ and $\bs{M}_{\tau_3}$ ($\bs{M}_{\tau_1}, \bs{M}_{\tau_2}, \bs{M}_{\tau_3}$ represent the reflection operation with respect to the axis which coincide with the 1D block labeled by $\tau_1, \tau_2,\tau_3$); for 0D block $\mu_2$, $D_2$ symmetry can be generated by reflection operations $\bs{M}_{\tau_1}$ and $\bs{M}_{\tau_2}$. According to this notation, the obstruction-free 0D block-states form the following group:
\begin{align}
\{\mathrm{OFBS}\}_{p4m,0}^{\mathrm{0D}}=\mathbb{Z}_2^9
\end{align}
where the group elements can be labeled by:
\[
[(\pm,\pm,\pm),(\pm,\pm,\pm),(\pm,\pm,\pm)]
\]
here three brackets represent the block-states at $\mu_1$, $\mu_2$ and $\mu_3$, respectively. 

Subsequently we consider the 1D block-state decoration. For $\tau_1$, $\tau_2$ and $\tau_3$, the total symmetry group is $\mathbb{Z}_2^f\times\mathbb{Z}_2$, so there are two possible 1D block-states: Majorana chain and 1D FSPT state, and all 1D block-states form a group:
\begin{align}
\{\mathrm{BS}\}_{p4m,0}^{\mathrm{1D}}=\mathbb{Z}_2^6
\end{align}
Below we discuss the decorations of these two root phases separately. 

\paragraph{Majorana chain decoration}Consider Majorana chain decoration on 1D blocks labeled by $\tau_1$, which leaves 4 dangling Majorana modes at each 0D block $\mu_1/\mu_3$, and 2 dangling Majorana modes at each 0D block $\mu_2$. Near $\mu_1$, Majorana modes have the following rotation and reflection symmetry (all subscripts are taken with modulo 4): 
\begin{align}
\bs{R}_{\mu_1}:~\gamma_j\mapsto\gamma_{j+1},~~\bs{M}_{\tau_2}:~\gamma_j\mapsto\gamma_{4-j}
\end{align}
The local fermion parity operator and its symmetry properties read:
\begin{align}
P_f=-\prod\limits_{j=1}^4\gamma_j,~~\bs{R}_{\mu_1},\bs{M}_{\tau_2}:~P_f\mapsto-P_f
\end{align}
Hence these four Majorana modes break the fermion parity. Thus Majorana chain decoration on $\tau_1$ does not contribute to nontrivial crystalline TSC because of the violation of the no-open-edge condition. It is similar for 1D blocks $\tau_2$ and $\tau_3$, so all types of Majorana chain decoration are obstructed.

\paragraph{1D FSPT state decoration}The 1D FSPT state decoration on $\tau_1$, $\tau_2$ and $\tau_3$ will leave 8 dangling Majorana modes ($\xi_j,\xi_j'$, $j=1,2,3,4$) at each 0D block labeled by $\mu_1/\mu_3$ and 4 dangling Majorana modes ($\eta_j,\eta_j'$, $j=1,2$) at each 0D block labeled by $\mu_2$. At $\mu_1/\mu_3$ (we discuss $\mu_1$ as an example), the corresponding 8 Majorana modes have the following rotation and reflection symmetry properties (all subscripts are taken under modulo 4, e.g., $\xi_5\equiv\xi_1$ and $\xi_5'\equiv\xi_1'$):
\begin{align}
\left.
\begin{aligned}
&\bs{R}_{\mu_1}:~\xi_j\mapsto\xi_{j+1},~\xi_j'\mapsto\xi_{j+1}'\\
&\bs{M}
_{\tau_1}:~\xi_j\mapsto\xi_{6-j},~\xi_j'\mapsto-\xi_{6-j}'
\end{aligned}
\right.,~~j=1,2,3,4.
\label{p4m Majorana}
\end{align}
We can define four complex fermions from these eight dangling Majorana modes:
\begin{align}
c_j^\dag=\frac{1}{2}\left(\xi_j+i\xi_j'\right),~j=1,2,3,4
\end{align}
And from the point group symmetry properties (\ref{p4m Majorana}), we can obtain the point group symmetry properties of the above complex fermions as:
\begin{align}
\begin{aligned}
&\bs{R}_{\mu_1}:~\left(c_1^\dag,c_2^\dag,c_3^\dag,c_4^\dag\right)\mapsto\left(c_2^\dag,c_3^\dag,c_4^\dag,c_1^\dag\right)\\
&\bs{M}
_{\tau_1}:~\left(c_1^\dag,c_2^\dag,c_3^\dag,c_4^\dag\right)\mapsto\Big(c_1,c_4,c_3,c_2\Big)
\end{aligned}
\label{p4m complex}
\end{align}
We denote the fermion number operators $n_j=c_j^\dag c_j, j=1,2,3,4$. Firstly we consider the Hamiltonian with Hubbard interaction ($U>0$) that can gap out these dangling Majorana modes:
\begin{align}
H_U=U\sum\limits_{j=1,2}\left(n_j-\frac{1}{2}\right)\left(n_{j+2}-\frac{1}{2}\right)
\label{HU}
\end{align}
And it can also be expressed in terms of Majorana modes with symmetry properties as shown in Eq. (\ref{p4m Majorana}):
\begin{align}
H_U=-\frac{U}{4}\left(\xi_1\xi_1'\xi_3\xi_3'+\xi_2\xi_2'\xi_4\xi_4'\right)
\end{align}
It is easy to verify that $H_U$ respects all symmetries. There is a 4-fold ground-state degeneracy from $(n_1,n_3)$ and $(n_2,n_4)$, which can be viewed as two spin-1/2 degrees of freedom: 
\begin{align}
\tau_{13}^\mu=\left(c_1^\dag,c_3^\dag\right)\sigma^\mu\left(
\begin{array}{ccc}
c_1\\
c_3
\end{array}
\right)
\end{align}
and
\begin{align}
\tau_{24}^\mu=\left(c_2^\dag,c_4^\dag\right)\sigma^\mu\left(
\begin{array}{ccc}
c_2\\
c_4
\end{array}
\right)
\end{align}
where $\sigma^\mu,\mu=x,y,z$ are Pauli matrices. In order to lift this ground-state degeneracy (GSD), we should further consider the interactions between these two spins. The symmetry properties of these two spins can be easily obtained from (\ref{p4m complex}):
\begin{align}
\begin{aligned}
&\bs{R}_{\mu_1}:~
\begin{aligned}
&\left(\tau_{13}^x,\tau_{13}^y,\tau_{13}^z\right)\mapsto\left(\tau_{24}^x,\tau_{24}^y,\tau_{24}^z\right)\\
&\left(\tau_{24}^x,\tau_{24}^y,\tau_{24}^z\right)\mapsto\left(\tau_{13}^x,-\tau_{13}^y,-\tau_{13}^z\right)
\end{aligned}\\
&\bs{M}
_{\tau_1}:
\begin{aligned}
&\left(\tau_{13}^x,\tau_{13}^y,\tau_{13}^z\right)\mapsto\left(-\tau_{13}^x,\tau_{13}^y,-\tau_{13}^z\right)\\
&\left(\tau_{24}^x,\tau_{24}^y,\tau_{24}^z\right)\mapsto\left(-\tau_{24}^x,-\tau_{24}^y,\tau_{24}^z\right)
\end{aligned}
\end{aligned}
\label{p4m spin}
\end{align}
Then we can further add a spin Hamiltonian ($J>0$):
\begin{align}
H_J=J\big(\tau_{13}^x\tau_{24}^x+\tau_{13}^y\tau_{24}^z-\tau_{13}^z\tau_{24}^y\big)
\label{HJ}
\end{align}
According to the symmetry properties of spin operations (\ref{p4m spin}), we can easily verify that the spin Hamiltonian $H_J$ is symmetric under all symmetries. We can also verify the symmetry properties in Majorana representations by expressing $H_J$ in terms of Majorana modes as:
\begin{align}
H_J=&-\frac{J}{4}\left(\xi_1\xi_3'-\xi_1'\xi_3\right)\left(\xi_2\xi_4'-\xi_2'\xi_4\right)\nonumber\\
&-\frac{J}{4}\left(\xi_1\xi_3+\xi_1'\xi_3'\right)\left(\xi_2\xi_2'-\xi_4\xi_4'\right)\nonumber\\
&+\frac{J}{4}\left(\xi_1\xi_1'-\xi_3\xi_3'\right)\left(\xi_2\xi_4+\xi_2'\xi_4'\right)
\end{align}
and it is invariant under the symmetry properties defined in Eq. (\ref{p4m Majorana}). The GSD is lifted by a symmetric Hamiltonian $H_U+H_J$, and the non-degenerate ground-state is:
\begin{align}
|\psi\rangle_{\mathrm{0D}}=-\frac{1}{2}\left(|\uparrow,\uparrow\rangle+i|\uparrow,\downarrow\rangle-i|\downarrow,\uparrow\rangle-|\downarrow,\downarrow\rangle\right)
\end{align}
where $\uparrow$ and $\downarrow$ represent spin-up and spin-down of two spin-1/2 degrees of freedom ($\vec{\tau}_{13}$ and $\vec{\tau}_{24}$), and the ground-state energy is $-3J$. It is easy to verify that this state is invariant under arbitrary symmetry actions because $|\psi\rangle_{\mathrm{0D}}$ is the eigenstate of the operators $\bs{R}_{\mu_1}$ and $\bs{M}_{\tau_1}$ as two generators of $D_4$ group at each $\mu_1$:
\begin{align}
\begin{aligned}
&\bs{R}_{\mu_1}|\psi\rangle_{\mathrm{0D}}=i|\psi\rangle_{\mathrm{0D}}\\
&\bs{M}_{\tau_1}|\psi\rangle_{\mathrm{0D}}=-|\psi\rangle_{\mathrm{0D}}
\end{aligned}
\end{align}
Thus the corresponding 8 Majorana modes are gapped out by interactions in a symmetric way. 

Next we consider the dangling Majorana modes from the 1D FSPT decorations on $\tau_1$ at $\mu_2$ with the rotation and reflection symmetry properties:
\begin{align}
\left.
\begin{aligned}
\bs{R}_{\mu_2}:~&\left(\eta_1,\eta_1',\eta_2,\eta_2'\right)\mapsto\left(\eta_2,\eta_2',\eta_1,\eta_1'\right)\\
\bs{M}_{\tau_1}:~&\left(\eta_1,\eta_1',\eta_2,\eta_2'\right)\mapsto\left(\eta_1,-\eta_1',\eta_2,-\eta_2'\right)
\end{aligned}
\right.
\label{p4m mu2 Majorana}
\end{align}
We can define two complex fermions from these four dangling Majorana modes:
\begin{align}
c^\dag=\frac{1}{2}(\eta_1+i\eta_2),~c'^\dag=\frac{1}{2}(\eta_1'+i\eta_2')
\end{align}
and from the symmetry properties (\ref{p4m mu2 Majorana}), we can obtain the point group symmetry properties of the above complex fermions:
\begin{align}
\begin{aligned}
&\bs{R}:~\Big(c^\dag,c'^\dag\Big)\mapsto\Big(ic,ic'\Big)\\
&\bs{M}:~\Big(c^\dag,c'^\dag\Big)\mapsto\Big(c^\dag,-c'^\dag\Big)
\end{aligned}
\label{p4m mu2 complex}
\end{align}
We denote the fermion number operators $n=c^\dag c$ and $n'=c'^\dag c'$. First we consider the Hamiltonian with Hubbard interaction ($U'>0$) that can gap out these dangling Majorana modes:
\begin{align}
H_U'=U'\left(n-\frac{1}{2}\right)\left(n'-\frac{1}{2}\right)
\end{align}
And it is easy to verify that $H_U'$ respects all symmetries according to the symmetry properties of defined complex fermions (\ref{p4m mu2 complex}). There is a 2-fold ground-state degeneracy from $(n,n')$ that can be viewed as a spin-1/2 degree of freedom:
\begin{align}
\tau^\mu=\left(c^\dag,c'^\dag\right)\sigma^\mu\left(
\begin{array}{ccc}
c\\
c'
\end{array}
\right)
\end{align}
In order to investigate whether the degenerate ground states can be gapped out or not, we focus on the projective Hilbert space spanned by two states $c^\dag|0\rangle$ and $c'^\dag|0\rangle$. In this projective Hilbert space, two generators of $D_2$ symmetry on each $\mu_2$, $\bs{R}_{\mu_2}$ and $\bs{M}_{\tau_1}$ can be represented by two $2\times2$ matrices:
\begin{align}
\begin{aligned}
&\bs{R}_{\mu_2}=\left(
\begin{array}{ccc}
0 & 1\\
1 & 0
\end{array}
\right)=\sigma^x\\
&\bs{M}_{\tau_1}=\left(
\begin{array}{ccc}
1 & 0\\
0 & -1
\end{array}
\right)=\sigma^z
\end{aligned}
\end{align}
It is obvious that these two generators are anticommute: 
$$\bs{R}_{\mu_2}\bs{M}_{\tau_1}=-\bs{M}_{\tau_1}\bs{R}_{\mu_2}$$
i.e., a sufficient condition shows that the projective Hilbert space is a projective representation of the symmetry group $D_2$ at each 0D block labeled by $\mu_2$. Hence, the two-fold ground-state degeneracy cannot be lifted. 

We demonstrate this conclusion in Majorana representation and elucidate that all possible mass terms are not compatible with symmetries. A mass term is formed by two Majorana operators, and all possible mass terms are:
\[
\eta_1\eta_2,~\eta_1\eta_1',~\eta_1\eta_2',~\eta_2\eta_1',~\eta_2\eta_2',~\eta_1'\eta_2'
\]
and their linear combinations. Under 2-fold rotation $\bs{R}_{\mu_2}$, these mass terms will be transformed to:
\[
-\eta_1\eta_2,~\eta_2\eta_2',~\eta_2\eta_1',~\eta_1\eta_2',~\eta_1\eta_1',~-\eta_1'\eta_2'
\]
so there are only two mass terms that are symmetric under $\bs{R}_{\mu_2}$: $\eta_1\eta_1'+\eta_2\eta_2'$ and $\eta_1\eta_2'+\eta_2\eta_1'$ and their linear combinations. Subsequently, under the reflection $\bs{M}_{\tau_1}$, these terms are not symmetric:
\[
-(\eta_1\eta_1'+\eta_2\eta_2'),~-(\eta_1\eta_2'+\eta_2\eta_1')
\]
Therefore, there is no symmetric mass term to lift the GSD. Accordingly, 1D FSPT state decoration on $\tau_1$ is \textit{obstructed} because of the degenerate ground state, and similar arguments can also be held on 1D blocks labeled by $\tau_2$ (and the obstruction also happens at 0D block $\mu_2$, as the center of $D_2$ symmetry). 1D FSPT state decoration on $\tau_3$ is \textit{obstruction-free} because this decoration leaves eight dangling Majorana modes at each 0D block labeled by $\mu_1$ and $\mu_3$, and both of them are centers of $D_4$ symmetry. 

\begin{figure*}
\centering
\includegraphics[width=\textwidth]{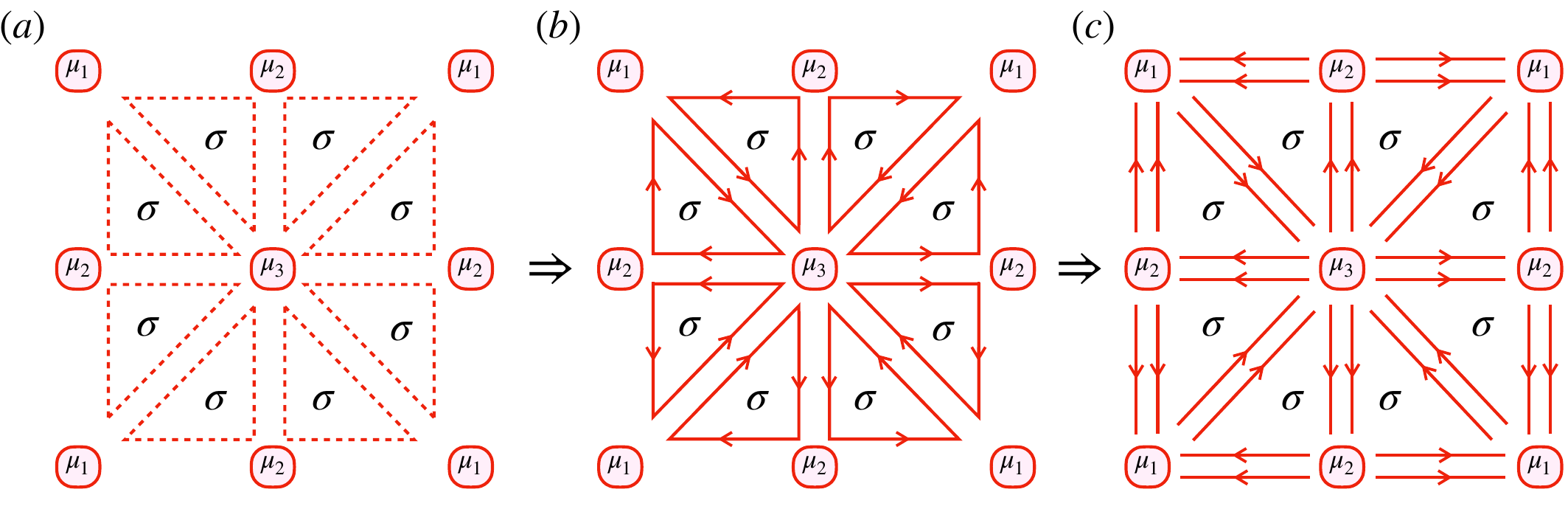}
\caption{Deformation of ``Majorana'' bubble construction. (a): 1D vacuum block-state that all 1D blocks are in vacuum state. Here $\mu_1$, $\mu_2$ and $\mu_3$ label different 0D blocks and $\sigma$ enclosed by a dashed triangle labels 2D blocks in one unit cell. (b): Decorate a ``Majorana'' bubble in each 2D block in a symmetric way. Each solid oriented triangle expresses a Majorana chain with anti-PBC. (c): Enlarge the ``Majorana'' bubbles, they can be deformed to the 1D block-state with 1D FSPT states protected by on-site $\mathbb{Z}_2$ symmetry on all 1D blocks. Here each double oriented lines expresses a 1D FSPT states protected by on-site $\mathbb{Z}_2$ symmetry that can be constructed by two Majorana chains.}
\label{deformation}
\end{figure*}

There is one exception: if we decorate a 1D FSPT phase on each 0D block labeled by $\tau_1$ and $\tau_2$ simultaneously, it leaves eight dangling Majorana modes at each 0D block $\mu_2$ ($\eta_j,\eta_j',j=1,2,3,4$), with the following rotation and reflection symmetry properties:
\begin{align}
\begin{aligned}
&\bs{R}_{\mu_2}:~\left\{\begin{aligned}
&(\eta_1,\eta_1',\eta_2,\eta_2')\mapsto(\eta_2,\eta_2',\eta_1,\eta_1')\\
&(\eta_3,\eta_3',\eta_4,\eta_4')\mapsto(\eta_4,\eta_4',\eta_3,\eta_3')
\end{aligned}\right.\\
&\bs{M}_{\tau_1}:~\left\{\begin{aligned}
&(\eta_1,\eta_1',\eta_2,\eta_2')\mapsto(\eta_1,-\eta_1',\eta_2,-\eta_2')\\
&(\eta_3,\eta_3',\eta_4,\eta_4')\mapsto(\eta_4,-\eta_4',\eta_3,-\eta_3')
\end{aligned}\right.
\end{aligned}
\label{p4m tau23 Majorana}
\end{align}
This situation is quite similar with aforementioned gapping situation at each 0D block labeled by $\mu_1$ or $\mu_3$, with lower point group symmetry ($D_2\in D_4$). Thus eight dangling Majorana modes at each 0D block $\mu_2$ from decorating a 1D FSPT state on each $\tau_1$ and $\tau_2$ can be gapped by previously discussed interactions $H_U+H_J$ [cf. Eqs. (\ref{HU}) and (\ref{HJ})] in a symmetric way, and the 1D FSPT state decoration on $\tau_1$ and $\tau_2$ simultaneously is \textit{obstruction-free}. We should note that this block-state has no free-fermion realization because as aforementioned, we should introduce some interactions to satisfy the no-open-edge condition, as non-interacting mass terms cannot gap them out. Hence the crystalline TSC realized here is an intrinsic interacting FSPT phase. In summary, all obstruction-free 1D block-states are:
\begin{itemize}
\item 1D FSPT state decoration on $\tau_1$ and $\tau_2$ simultaneously;
\item 1D FSPT state decoration on $\tau_3$.
\end{itemize}
and they form the following group:
\begin{align}
\{\mathrm{OFBS}\}_{p4m,0}^{\mathrm{1D}}=\mathbb{Z}_2^2
\end{align}
where the group elements can be labeled by:
\[
[m_1=m_2,m_3]
\]
here $m_j=0,1$ ($j=1,2,3$) represents the number of decorated 1D FSPT states on $\tau_j$, respectively. According to aforementioned discussions, a necessary condition of an obstruction-free block-state is $m_1=m_2$.

With all obstruction-free block-states, below we will discuss all possible trivializations. First we consider the 2D bubble equivalences: we decorate a ``Majorana bubble'' on each 2D block $\sigma$ (see Fig. \ref{deformation}), and then demonstrate that they can be deformed into double Majorana chains at each nearby 1D block, and this is exactly the definition of the nontrivial 1D FSPT phase protected by on-site $\mathbb{Z}_2$ symmetry. Fig. \ref{deformation}(b) shows that these ``Majorana bubbles'' can be deformed to double Majorana chains. For $p4m$ case, all 1D blocks are lying on the reflection axis, and reflection operation are acting on them internally: reflection operation (on-site $\mathbb{Z}_2$ symmetry on 1D blocks) exchanges two Majorana chains deformed from ``Majorana'' bubble constructions, and this is exactly the definition of the nontrivial 1D FSPT phase protected by on-site $\mathbb{Z}_2$ symmetry. Equivalently, we can say that 1D FSPT state decorations on all 1D blocks can be deformed to a trivial state via 2D bubble equivalence. Next, we further investigate whether 2D bubble equivalence can change the fermion parity of 0D blocks or not. We have already seen that 2D ``Majorana'' bubble equivalence leaves double Majorana chains on all 1D blocks. Correspondingly, it leaves 16 Majorana modes at each 0D block $\mu_1/\mu_3$ and 8 Majorana modes at each 0D block $\mu_2$ as the edge modes of double Majorana chains on 1D blocks. Apparently, these Majorana modes cannot be connected to Majorana chains with PBC surrounding the 0D blocks(with fermion parity odd): It is well-known that Majorana chain is not compatible with reflection symmetry; however, the Majorana chain with PBC surrounding 0D block must across at least one reflection axis. 
As a result, the overall effect of 2D ``Majorana'' bubble equivalence is deforming the 1D FSPT phase (protected by on-site $\mathbb{Z}_2$ symmetry) decorations on all 1D blocks to a trivial state.

\begin{figure}
\centering
\includegraphics[width=0.4\textwidth]{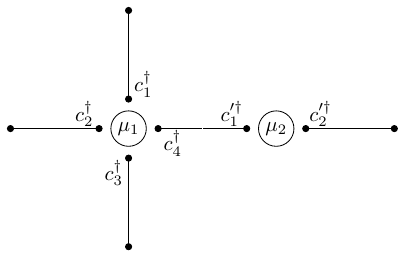}
\caption{1D bubble equivalence on $\tau_1$. Atomic insulators (\ref{mu1 atomic insulator}) and (\ref{mu2 atomic insulator}) are constructed by this procedure.}
\label{1D bubble equivalence}
\end{figure}

Subsequently we consider the 1D bubble equivalences. For instance, we decorate a pair of complex fermions [cf. Eq. (\ref{1D bubble})] on each 1D block $\tau_1$: Near each 0D block $\mu_1$, there are 4 complex fermions forming the following atomic insulator:
\begin{align}
|\psi\rangle_{p4m}^{\mu_1}=c_1^\dag c_2^\dag c_3^\dag c_4^\dag|0\rangle
\label{mu1 atomic insulator}
\end{align}
with two independent reflection properties:
\begin{align}
\begin{aligned}
&\bs{M}_{\tau_1}|\psi\rangle_{p4m}^{\mu_1}=c_1^\dag c_4^\dag c_3^\dag c_2^\dag|0\rangle=-|\psi\rangle_{p4m}^{\mu_1}\\
&\bs{M}_{\tau_3}|\psi\rangle_{p4m}^{\mu_1}=c_3^\dag c_4^\dag c_1^\dag c_2^\dag|0\rangle=|\psi\rangle_{p4m}^{\mu_1}
\end{aligned}
\end{align}
i.e., at 0D blocks $\mu_1$, 1D bubble construction on $\tau_1$ changes the reflection eigenvalue of $\bs{M}_{\tau_1}$, and leaves the reflection eigenvalue of $\bs{M}_{\tau_2}$ invariant. Near each 0D block $\mu_2$, there are two complex fermions forming another atomic insulator:
\begin{align}
|\psi\rangle_{p4m}^{\mu_2}=c_1'^\dag c_2'^\dag|0\rangle
\label{mu2 atomic insulator}
\end{align}
with two independent reflection properties:
\begin{align}
\begin{aligned}
&\bs{M}_{\tau_1}|\psi\rangle_{p4m}^{\mu_2}=c_1'^\dag c_2'^\dag|0\rangle=|\psi\rangle_{p4m}^{\mu_2}\\
&\bs{M}_{\tau_2}|\psi\rangle_{p4m}^{\mu_2}=c_2'^\dag c_1'^\dag|0\rangle=-|\psi\rangle_{p4m}^{\mu_2}
\end{aligned}
\end{align}
i.e., at 0D blocks $\mu_2$ 1D bubble construction on $\tau_1$ changes the reflection eigenvalue of $\bs{M}_{\tau_2}$, and leaves the reflection eigenvalue of $\bs{M}_{\tau_1}$ invariant. This 1D bubble equivalence is illustrated in Fig. \ref{1D bubble equivalence}. Similar 1D bubble constructions can be held on 1D blocks $\tau_2$ and $\tau_3$, and we summarize the effects of 1D bubble constructions as following:
\begin{enumerate}[1.]
\item 1D bubble construction on $\tau_1$: simultaneously changes the eigenvalues of $\bs{M}_{\tau_1}$ at $\mu_1$ and $\bs{M}_{\tau_2}$ at $\mu_2$;
\item 1D bubble construction on $\tau_2$: simultaneously changes the eigenvalues of $\bs{M}_{\tau_1}$ at $\mu_2$ and $\bs{M}_{\tau_2}$ at $\mu_3$;
\item 1D bubble construction on $\tau_3$: simultaneously changes the eigenvalues of $\bs{M}_{\tau_3}$ at $\mu_1$ and $\bs{M}_{\tau_3}$ at $\mu_3$;
\end{enumerate}

With all possible trivializations, we are ready to study the trivial states. Start from the original 0D trivial block-state:
\[
[(+,+,+),(+,+,+),(+,+,+)]
\]
If we take 1D bubble constructions on $\tau_j$ by $l_j$ times ($j=1,2,3$), the above trivial 0D block-state will be transformed to a new 0D block-state labeled by:
\begin{align}
&\left[\left(+,(-1)^{l_1},(-1)^{l_3}\right)\right.,\left(+,(-1)^{l_2},(-1)^{l_1}\right),\nonumber\\
&\left.\left(+,(-1)^{l_2},(-1)^{l_3}\right)\right]
\label{p4m spinless trivial state}
\end{align}
According to the definition of bubble equivalence, all these states should be trivial. It is easy to see that there are only three independent quantities ($l_j,j=1,2,3$) in Eq. (\ref{p4m spinless trivial state}). Together with the 2D ``Majorana'' bubble construction that deforms the vacuum 1D block-state to 1D FSPT states decorated on all 1D blocks, all these trivial states form the group:
\begin{align}
\{\mathrm{TBS}\}_{p4m,0}&=\{\mathrm{TBS}\}_{p4m,0}^{\mathrm{1D}}\times\{\mathrm{TBS}\}_{p4m,0}^{\mathrm{0D}}\nonumber\\
&=\mathbb{Z}_2\times\mathbb{Z}_2^3=\mathbb{Z}_2^4
\end{align}
where $\{\mathrm{TBS}\}_{p4m,0}^{\mathrm{1D}}$ represents the group of trivial states with non-vacuum 1D blocks (i.e., 1D FSPT phase decorations on all 1D blocks), and $\{\mathrm{TBS}\}_{p4m,0}^{\mathrm{0D}}$ represents the group of trivial states with non-vacuum 0D blocks.

Therefore, all independent nontrivial block-states with different dimensions are classifiied by:
\begin{align}
\begin{aligned}
&E_{p4m,0}^{\mathrm{1D}}=\{\mathrm{OFBS}\}_{p4m,0}^{\mathrm{1D}}/\{\mathrm{TBS}\}_{p4m,0}^{\mathrm{1D}}=\mathbb{Z}_2\\
&E_{p4m,0}^{\mathrm{0D}}=\{\mathrm{OFBS}\}_{p4m,0}^{\mathrm{0D}}/\{\mathrm{TBS}\}_{p4m,0}^{\mathrm{0D}}=\mathbb{Z}_2^6
\end{aligned},
\end{align}
where one $\mathbb{Z}_2$ is from the nontrivial 1D block-state, and other six $\mathbb{Z}_2$ are from the nontrivial 0D block-states.

With all nontrivial block-states, we consider the group structure of the ultimate classification. The physical meaning of the group structure is whether stacking of 1D block-state extends to 0D block-state or not. We argue that there is no stacking between block-states with different dimensions for p4m symmetry. In order to investigate the possible stacking, we consider two identical 1D block-states: for example, we decorate two copies of 1D FSPT states on each 1D block labeled by $\tau_3$, which leaves 16 dangling Majorana modes at each 0D block labeled by $\mu_1/\mu_3$. It is easy to verify that two copies of 1D FSPT states should be a trivial 1D block-state because the root phase has a $\mathbb{Z}_2$ structure. First of all, according to previous discussions, these decoration cannot be deformed to a Majorana chain surrounding the 0D block to change the corresponding fermion parity because the Majorana chain is not compatible with the reflection symmetry. Subsequently at each 0D block $\mu_1/\mu_3$, we can treat these 16 Majorana modes as 8 complex fermions: $c_j$ and $c_j'$ ($j=1,2,3,4$) form two atomic insulators:
\begin{align}
\begin{aligned}
&|\phi\rangle=a_1^\dag a_2^\dag a_3^\dag a_4^\dag|0\rangle\\
&|\phi'\rangle=a_1'^\dag a_2'^\dag a_3'^\dag a_4'^\dag|0\rangle
\end{aligned}
\end{align}
and the wavefunction of these 8 complex fermions is direct product of $|\phi\rangle$ and $|\phi'\rangle$:
\begin{align}
|\Phi\rangle=|\phi\rangle\otimes|\phi'\rangle
\end{align}
$|\phi\rangle$ and $|\phi'\rangle$ are eigenstates of two generators of $D_4$ symmetry, $\bs{M}_{\tau_1}$ and $\bs{M}_{\tau_3}$:
\begin{align}
\begin{aligned}
&\bs{M}_{\tau_1}|\phi\rangle=a_2^\dag a_1^\dag a_4^\dag a_3^\dag|0\rangle=|\phi\rangle\\
&\bs{M}_{\tau_1}|\phi'\rangle=a_2'^\dag a_1'^\dag a_4'^\dag a_3'^\dag|0\rangle=|\phi'\rangle\\
&\bs{M}_{\tau_3}|\phi\rangle=a_1^\dag a_4^\dag a_3^\dag a_2^\dag|0\rangle=-|\phi\rangle\\
&\bs{M}_{\tau_3}|\phi'\rangle=a_1'^\dag a_4'^\dag a_3'^\dag a_2'^\dag|0\rangle=-|\phi'\rangle\\
\end{aligned}
\end{align}
Then the eigenvalues of $|\Phi\rangle$ under $\bs{M}_{\tau_1}$ and $\bs{M}_{\tau_3}$ are trivial:
\begin{align}
\begin{aligned}
&\bs{M}_{\tau_1}|\Phi\rangle=|\Phi\rangle\\
&\bs{M}_{\tau_3}|\Phi\rangle=|\Phi\rangle
\end{aligned}
\end{align}
Therefore, stacking of 1D block-state cannot extend to 0D block-state, and the ultimate classification of 2D crystalline SPT phases with $p4m$ symmetry for spinless fermions is:
\begin{align}
\mathcal{G}_{p4m,0}=E_{p4m,0}^{\mathrm{1D}}\times E_{p4m,0}^{\mathrm{0D}}=\mathbb{Z}_2^{7}
\label{p4m spinless classification}
\end{align}

\subsubsection{Spin-1/2 fermions}

Now we turn to discuss systems with spin-1/2 fermions. We first consider the 0D block-state decoration. For each 0D block $\mu_j$ ($j=1,2,3$), the classification data can also be characterized by different 1D irreducible representations of the full symmetry group $\mathbb{Z}_2^f\times_{\omega_2}(\mathbb{Z}_n\rtimes\mathbb{Z}_2)$ ($n=2,4$, and the symbol ``$\times_{\omega_2}$'' means that the physical symmetry group is nontrivially extended by fermion parity $\mathbb{Z}_2^f$, which is characterized by a 2-cocycle $\omega_2$, see Sec. \ref{spinSec}):
\begin{align}
\mathcal{H}^1\left[\mathbb{Z}_2^f\times_{\omega_2}(\mathbb{Z}_n\rtimes\mathbb{Z}_2),U(1)\right]=\mathbb{Z}_2^2
\end{align}
To calculate this, we should firstly calculate the following two cohomologies:
\begin{align}
\left\{
\begin{aligned}
&\mathcal{H}^0(\mathbb{Z}_n\rtimes\mathbb{Z}_2,\mathbb{Z}_2)=\mathbb{Z}_2\\
&\mathcal{H}^1\left[\mathbb{Z}_n\rtimes\mathbb{Z}_2,U(1)\right]=\mathbb{Z}_2^2
\end{aligned}
\right.
\end{align}
But the 0-cocycle $n_0\in\mathcal{H}^0(\mathbb{Z}_n\rtimes\mathbb{Z}_2,\mathbb{Z}_2)$ does not contribute a nontrivial 0D block-state: a specific $n_0$ is obstructed if and only if $(-1)^{\omega_2\smile n_0}\in\mathcal{H}^2[\mathbb{Z}_4\rtimes\mathbb{Z}_2,U(1)]$ is a nontrivial 2-cocycle with $U(1)$ coefficient. From Refs. \cite{general2} and \cite{dihedral} we know that nontrivial 0-cocycle $n_0=1$ (fermion parity odd) leads to a nontrivial 2-cocycle $(-1)^{\omega_2\smile n_0}\in\mathcal{H}^2[\mathbb{Z}_4\rtimes\mathbb{Z}_2,U(1)]$, and the 0D block-states at $\mu_j$ with odd fermion parity are obstructed. Hence different $\mathbb{Z}_2$'s in the classification data represent the rotation and reflection eigenvalues at each $D_4$ or $D_2$ center. As a consequence, all obstruction-free 0D block-states form the following group:
\begin{align}
\{\mathrm{OFBS}\}_{p4m,1/2}^{\mathrm{0D}}=\mathbb{Z}_2^6
\end{align}

Then we demonstrate that there is no trivialization. For spinless fermions, we have demonstrated that the eigenvalue $-1$ of $\bs{M}_{\tau_1}/\bs{M}_{\tau_2}$ is trivialized by atomic insulator $|\psi\rangle_{p4m}^{\mu_1}/|\psi\rangle_{p4m}^{\mu_2}$ [cf. Eqs. (\ref{mu1 atomic insulator}) and (\ref{mu2 atomic insulator}), and Fig. \ref{1D bubble equivalence}]. Nevertheless, to fulfill the spin-1/2 condition ($\bs{M}_{\tau_1}^2=\bs{M}_{\tau_2}^2=-1$), there is an additional minus sign under reflections:
\begin{align}
\begin{aligned}
&\bs{M}_{\tau_1}|\psi\rangle_{p4m}^{\mu_1}=c_1^\dag c_4^\dag c_3^\dag (-c_2^\dag)|0\rangle=|\psi\rangle_{p4m}^{\mu_1}\\
&\bs{M}_{\tau_2}|\psi\rangle_{p4m}^{\mu_2}=c_1'^\dag(-c_2'^\dag)=|\psi\rangle_{p4m}^{\mu_2}
\end{aligned}
\end{align}
i.e., eigenvalues of $\bs{M}_{\tau_1}$ and $\bs{M}_{\tau_2}$ remain invariant. Equivalently, there is no trivialization for 0D block-states:
\begin{align}
\{\mathrm{TBS}\}_{p4m,1/2}^{\mathrm{0D}}=\mathbb{Z}_1
\end{align}

As the consequence,  the classification attributed to 0D block-states is:
\begin{align}
E_{p4m,1/2}^{\mathrm{0D}}=\mathbb{Z}_2^6
\end{align}

Subsequently we consider the 1D block-state decoration. For arbitrary 1D blocks, the total symmetry group is $\mathbb{Z}_4^f$, hence there is no nontrivial 1D block-state due to the trivial classification of the corresponding 1D FSPT phases, and the classification attributed to 1D block-state decorations is trivial:
\begin{align}
E_{p4m,1/2}^{\mathrm{1D}}=\{\mathrm{OFBS}\}_{p4m,1/2}^{\mathrm{1D}}=\mathbb{Z}_1
\end{align}

Therefore it is obvious that there is no stacking between 1D and 0D block-states because of the trivial contribution from 1D block-state. The ultimate classification with accurate group structure is:
\begin{align}
\mathcal{G}_{p4m,1/2}=E_{p4m,1/2}^{\mathrm{0D}}\times E_{p4m,1/2}^{\mathrm{1D}}=\mathbb{Z}_2^6
\end{align}

\subsection{Parallelogrammatic lattice: $p2$}
\begin{figure}
\centering
\includegraphics[width=0.46\textwidth]{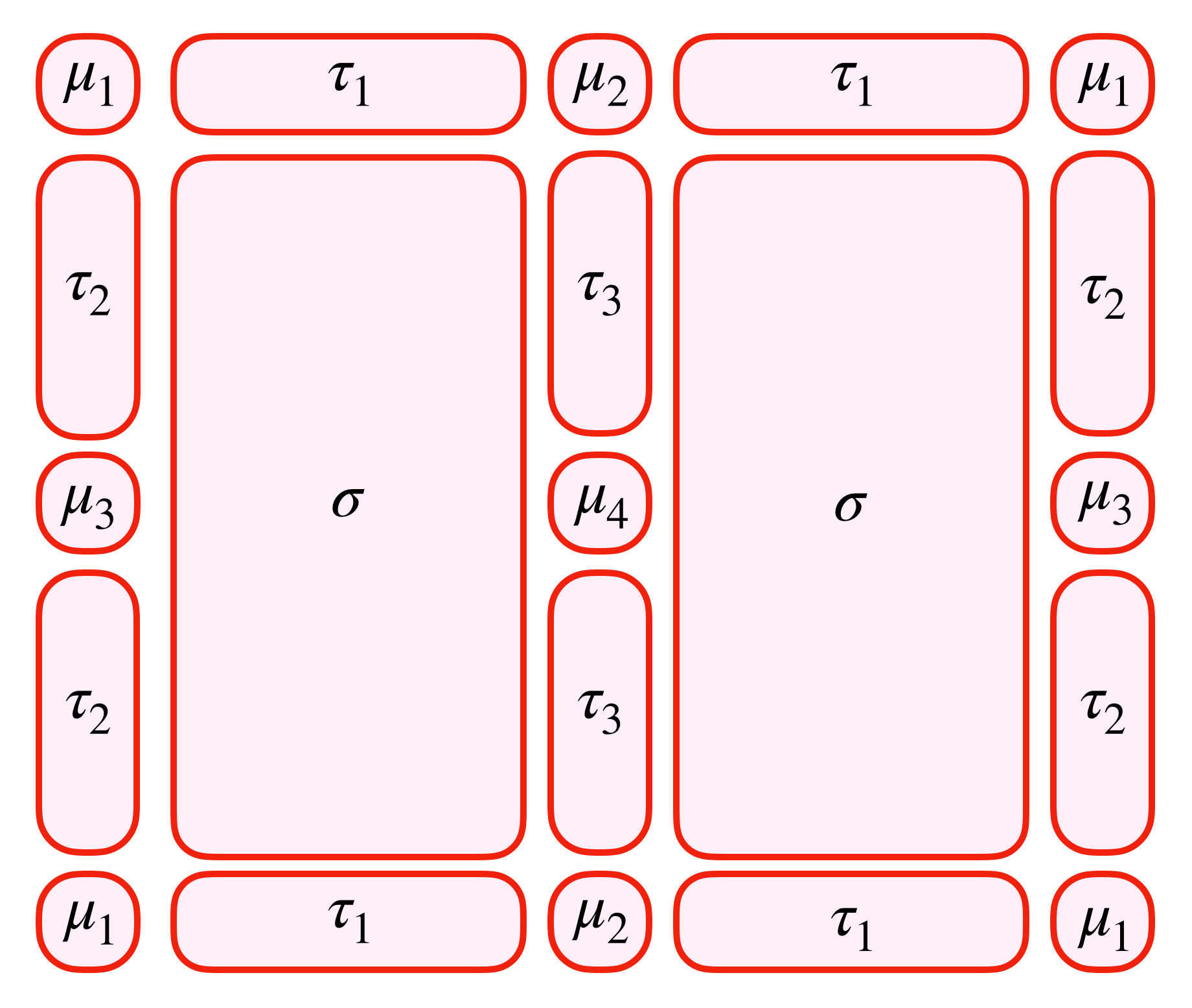}
\caption{\#2 wallpaper group $p2$ and its cell decomposition.}
\label{p2}
\end{figure}

For parallelogrammatic lattice, we demonstrate the crystalline TSC protected by $p2$ symmetry as an example. The corresponding point group of $p2$ is rotation group $C_2$. For 1D and 2D blocks, there is no on-site symmetry group, but the rotational subgroup $C_2$ acts on each 0D blocks internally, just identical with on-site $\mathbb{Z}_2$ symmetry, as seen Fig. \ref{p2}. Below we discuss systems with spinless and spin-1/2 fermions separately.

\subsubsection{Spinless fermions}
For spinless fermions, the total on-site symmetry of each 0D block labeled by $\mu_j, j=1,2,3,4$, is $\mathbb{Z}_2^f\times\mathbb{Z}_2$, and the classification data can be characterized by different 1D irreducible representations of the symmetry group $\mathbb{Z}_2^f\times\mathbb{Z}_2$:
\begin{align}
\mathcal{H}^1\left[\mathbb{Z}_2^f\times\mathbb{Z}_2,U(1)\right]=\mathbb{Z}_2^2
\end{align}

Here one $\mathbb{Z}_2$ is from the fermion parity, and the other is from the rotation eigenvalue $-1$. Thus at each 0D block, the block-state can be labeled by $(\pm,\pm)$ where one $\pm$ represents the fermion parity and the other represents rotation eigenvalue, respectively. According to this notation, the obstruction-free 0D block-states form the following group:
\begin{align}
\{\mathrm{OFBS}\}_{p2,0}^{\mathrm{0D}}=\mathbb{Z}_2^8
\end{align}
and the group elements can be labeled by (four brackets represent the block-states at $\mu_j,j=1,2,3,4$):
\[
[(\pm,\pm),(\pm,\pm),(\pm,\pm),(\pm,\pm)]
\]

Subsequently we consider the 1D block-state decorations. The unique possible 1D block-state is Majorana chain due to the absence of on-site symmetry on arbitrary 1D block, and all 1D block-states form a group:
\begin{align}
\{\mathrm{BS}\}_{p2,0}^{\mathrm{1D}}=\mathbb{Z}_2^3
\end{align}
Then we consider the possible obstructions: Majorana chain decoration on $\tau_1$ leaves 2 dangling Majorana modes at each 0D block labeled by $\mu_2$ which can be glued by an entanglement pair $i\gamma_1\gamma_2$. Nevertheless, this entanglement pair breaks $C_2$ symmetry:
\begin{align}
\bs{R}_{\mu_2}:~i\gamma_1\gamma_2\mapsto i\gamma_2\gamma_1=-i\gamma_1\gamma_2
\end{align}
hence this decoration is obstructed, and does not contribute nontrivial crystalline TSC because of the violation of the no-open-edge condition. It is similar for all other 1D blocks. As a consequence, 1D block-state decorations do not contribute any nontrivial crystalline TSC because all block-states are obstructed: 
\begin{align}
E_{p2,0}^{\mathrm{1D}}=\{\mathrm{OFBS}\}_{p2,0}^{\mathrm{1D}}=\mathbb{Z}_1
\end{align}

\begin{figure}
\centering
\includegraphics[width=0.48\textwidth]{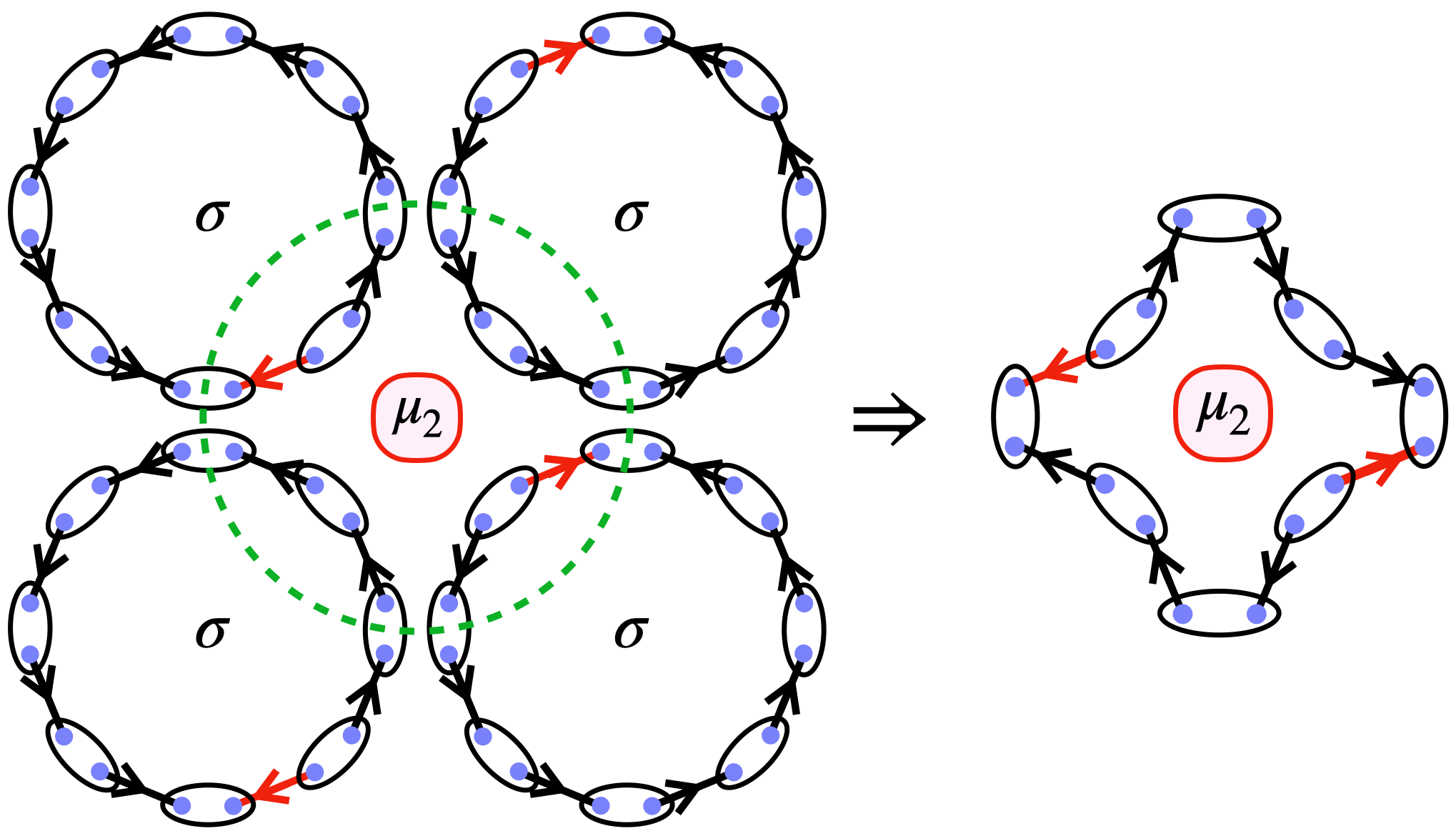}
\caption{2D bubble equivalence for \#2 wallpaper group $p2$. Near each 0D block ($\mu_2$ for example), Majorana modes surrounded by green dashed circle are deformed toward an enclosed Majorana chain surrounding the 0D block $\mu_2$. Left panel shows the bubble construction, and right panel is the deformed Majorana chain which is not Kasteleyn oriented, and the state has odd fermion parity.}
\label{antiPBC p2}
\end{figure}

With all obstruction-free block-states, we consider possible trivializations via bubble construction. First of all, we consider the 2D bubble equivalence: as illustrated in Fig. \ref{antiPBC p2}, we decorate a Majorana chain with anti-PBC on each 2D block that can be trivialized if it shrinks to a point. At each nearby 1D block, we can see that these ``Majorana'' bubbles can be deformed into double Majorana chains. 
Consequently, ``Majorana bubble'' construction has no effect on 1D blocks. At each nearby 0D block ($\mu_2$ as an example, see Fig. \ref{antiPBC p2}), these ``Majorana'' bubbles can be deformed into an alternative Majorana chain with \textit{odd} fermion parity surrounding it. Distinct from the $p4m$ case, this Majorana chain respects all symmetries of $p2$, so this ``Majorana'' bubble construction can change the fermion parities of all 0D blocks simultaneously.

Furthermore, consider 1D bubble equivalence on $\tau_1$: on each 1D block labeled by $\tau_1$, we decorate a pair of complex fermions [cf. Eq. (\ref{1D bubble})]: Near each 0D block $\mu_2$, there are 2 complex fermions forming an atomic insulator:
\begin{align}
|\psi\rangle_{p2}^{\mu_2}=c_1^\dag c_2^\dag|0\rangle
\end{align}
with rotation property:
\begin{align}
\bs{R}_{\mu_2}|\psi\rangle_{p2}^{\mu_2}=c_2^\dag c_1^\dag|0\rangle=-|\psi\rangle_{p2}^{\mu_2}
\end{align}
Hence the rotation eigenvalue $-1$ can be trivialized by atomic insulator $|\psi\rangle_{p2}^{\mu_2}$. Similar for $\mu_1$, and we can conclude that rotation eigenvalues at 0D blocks labeled by $\mu_1$ and $\mu_2$ are not independent. Similar bubble equivalences can be held on arbitrary 1D blocks $\tau_j$, $j=1,2,3,4$, and rotation eigenvalues at all 0D blocks are not independent. 

With all possible bubble constructions, we are ready to study the trivial states. Start from the original trivial state:
\[
[(+,+),(+,+),(+,+),(+,+)]
\]
if we take 2D bubble construction $l_0$ times, and take 1D bubble constructions on $\tau_j$ with $l_j$ times ($j=1,2,3$), above trivial state will be transformed to a new 0D block-state labeled by:
\begin{align}
&\left[((-1)^{l_0},(-1)^{l_1+l_2}),((-1)^{l_0},(-1)^{l_1+l_3}),\right.\nonumber\\
&\left.((-1)^{l_0},(-1)^{l_2}),((-1)^{l_0},(-1)^{l_3})\right]
\label{p2 spinless trivial state}
\end{align}
According to the definition of bubble equivalence, all these states should be trivial. Alternatively, all 0D block-states can be viewed as a vector of an 8-dimensional $\mathbb{Z}_2$-valued vector space, and all trivial 0D block-states with the form as Eq. (\ref{p2 spinless trivial state}) can be viewed as a 4-dimensional vector subspace generated by $l_0,l_1,l_2,l_3$. 
Hence all trivial 0D block-states form the group:
\begin{align}
\{\mathrm{TBS}\}_{p2,0}^{\mathrm{0D}}=\mathbb{Z}_2^4
\end{align}

Therefore, all independent nontrivial 0D block-states are labeled by different group elements of the following quotient group:
\begin{align}
E_{p2,0}^{\mathrm{0D}}=\{\mathrm{OFBS}\}_{p2,0}^{\mathrm{0D}}/\{\mathrm{TBS}\}_{p2,0}^{\mathrm{0D}}=\mathbb{Z}_2^4
\end{align}

It is obvious that there is no stacking between 1D and 0D block-states, and the ultimate classification with accurate group structure is:
\begin{align}
\mathcal{G}_{p2,0}=E_{p2,0}^{\mathrm{1D}}\times E_{p2,0}^{\mathrm{0D}}=\mathbb{Z}_2^4
\end{align}

\subsubsection{Spin-1/2 fermions}
For spin-1/2 fermions, first we consider the 0D block-state decoration, whose candidate states can be characterized by different 1D irreducible representations of the symmetry group $\mathbb{Z}_4^f$ (nontrivial $\mathbb{Z}_2^f$ extension of $\mathbb{Z}_2$ on-site symmetry):
\begin{align}
\mathcal{H}^1\left[\mathbb{Z}_4^f,U(1)\right]=\mathbb{Z}_4
\end{align}
All root phases are characterized by eigenvalues $\{i,-1,-i,1\}$ of $\mathbb{Z}_4^f$. So at each 0D block, the block-state can be labeled by $\nu\in\{i,-1,-i,1\}$. According to this notation, the obstruction-free 0D block-states form the following group:
\begin{align}
\{\mathrm{OFBS}\}_{p2,1/2}^{\mathrm{0D}}=\mathbb{Z}_4^4
\end{align}
and different group elements can be labeled by:
\[
[\nu_1,\nu_2,\nu_3,\nu_4]
\]
where $\nu_j$ labels the 0D block-state at $\mu_j$ (j=1,2,3,4). It is easy to see that there is no trivialization on 0D block-states (i.e., $\{\mathrm{TBS}\}_{p2,1/2}^{\mathrm{0D}}=\mathbb{Z}_1$), so the classification attributed to 0D block-state decoration is:
\begin{align}
E_{p2,1/2}^{\mathrm{0D}}=\{\mathrm{OFBS}\}_{p2,1/2}^{\mathrm{0D}}/\{\mathrm{TBS}\}_{p2,1/2}^{\mathrm{0D}}=\mathbb{Z}_4^4
\label{p2-0}
\end{align}

Subsequently consider the 1D block-state decorations. The unique possible 1D block-state is still the Majorana chain due to the absence of on-site symmetry on each 1D block. The Majorana chain decoration on $\tau_1$ leaves 2 dangling Majorana modes at each 0D block labeled by $\mu_2$ which can be glued by an entanglement pair $i\gamma_1\gamma_2$, and it respects rotational symmetry:
\begin{align}
\bs{R}_{\mu_2}:~i\gamma_1\gamma_2\mapsto-i\gamma_2\gamma_1=i\gamma_1\gamma_2
\end{align}
Hence Majorana chain decoration on $\tau_1$ is an obstruction-free block-state because of the satisfaction of the no-open-edge condition. It is similar for 1D blocks labeled by $\tau_2$ and $\tau_3$. Hence all obstruction-free 1D block-states form the following group:
\begin{align}
\{\mathrm{OFBS}\}_{p2,1/2}^{\mathrm{1D}}=\mathbb{Z}_2^3
\end{align}
We have demonstrated that the 2D ``Majorana'' bubble construction cannot change the parity of Majorana chains on each 1D block in Sec. \ref{general}, hence there is no trivialization (i.e., $\{\mathrm{TBS}\}_{p2,1/2}^{\mathrm{1D}}=\mathbb{Z}_1$), so the classification attributed to 1D block-state decorations is:
\begin{align}
E_{p2,1/2}^{\mathrm{1D}}=\{\mathrm{OFBS}\}_{p2,1/2}^{\mathrm{1D}}/\{\mathrm{TBS}\}_{p2,1/2}^{\mathrm{1D}}=\mathbb{Z}_2^3
\label{p2-1}
\end{align}

With the classification data Eqs. (\ref{p2-0}) and (\ref{p2-1}), we consider the group structure of the corresponding classification. Equivalently, we investigate whether 1D block-state extends 0D block-state or not. As an example, we decorate two copies of Majorana chains on each 1D block labeled by $\tau_1$ which leaves four dangling Majorana fermions at each 0D block labeled by $\mu_1/\mu_2$. Similar with Ref. \onlinecite{rotation}, these Majorana chains can be smoothly deformed to another assembly of Majorana chains surrounding 0D blocks labeled by $\mu_1$ and $\mu_2$ as follows (each yellow ellipse represents a physical site):
\begin{align}
\centering
\includegraphics[width=0.18\textwidth]{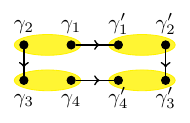}
\label{Majorana stacking}
\end{align}
with rotational symmetry properties: $\gamma_j\mapsto\gamma_j'$ and $\gamma_j'\mapsto-\gamma_j$. The gapped Hamiltonian corresponding to the graph in Eq. (\ref{Majorana stacking}) is:
\begin{align}
H=-i\gamma_1\gamma_1'-i\gamma_4\gamma_4'-i\gamma_2\gamma_3-i\gamma_2'\gamma_3'
\label{Majorana stacking H}
\end{align}
We can further define four complex fermions according to eight Majorana modes in Eq. (\ref{Majorana stacking}) as follows:
\begin{align}
\left.
\begin{aligned}
&c_1=(\gamma_2+i\gamma_1)/2~~~~~~c_2=(\gamma_3+i\gamma_4)/2\\
&c_1'=(\gamma_2'+i\gamma_1')/2~~~~~~c_2'=(\gamma_3'+i\gamma_4')/2
\end{aligned}
\right.
\end{align}
It is easy to find the ground state of Eq. (\ref{Majorana stacking H}):
\begin{align}
|\phi\rangle_{\mathrm{0D}}=(c_1^\dag-c_2^\dag&-ic_1'^\dag+ic_2'^\dag-c_1^\dag c_1'^\dag c_2'^\dag+c_2^\dag c_1'^\dag c_2'^\dag\nonumber\\
&+ic_1^\dag c_2^\dag c_1'^\dag-ic_2^\dag c_2^\dag c_1'^\dag)|0\rangle
\label{Majorana stacking 0D}
\end{align}
with the 2-fold rotation property:
\begin{align}
\bs{R}_{\mu_1}|\phi\rangle_{\mathrm{0D}}=i|\phi\rangle_{\mathrm{0D}}
\label{Majorana stacking sym.}
\end{align}
If a 0D block-state with eigenvalue $e^{i\pi q/2}$ under 2-fold rotation is attached to each 1D block-state near each 0D block labeled by $\mu_1$, the rotation eigenvalue $r$ of the obtained 0D block-state becomes:
\begin{align}
r=e^{i\pi/2+i\pi q},~q\in\mathbb{Z},
\end{align}
and there is no solution to the formula $r=1$. Therefore, near each 0D block labeled by $\mu_1/\mu_2$, 1D block-states extend 0D block-states, hence the 0D block-states at $\mu_1/\mu_2$ have the group structure $\mathbb{Z}_8$ as the nontrivial extension of $\mathbb{Z}_4$ and $\mathbb{Z}_2$ that should be attributed to 0D and 1D block-state decorations, respectively. 

Similar for other 1D and 0D block-states, we can obtain that the 0D block-states have the group structure $\mathbb{Z}_8$ for an arbitrary 0D block. Nevertheless, stacking between 1D and 0D block-states at different 0D blocks are not independent. For instance, if we decorate two copies of Majorana chain on 1D blocks $\tau_1$, these two Majorana chains extend the 0D block-states at both $\mu_1$ and $\mu_2$. It is not hard to verify that only three 0D blocks have independent stacking between 1D and 0D block-states, hence the ultimate classification with accurate group structure is:
\begin{align}
\mathcal{G}_{p2,1/2}=E_{p2,1/2}^{\mathrm{1D}}\times_{\omega_2}E_{p2,1/2}^{\mathrm{0D}}=\mathbb{Z}_4\times\mathbb{Z}_8^3
\end{align}
here the symbol $\times_{\omega_2}$ means that independent nontrivial 1D and 0D block-states $E_{p2,1/2}^{\mathrm{1D}}$ and $E_{p2,1/2}^{\mathrm{0D}}$ have nontrivial extension, characterized by the following short exact sequence:
\begin{align}
0\rightarrow E_{p2,1/2}^{\mathrm{1D}}\rightarrow G_{p2,1/2}\rightarrow E_{p2,1/2}^{\mathrm{0D}}\rightarrow 0
\end{align}

\subsection{Rhombic lattice: $cmm$}
\begin{figure}
\centering
\includegraphics[width=0.483\textwidth]{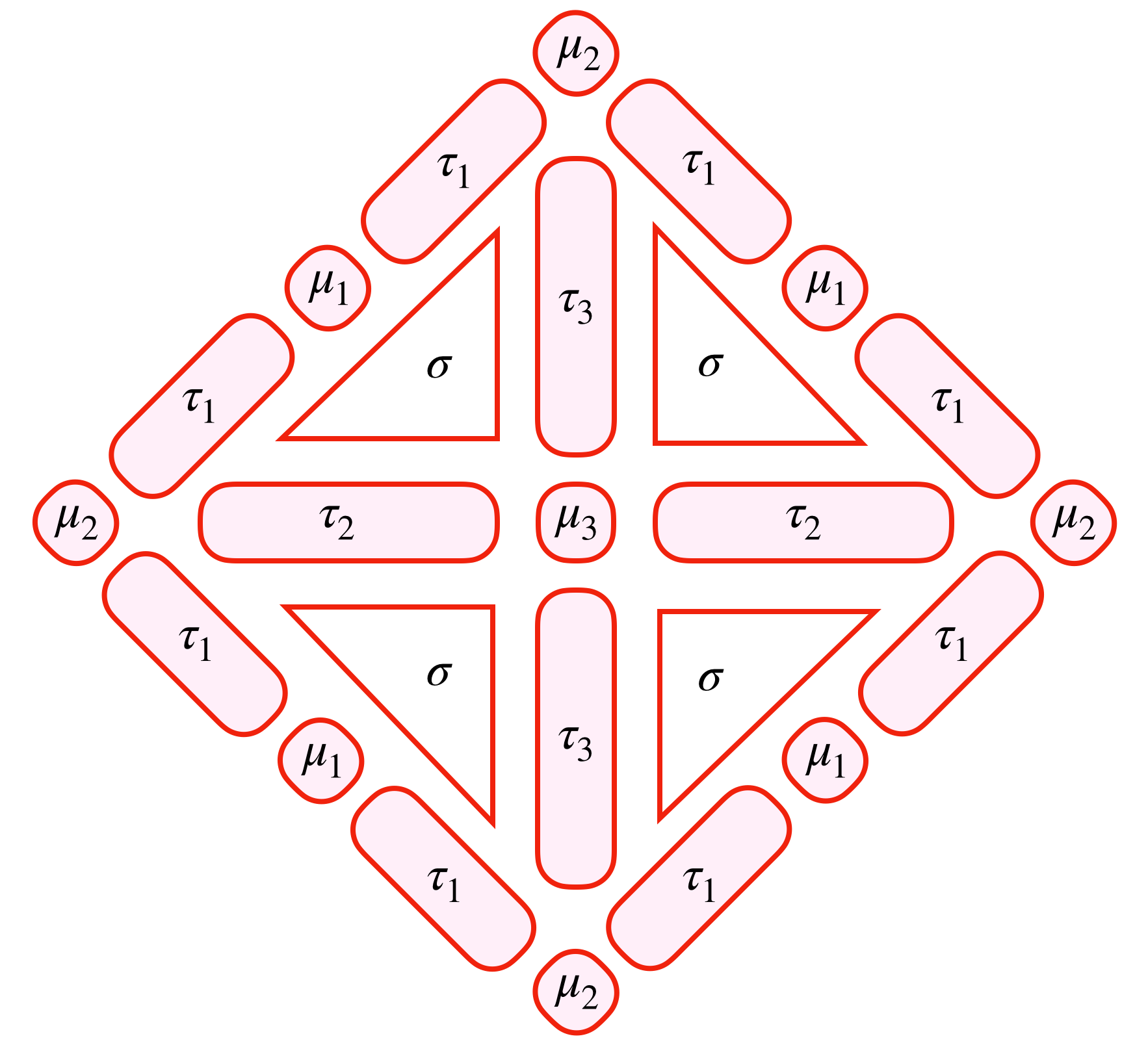}
\caption{\#9 wallpaper group $cmm$ and its cell decomposition.}
\label{cmm}
\end{figure}

For rhombic lattice, we demonstrate the crystalline TSC protected by $cmm$ symmetry as an example. The corresponding point group of $cmm$ is dihedral group $D_2$, and for 2D blocks $\sigma$ and 1D blocks $\tau_1$, there is no on-site symmetry; and for 1D blocks $\tau_2/\tau_3$ and 0D blocks $\mu_1$, the on-site symmetry is $\mathbb{Z}_2$ via the reflection symmetry acting internally; for 0D blocks $\mu_2$ and $\mu_3$, the on-site symmetry group is $\mathbb{Z}_2\rtimes\mathbb{Z}_2$ via the $D_2$ symmetry acting internally. The cell decomposition of $cmm$ is illustrated in Fig. \ref{cmm}.

\subsubsection{Spinless fermions}
For spinless fermions, consider the 0D block-state decoration: For 0D blocks $\mu_1$, the total symmetry group of each is $\mathbb{Z}_2^f\times\mathbb{Z}_2$, and candidate states can be characterized by different 1D irreducible representations of the symmetry group:
\begin{align}
\mathcal{H}^1\left[\mathbb{Z}_2^f\times\mathbb{Z}_2,U(1)\right]=\mathbb{Z}_2^2
\end{align}
So at each 0D block labeled by $\mu_1$, the block-state can be labeled by $(\pm,\pm)$,  and these two $\pm$'s represent the fermion and rotation eigenvalue, respectively. For 0D blocks $\mu_2$ and $\mu_3$, the classification data can be characterized by different irreducible representations of the full symmetry group $\mathbb{Z}_2^f\times(\mathbb{Z}_2\rtimes\mathbb{Z}_2)$:
\begin{align}
\mathcal{H}^1\left[\mathbb{Z}_2^f\times(\mathbb{Z}_2\rtimes\mathbb{Z}_2),U(1)\right]=\mathbb{Z}_2^3
\end{align}
So at each 0D block, the block-state can be labeled by $(\pm,\pm,\pm)$, and these three $\pm$'s represent the fermion parity and eigenvalues of two independent reflection generators $\bs{M}_{\tau_2}$ and $\bs{M}_{\tau_3}$, respectively. According to this notation, the obstruction-free 0D block-states form the group:
\begin{align}
\{\mathrm{OFBS}\}_{cmm,0}^{\mathrm{0D}}=\mathbb{Z}_2^8
\end{align}
where the group elements can be labeled by:
\begin{align}
[(\pm,\pm),(\pm,\pm,\pm),(\pm,\pm,\pm)]
\end{align}
here three brackets represent the block-states at $\mu_1$, $\mu_2$ and $\mu_3$, respectively. 

Subsequently we consider the 1D block-state decoration. For 1D block $\tau_1$, the total symmetry group is just fermion parity $\mathbb{Z}_2^f$, so the only nontrivial 1D block-state is Majorana chain; for 1D blocks $\tau_2$ and $\tau_3$, the total symmetry group is $\mathbb{Z}_2^f\times\mathbb{Z}_2$, so there are two possible 1D block-states: Majorana chain and 1D FSPT state (composed by double Majorana chains), so all 1D block-states form a group:
\begin{align}
\{\mathrm{BS}\}_{cmm,0}^{\mathrm{1D}}=\mathbb{Z}_2^5
\end{align}
Then we discuss the decorations of these two root phases separately.

\paragraph{Majorana chain decoration}First we consider the Majorana chain decoration on 1D blocks $\tau_1$ which leaves two/four dangling Majorana modes at each 0D block $\mu_1/\mu_2$. Near $\mu_1$, Majorana modes have following rotational symmetry properties:
\begin{align}
\bs{R}_{\mu_1}:~\gamma_1\leftrightarrow\gamma_2
\end{align}
with local fermion parity and its symmetry property:
\begin{align}
P_f=i\gamma_1\gamma_2,~\bs{R}_{\mu_1}:~P_f\mapsto-P_f
\end{align}
Hence these two Majorana modes break the fermion parity on 0D block $\mu_1$. Thus Majorana chain decoration on 1D block $\tau_1$ is obstructed because of the violation of the no-open-edge condition. 

Then we consider the Majorana chain decoration on 1D blocks $\tau_2$ that leaves two Majorana modes at each 0D block $\mu_2/\mu_3$. Near $\mu_2$, Majorana modes have following reflection symmetry properties:
\begin{align}
\bs{M}_{\tau_3}:~\gamma_1\leftrightarrow\gamma_2
\end{align}
with local fermion parity and its symmetry property:
\begin{align}
P_f=i\gamma_1\gamma_2,~\bs{M}_{\tau_3}:~P_f\mapsto-P_f
\end{align}
Hence these two Majorana modes break the fermion parity on 0D block $\mu_2$. Thus Majorana chain decoration on $\tau_2$ is obstructed because of the violation of the no-open-edge condition. Majorana chain decoration on 1D blocks $\tau_3$ is similar, hence all types of Majorana chain decorations are obstructed.

\paragraph{1D FSPT state decoration}First we consider the 1D FSPT state decoration on 1D blocks $\tau_2$ that leaves four dangling Majorana modes ($\xi_j,\xi_j',j=1,2$) at each 0D block $\mu_2/\mu_3$. Near $\mu_2/\mu_3$, the corresponding 4 Majorana modes have the symmetry properties ($j=1,2$):
\begin{align}
\begin{aligned}
&\bs{M}_{\tau_2}:~\xi_j\mapsto\xi_j,~\xi_j'\mapsto-\xi_j'\\
&\bs{M}_{\tau_3}:~\xi_1\leftrightarrow\xi_2,~\xi_1'\leftrightarrow\xi_2'
\end{aligned}
\end{align}
Similar with the 1D block-state decorations for $p4m$ case, these 4 Majorana modes cannot be gapped out because they form a projective representation of $D_2$ group at each corresponding 0D block, and a non-degenerate ground state is forbidden. Accordingly, the 1D FSPT state decoration on $\tau_2$ or $\tau_3$ is obstructed because of the degenerate ground state. 

There is one exception: if we decorate 1D FSPT phases on 1D blocks $\tau_2$ and $\tau_3$ simultaneously, it leaves eight dangling Majorana modes at each $\mu_2$ and $\mu_3$. Similar with the 0D block labeled by $\mu_2$ in $p4m$ case (see Fig. \ref{p4m}), these 8 dangling Majorana modes can be gapped symmetrically. As a consequence, the only nontrivial obstruction-free 1D block-state is 1D FSPT state decorations on $\tau_2$ and $\tau_3$ simultaneously, and all obstruction-free 1D block-states form a group:
\begin{align}
\{\mathrm{OFBS}\}_{cmm,0}^{\mathrm{1D}}=\mathbb{Z}_2
\end{align}
where the group elements can be labeled by $m_2=m_3$ ($m_2/m_3$ represents the number of decorated 1D FSPT states on $\tau_2/\tau_3$). 


With all obstruction-free block-states, we discuss all possible trivializations. First we consider the 2D bubble equivalences: we decorate a Majorana chain with anti-PBC on each 2D block and enlarge all ``Majorana bubbles'' near each 1D block labeled by $\tau_1$, and it can be deformed to double Majorana chains that can be trivialized because there is no on-site symmetry; near each 1D block labeled by $\tau_2/\tau_3$, it can also be deformed to double Majorana chains, nevertheless, these double Majorana chains cannot be trivialized because there is an on-site $\mathbb{Z}_2$ symmetry on each $\tau_2/\tau_3$. Equivalently, 1D FSPT state decorations on 1D blocks $\tau_2$ and $\tau_3$ can be deformed to a trivial state via 2D ``Majorana'' bubble equivalence. Furthermore, similar with the $p4m$ case, there is no effect on 0D blocks labeled by $\mu_2$ and $\mu_3$ by taking 2D ``Majorana'' bubble equivalence; nevertheless, similar with the $p2$ case, 2D ``Majorana bubble'' construction changes the fermion parity of each 0D block labeled by $\mu_1$. 

Subsequently we consider the 1D bubble equivalences. For instance, we decorate a pair of complex fermions [cf. Eq. (\ref{1D bubble})]: Near each 0D block $\mu_1$, there are 2 complex fermions forming the following atomic insulator:
\begin{align}
|\psi\rangle_{cmm}^{\mu_1}=c_1^\dag c_2^\dag|0\rangle
\label{cmm mu1 atomic insulator}
\end{align}
with rotation property:
\begin{align}
\bs{R}_{\mu_1}|\psi\rangle_{cmm}^{\mu_1}=c_2^\dag c_1^\dag|0\rangle=-|\psi\rangle_{cmm}^{\mu_1}
\end{align}
i.e., 1D bubble construction on $\tau_1$ changes the rotation eigenvalue at each 0D block $\mu_1$. Near each 0D block $\mu_2$, there are 4 complex fermions forming another atomic insulator:
\begin{align}
|\psi\rangle_{cmm}^{\mu_2}=c_1'^\dag c_2'^\dag c_3'^\dag c_4'^\dag|0\rangle
\label{cmm mu2 atomic insulator}
\end{align}
with two independent reflection symmetry properties ($D_2$ symmetry at 0D block $\mu_2$ can also be generated by two independent reflections $\bs{M}_{\tau_2}$ and $\bs{M}_{\tau_3}$):
\begin{align}
\begin{aligned}
&\bs{M}_{\tau_2}|\psi\rangle_{cmm}^{\mu_2}=c_3'^\dag c_4'^\dag c_1'^\dag c_2'^\dag|0\rangle=|\psi\rangle_{cmm}^{\mu_2}\\
&\bs{M}_{\tau_3}|\psi\rangle_{cmm}^{\mu_2}=c_4'^\dag c_3'^\dag c_2'^\dag c_1'^\dag|0\rangle=|\psi\rangle_{cmm}^{\mu_2}
\end{aligned}
\end{align}
i.e., 1D bubble construction on $\tau_1$ does not change anything on $\mu_2$. Similar 1D bubble constructions can be held on 1D blocks $\tau_2$ and $\tau_3$, and we summarize the effects of 1D bubble constructions as following:
\begin{enumerate}[1.]
\item 1D bubble construction on $\tau_1$: changes the eigenvalue of $\bs{R}_{\mu_1}$ at $\mu_1$;
\item 1D bubble construction on $\tau_2$: simultaneously changes the eigenvalues of $\bs{M}_{\tau_3}$ at $\mu_2$ and $\mu_3$;
\item 1D bubble construction on $\tau_3$: simultaneously changes the eigenvalues of $\bs{M}_{\tau_2}$ at $\mu_2$ and $\mu_3$;
\end{enumerate}

With all possible trivializations, we are ready to study the trivial states. Start from the original trivial 0D block-state (nothing is decorated on arbitrary 0D blocks):
\[
[(+,+),(+,+,+),(+,+,+)]
\]
If we take 2D ``Majorana bubble'' construction $l_0$ times, and take 1D bubble equivalences on $\tau_j$ with $l_j$ times ($j=1,2,3$), above trivial state will be deformed to a new 0D block-state labeled by:
\begin{align}
&\left[((-1)^{l_0},(-1)^{l_1}),(+,(-1)^{l_3},(-1)^{l_2}),\right.\nonumber\\
&\left.(+,(-1)^{l_3},(-1)^{l_2})\right]
\label{cmm spinless trivial state}
\end{align}
According to the definition of bubble equivalence, all these states should be trivial. It is easy to see that there are only four independent quantities ($l_j=0,1,2,3$) in Eq. (\ref{cmm spinless trivial state}), hence all these trivial states form the following group:
\begin{align}
\{\mathrm{TBS}\}_{cmm,0}&=\{\mathrm{TBS}\}_{cmm,0}^{\mathrm{1D}}\times\{\mathrm{TBS}\}_{cmm,0}^{\mathrm{0D}}\nonumber\\
&=\mathbb{Z}_2\times\mathbb{Z}_2^3=\mathbb{Z}_2^4
\end{align}
here $\{\mathrm{TBS}\}_{cmm,0}^{\mathrm{1D}}$ represents the group of trivial states, and $\{\mathrm{TBS}\}_{cmm,0}^{\mathrm{0D}}$ represents the group of trivial states with non-vacuum 0D blocks.

Therefore, all independent nontrivial block-states with different dimensions form the following groups:
\begin{align}
\begin{aligned}
&E_{cmm,0}^{1\mathrm{D}}=\{\mathrm{OFBS}\}_{cmm,0}^{1\mathrm{D}}/\{\mathrm{TBS}\}_{cmm,0}^{1\mathrm{D}}=\mathbb{Z}_1\\
&E_{cmm,0}^{0\mathrm{D}}=\{\mathrm{OFBS}\}_{cmm,0}^{0\mathrm{D}}/\{\mathrm{TBS}\}_{cmm,0}^{0\mathrm{D}}=\mathbb{Z}_2^5
\end{aligned}
\end{align}
and together form the classification:
\begin{align}
\mathcal{G}_{cmm,0}=E_{cmm,0}^{\mathrm{0D}}=\mathbb{Z}_2^5
\end{align}
here all $\mathbb{Z}_2$'s are from the nontrivial 0D block-states. It is obvious that there is no nontrivial group extension because of the absence of nontrivial 1D block-state, and the group structure of $\mathcal{G}_{cmm,0}$ has already been correct.

\subsubsection{Spin-1/2 fermions}
Now we turn to discuss systems with spin-1/2 fermions. First we consider the 0D block-state decorations: for 0D blocks labeled by $\mu_1$, the 2-fold rotational symmetry acts on each of them internally, hence the total symmetry is $\mathbb{Z}_4^f$: nontrivial $\mathbb{Z}_2^f$ extension of on-site $\mathbb{Z}_2$ symmetry. And all different 0D block-states which can be characterized by different 1D irreducible representations of the corresponding symmetry group are:
\begin{align}
\mathcal{H}^1\left[\mathbb{Z}_4^f,U(1)\right]=\mathbb{Z}_4,
\end{align}
and there is no trivialization on them. Furthermore, for 0D blocks labeled by $\mu_2$ and $\mu_3$, the dihedral group symmetry $D_2$ acts on each of them internally, and similar with the $p4m$ case, the classification of corresponding 0D block-states can be characterized by different 1D irreducible representations of the full symmetry group:
\begin{align}
\mathcal{H}^1\left[\mathbb{Z}_2^f\times_{\omega_2}(\mathbb{Z}_2\rtimes\mathbb{Z}_2),U(1)\right]=\mathbb{Z}_2^2
\label{cmm classification data}
\end{align}
Here different $\mathbb{Z}_2$'s represent the rotation and reflection eigenvalues at each $D_2$ center. As a consequence, all obstruction-free 0D block-states form the following group:
\begin{align}
\{\mathrm{OFBS}\}_{cmm,1/2}^{\mathrm{0D}}=\mathbb{Z}_4\times\mathbb{Z}_2^4,
\end{align}
and there is no trivialization on them: For spinless fermions, the rotation eigenvalue $-1$ at 0D block $\mu_1$ is changed by atomic insulator $|\psi\rangle_{cmm}^{\mu_1}$ [cf. Eq. (\ref{cmm mu1 atomic insulator})]. Nevertheless, for spin-1/2 fermions, there is an addtional minus sign under rotation to fulfill the condition $\bs{R}_{\mu_1}^2=-1$:
\begin{align}
\bs{R}_{\mu_1}|\psi\rangle_{cmm}^{\mu_1}=c_2^\dag(-c_1^\dag)=|\psi\rangle_{cmm}^{\mu_1}
\end{align}
i.e., $|\psi\rangle_{cmm}^{\mu_1}$ does not change the rotation eigenvalue $-1$ at 0D block $\mu_1$. Similar for all other 0D blocks.

As a consequence, the classification attributed to 0D block-state decorations is:
\begin{align}
E_{cmm,1/2}^{\mathrm{0D}}=\mathbb{Z}_4\times\mathbb{Z}_2^4
\label{cmm0}
\end{align}

Subsequently we investigate the 1D block-state decoration. On $\tau_1$, the unique possible 1D block-state is Majorana chain because of the absence of the on-site symmetry; on $\tau_2$ and $\tau_3$, the total symmetry group is $\mathbb{Z}_4^f$, hence there is no candidate block-state due to the trivial classification of the corresponding 1D FSPT phases. The Majorana chain decoration on $\tau_1$ leaves 2 dangling Majorana modes at each $\mu_1$, and 4 dangling Majorana modes at each $\mu_2$. At $\mu_1$, the 2 dangling Majorana modes which can be gapped out by an entanglement pair without breaking any symmetry are:
\begin{align}
\bs{R}_{\mu_1}:~i\gamma_1\gamma_2\mapsto-i\gamma_2\gamma_1=i\gamma_1\gamma_2
\end{align}
at $\mu_2$, the 4 Majorana modes have the following reflection symmetry properties ($D_2$ symmetry can also be generated by two independent reflections $\bs{M}_{\tau_2}$ and $\bs{M}_{\tau_3}$):
\begin{align}
\left.
\begin{aligned}
\bs{M}_{\tau_2}:~&(\eta_1,\eta_2,\eta_3,\eta_4)\mapsto(\eta_2,-\eta_1,\eta_4,-\eta_3)\\
\bs{M}_{\tau_3}:~&(\eta_1,\eta_2,\eta_3,\eta_4)\mapsto(\eta_4,\eta_3,-\eta_2,-\eta_1)
\end{aligned}
\right.
\label{cmm spin-1/2 symmetry at mu2}
\end{align}
Consider the following Hamiltonian containing two entanglement pairs of these four Majorana modes:
\begin{align}
H_{\mu_2}=-i\eta_1\eta_3-i\eta_2\eta_4
\end{align}
It is easy to verify that $H_{\mu_2}$ is invariant under the symmetry actions (\ref{cmm spin-1/2 symmetry at mu2}). As a consequence, all obstruction-free 1D block-states form the following group:
\begin{align}
\{\mathrm{OFBS}\}_{cmm,1/2}^{\mathrm{1D}}=\mathbb{Z}_2,
\end{align}
and it is easy to see that there is no trivialization (i.e., $\{\mathrm{TBS}\}_{cmm,1/2}^{\mathrm{0D}}=\mathbb{Z}_1$). So the classification attributed to 1D block-state decorations is:
\begin{align}
E_{cmm,1/2}^{\mathrm{1D}}=\mathbb{Z}_2
\label{cmm1}
\end{align}

With the classification data as Eqs. (\ref{cmm0}) and (\ref{cmm1}), we consider the group structure of the corresponding classification. Equivalently, we investigate if 1D block-state extends 0D block-state. The only possible case of stacking should happen on 1D blocks labeled by $\tau_1$ because other 1D blocks have no nontrivial block-state, similar with $p4m$ and $p2$ cases. We decorate two copies of Majorana chains on $\tau_1$ that leave 2 dangling Majorana modes at each 0D block labeled by $\mu_1$ and 4 dangling Majorana modes at each 0D block labeled by $\mu_2$. At $\mu_1$, these Majorana chains can be smoothly deformed to the state described by Eqs. (\ref{Majorana stacking}) and (\ref{Majorana stacking 0D}), with the symmetry properties as Eq. (\ref{Majorana stacking sym.}). So similar with $p2$ case, near each 0D block labeled by $\mu_1$, 1D block-states extend 0D block-state, and 0D block-states at $\mu_1$ have the group structure $\mathbb{Z}_8$ as a consequence. At $\mu_2$, these Majorana chains can be smoothly deformed to two copies of the state described by Eqs. (\ref{Majorana stacking}) and (\ref{Majorana stacking 0D}), and have eigenvalue $-1$ under 2-fold rotational symmetry. The classification data of 0D block-states at $\mu_2$ is determined by Eq. (\ref{cmm classification data}), hence if a 0D block-state with eigenvalue $-1$ under 2-fold rotation is attached to each 1D block-state near each 0D block labeled by $\mu_2$, the rotation eigenvalue $s$ of the obtained 0D block-state becomes:
\begin{align}
s=(-1)\times(-1)=1
\end{align}
Therefore, near 0D block $\mu_2$ there is an appropriate 1D block-state which itself forms a $\mathbb{Z}_2$ structure under stacking, and there is no stacking between 1D and 0D block-states at $\mu_2$ as a consequence. Finally, the ultimate classification with accurate group structure is:
\begin{align}
\mathcal{G}_{cmm,1/2}=E_{cmm,1/2}^{\mathrm{0D}}\times_{\omega_2}E_{cmm,1/2}^{\mathrm{1D}}=\mathbb{Z}_8\times\mathbb{Z}_2^4
\end{align}
here the symbol ``$\times_{\omega_2}$'' means that 1D and 0D block-states $E_{cmm,1/2}^{\mathrm{1D}}$ and $E_{cmm,1/2}^{\mathrm{0D}}$ have nontrivial extension, and described by the following short exact sequence:
\begin{align}
0\rightarrow E_{cmm,1/2}^{\mathrm{1D}}\rightarrow G_{cmm,1/2}\rightarrow E_{cmm,1/2}^{\mathrm{0D}}\rightarrow0
\end{align}

\subsection{Rectangle lattice: $pgg$}
\begin{figure}
\centering
\includegraphics[width=0.4\textwidth]{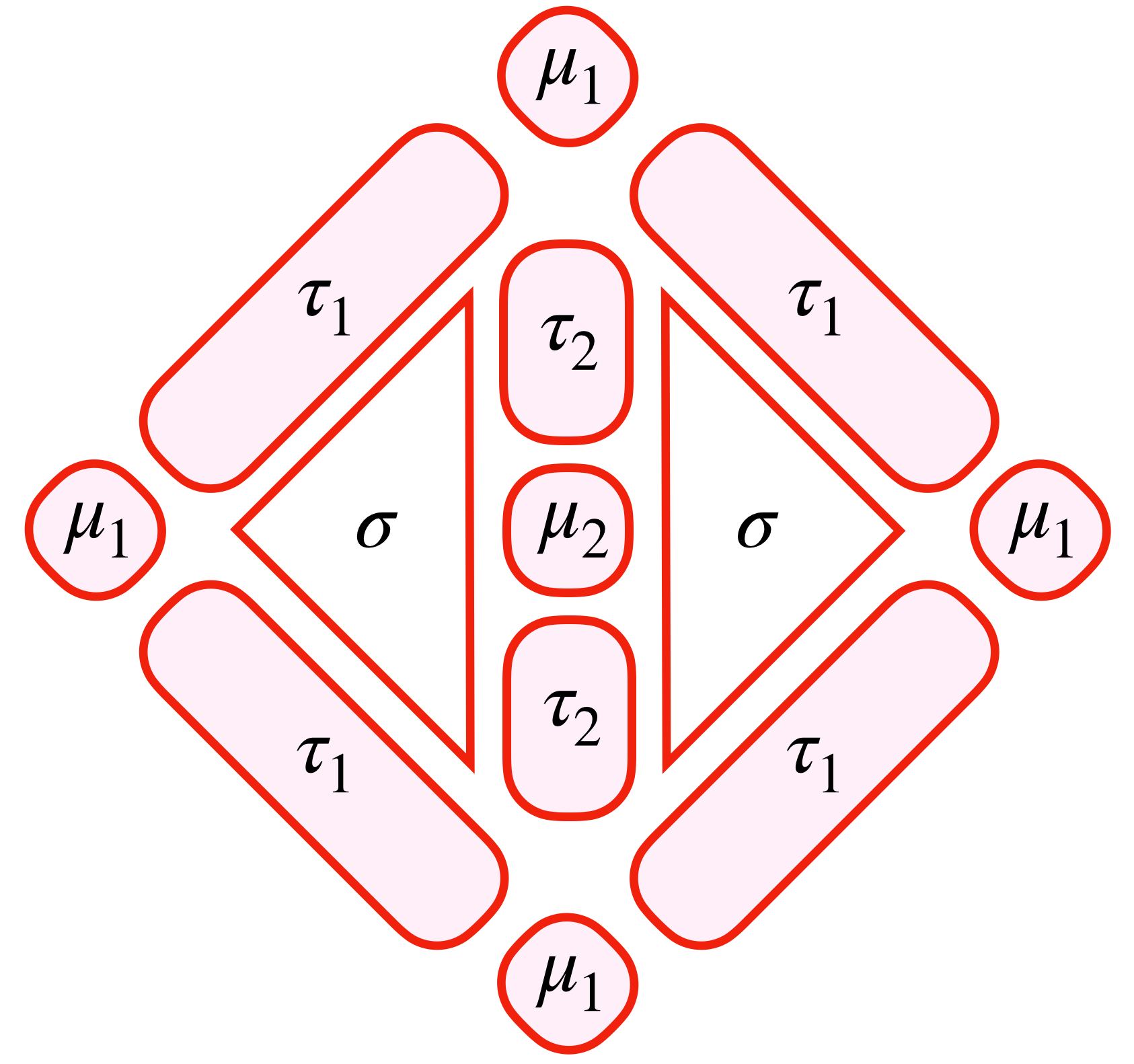}
\caption{\#8 wallpaper group $pgg$ and its cell decomposition.}
\label{pgg}
\end{figure}

For rectangle lattice, we demonstrate the crystalline TSC protected by $pgg$ symmetry as an example. $pgg$ is a non-symorphic wallpaper group and the corresponding point group is dihedral group. The corresponding point group for this case is 2-fold dihedral group $D_2$. For 2D blocks $\sigma$ and 1D blocks $\tau_1$ and $\tau_2$, there is no on-site symmetry, and for 0D blocks $\mu_1$ and $\mu_2$, the on-site symmetry is $\mathbb{Z}_2$ because 2-fold rotational symmetry $C_2$ acts on the 0D blocks internally, as seen Fig. \ref{pgg}.

\subsubsection{Spinless fermions}

First we investigate the 0D block-state decoration. For an arbitrary 0D block, the total symmetry group is an on-site $\mathbb{Z}_2$ symmetry together with the fermion parity: $\mathbb{Z}_2^f\times\mathbb{Z}_2$, and the classification data can be characterized by different 1D irreducible representations of the symmetry group:
\begin{align}
\mathcal{H}^1\left[\mathbb{Z}_2^f\times\mathbb{Z}_2,U(1)\right]=\mathbb{Z}_2^2
\end{align}
these two $\mathbb{Z}_2$'s represent the fermion parity and eigenvalues of 2-fold rotational symmetry on each 0D block, respectively. So at each 0D block, the block-state can be labeled by $(\pm,\pm)$. According to this notation, the obstruction-free 0D block-states form the following group:
\begin{align}
\{\mathrm{OFBS}\}_{pgg,0}^{\mathrm{0D}}=\mathbb{Z}_2^4
\end{align}
and the group elements can be labeled by (two brackets represent the block-states at $\mu_1$ and $\mu_2$):
\[
[(\pm,\pm),(\pm,\pm)]
\]

Subsequently we investigate the 1D block-state decoration. Due to the absence of the on-site symmetry, the unique possible 1D block-state is Majorana chain. So all 1D block-states form a group:
\begin{align}
\{\mathrm{BS}\}_{pgg,0}^{\mathrm{1D}}=\mathbb{Z}_2^2
\end{align}
Then we discuss the possible obstructions: we discuss the 1D block-state decorations on $\tau_1$ and $\tau_2$ separately.

\paragraph{Majorana chain decoration on $\tau_1$}Majorana chain decoration on $\tau_1$ leaves 4 dangling Majorana modes at each corresponding 0D blocks $\mu_1$ with the following rotational symmetry properties:
\begin{align}
\bs{R}_{\mu_1}:~\gamma_j\mapsto\gamma_{j+2}
\label{pgg sym}
\end{align}
Here all subscripts are taken with modulo 4 (i.e., $\gamma_5$ represents the Majorana mode labeled by $\gamma_1$). Consider the following Hamiltonian near each 0D block $\mu_1$:
\begin{align}
H=i\gamma_1\gamma_2+i\gamma_3\gamma_4
\end{align}
it is obvious that $H$ is symmetric under (\ref{pgg sym}), and it can gap out the four Majorana modes at each $\mu_1$.

\paragraph{Majorana chain decoration on $\tau_2$}Majorana chain decoration on $\tau_2$ leaves 2 dangling Majorana modes at each corresponding 0D block which can be gapped out by an entanglement pair. Nevertheless this entanglement pair breaks the fermion parity, and the no-open-edge condition is violated.

As a consequence, all obstruction-free 1D block-states form the following group:
\begin{align}
\{\mathrm{OFBS}\}_{pgg,0}^{\mathrm{1D}}=\mathbb{Z}_2,
\end{align}
and we have demonstrated that the 2D ``Majorana'' bubble cannot change the parity of Majorana chains of each 1D block in Sec. \ref{general}, hence there is no trivialization (i.e., $\{\mathrm{TBS}\}_{pgg,0}^{\mathrm{1D}}=\mathbb{Z}_1$). 
Therefore, all independent nontrivial 1D block-states are labeled by different group elements of the following group:
\begin{align}
E_{pgg,0}^{\mathrm{1D}}=\{\mathrm{OFBS}\}_{pgg,0}^{\mathrm{1D}}/\{\mathrm{TBS}\}_{pgg,0}^{\mathrm{1D}}=\mathbb{Z}_2
\end{align}

With all obstruction-free block-states, we consider possible trivializations via bubble construction. First of all, we consider the 2D bubble equivalence: we decorate a Majorana chain with anti-PBC on each 2D block that can be trivialized if it shrinks to a point. Similar with the $p2$ case, by some proper local unitary transformations, this assembly of bubbles can be deformed to an assembly of Majorana chains with odd fermion parity surrounding each of 0D block, and the fermion parities of all 0D blocks are changed simultaneously. Equivalently, the fermion parities of 0D blocks labeled by $\mu_1$ and $\mu_2$ are not independent. 

Then we study the role of rotation symmetry. Consider the 1D bubble equivalence on $\tau_2$: we decorate a pair of complex fermions [cf. Eq. (\ref{1D bubble})]: Near $\mu_2$, there are 2 complex fermions which form an atomic insulator:
\begin{align}
|\psi\rangle_{pgg}^{\mu_2}=c_1^\dag c_2^\dag|0\rangle
\label{pgg mu1 atomic insulator}
\end{align}
with rotation property as ($\bs{R}_{\mu_2}$ represents the rotation operation centred at the 0D block labeled by $\mu_2$):
\begin{align}
\bs{R}_{\mu_2}|\psi\rangle_{pgg}^{\mu_2}=c_2^\dag c_1^\dag|0\rangle=-|\psi\rangle_{pgg}^{\mu_2}
\end{align}
i.e., $|\psi\rangle_{pgg}^{\mu_2}$ can trivialize the rotation eigenvalue $-1$ at each 0D block labeled by $\mu_2$, similar for the 0D block labeled by $\mu_1$. Hence the rotation eigenvalues at $\mu_1$ and $\mu_2$ are not independent; and we further consider the 1D bubble equivalence on $\tau_1$: Near each 0D block labeled by $\mu_1$, there are 4 complex fermions which form another atomic insulator:
\begin{align}
|\psi\rangle_{pgg}^{\mu_1}=c_1'^\dag c_2'^\dag c_3'^\dag c_4'^\dag |0\rangle
\label{pgg mu2 atomic insulator}
\end{align}
with rotation property as ($\bs{R}_{\mu_1}$ represents the rotation operation centred at the 0D block labeled by $\mu_1$):
\begin{align}
\bs{R}_{\mu_1}|\psi\rangle_{pgg}^{\mu_1}=c_3'^\dag c_4'^\dag c_1'^\dag c_2'^\dag=|\psi\rangle_{pgg}^{\mu_1}
\end{align}
So there is no trivialization from this bubble construction. 

With all possible bubble constructions, we are ready to study the trivial states. Start from the original trivial state:
\[
[(+,+),(+,+)]
\]
if we take 2D bubble construction $l_0$ times and 1D bubble construction on $\tau_2$ with $l_2$ times, above trivial state will be deformed to a new 0D block-state labeled by:
\begin{align}
[((-1)^{l_0},(-1)^{l_2}),((-1)^{l_0},(-1)^{l_2})]
\label{pgg spinless trivial state}
\end{align}
According to the definition of bubble equivalence, all these states should be trivial. It is easy to see that there are only two independent quantities in the state (\ref{pgg spinless trivial state}), hence all trivial states form the group:
\begin{align}
\{\mathrm{TBS}\}_{pgg,0}^{\mathrm{0D}}=\mathbb{Z}_2^2
\end{align}
Therefore, all independent nontrivial 0D block-states are labeled by different group elements of the following quotient group:
\begin{align}
E_{pgg,0}^{\mathrm{0D}}=\{\mathrm{OFBS}\}_{pgg,0}^{\mathrm{0D}}/\{\mathrm{TBS}\}_{pgg,0}^{\mathrm{0D}}=\mathbb{Z}_2^2
\end{align}

It is straightforward to see that there is no stacking between 1D and 0D block-states, and the ultimate classification with accurate group structure is:
\begin{align}
\mathcal{G}_{pgg,0}=E_{pgg,0}^{\mathrm{0D}}\times E_{pgg,0}^{\mathrm{1D}}=\mathbb{Z}_2^3
\end{align}

\subsubsection{Spin-1/2 fermions}
First we investigate the 0D block-state decorations. All 0D blocks are 2-fold rotation centers, hence the total symmetry group of each 0D block is $\mathbb{Z}_4^f$, and different 0D block-states which can be characterized by different 1D irreducible representations of the corresponding symmetry group are:
\begin{align}
\mathcal{H}^1\left[\mathbb{Z}_4^f,U(1)\right]=\mathbb{Z}_4
\end{align}
All root phases at each 0D block are characterized by group elements of $\{1,i,-1,-i\}$. So at each 0D block, the block-state can be labeled by $\nu\in\{1,i,-1,-i\}$. According to this notation, the obstruction-free 0D block-states form the following group:
\begin{align}
\{\mathrm{OFBS}\}_{pgg,1/2}^{\mathrm{0D}}=\mathbb{Z}_4^2
\end{align}
and different group elements can be labeled by:
\[
[\nu_1,\nu_2]
\]
where $\nu_1$ and $\nu_2$ label the 0D block-state at $\mu_1$ and $\mu_2$. It is easy to see that there is no trivialization on 0D block-states (i.e., $\{\mathrm{TBS}\}_{pgg,1/2}^{\mathrm{0D}}=\mathbb{Z}_1$), so the classification attributed to 0D block-state decorations is:
\begin{align}
E_{pgg,1/2}^{\mathrm{0D}}=\{\mathrm{OFBS}\}_{pgg,1/2}^{\mathrm{0D}}/\{\mathrm{TBS}\}_{pgg,1/2}^{\mathrm{0D}}=\mathbb{Z}_4^2
\end{align}

Subsequently we investigate the 1D block-state decoration. The unique possible 1D block-state is Majorana chain because of the absence of on-site symmetry. 
\paragraph{Majorana chain deocration on $\tau_1$}Majorana chain decoration on $\tau_1$ leaves 4 dangling Majorana modes at each  0D blocks labeled by $\mu_1$ with identical symmetry properties with the spinless fermions [cf. Eq. (\ref{pgg sym})], hence these 4 Majorana modes can be gapped out by some entanglement pairs in a symmetric way, and the no-open-edge condition is satisfied.
\paragraph{Majorana chain decoration on $\tau_2$}Majorana chain decoration on $\tau_2$ leaves 2 dangling Majorana modes at each 0D block $\mu_2$ which can be gapped out by an entanglement pair in a symmetric way. Therefore the no-open-edge condition is satisfied. Consequently, all obstruction-free 1D block-states form the following group:
\begin{align}
\{\mathrm{OFBS}\}_{pgg,1/2}^{\mathrm{1D}}=\mathbb{Z}_2^2
\end{align}
Then we demonstrate that there is no trivialization from bubble constructions: For spinless fermions, eigenvalue $-1$ of rotation $\bs{R}_{\mu_1}$ is changed by atomic insulator $|\psi\rangle_{pgg}^{\mu_1}$ [cf. Eq. (\ref{pgg mu1 atomic insulator})]. Nevertheless, for spin-1/2 fermions, there is an additional minus sign under rotation to fulfill the condition $\bs{R}_{\mu_1}^2=-1$:
\begin{align}
\bs{R}_{\mu_1}|\psi\rangle_{pgg}^{\mu_1}=c_2^\dag(-c_1^\dag)|0\rangle=|\psi\rangle_{pgg}^{\mu_1}
\end{align}
i.e., $|\psi\rangle_{pgg}^{\mu_1}$ does not change the eigenvalue of $\bs{R}_{\mu_1}$.

As the consequence, the classification of 2D FSPT phases with $pgg$ symmetry attributed to 1D block-state decoration is:
\begin{align}
E_{pgg,1/2}^{\mathrm{1D}}=\{\mathrm{OFBS}\}_{pgg,1/2}^{\mathrm{1D}}/\{\mathrm{TBS}\}_{pgg,1/2}^{\mathrm{1D}}=\mathbb{Z}_2^2
\end{align}
Then we study the possible stacking between 1D and 0D block-states. If we decorate two Majorana chains on each 1D block labeled by $\tau_1$, similar with $cmm$ case, there is no stacking between 1D and 0D block-states; if we decorate two Majorana chains on each 1D block labeled by $\tau_2$, similar with $p2$ case, it can be smoothly deformed to an assembly of 0D root phases at 0D blocks $\mu_2$. Therefore, the ultimate classification with accurate group structure is:
\begin{align}
\mathcal{G}_{pgg,1/2}=E_{pgg,1/2}^{\mathrm{1D}}\times_{\omega_2}E_{pgg,1/2}^{\mathrm{0D}}=\mathbb{Z}_2\times\mathbb{Z}_4\times\mathbb{Z}_8
\end{align}
here the symbol ``$\times_{\omega_2}$'' means that 1D and 0D block-states $E_{cmm,1/2}^{\mathrm{1D}}$ and $E_{cmm,1/2}^{\mathrm{0D}}$ have nontrivial extension, described by the following short exact sequence:
\begin{align}
0\rightarrow E_{pgg,1/2}^{\mathrm{1D}}\rightarrow G_{pgg,1/2}\rightarrow E_{pgg,1/2}^{\mathrm{0D}}\rightarrow0
\end{align}

\subsection{Hexagonal lattice: $p6m$}
\begin{figure}
\centering
\includegraphics[width=0.46\textwidth]{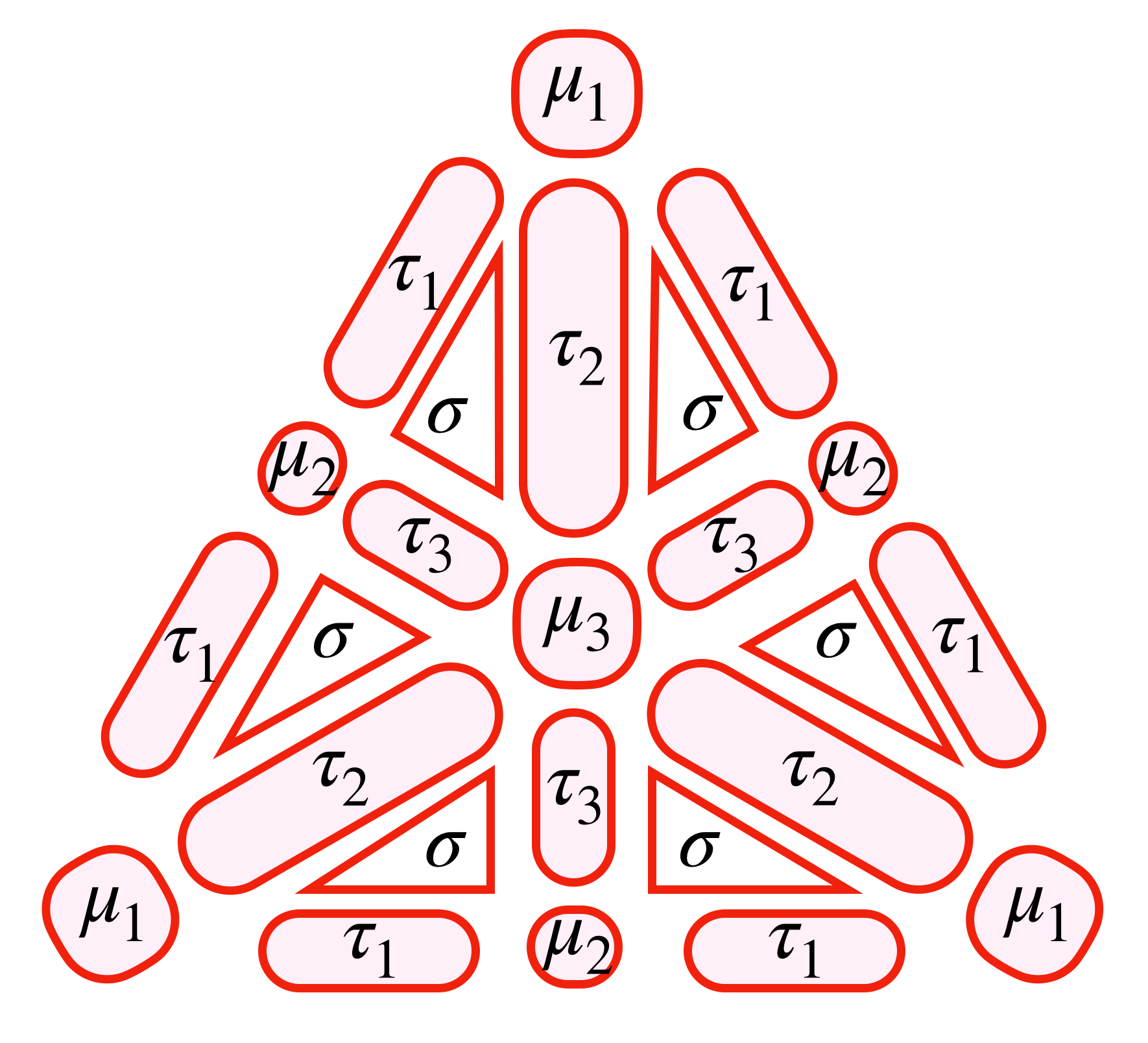}
\caption{\#17 wallpaper group $p6m$ and its cell decomposition.}
\label{p6m}
\end{figure}

For hexagonal lattice, we demonstrate the crystalline TSC protected by $p6m$ symmetry as an example. The corresponding point group of $p6m$ is dihedral group $D_6$, and for 2D blocks labeled by $\sigma$, there is no on-site symmetry; for arbitrary 1D block, the on-site symmetry is $\mathbb{Z}_2$ which is attributed to the reflection symmetry acting internally; for 0D blocks $\mu_1$, the on-site symmetry group is $\mathbb{Z}_6\rtimes\mathbb{Z}_2$ which is attributed to the $D_6$ group acting internally; for 0D blocks $\mu_2$, the on-site symmetry is $\mathbb{Z}_2\rtimes\mathbb{Z}_2$ which is attributed to the $D_2\subset D_6$ acting internally; for 0D blocks $\mu_3$, the on-site symmetry is $\mathbb{Z}_3\rtimes\mathbb{Z}_2$, which is attributed to the $D_3\subset D_6$ acting internally. The cell decomposition is shown in Fig. \ref{p6m}.

\subsubsection{Spinless fermions}
Consider the 0D block-state decorations, for $\mu_j$, $j=1,2,3$, the classification data can be characterized by different 1D irreducible representations of the full symmetry groups, respectively:
\begin{align}
\left.
\begin{aligned}
&\mathcal{H}^1\left[\mathbb{Z}_2^f\times(\mathbb{Z}_6\rtimes\mathbb{Z}_2),U(1)\right]=\mathbb{Z}_2^3\\
&\mathcal{H}^1\left[\mathbb{Z}_2^f\times(\mathbb{Z}_2\rtimes\mathbb{Z}_2),U(1)\right]=\mathbb{Z}_2^3\\
&\mathcal{H}^1\left[\mathbb{Z}_2^f\times(\mathbb{Z}_3\rtimes\mathbb{Z}_2),U(1)\right]=\mathbb{Z}_2^2
\end{aligned}
\right.
\label{p6m classification data}
\end{align}
Similar to the $p4m$ case, the 0D block-states at $\mu_1$ and $\mu_2$ can be labeled by $(\pm,\pm,\pm)$ where the three $\pm$'s represent the fermion parity and eigenvalues of two independent reflection generators $\bs{M}_{\tau_1}$ and $\bs{M}_{\tau_2}$, respectively. However, the 0D block-states at $\mu_3$ should be labeled by $(\pm,\pm)$ where the two $\pm$'s represent the fermion parity and reflection symmetry eigenvalues of $\bs{M}_{\tau_3}$.(We note that there is only one independent reflection eigenvalue for $D_n$ symmetry group with odd $n$.) According to this notation, the obstruction-free 0D block-states form the following group:
\begin{align}
\{\mathrm{OFBS}\}_{p6m}^{\mathrm{0D}}=\mathbb{Z}_2^8
\end{align}
and the group elements can be labeled by (three brackets represent the block-states at $\mu_1$, $\mu_2$ and $\mu_3$):
\[
[(\pm,\pm,\pm),(\pm,\pm,\pm),(\pm,\pm)]
\]

Subsequently we investigate the 1D block-state decoration. For all 1D blocks, the total symmetry group is $\mathbb{Z}_2^f\times\mathbb{Z}_2$, and the candidate 1D block-state is Majorana chain and 1D FSPT state. So all 1D block-states form a group:
\begin{align}
\{\mathrm{BS}\}_{p6m,0}^{\mathrm{1D}}=\mathbb{Z}_2^6
\end{align}
Then we discuss the decorations of these two root phases separately.

\paragraph{Majorana chain decoration}Consider Majorana chain decorations on 1D blocks labeled by $\tau_1$, which leave 6 dangling Majorana modes at each $\mu_1$ and 2 dangling Majorana modes at each $\mu_2$. Near each 0D block $\mu_1$, six dangling Majorana modes have the following rotational symmetry properties (all subscripts are taken with modulo 6):
\begin{align}
\bs{R}_{\mu_1}:~\gamma_j\mapsto\gamma_{j+1},~~j=1,...,6.
\end{align}
Then we consider the local fermion parity and its rotational symmetry property:
\begin{align}
P_f=i\prod\limits_{j=1}^6\gamma_j,~~\bs{R}_{\mu_1}:~P_f\mapsto-P_f
\end{align}
Thus these 6 dangling Majorana modes break fermion parity symmetry and a non-degenerate ground state is forbidden. The corresponding Majorana chain decoration on 1D blocks $\tau_1$ is obstructed because of the violation of the no-open-edge condition. On $\tau_2$, the Majorana chain decoration leaves 6 dangling Majorana modes at each $\mu_1$ and 3 dangling Majorana mdoes at each $\mu_3$. It is well-known that odd number of Majorana modes cannot be gapped out, hence Majorana chain decoration on $
\tau_2$ is obstructed. On $\tau_3$, Majorana chain decoration leaves 2 dangling Majorana modes at each $\mu_2$ and 3 dangling Majorana modes at each $\mu_3$. Similar with the $\tau_2$ case, Majorana chain decoration is obstructed. Note that if we consider all 1D blocks together and decorate a Majorana chain on each, it leaves 12 dangling Majorana modes at each $\mu_1$, 4 dangling Majorana modes at each $\mu_2$  and 6 dangling Majorana modes at each $\mu_3$. Consider Majorana modes at each $\mu_2$, with the following rotation and reflection symmetry properties (all subscripts are taken with modulo 4):
\begin{align}
\bs{R}_{\mu_3}:~\gamma_j'\mapsto\gamma_{j+2}',~~\bs{M}_{\tau_3}:~\gamma_j'\mapsto\gamma_{6-j}'
\end{align}
Then we consider the local fermion parity and its rotation and reflection symmetry properties:
\begin{align}
P_f'=-\prod\limits_{j=1}^4\gamma_j',~~
\left\{
\begin{aligned}
&\bs{R}_{\mu_3}:~P_f'\mapsto P_f'\\
&\bs{M}_{\tau_3}:~P_f'\mapsto-P_f'
\end{aligned}
\right.
\end{align}
Thus these Majorana modes cannot be gapped in a symmetric way, and Majorana chain decoration on all 1D blocks is obstructed. As a consequence, Majorana chain decoration does not contribute a nontrivial crystalline TSC.

\paragraph{1D FSPT state decoration}1D FSPT state decoration on $\tau_1$ leaves 12 dangling Majorana modes at each $\mu_1$ and 4 dangling Majorana modes at each $\mu_2$. Similar with the $p4m$ and $cmm$ cases, four Majorana modes at each $\mu_2$ form a projective representation of $D_2$ symmetry group, and a non-degenerate ground-state is forbidden. Thus the 1D FSPT state decoration on $\tau_1$ is \textit{obstructed}. 

1D FSPT state decoration on $\tau_2$ leaves 12 dangling Majorana modes at each $\mu_1$ and 6 dangling Majorana modes at each $\mu_3$. Consider the Majorana modes at each $\mu_3$, with the following rotation and reflection symmetry properties (all subscripts are taken with modulo 3):
\begin{align}
\left.
\begin{aligned}
\bs{R}_{\mu_3}:~&\eta_j\mapsto\eta_{j+1},~\eta_j'\mapsto\eta_{j+1}'\\
\bs{M}_{\tau_3}:~&\eta_j\mapsto-\eta_{5-j},~\eta_j'\mapsto\eta_{5-j}'
\end{aligned}
\right.,~j=1,2,3.
\end{align}
Then we consider the local fermion parity with its rotation and reflection symmetry properties:
\begin{align}
P_f^{\tau_2}=i\prod\limits_{j=1}^3\eta_j\eta_j',~~
\left\{
\begin{aligned}
&\bs{R}_{\mu_3}:~P_f^{\tau_2}\mapsto P_f^{\tau_2}\\
&\bs{M}_{\tau_3}:~P_f^{\tau_2}\mapsto-P_f^{\tau_2}
\end{aligned}
\right.
\end{align}
Hence these 6 Majorana modes break fermion parity symmetry and cannot be gapped out in a symmetric way. The corresponding 1D FSPT state decoration is \textit{obstructed} because of the violation of the no-open-edge condition. 

1D FSPT state decoration on $\tau_3$ leaves 4 dangling Majorana modes at each $\mu_2$ and 6 dangling Majorana modes at each $\mu_3$. Similar with the 1D FSPT state decoration on $\tau_2$ case, 6 Majorana modes at each $\mu_3$ cannot be gapped out in a symmetric way: consider the Majorana modes as the edge modes of decorated Majorana chains on $\tau_3$ at each $\mu_3$, with the following rotation and reflection symmetry properties: (all subscripts are taken with modulo 3):
\begin{align}
\bs{R}_{\mu_3}:~&\zeta_j\mapsto\zeta_{j+1},~\zeta_j'\mapsto\zeta_{j+1}'\\
\bs{M}_{\tau_3}:~&\zeta_j\mapsto-\zeta_{5-j},~\zeta_j'\mapsto\zeta_{5-j}'
\end{align}
with the local fermion parity and its rotation and reflection symmetry properties:
\begin{align}
P_f^{\tau_3}=i\prod\limits_{j=1}^3\zeta_j\zeta_j',~~
\left\{
\begin{aligned}
&\bs{R}_{\mu_3}:~P_f^{\tau_3}\mapsto P_f^{\tau_3}\\
&\bs{M}_{\tau_3}:~P_f^{\tau_3}\mapsto-P_f^{\tau_3}
\end{aligned}
\right.
\end{align}
Thus 1D FSPT state decoration on $\tau_3$ is \textit{obstructed}, and it does not contribute nontrivial crystalline TSC because of the violation of the no-open-edge condition. 

Note that if we consider 1D blocks labeled by $\tau_2$ and $\tau_3$ together and decorate a 1D FSPT state on each of them, this decoration leaves 12 dangling Majorana modes at each $\mu_1$ and $\mu_3$, and 4 dangling Majorana modes at each $\mu_2$. For the Majorana modes as the edge modes of the decorated 1D FSPT states at each $\mu_2$, as aforementioned, they can be gapped out in a symmetric way; For Majorana modes as the edge modes of the decorated 1D FSPT states at each $\mu_1/\mu_3$, the local fermion parity is the product of $P_f^{\tau_2}$ and $P_f^{\tau_3}$, with the following symmetry properties:
\begin{align}
P_f''=P_f^{\tau_2}P_f^{\tau_3},~~
\left\{
\begin{aligned}
&\bs{R}_{\mu_3}:~P_f''\mapsto P_f''\\
&\bs{M}_{\tau_3}:~P_f''\mapsto P_f''
\end{aligned}
\right.
\end{align}
Hence any symmetry operations commute with the fermion parity. Furthermore, there is no nontrivial projective representation of the $D_3$ group acting internally (identical with the internal symmetry group $\mathbb{Z}_3\rtimes\mathbb{Z}_2$), it can be obtained by calculating the following 2-cohomology of the symmetry group:
\begin{align}
\mathcal{H}^2\left[\mathbb{Z}_3\rtimes\mathbb{Z}_2,U(1)\right]=\mathbb{Z}_1
\end{align}
Therefore, these 12 dangling Majorana modes form a linear representation of the symmetry group, and can be gapped out by some proper interactions in symmetry way. Nevertheless, four Majorana modes at each 0D block labeled by $\mu_2$ form a projective representation of the $D_2$ symmetry group that forbids the non-degenerate ground-state, so the 1D FSPT state decoration on $\tau_2$ and $\tau_3$ is still \textit{obstructed} because of the violation of no-open-edge condition at each 0D block $\mu_2$. 

There is one exception: If we decorate a 1D FSPT phase on each 1D block (including $\tau_j,j=1,2,3$), the dangling Majorana modes at each 0D block can be gapped out in a symmetric way. In the aforementioned discussions we have elucidated that at each $\mu_3$, there are 12 dangling Majorana modes via 1D FSPT state decorations that can be gapped in a symmetric way; and at each $\mu_2$, there are 8 dangling Majorana modes and similar with the $p4m$ and $cmm$ case, they can be gapped out in a symmetric way because they form a linear representation of the corresponding symmetry group. Near each 0D block labeled by $\mu_1$, this decoration leaves 24 dangling Majorana modes as the edge modes of decorated 1D FSPT phases. Consider half of them from 1D FSPT state decorations on $\tau_1$ with the following rotation and reflection symmetry properties (all subscripts are taken with modulo 6):
\begin{align}
\left.
\begin{aligned}
\bs{R}_{\mu_1}:~&\gamma_j\mapsto\gamma_{j+1},~\gamma_j'\mapsto\gamma_{j+1}'\\
\bs{M}_{\tau_1}:~&\gamma_j\mapsto\gamma_{8-j},~\gamma_j'\mapsto\gamma_{8-j}'
\end{aligned}
\right.
\end{align}
Then we consider the local fermion parity and its rotation and reflection symmetry properties:
\begin{align}
P_f^{\tau_1}=-\prod\limits_{j=1}^6\gamma_j\gamma_j',~~\bs{R}_{\mu_1},\bs{M}_{\tau_1}:~P_f^{\tau_1}\mapsto P_f^{\tau_1}
\end{align}
Hence arbitrary symmetry actions commute with the fermion parity of these 12 Majorana modes, and they form either a linear representation or a projective representation of the $D_6$ symmetry. Similar arguments can be held for other 12 Majorana modes. We should note that there is only one nontrivial projective representation of the $D_6$ symmetry group acting internally (i.e., $\mathbb{Z}_6\rtimes\mathbb{Z}_2$ on-site symmetry) that can easily to be verified by the following 2-cohomology:
\begin{align}
\mathcal{H}^2\left[\mathbb{Z}_6\rtimes\mathbb{Z}_2,U(1)\right]=\mathbb{Z}_2
\end{align}
So these 24 Majorana modes together can always form a linear representation of the $D_6$ symmetry at each 0D block labeled by $\mu_1$, and they can be gapped out in a symmetric way. Thus the 1D FSPT state decorations on all 1D blocks simultaneously is \textit{obstruction-free}, and all obstruction-free 1D block-states form the following group:
\begin{align}
\{\mathrm{OFBS}\}_{p6m,0}^{\mathrm{1D}}=\mathbb{Z}_2
\end{align}
and the group elements can be labeled by $m_1=m_2=m_3$. Here $m_j=0,1$ ($j=1,2,3$) represents the number of decorated 1D FSPT states on $\tau_j$, respectively. According to aforementioned discussions, a necessary condition of an obstruction-free block-state is $m_1=m_2=m_3$. 


With all obstruction-free block-states, subsequently we discuss all possible trivializations. First we consider the 2D bubble equivalence: Similar with the $p4m$ case, ``Majorana bubbles'' can be deformed to double Majorana chains at each nearby 1D block, and this is exactly the definition of the nontrivial 1D FSPT phase protected by on-site $\mathbb{Z}_2$ symmetry (by reflection symmetry acting internally). As a consequence, 1D FSPT state decorations on all 1D blocks can be deformed to a trivial state via 2D ``Majorana'' bubble equivalences. Furthermore, repeatedly similar with the $p4m$ case, ``Majorana bubble'' constructions have no effect on 0D blocks. 

\begin{figure*}
\centering
\includegraphics[width=0.88\textwidth]{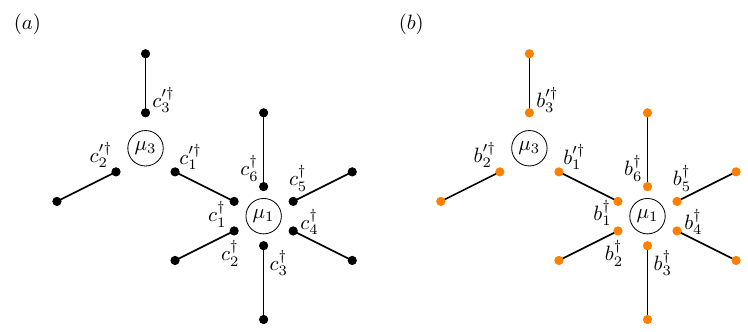}
\caption{1D bubble construction on $\tau_2$. (a): 1D type-\1 (fermionic) bubble construction, where complex fermions are indicated by solid black dots. Atomic insulators (\ref{p6m mu1 atomic}) and (\ref{p6m mu3 atomic}) are created by this procedure. (b). 1D type-\2 (bosonic) bubble construction, where bosonic modes are indicated by solid orange dots. Bosonic states (\ref{p6m mu1 bosonic}) and (\ref{p6m mu3 bosonic}) are created by this procedure.}
\label{p6m 1D bubble}
\end{figure*}

Subsequently we consider the 1D bubble equivalences. For example, on each 1D block labeled by $\tau_2$, we decorate a pair of complex fermions [cf. Eq. (\ref{1D bubble}) and Fig. \ref{p6m 1D bubble}(a)]: Near each 0D block labeled by $\mu_1$, there are 6 complex fermions which form an atomic insulator with even fermion parity:
\begin{align}
|\psi\rangle_{p6m}^{\mu_1}=\prod\limits_{j=1}^6c_j^\dag|0\rangle
\label{p6m mu1 atomic}
\end{align}
hence $|\psi\rangle_{p6m}^{\mu_1}$ cannot change the fermion parity of the 0D block labeled by $\mu_1$; Near each 0D block labeled by $\mu_3$, there are three complex fermions which form another atomic insulator with odd fermion parity:
\begin{align}
|\psi\rangle_{p6m}^{\mu_3}=c_1'^\dag c_2'^\dag c_3'^\dag|0\rangle
\label{p6m mu3 atomic}
\end{align}
and it can change the fermion parity at each 0D block labeled by $\mu_3$. Then we consider the symmetry properties of these atomic insulators: the eigenvalues of $|\psi\rangle_{p6m}^{\mu_1}$ at $\mu_1$ under two independent reflection operations are:
\begin{align}
\begin{aligned}
&\bs{M}_{\tau_1}|\psi\rangle_{p6m}^{\mu_1}=c_6^\dag c_5^\dag c_4^\dag c_3^\dag c_2^\dag c_1^\dag=-|\psi\rangle_{p6m}^{\mu_1}\\
&\bs{M}_{\tau_2}|\psi\rangle_{p6m}^{\mu_1}=c_1^\dag c_6^\dag c_5^\dag c_4^\dag c_3^\dag c_2^\dag=|\psi\rangle_{p6m}^{\mu_1}
\end{aligned}
\end{align}
i.e., 1D bubble construction on $\tau_2$ can change the eigenvalue of $\bs{M}_{\tau_1}$ and leave the eigenvalue of $\bs{M}_{\tau_2}$ invariant. The eigenvalue of $|\psi\rangle_{p6m}^{\mu_3}$ at $\mu_3$ under reflection $\bs{M}_{\tau_2}$ is:
\begin{align}
\bs{M}_{\tau_2}|\psi\rangle_{p6m}^{\mu_3}=c_1'^\dag c_3'^\dag c_2'^\dag|0\rangle=-|\psi\rangle_{p6m}^{\mu_3}
\end{align}
i.e., 1D bubble construction on $\tau_2$ can change the eigenvalue of $\bs{M}_{\tau_2}$. Similar 1D bubble constructions can be held on other 1D blocks, and we summarize the effects of 1D bubble constructions as following:
\begin{enumerate}[1.]
\item 1D bubble construction on $\tau_1$: simultaneously changes the eigenvalues of $\bs{M}_{\tau_2}$ at $\mu_1$ and $\bs{M}_{\tau_3}$ at $\mu_2$;
\item 1D bubble construction on $\tau_2$: simultaneously changes the eigenvalues of $\bs{M}_{\tau_1}$ at $\mu_1$, $\bs{M}_{\tau_2}$ at $\mu_3$ and the fermion parity of $\mu_3$;
\item 1D bubble construction on $\tau_3$: simultaneously changes the eigenvalues of $\bs{M}_{\tau_1}$ at $\mu_2$, $\bs{M}_{\tau_2}$ at $\mu_3$ and the fermion parity of $\mu_3$.
\end{enumerate}
There is another type of 1D bubble construction on $\tau_2$ and $\tau_3$ (we denote the above ``fermionic'' 1D bubble construction by ``type-\1'' and this ``bosonic'' 1D bubble construction by ``type-\2''): we decorate an bosonic 1D bubble [cf. Eq. (\ref{1D bubble}) and Fig. \ref{p6m 1D bubble}(b)] on each $\tau_2$ (here both yellow and red dots represent the 0D bosonic mode with reflection eigenvalue $-1$), near $\mu_1$, there are six 0D bosonic modes, each of them carries reflection eigenvalue $-1$ (six bosonic modes changes nothing):
\begin{align}
|\phi\rangle_{p6m}^{\mu_1}=b_1^\dag b_2^\dag b_3^\dag b_4^\dag b_5^\dag b_6^\dag|0\rangle
\label{p6m mu1 bosonic}
\end{align}
Near $\mu_3$, there are three 0D bosonic modes, each of them carries reflection eigenvalues $-1$: 
\begin{align}
|\phi\rangle_{p6m}^{\mu_3}=b_1'^\dag b_2'^\dag b_3'^\dag|0\rangle
\label{p6m mu3 bosonic}
\end{align}
Hence $|\phi\rangle_{p6m}^{\mu_3}$ changes the reflection eigenvalue by $-1$ at $\mu_3$. Similar for 1D bubble constructions on $\tau_3$.

With all possible bubble constructions, we are ready to investigate the trivial states. Start from the original 0D trivial block-state (nothing is decorated on arbitrary 0D blocks):
\[
[(+,+,+),(+,+,+),(+,+)]
\]
if we take type-\1 1D bubble constructions on $\tau_j$ with $l_j$ times ($j=1,2,3$), and type-\2 1D bubble constructions on $\tau_2$ and $\tau_3$ with $l_2'$ and $l_3'$ times, above trivial state will be deformed to a new block-state labeled by:
\begin{align}
&\left[(+,(-1)^{l_2},(-1)^{l_1}),(+,(-1)^{l_3},(-1)^{l_1}),\right.\nonumber\\
&\left.((-1)^{l_2+l_3},(-1)^{l_2+l_3+l_2'+l_3'})\right]
\label{p6m spinless trivial state}
\end{align}

According to the definition of bubble equivalence, all these states should be trivial. Alternatively, all 0D block-states can be viewed as vectors of an 8-dimensional $\mathbb{Z}_2$-valued vector space $V$, and all trivial 0D block-states with the form as Eq. (\ref{p6m spinless trivial state}) can be viewed as vectors of the subspace of $V$. The dimension of this subspace is 4 because there are only 4 independent indices in $l_1$, $l_2$, $l_3$, and $l_2'+l_3'$. Together with the 2D bubble equivalence, all trivial states form the group:
\begin{align}
\{\mathrm{TBS}\}_{p6m,0}&=\{\mathrm{TBS}\}_{p6m,0}^{\mathrm{1D}}\times\{\mathrm{TBS}\}_{p6m,0}^{\mathrm{0D}}\nonumber\\ 
&=\mathbb{Z}_2 \times \mathbb{Z}_2^4=\mathbb{Z}_2^5
\end{align}
here $\{\mathrm{TBS}\}_{p6m,0}^{\mathrm{1D}}$ represents the group of trivial states with non-vacuum 1D blocks (i.e., 1D FSPT phase decorations on all 1D blocks simultaneously), and $\{\mathrm{TBS}\}_{p6m,0}^{\mathrm{0D}}$ represents the group of trivial states with non-vacuum 0D blocks.

Therefore, all independent nontrivial block-states are labeled by the group elements of the following quotient groups:
\begin{align}
\begin{aligned}
&E_{p6m,0}^{\mathrm{1D}}=\{\mathrm{OFBS}\}_{p6m,0}^{\mathrm{1D}}/\{\mathrm{TBS}\}_{p6m,0}^{\mathrm{1D}}=\mathbb{Z}_1\\
&E_{p6m,0}^{\mathrm{0D}}=\{\mathrm{OFBS}\}_{p6m,0}^{\mathrm{0D}}/\{\mathrm{TBS}\}_{p6m,0}^{\mathrm{0D}}=\mathbb{Z}_2^4
\end{aligned}
\end{align}
here all $\mathbb{Z}_2$'s are from the nontrivial 0D block-states. It is obvious that there is no nontrivial group extension because of the absence of nontrivial 1D block-state. Therefore, the ultimate classification of 2D crystalline FSPT phases with $p6m$ symmetry for spinless fermions is:
\begin{align}
\mathcal{G}_{p6m,0}=E_{p6m,0}^{\mathrm{1D}}\times E_{p6m,0}^{\mathrm{0D}}=\mathbb{Z}_2^4
\end{align}

\subsubsection{Spin-1/2 fermions}
Consider the 0D block-state decoration, and similar with the $p4m$ case, the classification data can also be characterized by different 1D irreducible representations of alternative symmetry groups:
\begin{align}
\left.
\begin{aligned}
&\mathcal{H}^1\left[\mathbb{Z}_2^f\times_{\omega_2}(\mathbb{Z}_6\rtimes\mathbb{Z}_2),U(1)\right]=\mathbb{Z}_2^2\\
&\mathcal{H}^1\left[\mathbb{Z}_2^f\times_{\omega_2}(\mathbb{Z}_2\rtimes\mathbb{Z}_2),U(1)\right]=\mathbb{Z}_2^2\\
&\mathcal{H}^1\left[\mathbb{Z}_2^f\times_{\omega_2}(\mathbb{Z}_3\rtimes\mathbb{Z}_2),U(1)\right]=\mathbb{Z}_4
\end{aligned}
\right.
\end{align}
For $D_6$ and $D_2$ centers, the physical meanings of two $\mathbb{Z}_2$'s in the classification data are rotation and reflection eigenvalues, respectively. Furthermore, the group structure of the classification of 0D FSPT phases protected by $\mathbb{Z}_3\rtimes\mathbb{Z}_2$ on-site symmetry for systems with spin-1/2 fermions is $\mathbb{Z}_4$. Equivalently, we can label different 0D block-states by the group elements of the 4-fold cyclic group:
\begin{align}
\mathbb{Z}_4=\left\{1,i,-1,-i\right\}
\label{Z4}
\end{align}
So the 0D block-states at $\mu_1$ and $\mu_2$ can be labeled by $(\pm,\pm)$, here these two $\pm$'s represent the 2-fold rotation and reflection symmetry eigenvalues (alternatively, they can also represent the eigenvalues of two independent reflection operations because even-fold dihedral group can also be generated by two independent reflections); the 0D block-states at $\mu_3$ can be labeled by $\nu\in\left\{1,i,-1,-i\right\}$ as the eigenvalues of $\mathbb{Z}_4^f$ symmetry. According to this notation, all obstruction-free 0D block-states form the following group:
\begin{align}
\{\mathrm{OFBS}\}_{p6m,1/2}^{\mathrm{0D}}=\mathbb{Z}_2^4\times\mathbb{Z}_4
\end{align}
and the group elements can be labeled by (three brackets represent the block-states at $\mu_1$, $\mu_2$ and $\mu_3$):
\[
[(\pm,\pm),(\pm,\pm),\nu]
\]

Then we investigate the possible trivializations. Consider the 1D bubble equivalence on 1D blocks labeled by $\tau_1$: on each $\tau_1$, the total on-site symmetry is $\mathbb{Z}_4^f$: nontrivial $\mathbb{Z}_2^f$ extension of the on-site symmetry $\mathbb{Z}_2$. Next we decorate an Eq. (\ref{1D bubble}) onto each of them, here the yellow/red dots represent the 0D FSPT modes protected by $\mathbb{Z}_4^f$ symmetry which are labeled by $i~\&~-i\in\mathbb{Z}_4$, cf. Eq. (\ref{Z4}), and they can be trivialized if they shrink to a point. Near each 0D block labeled by $\mu_3$, there are three 0D FSPT modes labeled by $i\in\mathbb{Z}_4$ and they can change the label of 0D block-state decorated at each 0D block $\mu_3$ by $-i\in\mathbb{Z}_4$. Therefore, the 0D block-state on each $\mu_3$ can be trivialized by this bubble construction. Near 0D block $\mu_1$, this 1D bubble construction changes nothing because there is no $\mathbb{Z}_4^f$ on-site symmetry on $\mu_1$. Similar 1D bubble construction can be held on $\tau_3$.

With all possible bubble constructions, we are ready to investigate the trivial states. Start from the original trivial state (nothing decorated on arbitrary 0D block):
\[
[(+,+),(+,+),1]
\]
if we take above 1D bubble constructions on $\tau_2$ and $\tau_3$ with $l_2$ and $l_3$ times, above trivial state will be deformed to a new 0D block-state labeled by:
\begin{align}
[(+,+),(+,+),(-i)^{3(l_2+l_3)}]
\label{p6m spin-1/2 trivial state}
\end{align}
According to the definition of bubble equivalence, all these states should be trivial and all trivial states form the group:
\begin{align}
\{\mathrm{TBS}\}_{p6m,1/2}^{\mathrm{0D}}=\mathbb{Z}_4
\end{align}
Therefore, all independent nontrivial 0D block-states are labeled by different group elements of the following quotient group:
\begin{align}
E_{p6m,1/2}^{\mathrm{0D}}=\{\mathrm{OFBS}\}_{p6m,1/2}^{\mathrm{0D}}/\{\mathrm{TBS}\}_{p6m,1/2}^{\mathrm{0D}}=\mathbb{Z}_2^4
\end{align}

Subsequently we consider the 1D block-state decoration. For arbitrary 1D blocks, the total on-site symmetry on them is $\mathbb{Z}_4^f$: nontrivial $\mathbb{Z}_2^f$ extension of $\mathbb{Z}_2$ on-site symmetry, hence there is no nontrivial 1D block-state due to the trivial classification of the corresponding 1D FSPT phases, and the classification attributed to 1D block-state decorations is trivial: 
\begin{align}
E_{p6m,1/2}^{\mathrm{1D}}=\{\mathrm{OFBS}\}_{p6m,1/2}^{\mathrm{1D}}=\mathbb{Z}_1
\end{align}
Therefore, it is obvious that there is no stacking between 1D and 0D block-states, and the ultimate classification with accurate group structure is:
\begin{align}
\mathcal{G}_{p6m,1/2}=\mathbb{Z}_2^4
\end{align}

\section{Construction and classification of crystalline TI\label{insulator}}
So far we have discussed the construction and classification of crystalline TSC in 2D interacting fermionic systems. 
In this section, we will discuss the crystalline TI with additional $U^f(1)$ symmetry by generalizing the real-space construction highlighted in Sec. \ref{general}. In particular, all block-states decorations will admit an additional $U^f(1)$ internal symmetry. Below we demonstrate that 1D block-state decoration has no contribution and all nontrivial crystalline TI in 2D interacting fermionic systems can be constructed by 0D block-state decoration. 

For 1D blocks, there are two different cases: symmetry group with/without the reflection symmetry operation. Since bosonic and fermionic systems can be mapped to each other by Jordan-Wigner transformation, the classification data of 1D SPT phases for bosonic and fermionic systems are identical: by calculating the different projective representations of the symmetry group. (However, the group structure of the classification data could be different in general as stacking operation has different physical meaning for boson and fermion systems.)

For symmetry groups without reflection symmetry operation, the on-site symmetry group of an arbitrary 1D blocks should be $U^f(1)$ charge conservation only, and the corresponding classification for 2D systems with spinless/spin-1/2 fermions can be calculated by the following group cohomology:
\begin{align}
\mathcal{H}^2[U^f(1),U(1)]=\mathbb{Z}_1
\end{align}
Thus, there is no nontrivial 1D block-state for this case. 

For the symmetry group with reflection symmetry operation, the on-site symmetry group of some 1D blocks should be $U^f(1)$ charge conservation and $\mathbb{Z}_2$ symmetry via reflection symmetry acting internally. The corresponding classification for 2D systems with spinless/spin-1/2 fermions can be calculated by the following group cohomology:
\begin{align}
\mathcal{H}^2[U^f(1)\times\mathbb{Z}_2,U(1)]=\mathbb{Z}_1
\end{align}
Again, there is also no nontrivial 1D block-state for this case. 
 

Below we will again study five representative cases  belonging to different crystallographic systems:
\begin{enumerate}
\item square lattice: $p4m$;
\item parallelogrammatic lattice: $p2$.
\item rhombic lattice: $cmm$;
\item rectangle lattice: $pgg$;
\item hexagonal lattice: $p6m$;
\end{enumerate}
and all other cases are assigned in Supplementary Materials \cite{supplementary}. The classification results are summarized in Table \ref{insulator U(1)}.

\subsection{Square lattice: $p4m$}
According to the cell decomposition (see Fig. \ref{p4m}), for 0D blocks labeled by $\mu_j$ ($j=1,2,3$), different 0D block-states are characterized by different irreducible representations of the corresponding on-site symmetry group ($n=2,4$):
\begin{align}
\mathcal{H}^1[U^f(1)\times(\mathbb{Z}_n\rtimes\mathbb{Z}_2),U(1)]=\mathbb{Z}\times\mathbb{Z}_2^2
\end{align}
For systems with spinless fermions, the 0D block-states at $\mu_j$ ($j=1,2,3$) can be labeled by $(n_j,\pm,\pm)$, where $n_j\in\mathbb{Z}$ represents the $U^f(1)$ charge carried by complex fermions decorated on $\mu_j$ and two $\pm$'s represent the eigenvalues of two independent reflection generators. According to this notation, the obstruction-free 0D block-states form the following group:
\begin{align}
\{\mathrm{OFBS}\}_{p4m,0}^{U(1)}=\mathbb{Z}^3\times\mathbb{Z}_2^6
\end{align}
and different group elements can be labeled by (three brackets represent the block-states at $\mu_1$, $\mu_2$ and $\mu_3$):
\begin{align}
[(n_1,\pm,\pm),(n_2,\pm,\pm),(n_3,\pm,\pm)]
\end{align}
Nevertheless, we should further consider possible trivializations. For systems with spinless fermions, we first consider the 1D bubble equivalence on 1D blocks labeled by $\tau_1$: we decorate a 1D ``particle-hole'' bubble on each $\tau_1$ that can be trivialized if we shrink them to a point. Near each 0D block labeled by $\mu_1$, there are four particles forming the following atomic insulator:
\begin{align}
|\phi\rangle_{p4m}^{\mu_1}=p_1^\dag p_2^\dag p_3^\dag p_4^\dag|0\rangle
\end{align}
it has eigenvalues under independent reflections:
\begin{align}
&\bs{M}_{\tau_1}|\phi\rangle_{p4m}^{\mu_1}=p_1^\dag p_4^\dag p_3^\dag p_2^\dag=-|\phi\rangle_{p4m}^{\mu_1}\\
&\bs{M}_{\tau_3}|\phi\rangle_{p4m}^{\mu_1}=p_3^\dag p_4^\dag p_1^\dag p_2^\dag=|\phi\rangle_{p4m}^{\mu_1}
\end{align}
i.e., eigenvalue $-1$ of $\bs{M}_{\tau_1}$ at each 0D block $\mu_1$ can be trivialized by the atomic insulator $|\phi\rangle_{p4m}^{\mu_1}$. Near $\mu_2$, there are two holes forming another atomic insulator:
\begin{align}
|\phi\rangle_{p4m}^{\mu_2}=h_1^\dag h_2^\dag|0\rangle
\end{align}
it has eigenvalues under independent reflections:
\begin{align}
&\bs{M}_{\tau_1}|\phi\rangle_{p4m}^{\mu_2}=h_1^\dag h_2^\dag|0\rangle=|\phi\rangle_{p4m}^{\mu_2}\\
&\bs{M}_{\tau_2}|\phi\rangle_{p4m}^{\mu_2}=h_2^\dag h_1^\dag|0\rangle=-|\phi\rangle_{p4m}^{\mu_2}
\end{align}
i.e., eigenvalues $-1$ of the reflection $\bs{M}_{\tau_2}$ at each 0D block $\mu_2$ can be trivialized by atomic insulator $|\phi\rangle_{p4m}^{\mu_2}$. Therefore, aforementioned 1D bubble construction leads to the dependence of reflection eigenvalues at $\mu_1$ and $\mu_2$ (can be changed simultaneously). Similar 1D bubble construction can be held on $\tau_2$ and $\tau_3$ as well. 

Now we move to the $U^f(1)$ charge sector. As shown in Fig. \ref{p4m}, we note that within a specific unit cell, there is one 0D block labeled by $\mu_1$ and $\mu_3$, two 0D blocks labeled by $\mu_2$. Consider the 1D bubble construction on $\tau_1$: it adds four $U^f(1)$ charges at each 0D block $\mu_1$ and removes two $U^f(1)$ charges at each 0D block $\mu_2$, hence the $U^f(1)$ charge at $\mu_1$ and $\mu_2$ are not independent. Similar arguments are also applied to 1D blocks labeled by $\tau_2$ and $\tau_3$.

With the help of above discussions, we consider the 1D bubble equivalence. Start from the trivial state:
\begin{align}
[(0,+,+),(0,+,+),(0,+,+)]
\label{p4m original trivial state}
\end{align}
Taking aforementioned 1D bubble constructions on $\tau_j$ with $l_j\in\mathbb{Z}$ times, it will lead to a new 0D block-state labeled by:
\begin{align}
&\left[\left(4l_1+4l_3,(-1)^{l_1},(-1)^{l_3}\right),\right.\nonumber\\
&\left(-2l_1+2l_2,(-1)^{l_2},(-1)^{l_1}\right),\nonumber\\
&\left.\left(-4l_2-4l_3,(-1)^{l_2},(-1)^{l_3}\right)\right],
\label{p4m trivial state}
\end{align}
which should be trivial. Alternatively, all 0D block-states can be viewed as vectors of a 9-dimensional vector space $V$, where the $U^f(1)$ charge components are $\mathbb{Z}$-valued and all other components are $\mathbb{Z}_2$-valued attributed to rotation and reflection eigenvalues. Then all trivial 0D block-states with the form as Eq. (\ref{p4m trivial state}) can be viewed as a vector subspace $V'$ of $V$. It is easy to see that there are only three independent quantities in Eq. (\ref{p4m trivial state}): $l_1$, $l_2$ and $l_3$, so the dimension of the vector subspace $V'$ should be 3. For the $U^f(1)$ charge  sector, we have the following relationship:
\begin{align}
-(4l_1+4l_3)-2(-2l_1+2l_2)=-4l_2-4l_3
\end{align}
i.e., there are only two independent quantities which serves a $2\mathbb{Z}\times4\mathbb{Z}$ trivialization. The remaining one degree of freedom of the vector subspace $V'$ should be attributed to the eigenvalues of point group symmetry action with ${(-1)}^{l_1}={(-1)}^{l_2}={(-1)}^{(-l_3)}$ which serve a $\mathbb{Z}_2$ trivialization. Therefore, all trivial states with the form as shown in Eq. (\ref{p4m trivial state}) compose the following group:
\begin{align}
\{\mathrm{TBS}\}_{p4m,0}^{U(1)}=2\mathbb{Z}\times4\mathbb{Z}\times\mathbb{Z}_2
\end{align}
and different independent nontrivial 0D block-states can be labeled by different group elements of the following quotient group:
\begin{align}
\mathcal{G}_{p4m,0}^{U(1)}&=\{\mathrm{OFBS}\}_{p4m,0}^{U(1)}/\{\mathrm{TBS}\}_{p4m,0}^{U(1)}\nonumber\\
&=\mathbb{Z}\times\mathbb{Z}_8\times\mathbb{Z}_4\times\mathbb{Z}_2^3
\end{align}

For systems with spin-1/2 fermions, the classification data of the corresponding 0D block-states can be characterized by different irreducible representations of the corresponding on-site symmetry group ($n=2,4$):
\begin{align}
\mathcal{H}^1[U^f(1)\times_{\omega_2}(\mathbb{Z}_n\rtimes\mathbb{Z}_2),U(1)]=2\mathbb{Z}\times\mathbb{Z}_2^2
\end{align}
The precise meaning of $\omega_2$ are refer to Sec. \ref{spinSec}).
To calculate this classification data, we should firstly calculate the following two cohomologies \cite{supplementary}:
\begin{align}
\left\{
\begin{aligned}
&n_0\in\mathcal{H}^0(\mathbb{Z}_n\rtimes\mathbb{Z}_2,\mathbb{Z})=\mathbb{Z}\\
&\nu_1\in\mathcal{H}^1\left[\mathbb{Z}_n\rtimes\mathbb{Z}_2,U(1)\right]=\mathbb{Z}_2^2
\end{aligned}
\right.
\end{align}
Here $\mathbb{Z}$ represents the $U^f(1)$ charge carried by complex fermions, and two $\mathbb{Z}_2$'s represent the rotation and reflection eigenvalues. We demonstrate that the odd number of the $U^f(1)$ charge at each 0D block is not allowed: a specific $n_0$ is obstructed if and only if $(-1)^{\omega_2\smile n_0}\in\mathcal{H}^2[\mathbb{Z}_n\rtimes\mathbb{Z}_2,U(1)]$ is a nontrivial 2-cocycle with $U(1)$-coefficient. From Refs. \onlinecite{general2} and \onlinecite{dihedral} we know that for cases without $U^f(1)$ charge conservation, nontrivial 0-cocycle $n_0=1\in\mathcal{H}^0(\mathbb{Z}_n\rtimes\mathbb{Z}_2,\mathbb{Z}_2)$ leads to nontrivial 2-cocycle $(-1)^{\omega_2\smile n_0}\in\mathcal{H}^2[\mathbb{Z}_n\rtimes\mathbb{Z}_2,U(1)]$. So for $U^f(1)$ charge conserved cases, odd $n_0\in\mathcal{H}^0(\mathbb{Z}_n\rtimes\mathbb{Z}_2,\mathbb{Z})$ lead to nontrivial 2-cocycle $(-1)^{\omega_2\smile n_0}\in\mathcal{H}^2[\mathbb{Z}_n\rtimes\mathbb{Z}_2,U(1)]$. As a consequence, for systems with spin-1/2 fermions, we can only decorate even number of complex fermions on each 0D block and all obstruction-free block-states form a group:
\begin{align}
\{\mathrm{OFBS}\}_{p4m,1/2}^{U(1)}=(2\mathbb{Z})^3\times\mathbb{Z}_2^6
\end{align}

Then we consider the possible trivializations via 1D bubble constructions. Similar to the TSC case, since the reflection properties of $|\phi\rangle_{p4m}^{\mu_1}$ and $|\phi\rangle_{p4m}^{\mu_2}$ at $\mu_1$ and $\mu_2$ are changed by an additional $-1$, there is no trivialization. The discussion of $U^f(1)$ charge sector is identical to the spinless case: start from the original trivial state (\ref{p4m original trivial state}), take above 1D bubble constructions on $\tau_j$ with $l_j\in\mathbb{Z}$ times, it leads to a new 0D block-state labeled by:
\begin{align}
&\left[\left(4l_1+4l_3,0,0\right),\right.\nonumber\\
&\left(-2l_1+2l_2,0,0\right),\nonumber\\
&\left.\left(-4l_2-4l_3,0,0\right)\right]
\label{p4m U(1) spin-1/2 trivial state}
\end{align}
Again, all states with the form (\ref{p4m U(1) spin-1/2 trivial state}) are trivial, forming the following group:
\begin{align}
\{\mathrm{TBS}\}_{p4m,1/2}^{U(1)}=2\mathbb{Z}\times4\mathbb{Z}
\end{align}
Different independent nontrivial 0D block-states can be labeled by different group elements of the following quotient group:
\begin{align}
\mathcal{G}_{p4m,1/2}^{U(1)}&=\{\mathrm{OFBS}\}_{p4m,1/2}^{U(1)}/\{\mathrm{TBS}\}_{p4m,1/2}^{U(1)}\nonumber\\
&=2\mathbb{Z}\times\mathbb{Z}_2^7
\end{align}
Here $2\mathbb{Z}$ means that we can only decorate even number of complex fermions on each 0D block.

\subsection{Parallelogrammatic lattice: $p2$}
Similar to the $p4m$ case, different 0D block-states are characterized by different irreducible representations of the symmetry group:
\begin{align}
\mathcal{H}^1[U^f(1)\times\mathbb{Z}_2,U(1)]=\mathbb{Z}\times\mathbb{Z}_2
\end{align}
Here $\mathbb{Z}$ represents the $U^f(1)$ charge and $\mathbb{Z}_2$ represents the rotation eigenvalues. So 0D block-states at $\mu_j$ ($j=1,2,3,4$) can be labeled by $(n_j,\pm)$, here $n_j\in\mathbb{Z}$ represents the $U^f(1)$ charge carried by complex fermions on $\mu_j$ and $\pm$ represents the eigenvalue of 2-fold rotation operation. According to this notation, all obstruction-free 0D block-states form the following group:
\begin{align}
\{\mathrm{OFBS}\}_{p2,0}^{U(1)}=\mathbb{Z}^4\times\mathbb{Z}_2^4
\end{align}

We should further consider possible trivializations: for systems with spinless fermions, consider the 1D bubble equivalence on 1D blocks labeled by $\tau_1$: we decorate a 1D ``particle-hole'' bubble on each $\tau_1$. Near each 0D block labeled by $\mu_1$, there are two particles forming the following atomic insulator:
\begin{align}
|\xi\rangle_{p2}^{\mu_1}=p_1^\dag p_2^\dag|0\rangle
\end{align}
with following rotation property:
\begin{align}
\bs{R}_{\mu_1}|\xi\rangle_{p2}^{\mu_1}=p_2^\dag p_1^\dag|0\rangle=-|\xi\rangle_{p2}^{\mu_1}
\end{align}
i.e., rotation eigenvalue $-1$ at each 0D block $\mu_1$ can be trivialized by atomic insulator $|\xi\rangle_{p2}^{\mu_1}$. Near $\mu_2$, there are two holes forming another atomic insulator:
\begin{align}
|\xi\rangle_{p2}^{\mu_2}=h_1^\dag h_2^\dag|0\rangle
\end{align}
with rotation property:
\begin{align}
\bs{R}_{\mu_2}|\xi\rangle_{p2}^{\mu_2}=h_2^\dag h_1^\dag|0\rangle=-|\xi\rangle_{p2}^{\mu_2}
\end{align}
i.e., rotation eigenvalue $-1$ at each 0D block $\mu_2$ can be trivialized by atomic insulator $|\xi\rangle_{p2}^{\mu_2}$. Therefore, aforementioned 1D bubble construction leads to the dependence of rotation eigenvalues at $\mu_1$ and $\mu_2$. 

Now we move to the $U^f(1)$ charge sector. 
Repeatedly consider the aforementioned 1D bubble construction on $\tau_1$: it adds two $U^f(1)$ charges at each 0D block $\mu_1$ and removes two $U^f(1)$ charges at each 0D block $\mu_2$, hence the $U^f(1)$ charge at $\mu_1$ and $\mu_2$ are not independent. 
We summarize effects of all possible 1D bubble constructions:
\begin{enumerate}
\item 1D bubble construction on $\tau_1$: Add two $U^f(1)$ charges on $\mu_1$, eliminate two $U^f(1)$ charges on $\mu_2$, and simultaneously change the rotation eigenvalues of $\mu_1$ and $\mu_2$;
\item 1D bubble construction on $\tau_2$: Add two $U^f(1)$ charges on $\mu_1$, eliminate two $U^f(1)$ charges on $\mu_3$, and simultaneously change the rotation eigenvalues of $\mu_1$ and $\mu_3$;
\item 1D bubble construction on $\tau_3$: Add two $U^f(1)$ charges on $\mu_2$, eliminate two $U^f(1)$ charges on $\mu_4$, and simultaneously change the rotation eigenvalues of $\mu_2$ and $\mu_4$;
\end{enumerate}

With the help of above discussions, we consider the 1D bubble equivalence. Start from the original trivial state:
\begin{align}
[(0,+),(0,+),(0,+),(0,+)]
\label{p2 original trivial state}
\end{align}
Taking aforementioned 1D bubble constructions on $\tau_j$ with $l_j\in\mathbb{Z}$ times ($j=1,2,3$), this trivial state will be deformed to a new 0D block-state labeled by:
\begin{align}
&\left[(2l_1+2l_2,(-1)^{l_1+l_2}),(-2l_1+2l_3,(-1)^{l_1+l_3}),\right.\nonumber\\
&\left.(-2l_2,(-1)^{l_2}),(-2l_3,(-1)^{l_3})\right]
\label{p2 U(1) spinless trivial state}
\end{align}
According to the definition of bubble equivalence, this state should be trivial. Alternatively, all 0D block-states can be viewed as vectors of an 8-dimensional vector space $V$, where the complex fermion components are $\mathbb{Z}$-valued, and all other components are $\mathbb{Z}_2$-valued. Then all trivial 0D block-states with the form as Eq. (\ref{p2 U(1) spinless trivial state}) can be viewed as a vector space $V'$ of $V$. It is easy to see that there are only three independent quantities in Eq. (\ref{p2 U(1) spinless trivial state}): $l_1$, $l_2$ and $l_3$. So the dimension of the vector subspace $V'$ should be 3. For the $U^f(1)$ charge sector, there are 3 independent quantities in the following 4 variables:
\[
2l_1+2l_2,-2l_1+2l_3,-2l_2,-2l_3
\]
Thus all 1D bubble constructions serve a $(2\mathbb{Z})^3$ trivialization in $U^f(1)$ charge sector, and all trivial states form the following group:
\begin{align}
\{\mathrm{TBS}\}_{p2,0}^{U(1)}=(2\mathbb{Z})^3
\end{align}
and different independent nontrivial 0D block-states can be labeled by different group elements of the following quotient group:
\begin{align}
\mathcal{G}_{p2,0}^{U(1)}&=\{\mathrm{OFBS}\}_{p2,0}^{U(1)}/\{\mathrm{TBS}\}_{p2,0}^{U(1)}\nonumber\\
&=\mathbb{Z}^4\times\mathbb{Z}_2^4/(2\mathbb{Z})^3=\mathbb{Z}\times\mathbb{Z}_2^7
\end{align}

For systems with spin-1/2 fermions, 0D obstruction-free block-states are identical with spinless case:
\begin{align}
\{\mathrm{OFBS}\}_{p2,1/2}^{U(1)}=\mathbb{Z}^4\times\mathbb{Z}_2^4
\end{align}
then repeatedly consider the aforementioned 1D bubble constructions: rotation properties of $|\xi\rangle_{p2}^{\mu_1}$ and $|\xi\rangle_{p2}^{\mu_2}$ at $\mu_1$ and $\mu_2$ are changed by an additional $-1$ and it leads to no trivialization. Furthermore, it is easy to verify that the complex fermion decorations for spinless and spin-1/2 fermions are identical. So again we start from the original trivial state (\ref{p2 original trivial state}), take above 1D bubble constructions on $\tau_j$ with $l_j\in\mathbb{Z}$ times ($j=1,2,3$), and it will lead to a new 0D block-state labeled by:
\begin{align}
&\left[(2l_1+2l_2,+),(-2l_1+2l_3,+),\right.\nonumber\\
&\left.(-2l_2,+),(-2l_3,+)\right]
\end{align} 
Similar with the spinless case, all states with this form are trivial, forming the following group:
\begin{align}
\{\mathrm{TBS}\}_{p2,1/2}^{U(1)}=(2\mathbb{Z})^3
\end{align}
and different independent nontrivial 0D block-states can be labeled by different group elements of the following quotient group
\begin{align}
\mathcal{G}_{p2,1/2}^{U(1)}&=\{\mathrm{OFBS}\}_{p2,1/2}^{U(1)}/\{\mathrm{TBS}\}_{p2,1/2}^{U(1)}\nonumber\\
&=\mathbb{Z}^4\times\mathbb{Z}_2^4/(2\mathbb{Z})^3=\mathbb{Z}\times\mathbb{Z}_4^3\times\mathbb{Z}_2
\end{align}

We notice that the classifications of 2D crystalline TI protected by $p2$ symmetry for both spinless and spin-1/2 fermions are identical. Now we give a comprehension of this issue: for both spinless and spin-1/2 fermions ($\bs{R}^2=1$ and $\bs{R}^2=-1$, respectively), the group structures of the total symmetry groups are identical: direct product of $U^f(1)$ charge conservation and 2-fold rotation symmetry: $U^f(1)\times C_2$. We explicitly formulate the $U^f(1)$ charge conservation and $C_2$ rotation symmetry as:
\begin{align}
C_2=\left\{E,\bs{R}\right\},~~U^f(1)=\left\{e^{i\theta}\big|\theta\in[0,2\pi)\right\}
\end{align}
For systems with spinless fermions, $\bs{R}^2=1$. Nevertheless, we can twist the group elements of $C_2$ by a $U^f(1)$ phase factor as:
\begin{align}
\bs{R}'=\bs{R}e^{i\pi/2},~~e^{i\pi/2}\in U^f(1)
\label{twist}
\end{align}
then we reformulate the total symmetry group with the twisted operators:
\begin{align}
C_2=\left\{E,\bs{R}'\right\},~~U^f(1)=\left\{e^{i\theta}\big|\theta\in[0,2\pi)\right\}
\end{align}
But $\bs{R}'^2=-1$ for this case. Therefore, the symmetry groups for both spinless and spin-1/2 fermions are identical, and can be deformed to each other by Eq. (\ref{twist}). We stress that such a statement is true for all wallpaper group with a single reflection axis.

\subsection{Rhombic lattice: $cmm$}
Repeatedly consider the cell decomposition of $cmm$ as illustrated in Fig. \ref{cmm}. For 0D blocks labeled by $\mu_1$, different 0D block-states are characterized by different irreducible representations of the symmetry group as:
\begin{align}
\mathcal{H}^1[U^f(1)\times\mathbb{Z}_2,U(1)]=\mathbb{Z}\times\mathbb{Z}_2
\end{align}
Here $\mathbb{Z}$ represents the $U^f(1)$ charge and $\mathbb{Z}_2$ represents the rotation eigenvalue $-1$. For 0D blocks labeled by $\mu_2$ and $\mu_3$, different 0D block-states are characterized by different irreducible representations of the symmetry group as:
\begin{align}
\mathcal{H}^1[U^f(1)\times(\mathbb{Z}_2\rtimes\mathbb{Z}_2),U(1)]=\mathbb{Z}\times\mathbb{Z}_2^2
\end{align}
Here $\mathbb{Z}$ represents the $U^f(1)$ charge and two $\mathbb{Z}_2$'s represent the two independent reflection eigenvalue $-1$. Thus 0D block-states at $\mu_1$ can be labeled by $(n_1,\pm)$, here $n_1\in\mathbb{Z}$ represents the $U^f(1)$ charge at each $\mu_1$ and $\pm$ represents the eigenvalues of 2-fold rotation operation $\bs{R}_{\mu_1}$; and 0D block-states at $\mu_2$/$\mu_3$ can be labeled by $(n_2/n_3,\pm,\pm)$, here $n_2/n_3\in\mathbb{Z}$ represents the $U^f(1)$ charge at each $\mu_2/\mu_3$, and two $\pm$'s represent the eigenvalues of two independent reflection generators $\bs{M}_{\tau_2}$ and $\bs{M}_{\tau_3}$. According to this notation, all obstruction-free 0D block-states form the following group:
\begin{align}
\{\mathrm{OFBS}\}_{cmm,0}^{U(1)}=\mathbb{Z}^3\times\mathbb{Z}_2^5
\end{align}

We should further consider possible trivializations: for systems with spinless fermions, consider the 1D bubble equivalence on 1D blocks labeled by $\tau_1$: we decorate a 1D ``particle-hole'' bubble [cf. Eq. (\ref{1D bubble}), here yellow and red dots represent particle and hole, respectively] on each $\tau_1$, and they can be trivialized if we shrink them to a point. Near each 0D block labeled by $\mu_1$, there are two particles forming atomic insulator:
\begin{align}
|\xi\rangle_{cmm}^{\mu_1}=p_1^\dag p_2^\dag|0\rangle
\end{align}
Near $\mu_2$, there are four holes forming another atomic insulator: 
\begin{align}
|\xi\rangle_{cmm}^{\mu_2}=h_1^\dag h_2^\dag h_3^\dag h_4^\dag|0\rangle
\end{align}
Similar with the crystalline TSC, rotation eigenvalue at each $\mu_1$ can be changed by $|\xi\rangle_{cmm}^{\mu_1}$. Then we consider the 1D bubble equivalence on 1D blocks labeled by $\tau_2$: we decorate an identical 1D ``particle-hole'' bubble as aforementioned on each $\tau_2$. Near each 0D block labeled by $\mu_2$, there are two particles forming the following atomic insulator:
\begin{align}
|\eta\rangle_{cmm}^{\mu_2}=p_1'^\dag p_2'^\dag|0\rangle
\end{align}
Near $\mu_3$, there are two holes forming another atomic insulator:
\begin{align}
|\eta\rangle_{cmm}^{\mu_3}=h_1'^\dag h_2'^\dag|0\rangle
\end{align}
Similar with the crystalline TSC, eigenvalue of $\bs{M}_{\tau_3}$ at each $\mu_2/\mu_3$ can be changed by $|\eta\rangle_{cmm}^{\mu_2}/|\eta\rangle_{cmm}^{\mu_3}$. The bubble construction on $\tau_3$ can be understood in a similar way.

Subsequently we consider the $U^f(1)$ charge sector. First of all, as shown in Fig. \ref{cmm}, we should identify that within a specific unit cell, there are two 0D blocks labeled by $\mu_1$ and one 0D block labeled by $\mu_2/\mu_3$. Repeatedly consider above 1D bubble construction on $\tau_1$: it adds two complex fermions on each 0D block $\mu_1$ and removes four complex fermions at each 0D block $\mu_2$ (by adding four holes), hence the numbers of complex fermions at $\mu_1$ and $\mu_2$ are not independent. 
Then we consider aforementioned 1D bubble equivalence on 1D blocks $\tau_2$/$\tau_3$: it adds two complex fermions at each 0D block $\mu_2$ and adding two holes at each 0D block $\mu_3$, hence the $U^f(1)$ charge at $\mu_2$ and $\mu_3$ are not independent. 

With the help of above discussions, we consider the 0D block-state decorations. Start from the original trivial state (nothing is decorated on all blocks):
\begin{align}
[(0,+),(0,+,+),(0,+,+)]
\label{cmm original trivial state}
\end{align}
Taking aforementioned 1D bubble construction on $\tau_j$ with $l_j\in\mathbb{Z}$ times ($j=1,2,3$), it will lead to a new 0D block-state labeled by:
\begin{align}
&\left[\left(2l_1,(-1)^{l_1}\right),\left(-2l_1+2l_2+2l_3,(-1)^{l_3},(-1)^{l_2}\right)\right.\nonumber\\
&\left.\left(-2l_2-2l_3,(-1)^{l_3},(-1)^{l_2}\right)\right]
\label{cmm U(1) spinless trivial state}
\end{align}
According to the definition of bubble equivalence, all states with this form should be trivial. Alternatively, all 0D block-states can be viewed as vectors of an 8-dimensional vector space $V$, where the complex fermion components are $\mathbb{Z}$-valued and all other components are $\mathbb{Z}_2$-valued. Then all trivial 0D block-states with the form as Eq. (\ref{cmm U(1) spinless trivial state}) can be viewed as a vector subspace $V'$ of $V$. It is easy to see that there are only three independent quantities in Eq. (\ref{cmm U(1) spinless trivial state}): $l_1$, $l_2$ and $l_3$. So the dimensionality of the vector subspace $V'$ should be 3. For the $U^f(1)$ charge sector, we have the following relationship:
\begin{align}
-2l_1-(-2l_1+2l_2+2l_3)=-2l_2-2l_3
\end{align}
i.e., there are only two independent quantities which serve a $(2\mathbb{Z})^2$ trivialization. The remaining one degree of freedom of the vector subspace $V'$ should be attributed to the eigenvalues of point group symmetry action with $(-1)^{l_2}=(-1)^{(-l_3)}$ which serves a $\mathbb{Z}_2$ trivialization. Therefore, all trivial states (\ref{cmm U(1) spinless trivial state}) form the following group:
\begin{align}
\{\mathrm{TBS}\}_{cmm,0}^{U(1)}=(2\mathbb{Z})^2\times\mathbb{Z}_2
\end{align}
and different independent nontrivial 0D block-states can be labeled by different group elements of the following quotient group:
\begin{align}
\mathcal{G}_{cmm,0}^{U(1)}&=\{\mathrm{OFBS}\}_{cmm,0}^{U(1)}/\{\mathrm{TBS}\}_{cmm,0}^{U(1)}\nonumber\\
&=\mathbb{Z}^3\times\mathbb{Z}_2^5/(2\mathbb{Z})^2\times\mathbb{Z}_2=\mathbb{Z}\times\mathbb{Z}_4^2\times\mathbb{Z}_2^2
\end{align}

For systems with spin-1/2 fermions, like the cases without $U^f(1)$ charge conservation, the classification data of the 0D block-states of 0D blocks labeled by $\mu_2$ and $\mu_3$ can be characterized by different irreducible representations of the corresponding on-site symmetry group, 
\begin{align}
\mathcal{H}^1[U^f(1)\times_{\omega_2}(\mathbb{Z}_2\rtimes\mathbb{Z}_2),U(1)]=2\mathbb{Z}\times\mathbb{Z}_2^2
\end{align}
Here $2\mathbb{Z}$ represents the $U^f(1)$ charge carried by complex fermion, and two $\mathbb{Z}_2$'s represent the two independent reflection eigenvalues (similar with the $p4m$ case, we can only decorate even number of $U^f(1)$ charge on each 0D block). 
and all obstruction-free 0D block-states form the following group:
\begin{align}
\{\mathrm{OFBS}\}_{cmm,1/2}^{U(1)}=\mathbb{Z}\times(2\mathbb{Z})^2\times\mathbb{Z}_2^5
\end{align}
Then we discuss possible trivializations. Repeatedly consider aforementioned 1D bubble constructions, and now the rotation properties of $|\xi\rangle_{cmm}^{\mu_1}$, $|\xi\rangle_{cmm}^{\mu_2}$, $|\eta\rangle_{cmm}^{\mu_2}$ and $|\eta\rangle_{cmm}^{\mu_3}$ at $\mu_j, j=1,2,3$ are changed by an additional $-1$; the reflection properties of $|\eta\rangle_{cmm}^{\mu_2}$ and $|\eta\rangle_{cmm}^{\mu_3}$ at $\mu_2$ and $\mu_3$ are also changed by an additional $-1$. All of them lead to no trivialization. Furthermore, it is easy to see that all arguments about the $U^f(1)$ charge sector are identical. Again we start from the original trivial state (\ref{cmm original trivial state}), and take above 1D bubble constructions on $\tau
_j$ with $l_j$ times ($j=1,2,3$), it will lead to a new 0D block-state labeled by:
\begin{align}
&\left[\left(2l_1,+\right),\left(-2l_1+2l_2+2l_3,+,+\right)\right.\nonumber\\
&\left.\left(-2l_2-2l_3,+,+\right)\right]
\label{cmm U(1) spin-1/2 trivial state}
\end{align}
Similar with the spinless case, all states with this form are trivial, forming the following group:
\begin{align}
\{\mathrm{TBS}\}_{cmm,1/2}^{U(1)}=(2\mathbb{Z})^2
\end{align}
and all different independent nontrivial 0D block-states can be labeled by different group elements of the following quotient group:
\begin{align}
&\mathcal{G}_{cmm,1/2}^{U(1)}=\{\mathrm{OFBS}\}_{cmm,1/2}^{U(1)}/\{\mathrm{TBS}\}_{cmm,1/2}^{U(1)}\nonumber\\
&=\mathbb{Z}\times(2\mathbb{Z})^2\times\mathbb{Z}_2^5/(2\mathbb{Z})^2=2\mathbb{Z}\times\mathbb{Z}_4\times\mathbb{Z}_2^4
\end{align}
We should notice that the group structure of the classification should be $2\mathbb{Z}\times\mathbb{Z}_2^6$ rather than $\mathbb{Z}\times\mathbb{Z}_2^5$: two independent quantities are $l_1$ and $l_2+l_3$, hence the classification contributed from complex fermion decorations on $\mu_1$ should be $\mathbb{Z}/2\mathbb{Z}=\mathbb{Z}_2$. Equivalently, 0D block-state $(1,+)$ at $\mu_1$ is nontrivial. 

\subsection{Rectangle lattice: $pgg$}
Repeatedly consider the cell decomposition of $pgg$ as illustrated in Fig. \ref{pgg}. For an arbitrary 0D block, different 0D block-states are characterized by different irreducible representations of symmetry group as:
\begin{align}
\mathcal{H}^1[U^f(1)\times\mathbb{Z}_2,U(1)]=\mathbb{Z}\times\mathbb{Z}_2
\end{align}
Here $\mathbb{Z}$ represents the $U^f(1)$ charge and $\mathbb{Z}_2$ represents the eigenvalues of 2-fold rotational symmetry operation. So the 0D block-state decorated on $\mu_j~(j=1,2)$ can be labeled by $(n_j,\pm)$, where $n_j\in\mathbb{Z}$ represents the $U^f(1)$ charge carried by complex fermions on $\mu_j$ and $\pm$ represents the eigenvalues of 2-fold rotational symmetry on $\mu_j$. According to this notation, all obstruction-free 0D block-states form the following group:
\begin{align}
\{\mathrm{OFBS}\}_{pgg,0}^{U(1)}=\mathbb{Z}^2\times\mathbb{Z}_2^2
\end{align}
We should further consider the possible trivialization. For systems with spinless fermions, consider the 1D bubble equivalence on $\tau_2$: we decorate a 1D ``particle-hole'' bubble [cf. Eq. (\ref{1D bubble})] on each $\tau_2$ that can be trivialized if we shrink them to a point. Near each 0D block labeled by $\mu_1$, there are two particles that form an atomic insulator:
\begin{align}
|\phi\rangle_{pgg}^{\mu_1}=p_1^\dag p_2^\dag|0\rangle
\end{align}
with rotation property as:
\begin{align}
\bs{R}_{\mu_1}|\phi\rangle_{pgg}^{\mu_1}=p_2^\dag p_1^\dag|0\rangle=-|\phi\rangle_{pgg}^{\mu_1}
\end{align}
i.e., rotation eigenvalue $-1$ can be trivialized by the atomic insulator $|\phi\rangle_{pgg}^{\mu_1}$ at each 0D block labeled by $\mu_1$. Near each 0D block labeled by $\mu_2$, there are two holes that form another atomic insulator:
\begin{align}
|\phi\rangle_{pgg}^{\mu_2}=h_1^\dag h_2^\dag|0\rangle
\end{align}
with rotation property as:
\begin{align}
\bs{R}_{\mu_2}|\phi\rangle_{pgg}^{\mu_2}=h_2^\dag h_1^\dag|0\rangle=-|\phi\rangle_{pgg}^{\mu_2}
\end{align}
i.e., rotation eigenvalue $-1$ can be trivialized by the atomic insulator $|\phi\rangle_{pgg}^{\mu_2}$ at each 0D block labeled by $\mu_2$. Thus the 1D bubble construction on $\tau_2$ can change the rotation eigenvalues of $\mu_1$ and $\mu_2$ simultaneously, which lead to the dependence of rotation eigenvalues of $\mu_1$ and $\mu_2$. 

Subsequently we consider the $U^f(1)$ charge sector: consider 1D bubble equivalence on 1D blocks $\tau_2$ [cf. Eq. (\ref{1D bubble})]: it adds two $U^f(1)$ charges at each 0D block $\mu_1$ and removes two $U^f(1)$ charges at each 0D block $\mu_2$, hence the numbers of $U^f(1)$ charges at $\mu_1$ and $\mu_2$ are not independent. 

With the help of above discussions, we consider the 0D block-state decorations. Start from the trivial state:
\begin{align}
[(0,+),(0,+)]
\end{align}
Taking aforementioned 1D bubble construction on $\tau_2$ with $n\in\mathbb{Z}$ times will obtain the group containing all trivial states:
\begin{align}
\{\mathrm{TBS}\}_{pgg,0}^{U(1)}&=\left\{\big[(2n,(-1)^n),(-2n,(-1)^n)\big]\Big|n\in\mathbb{Z}\right\}\nonumber\\
&=2\mathbb{Z}
\end{align}
Therefore, the ultimate classification of crystalline TSC protected by $pgg$ symmetry for 2D systems with spinless fermions is:
\begin{align}
\mathcal{G}_{pgg,0}^{U(1)}&=\{\mathrm{OFBS}\}_{pgg,0}^{U(1)}/\{\mathrm{TBS}\}_{pgg,0}^{U(1)}\nonumber\\
&=\mathbb{Z}^2\times\mathbb{Z}_2^2/2\mathbb{Z}=\mathbb{Z}\times\mathbb{Z}_4\times\mathbb{Z}_2
\end{align}

For systems with spin-1/2 fermions, 0D obstruction-free block-states are identical with spinless case:
\begin{align}
\{\mathrm{OFBS}\}_{pgg,1/2}^{U(1)}=\mathbb{Z}^2\times\mathbb{Z}_2^2
\end{align}
Then repeatedly consider the aforementioned 1D bubble constructions: the rotation properties of $|\phi\rangle_{pgg}^{\mu_1}$ and $|\phi\rangle_{pgg}^{\mu_2}$ are changed by an additional $-1$, which leads to no trivialization. It is easy to verify that the complex fermion decorations for spinless and spin-1/2 fermions are identical. Repeatedly consider the 1D bubble construction on $\tau_2$ and it will lead to the following group containing all trivial states:
\begin{align}
\{\mathrm{TBS}\}_{pgg,1/2}^{U(1)}=\left\{\big[(2n,+),(-2n,+)\big]\Big|n\in\mathbb{Z}\right\}=2\mathbb{Z}
\end{align}
Therefore, the ultimate classification of crystalline topological phases protected by $pgg$ symmetry for 2D systems with spin-1/2 fermions is:
\begin{align}
\mathcal{G}_{pgg,1/2}^{U(1)}&=\{\mathrm{OFBS}\}_{pgg,1/2}^{U(1)}/\{\mathrm{TBS}\}_{pgg,1/2}^{U(1)}\nonumber\\
&=\mathbb{Z}^2\times\mathbb{Z}_2^2/2\mathbb{Z}=\mathbb{Z}\times\mathbb{Z}_4\times\mathbb{Z}_2
\end{align}

\subsection{Hexagonal lattice: $p6m$}
Repeatedly consider the cell decomposition of $p6m$ as illustrated in Fig. \ref{p6m}. For 0D blocks labeled by $\mu_1$ and $\mu_2$, different 0D block-states are characterized by different irreducible representations of the symmetry group as $n=6,2$:
\begin{align}
\mathcal{H}^1[U^f(1)\times(\mathbb{Z}_n\rtimes\mathbb{Z}_2),U(1)]=\mathbb{Z}\times\mathbb{Z}_2^2
\end{align}
Here $\mathbb{Z}$ represents the $U^f(1)$ charge. 
For $\mu_1$, two $\mathbb{Z}_2$'s represent the reflection eigenvalues of $\bs{M}_{\tau_1}$ and $\bs{M}_{\tau_2}$, respectively; for $\mu_2$, two $\mathbb{Z}_2$'s represent the reflection eigenvalues of $\bs{M}_{\tau_1}$ and $\bs{M}_{\tau_3}$, respectively.

For 0D blocks labeled by $\mu_3$, different 0D block-states are characterized by different irreducible representations of the symmetry group as:
\begin{align}
\mathcal{H}^1[U^f(1)\times(\mathbb{Z}_3\rtimes\mathbb{Z}_2),U(1)]=\mathbb{Z}\times\mathbb{Z}_2
\end{align}
Here $\mathbb{Z}$ represents the $U^f(1)$ charge and $\mathbb{Z}_2$ represents the eigenvalue $-1$ of the reflection $\bs{M}_{\tau_2}$. 

Therefore, the 0D block-states on $\mu_1$ and $\mu_2$ can be labeled by $(n_1/n_2,\pm,\pm)$, where $n_1/n_2$ represents the $U^f(1)$ charges on $\mu_1/\mu_2$ and two $\pm$'s represent the reflection eigenvalues; the 0D block-states on $\mu_3$ can be labeled by $(n_3,\pm)$, where $n_3$ represents the $U^f(1)$ charges on $\mu_3$ and $\pm$ represents the eigenvalue of reflection operation. According to this notation, all obstruction-free 0D block-states form the following group:
\begin{align}
\{\mathrm{OFBS}\}_{p6m,0}^{U(1)}=\mathbb{Z}^3\times\mathbb{Z}_2^5
\end{align}

\begin{figure}
\centering
\includegraphics[width=0.44\textwidth]{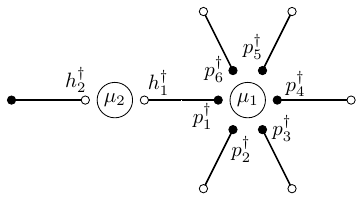}
\caption{1D particle-hole bubble construction on $\tau_1$, where particle/hole is labeled by solid/hollow dot. Atomic insulators (\ref{p6m mu1 atomic insulator}) and (\ref{p6m mu2 atomic insulator}) are created by this procedure.}
\label{p6m particle-hole 1D bubble}
\end{figure}

We should further consider possible trivializations: for systems with spinless fermions, consider the 1D bubble equivalence on 1D blocks labeled by $\tau_1$: we decorate a 1D ``particle-hole'' bubble [cf. Eq. (\ref{1D bubble}) and Fig. \ref{p6m particle-hole 1D bubble}] at each $\tau_1$. Near each 0D block labeled by $\mu_1$, there are six particles forming the following atomic insulator:
\begin{align}
|\xi\rangle_{p6m}^{\mu_1}=p_1^\dag p_2^\dag p_3^\dag p_4^\dag p_5^\dag p_6^\dag|0\rangle
\label{p6m mu1 atomic insulator}
\end{align}
There are two holes forming another atomic insulator:
\begin{align}
|\xi\rangle_{p6m}^{\mu_2}=h_1^\dag h_2^\dag|0\rangle
\label{p6m mu2 atomic insulator}
\end{align}
Similar with the crystalline TSC, eigenvalue of $\bs{M}_{\tau_2}$ can be changed by Eq. (\ref{p6m mu1 atomic insulator}) at each $\mu_1$, and eigenvalue of $\bs{M}_{\tau_3}$ can be changed by Eq. (\ref{p6m mu2 atomic insulator}) at each $\mu_2$. 

Then we consider the 1D bubble equivalence on 1D blocks labeled by $\tau_3$: we decorate an identical 1D ``particle-hole'' bubble as aforementioned on each $\tau_3$. Near each 0D block labeled by $\mu_2$, there are two particles forming the following atomic insulator:
\begin{align}
|\eta\rangle_{p6m}^{\mu_2}=p_1'^\dag p_2'^\dag|0\rangle
\end{align}
Near each 0D block $\mu_3$, there are three holes forming another atomic insulator:
\begin{align}
|\eta\rangle_{p6m}^{\mu_3}=h_1'^\dag h_2'^\dag h_3'^\dag|0\rangle
\end{align}
Similar with the crystalline TSC, eigenvalue of $\bs{M}_{\tau_1}$ can be changed by $|\eta\rangle_{p6m}^{\mu_2}$ at each $\mu_2$, and eigenvalue of $\bs{M}_{\tau_2}$ can be changed by $|\eta\rangle_{p6m}^{\mu_3}$ at each $\mu_3$. 

1D bubble construction on $\tau_2$ can be understood in a similar way, which changes the eigenvalue of $\bs{M}_{\tau_1}$ at $\mu_1$ and the eigenvalue of $\bs{M}_{\tau_2}$ at $\mu_3$. In addition, we also need to consider an alternative 1D bubble equivalence on 1D blocks $\tau_2$ (we label above 1D ``particle-hole'' bubble construction by ``type-\1'', and label this 1D ``bosonic'' bubble construction by ``type-\2''): we decorate an alternative 1D bubble on each 1D block labeled by $\tau_2$ [cf. Eq. (\ref{1D bubble}) and Fig. \ref{p6m 1D bubble}(b)], here both yellow and red dots represent the 0D bosonic modes carry eigenvalues $-1$ of reflection symmetry. According to this 1D bubble construction, the reflection eigenvalue at each 0D block $\mu_3$ is changed by $-1$ while the reflection eigenvalue at each 0D block $\mu_2$ remains invariant. Another type-\2 1D bubble construction can also be constructed on $\tau_3$.

Subsequently we consider the $U^f(1)$ charge sector. First of all, we should identify that within a specific unit cell, there is one 0D block labeled by $\mu_1$, two 0D blocks labeled by $\mu_3$ and three 0D blocks labeled by $\mu_2$. Repeatedly consider the aforementioned 1D bubble construction on $\tau_1$: it adds six $U^f(1)$ charges at each 0D block $\mu_1$ and removes two $U^f(1)$ charges at each 0D block $\mu_2$, hence the number of $U^f(1)$ charges at $\mu_1$ and $\mu_2$ are not independent. Similar argument can also be applied for $\tau_2$ and $\tau_3$.

With the help of above discussions, we consider the 0D block-state decorations. Start from the original trivial state:
\begin{align}
[(0,+,+),(0,+,+),(0,+)]
\label{p6m original trivial state}
\end{align}
Taking aforementioned type-\1 1D bubble constructions on $\tau_j$ with $l_j$ times ($j=1,2,3$), and type-\2 1D bubble constructions on $\tau_2/\tau_3$ with $l_2'/l_3'$ times, it will lead to a new 0D block-state labeled by:
\begin{align}
&\left[\left(6l_1+6l_2,(-1)^{l_2},(-1)^{l_1}\right),\right.\nonumber\\
&\left(-2l_1+2l_3,(-1)^{l_3},(-1)^{l_1}\right),\nonumber\\
&\left.\left(-3l_2-3l_3,(-1)^{l_2+l_3+l_2'+l_3'}\right)\right]
\label{p6m U(1) spinless trivial state}
\end{align}
According to the definition of bubble equivalence, all states with this form should be trivial. Alternatively, all 0D block-states can be viewed as vectors of an 8-dimensional vector space $V$, where the complex fermion components are $\mathbb{Z}$-valued and all other components are $\mathbb{Z}_2$-valued. Then all trivial 0D block-states with the form as Eq. (\ref{p6m U(1) spinless trivial state}) can be viewed as a vector subspace $V'$ of $V$. 
As a consequence, there are only four independent quantities in Eq. (\ref{p6m U(1) spinless trivial state}): $l_1$, $l_2$, $l_3$ and $l_2'+l_3'$. So the dimensionality of the vector subspace $V'$ should be 4. For the $U^f(1)$ charge sector, we have the following relationship:
\begin{align} 
-(6l_1+6l_2)-3(-2l_1+2l_3)=2(-3l_2-3l_3)
\end{align}
i.e., there are only two independent quantities that serve a $2\mathbb{Z}\times3\mathbb{Z}$ trivialization. The remaining two degrees of freedom of the vector subspace $V'$ should be attributed to the eigenvalues of point group symmetry actions labeled by ${(-1)}^{l_1}={(-1)}^{(-l_2)}={(-1)}^{l_3}$ and ${(-1)}^{l_2^\prime+l_3^\prime}$ which serve a $\mathbb{Z}_2^2$ trivialization. Therefore, all trivial states with form as shown in Eq. (\ref{p6m U(1) spinless trivial state}) compose the group:
\begin{align}
\{\mathrm{TBS}\}_{p6m,0}^{U(1)}=2\mathbb{Z}\times3\mathbb{Z}\times\mathbb{Z}_2^2
\end{align}
hence different independent nontrivial 0D block-states can be labeled by different group elements of the following quotient group:
\begin{align}
\mathcal{G}_{p6m,0}^{U(1)}&=\{\mathrm{OFBS}\}_{p6m,0}^{U(1)}/\{\mathrm{TBS}\}_{p6m,0}^{U(1)}\nonumber\\
&=\mathbb{Z}^3\times\mathbb{Z}_2^5/2\mathbb{Z}\times3\mathbb{Z}\times\mathbb{Z}_2^2=\mathbb{Z}\times\mathbb{Z}_{12}\times\mathbb{Z}_2^2
\end{align}

For systems with spin-1/2 fermions, like the cases without $U^f(1)$ charge conservation, the classification data of the corresponding 0D block-states on $\mu_1$, $\mu_2$ and $\mu_3$ can be characterized by different irreducible representations of the corresponding on-site symmetry group,
\begin{align}
\begin{aligned}
&\mathcal{H}^1\left[U^f(1)\times_{\omega_2}(\mathbb{Z}_6\rtimes\mathbb{Z}_2),U(1)\right]=2\mathbb{Z}\times\mathbb{Z}_2^2\\
&\mathcal{H}^1\left[U^f(1)\times_{\omega_2}(\mathbb{Z}_2\rtimes\mathbb{Z}_2),U(1)\right]=2\mathbb{Z}\times\mathbb{Z}_2^2
\end{aligned}
\end{align}
Here each $2\mathbb{Z}$ represents the $U^f(1)$ charge carried by complex fermion, and different $\mathbb{Z}_2$'s represent the rotation and reflection eigenvalues at each 0D block labeled by $\mu_1$ and $\mu_2$ (similar with the $p4m$ case, we can only decorate even number of $U^f(1)$ charge on each 0D block). 
and all obstruction-free 0D block-states form the following group:
\begin{align}
\{\mathrm{OFBS}\}_{p6m,1/2}^{U(1)}=\mathbb{Z}\times(2\mathbb{Z})^2\times\mathbb{Z}_2^5
\end{align}
then repeatedly consider the aforementioned 1D bubble constructions, the reflection properties of the atomic insulators: $|\xi\rangle_{p6m}^{\mu_1}$, $|\xi\rangle_{p6m}^{\mu_2}$, $|\eta\rangle_{p6m}^{\mu_1}$, $|\eta\rangle_{p6m}^{\mu_2}$ and $|\eta\rangle_{p6m}^{\mu_3}$ are changed by an additional $-1$, and all of them lead to no trivialization. Other 1D bubble constructions are identical. So again we start from the original trivial state (\ref{p6m original trivial state}), take above type-\1 1D bubble constructions on $\tau_j$ with $l_j$ times ($j=1,2,3$), and type-\2 1D bubble constructions on $\tau_2/\tau_3$ with $l_2'/l_3'$ times, it will lead to a new 0D block-state labeled by:
\begin{align}
&\left[\left(6l_1+6l_2,+,+\right),\right.\nonumber\\
&\left(-2l_1+2l_3,+,+\right),\nonumber\\
&\left.\left(-3l_2-3l_3,(-1)^{l_2'+l_3'}\right)\right]
\label{p6m U(1) spin-1/2 trivial state}
\end{align}
The $U^f(1)$ charge sector is identical with spinless case, and there is one independent nonzero reflection eigenvalue $(-1)^{l_2'+l_3'}$. Therefore, all trivial states with form as shown in Eq. (\ref{p6m U(1) spin-1/2 trivial state}) compose the following group:
\begin{align}
\{\mathrm{TBS}\}_{p6m,1/2}^{U(1)}=2\mathbb{Z}\times3\mathbb{Z}\times\mathbb{Z}_2
\end{align}
and different independent nontrivial 0D block-states can be labeled by different group elements of the following group:
\begin{align}
\mathcal{G}_{p6m,1/2}^{U(1)}&=\{\mathrm{OFBS}\}_{p6m,1/2}^{U(1)}/\{\mathrm{TBS}\}_{p6m,1/2}^{U(1)}\nonumber\\
&=\mathbb{Z}\times(2\mathbb{Z})^2\times\mathbb{Z}_2^5/(2\mathbb{Z}\times3\mathbb{Z}\times\mathbb{Z}_2)\nonumber\\
&=2\mathbb{Z}\times\mathbb{Z}_6\times\mathbb{Z}_2^3
\end{align}

\section{Generalized crystalline equivalence principle\label{principle}}
In this section, we discuss how to generalize the crystalline equivalence principle that is rigorously proven for interacting bosonic systems \cite{correspondence}.
By comparing the classification results of the topological crystalline TSC summarized in Table \ref{spinless}, Table \ref{spin-1/2} and the classification results of crystalline TI summarized in Table \ref{insulator U(1)} with the classification results of the 2D FSPT phases protected by the corresponding on-site symmetry\cite{resolution, QingruiTI}, 
we verify the fermionic crystalline equivalence principle for all TSC and TI(for both spinless and spin-1/2 cases)constructed in this paper,.  

In particular, we should map the space group symmetry to on-site symmetry according to the following rules:
\begin{enumerate}
\item Subgroup of translational symmetry along a particular direction should be mapped to the on-site symmetry group $\mathbb{Z}$. Equivalently, the total translational subgroup should be mapped to the on-site symmetry group $\mathbb{Z}^2$;
\item $n$-fold rotational symmetry subgroup should be mapped to the on-site symmetry group $\mathbb{Z}_n$;
\item Reflection symmetry subgroup should be mapped to the time-reversal symmetry group $\mathbb{Z}_2^T$ which is antiunitary.
\item Spinless (spin-1/2) fermionic systems should be mapped
into spin-1/2 (spinless) fermionic systems.
\end{enumerate}
The additional twist on spinless and spin-$1/2$ fermions can be naturally interpreted as the spin rotation of fermions: a $2\pi$ rotation of a fermion around a specific axis results in a $-1$ phase factor \cite{supplementary}.

Apparently, bosonic/fermionic crystalline SPT phases will be mapped to the corresponding on-site symmetry bosonic/fermionic SPT phases.
In fact, there is even a more precise one to one mapping between crystalline SPT phases and the corresponding on-site symmetry SPT phases. It is well known that the 2D fermionic SPT states protected by on-site symmetry have a layered structure: they can be constructed by decorating (subject to certain obstructions)  
1D Majorana chains to 1D symmetry domain walls, 0D complex fermion modes to intersection points of domain walls, in addition to the bosonic SPT layer. We find that all crystalline SPT states constructed via 1D block state with Majorana chain decorations will be mapped to on-site symmetry SPT states with Majorana chain decorations to 1D symmetry domain wall, while all crystalline SPT states constructed via 1D block state with FSPT decorations or 0D complex fermion decoration will be mapped to on-site symmetry SPT states with 0D complex fermion modes decoration to intersection points of domain walls. For crystalline topological insulator with an additional $U^f(1)$ charge conservation symmetry, there would be no Majorana chain decoration for the corresponding on-site symmetry SPT phases, that's why there is also no 1D block state in our crystalline SPT states construction.


\section{Conclusion and discussion\label{conclusion}}
In this paper, we derive the classification of crystalline TSC and TI in 2D interacting femionic systems by using the explicit real-space constructions. For a 2D system with a specific wallpaper group symmetry, we first decompose the system into an assembly of unit cells. Then according to the so-called \textit{extensive trivialization} scheme, we can further decompose each unit cell into an assembly of lower-dimensional blocks. After cell decompositions, we can decorate some lower-dimensional block-states on them, and investigate the \textit{obstruction} and \textit{trivialization} for all block states by checking the no-open-edge condition and bubble equivalence. An obstruction/trivialization free decoration corresponds to a nontrivial crystalline SPT phase. We further investigated the group structures of the classification data by considering the possible stacking between 1D and 0D block-states. Finally, with the complete classification results, we compare our results with classification of 2D FSPT phases protected by the corresponding on-site symmetry, we verify the crystalline equivalence principle for generic 2D interacting fermionic systems. 

We believe that the real-space construction scheme for crystalline SPT is also applicable to 3D interacting fermionic systems, with similar procedures discussed in this work. We conjecture that the crystalline equivalence principle is also correct for 3D crystalline FSPT phases as well.
In future works, we will try to construct and fully classify the crystalline TSC/TI in 3D interacting fermionic systems. 


We stress that the method in this paper can also be applied to cases with mixture of internal and space group symmetries, i.e. when considering about the lower-dimensional block-states, we should also include the internal symmetry together with the space group symmetry acting internally that leads to different lower-dimensional root phases and bubbles. Then based on these root phases, we can further discuss possible obstructions and trivializations by using the general paradigms highlighted in Sec. \ref{general}.

Moreover, we also predict an intriguing fermionic crystalline TSC (that cannot be realized in both free-fermion and interacting bosonic systems) with $p4m$ wall paper group symmetry. 
The iron-based superconductor could be a natural strongly correlation electron system to realize such a new phase, especially the monolayer iron selenide/pnictide \cite{monolayer}. Since the spin-orbit interaction in FeSe is relatively small [distinct from Fe(Se,Te) because of the absence of tellurium], we can effectively treat fermions in this system as spinless.

\begin{acknowledgements}
We thank Qingrui Wang, Zheng-Xin Liu and Meng Cheng for enlightening discussions. This work is supported by Direct Grant No. 4053409 from The Chinese University of Hong Kong and funding from Hong Kong's Research Grants Council (GRF No.14306918, ANR/RGC Joint Research Scheme No. A-CUHK402/18). SY is supported by NSFC (Grant No. 11804181) and the National Key R\&D Program of China (Grant No. 2018YFA0306504).
\end{acknowledgements}

\providecommand{\noopsort}[1]{}\providecommand{\singleletter}[1]{#1}%
%


\begin{thebibliography}{91}
\bibitem [{mon()}]{monolayer}%
  \BibitemOpen
\bibfield  {journal} {  }\href@noop {} {\bibinfo  {journal} {The space group of
  3D iron selenide/pnictide is $P4/nmm$, and according to recent STM
  experiments, there is no nematicity for monolayer iron selenide/pnictide
  grown on STO substrate by MBE, distinguished from the 3D crystals. So the
  corresponding wallpaper group of monolayer iron selenide/pnictide is $p4m$}\
  }\BibitemShut {NoStop}%
\end{thebibliography}

\pagebreak

\clearpage

\appendix
\setcounter{equation}{0}
\newpage

\renewcommand{\thesection}{S-\arabic{section}} \renewcommand{\theequation}{S%
\arabic{equation}} \setcounter{equation}{0} \renewcommand{\thefigure}{S%
\arabic{figure}} \setcounter{figure}{0}

\centerline{\textbf{Supplemental Materials}}

\maketitle

\section{Classifications of FSPT states in 0D}
In the main text, we have characterized the 0D block-states by different 1D irreducible representations of the full symmetry group, for both crystalline TSC and TI. In Ref. \cite{Sgeneral2}, the crystalline TSC cases are systematically studied, hence we mainly focus on the crystalline TI in this section. 

It is well-known that the 0D BSPT states with symmetry group $G$ are classified by 1D linear representations of $G$: $\mathcal{H}^1(G,U_T(1))$ \cite{Scohomology}. This is because the SPT state should be both symmetric and non-degenerate. In 0D, there are essentially no difference between bosonic and fermionic systems, except that there is an additional $U^f(1)$ for fermionic systems with charge conservation. We can treat a fermionic system with total fermionic symmetry group $G_f=U^f(1)\rtimes_{\omega_2}G$ as a bosonic system with total symmetry group $G_f$. Therefore, we can conclude:
\begin{itemize}
\item 0D FSPT phases with symmetry group $U^f(1)\rtimes_{\omega_2}G$ are classified by 1D irreducible representations of $G_f$. i.e., $\mathcal{H}^1(G_f,U_T(1))$.
\end{itemize}
Equivalently, we can unpack the above result and show that:
\begin{itemize}
\item 0D FSPT phases with symmetry group $U^f(1)\rtimes_{\omega_2}G$ are classified by 0-cocycle $n_0$ and 1-cocycle $\nu_1$, with some symmetry conditions and consistency equations.
\end{itemize}
Here $n_0\in\mathcal{H}^0(G,\mathbb{Z})$ is the number of $U^f(1)$ charge carried by complex fermions, and $\nu_1\in\mathcal{H}^1(G,U_T(1))$ is the usual 0D BSPT classification.

In general, for a given 1D representation $\tilde{U}$ of $G_f$, we can always separate $\tilde{U}\left(e^{i\theta n_0},g\right)$ [with $g\in G$ and $e^{i\theta n_0}\in U^f(1)$, $\theta\in[0,2\pi)$] into three parts:
\begin{align}
\tilde{U}\left(e^{i\theta n_0},g\right)=e^{i\theta n_0}\nu_1(g)K^{s_1(g)}
\end{align}
where $\nu_1(g)$ is a $U(1)$ phase factor, $K$ is the complex conjugation operator and $s_1(g)$ represents whether $g$ contains time-reversal or not [if so, $s_1(g)=1$, otherwise $s_1(g)=0$]. Using the multiplication rule of $G_f$ (defined in the main text), the representation condition $\tilde{U}\left(e^{i\theta n_0},g\right)\tilde{U}\left(e^{i\theta' n_0},h\right)=\tilde{U}\left[\left(e^{i\theta n_0},g\right)\cdot\left(e^{i\theta' n_0},h\right)\right]$ becomes:
\begin{align}
\nu_1(g)\nu_1(h)^{1-2s_1(g)}=e^{2\pi i \omega_2(g,h)n_0}\nu_1(gh)
\end{align}
and can be summarized as:
\begin{align}
(d\nu_1)(g,h):=\frac{\nu_1(h)^{1-2s_1(g)}\nu_1(g)}{\nu_1(gh)}=e^{2\pi i(\omega_2\smile n_0)(g,h)}
\end{align}
which means that the cocycle equation of $\nu_1$ is twisted by $\omega_2\smile n_0$. 

\section{Point group protected TI in 2D interacting fermion systems\label{App2}}
In Ref. \onlinecite{Srotation,Sdihedral}, the authors systematically constructed and classified the point group SPT phases without $U^f(1)$ charge conservation in 2D interacting fermionic systems. In this section, we construct and classify the 2D point group FSPT phases with $U^f(1)$ charge conservation. As demonstrated in the main text, we do not need to investigate the 1D block-state decoration because of the absence of 1D root phase. 

Here we explicit construct and classify the point group SPT phases with $U^f(1)$ charge conservation in 2D interacting fermionic systems by real-space construction. We demonstrate the $D_4$ case as an example. Inherit the definitions and terminologies of Ref. \onlinecite{Sdihedral}, we decorate some block-states on the corresponding lower-dimensional blocks. We only need to consider the 0D block-state decoration: different 0D block-states are characterized by different irreducible representations of the symmetry group ($\mathbb{Z}_4\rtimes\mathbb{Z}_2$, by $D_4$ acting internally on the 0D block):
\begin{align}
\mathcal{H}^1[U(1)\times(\mathbb{Z}_4\rtimes\mathbb{Z}_2),U(1)]=\mathbb{Z}\times\mathbb{Z}_2^2
\end{align}

\begin{figure}[t]
\centering
\includegraphics[width=0.48\textwidth]{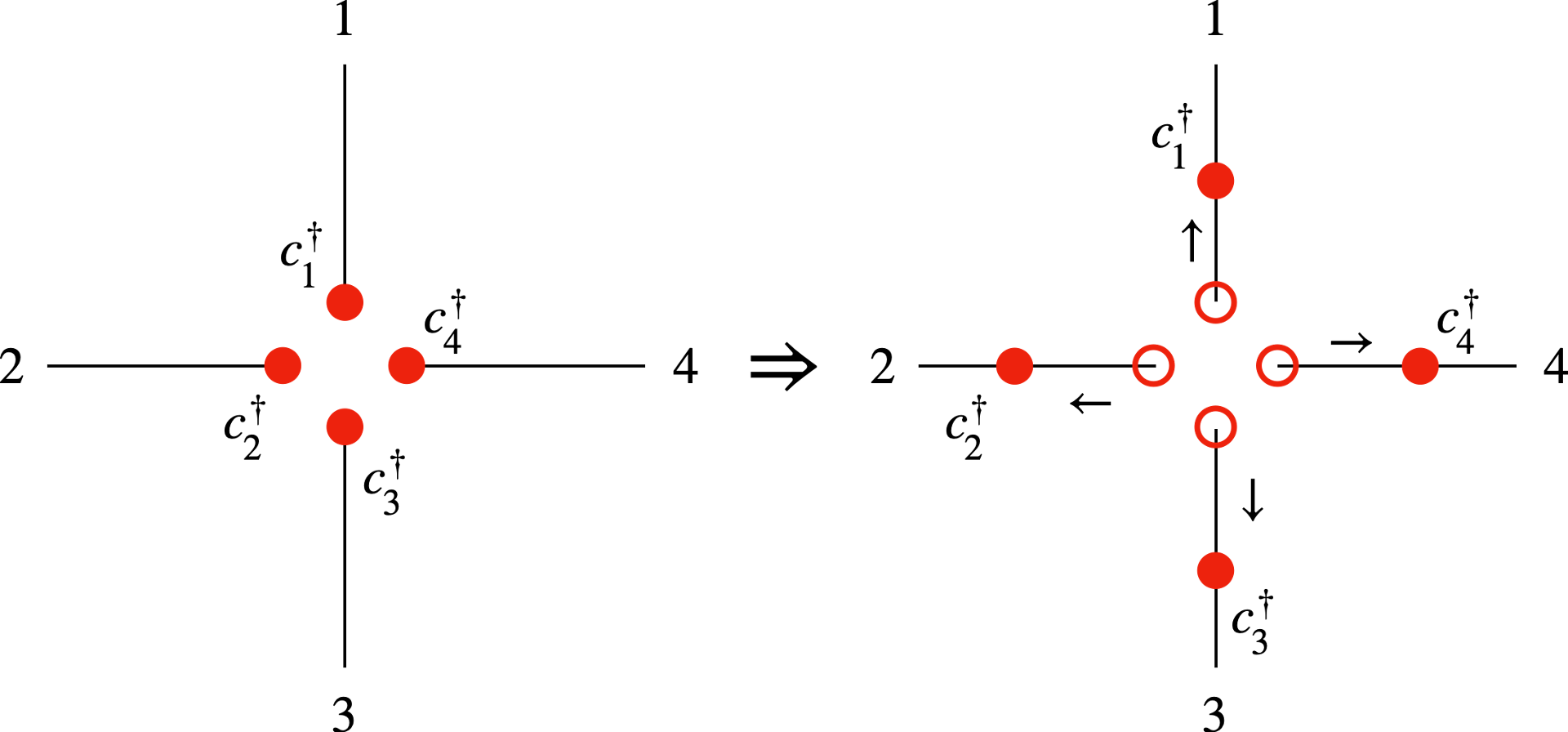}
\caption{Trivialization of central complex fermions for 2D crystalline topological phases protected by $D_4$ symmetry, with $U^f(1)$ charge conservation.}
\label{insulator trivialization}
\end{figure}

Here $\mathbb{Z}$ represents the complex fermion, the first $\mathbb{Z}_2$ represents the rotation eigenvalue $-1$ and another $\mathbb{Z}_2$ represents the reflection eigenvalue $-1$. We demonstrate that the 0D block-state with four complex fermions can be trivialized. We label these four complex fermions as $\left(c_1^\dag,c_2^\dag,c_3^\dag,c_4^\dag\right)$. As illustrated in Fig. \ref{insulator trivialization}, we can manipulate these complex fermions in a symmetric way and trivialize the corresponding 0D block-state: move the complex fermion labeled by $c_j$ outward the origin along the 1D block labeled by $j$ ($j=1,2,3,4$) toward infinite far away from the origin, and the system is trivialized to a vacuum under this symmetric manipulation. As a consequence, the classification attributed to the complex fermion decoration on 0D block is trivialized from $\mathbb{Z}$ to $\mathbb{Z}_4$. Then we consider the trivializations beyond this. For systems with spinless fermions, we repeatedly consider Fig. \ref{insulator trivialization}, we decorate 4 complex fermions at the rotation center (which can be trivialized as aforementioned), they can form an atomic insulator $|\psi\rangle_{D_4}^{U(1)}=c_1^\dag c_2^\dag c_3^\dag c_4^\dag|0\rangle$, with the following rotation and reflection symmetry properties:
\begin{align}
\renewcommand\arraystretch{1.2}
\begin{aligned}
&\bs{R}|\psi\rangle_{D_4}^{U(1)}=c_2^\dag c_3^\dag c_4^\dag c_1^\dag|0\rangle=-|\psi\rangle_{D_4}^{U(1)}\\
&\bs{M}|\psi\rangle_{D_4}^{U(1)}=c_1^\dag c_4^\dag c_3^\dag c_2^\dag|0\rangle=-|\psi\rangle_{D_4}^{U(1)}
\label{D4 insulator}
\end{aligned}
\end{align}
i.e., eigenvalue $-1$ of rotation and reflection symmetry can be trivialized by $|\psi\rangle_{D_4}^{U(1)}$, and the classification for systems with spinless fermions is:
\begin{align}
\mathcal{G}_{D_4,0}^{U(1)}=\mathbb{Z}_8
\end{align}
For systems with spin-1/2 fermions, The minus signs in Eq. (\ref{D4 insulator}) disappear, hence there is no trivialization beyond Fig. \ref{insulator trivialization}. But there is an obstruction: a specific $n_0$ is obstructed if and only if $(-1)^{\omega_2\smile n_0}\in\mathcal{H}^2[\mathbb{Z}_2\rtimes\mathbb{Z}_2,U(1)]$ is a nontrivial 2-cocycle with $U(1)$-coefficient. From Refs. \cite{Sgeneral2} and \cite{Sdihedral} we know that for cases without $U^f(1)$ charge conservation, nontrivial 0-cocycle $n_0=1$, $n_0\in\mathcal{H}^0(\mathbb{Z}_2\rtimes\mathbb{Z}_2,\mathbb{Z}_2)$ leads to nontrivial 2-cocycle $(-1)^{\omega_2\smile n_0}\in\mathcal{H}^2[\mathbb{Z}_2\rtimes\mathbb{Z}_2,U(1)]$. So for $U^f(1)$ charge conserved cases, all odd $n_0\in\mathcal{H}^0[\mathbb{Z}_2\rtimes\mathbb{Z}_2,\mathbb{Z}]$ lead to nontrivial 2-cocycle $(-1)^{\omega_2\smile n_0}\in\mathcal{H}^2[\mathbb{Z}_2\rtimes\mathbb{Z}_2,U(1)]$. As a consequence, for systems with spin-1/2 fermions, we can only decorate even $U^f(1)$ charges on the center of $D_4$ symmetry. Therefore, the classification for systems with spin-1/2 fermions is:
\begin{align}
\mathcal{G}_{D_4,1/2}^{U(1)}=\mathbb{Z}_2\times\mathbb{Z}_2^2
\end{align}

Similar discussions can be held for other point group FSPT phases, and we summarize all results of classification in Table \ref{point}.

\begin{table}[t]
\renewcommand\arraystretch{1.5}
\begin{tabular}{|c|c|c|c|c|}
\hline
\diagbox{$~G$}{spin}&$~\text{spinless}~$ & spin-1/2\\
\hline
$C_{2m-1}$&$\mathbb{Z}_{2m-1}^2$&$\mathbb{Z}_{2m-1}^2$\\
\hline
$C_{2m}$&$\mathbb{Z}_{2m}^2$&$\mathbb{Z}_{2m}^2$\\
\hline
$D_{2m-1}$&$\mathbb{Z}_{2m-1}$&$\mathbb{Z}_{2m-1}$\\
\hline
$D_{2m}$&$\mathbb{Z}_{4m}$ & $\left\{
\begin{aligned}
&\mathbb{Z}_{m}\times\mathbb{Z}_2^2,~m\in\mathrm{even}\\
&\mathbb{Z}_{2m}\times\mathbb{Z}_2,~m\in\mathrm{odd}
\end{aligned}
\right.$\\
\hline
\end{tabular}
\caption{The classification of interacting 2D FSPT phases with point group symmetry and $U^f(1)$ charge conservation, $m=1,2,3$.}
\label{point}
\end{table}
Finally, we compare our results with the classification of SPT phases protected by corresponding internal group and find a perfect agreement.

 
\section{Other cases of crystalline topological phases with wallpaper group symmetry\label{App3}}
In the main text, we explicitly constructed and classified five examples of 2D interacting fermionic crystalline topological phases protected by wallpaper group which belongs to different crystal systems. In this section we summarize all other cases, and the results of classification are summarized in Table \1 and \2 in the main text. Furthermore, we also discuss about the systems with $U^f(1)$ charge conservation whose general principles are highlighted in the main text, and all results of classification are summarized in Table \3 in the main text.

\begin{figure}
\centering
\includegraphics[width=0.34\textwidth]{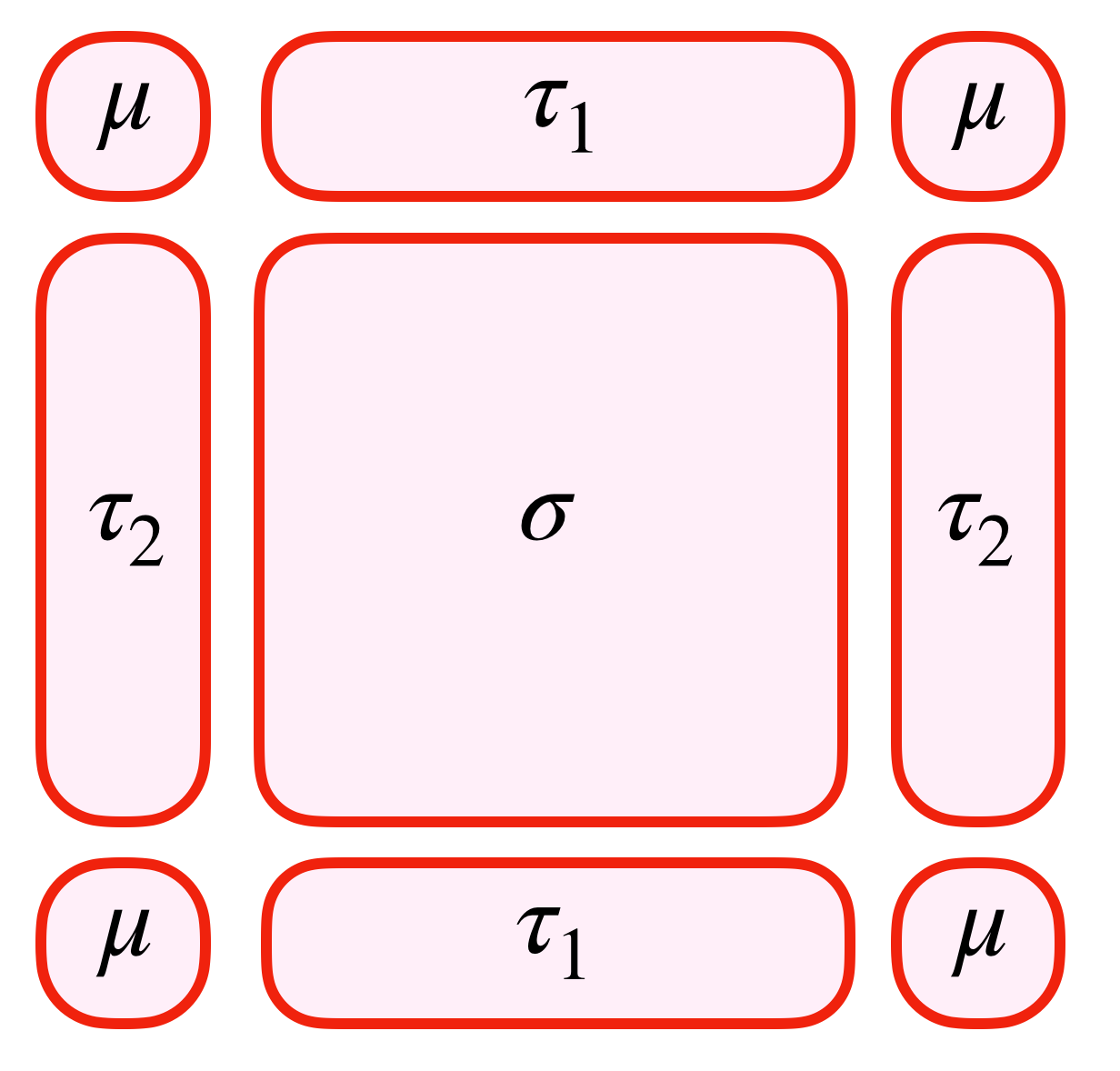}
\caption{\#1 wallpaper group $p1$ and its cell decomposition.}
\label{p1}
\end{figure}

\subsection{$p1$}
There is no more on-site symmetry on each block for arbitrary dimensions (cf. Fig. \ref{p1}, and we use the same label for $p$-dimensional blocks who can be related by symmetry actions), so the only significant decoration is Majorana chain decorations on 1D blocks and complex fermion decoration on 0D blocks ($G_{\mathrm{0D}}=\mathbb{Z}_2$). Spin of fermions is irrelevant for this case because there is no symmetry operation which rotates fermions in the system. Consider the Majorana chain decorations on $\tau_1$ which leave two dangling Majorana modes at each 0D block. Add the term $i\gamma_1\gamma_2$ at each 0D block in order to glue these dangling Majorana modes toward an entanglement pair and the no-open-edge condition \cite{Srealspace} is satisfied as a consequence. Similar for 1D blocks labeled by $\tau_2$. So the eventual classification is $G_{\mathrm{1D}}=\mathbb{Z}_2^3$, which is attributed to complex fermion decoration:
\begin{align}
E_{p1}^{\mathrm{0D}}=\mathbb{Z}_2
\end{align}
and Majorana chain decoration:
\begin{align}
E_{p1}^{\mathrm{1D}}=\mathbb{Z}_2^2
\end{align}
We demonstrate that there is no nontrivial extension between the Majorana chain decorations (1D) and complex fermion decorations (0D). Consider two Majorana chains decorated on $\tau_1$ which is essentially a trivial 1D block-state, firstly we consider two Majorana pairs [cf. Eq. (\ref{p1 deformation})] with the following Hamiltonian,
\begin{align}
H=i\left(\gamma_j^2\gamma_{j+1}^1+\gamma_j'^2\gamma_{j+1}'^1\right)
\end{align}

\begin{widetext}
\begin{align}
\begin{tikzpicture}
\tikzstyle{xjp}=[rectangle,draw=none]
\filldraw[fill=yellow!80!white, draw=yellow] (-2.5,1.5) ellipse (20pt and 5pt);
\filldraw[fill=yellow!80!white, draw=yellow] (-0.5,1.5) ellipse (20pt and 5pt);
\draw[->,thick] (-2,1.5) -- (-1.45,1.5);
\draw[thick] (-2,1.5) -- (-1,1.5);
\filldraw[fill=black, draw=black] (-3,1.5)circle (2pt);
\filldraw[fill=black, draw=black] (0,1.5)circle (2pt);
\filldraw[fill=black, draw=black] (-1,1.5)circle (2pt);
\filldraw[fill=black, draw=black] (-2,1.5)circle (2pt);
\filldraw[fill=yellow!80!white, draw=yellow] (-2.5,1) ellipse (20pt and 5pt);
\filldraw[fill=yellow!80!white, draw=yellow] (-0.5,1) ellipse (20pt and 5pt);
\draw[->,thick] (-2,1) -- (-1.45,1);
\draw[thick] (-2,1) -- (-1,1);
\filldraw[fill=black, draw=black] (-3,1)circle (2pt);
\filldraw[fill=black, draw=black] (0,1)circle (2pt);
\filldraw[fill=black, draw=black] (-1,1)circle (2pt);
\filldraw[fill=black, draw=black] (-2,1)circle (2pt);
\draw[thick] (-3.5,1.5) -- (-3,1.5);
\draw[thick] (-3.5,1) -- (-3,1);
\draw[thick] (0,1.5) -- (0.5,1.5);
\draw[thick] (0.5,1) -- (0,1);
\filldraw[fill=yellow!80!white, draw=yellow] (-2.5+4.75,1) ellipse (20pt and 5pt);
\filldraw[fill=yellow!80!white, draw=yellow] (-0.5+4.75,1) ellipse (20pt and 5pt);
\filldraw[fill=yellow!80!white, draw=yellow] (-2.5+4.75,1.5) ellipse (20pt and 5pt);
\filldraw[fill=yellow!80!white, draw=yellow] (-0.5+4.75,1.5) ellipse (20pt and 5pt);
\draw[->,thick] (-3+4.75,1.5) -- (-3+4.75,1.2);
\draw[thick] (-3+4.75,1) -- (-3+4.75,1.5);
\draw[->,thick] (-2+4.75,1) -- (-2+4.75,1.3);
\draw[thick] (-2+4.75,1) -- (-2+4.75,1.5);
\draw[->,thick] (-1+4.75,1.5) -- (-1+4.75,1.2);
\draw[thick] (-1+4.75,1) -- (-1+4.75,1.5);
\draw[->,thick] (-0+4.75,1) -- (-0+4.75,1.3);
\draw[thick] (-0+4.75,1) -- (-0+4.75,1.5);
\filldraw[fill=black, draw=black] (-3+4.75,1)circle (2pt);
\filldraw[fill=black, draw=black] (0+4.75,1)circle (2pt);
\filldraw[fill=black, draw=black] (-1+4.75,1)circle (2pt);
\filldraw[fill=black, draw=black] (-2+4.75,1)circle (2pt);
\filldraw[fill=black, draw=black] (-3+4.75,1.5)circle (2pt);
\filldraw[fill=black, draw=black] (0+4.75,1.5)circle (2pt);
\filldraw[fill=black, draw=black] (-1+4.75,1.5)circle (2pt);
\filldraw[fill=black, draw=black] (-2+4.75,1.5)circle (2pt);
\path (1,1.25) node [style=xjp]{$\Longleftrightarrow$};
\path (5.5,1.25) node [style=xjp]{$\Longleftrightarrow$};
\filldraw[fill=yellow!80!white, draw=yellow] (-2.5+9.25,1) ellipse (20pt and 5pt);
\filldraw[fill=yellow!80!white, draw=yellow] (-0.5+9.25,1) ellipse (20pt and 5pt);
\filldraw[fill=yellow!80!white, draw=yellow] (-2.5+9.25,1.5) ellipse (20pt and 5pt);
\filldraw[fill=yellow!80!white, draw=yellow] (-0.5+9.25,1.5) ellipse (20pt and 5pt);
\filldraw[fill=black, draw=black] (-3+9.25,1)circle (2pt);
\filldraw[fill=black, draw=black] (0+9.25,1)circle (2pt);
\filldraw[fill=black, draw=black] (-1+9.25,1)circle (2pt);
\filldraw[fill=black, draw=black] (-2+9.25,1)circle (2pt);
\filldraw[fill=black, draw=black] (-3+9.25,1.5)circle (2pt);
\filldraw[fill=black, draw=black] (0+9.25,1.5)circle (2pt);
\filldraw[fill=black, draw=black] (-1+9.25,1.5)circle (2pt);
\filldraw[fill=black, draw=black] (-2+9.25,1.5)circle (2pt);
\draw[->,thick] (-3+9.25,1.5) -- (-2.45+9.25,1.5);
\draw[thick] (-3+9.25,1.5) -- (-2+9.25,1.5);
\draw[->,thick] (-1+9.25,1.5) -- (-0.45+9.25,1.5);
\draw[thick] (-1+9.25,1.5) -- (-0+9.25,1.5);
\draw[->,thick] (-3+9.25,1) -- (-2.45+9.25,1);
\draw[thick] (-3+9.25,1) -- (-2+9.25,1);
\draw[->,thick] (-1+9.25,1) -- (-0.45+9.25,1);
\draw[thick] (-1+9.25,1) -- (-0+9.25,1);
\path (-3,1.9) node [style=xjp]{$\gamma_j^1$};
\path (-3,0.6) node [style=xjp]{$\gamma_j'^1$};
\path (-2,1.9) node [style=xjp]{$\gamma_j^2$};
\path (-2,0.6) node [style=xjp]{$\gamma_j'^2$};
\path (-1,1.9) node [style=xjp]{$\gamma_{j+1}^1$};
\path (-1,0.6) node [style=xjp]{$\gamma_{j+1}'^1$};
\path (0,1.9) node [style=xjp]{$\gamma_{j+1}^2$};
\path (0,0.6) node [style=xjp]{$\gamma_{j+1}'^2$};
\path (-3+4.75,1.9) node [style=xjp]{$\gamma_j^1$};
\path (-3+4.75,0.6) node [style=xjp]{$\gamma_j'^1$};
\path (-2+4.75,1.9) node [style=xjp]{$\gamma_j^2$};
\path (-2+4.75,0.6) node [style=xjp]{$\gamma_j'^2$};
\path (-1+4.75,1.9) node [style=xjp]{$\gamma_{j+1}^1$};
\path (-1+4.75,0.6) node [style=xjp]{$\gamma_{j+1}'^1$};
\path (0+4.75,1.9) node [style=xjp]{$\gamma_{j+1}^2$};
\path (0+4.75,0.6) node [style=xjp]{$\gamma_{j+1}'^2$};
\path (-3+9.25,1.9) node [style=xjp]{$\gamma_j^1$};
\path (-3+9.25,0.6) node [style=xjp]{$\gamma_j'^1$};
\path (-2+9.25,1.9) node [style=xjp]{$\gamma_j^2$};
\path (-2+9.25,0.6) node [style=xjp]{$\gamma_j'^2$};
\path (-1+9.25,1.9) node [style=xjp]{$\gamma_{j+1}^1$};
\path (-1+9.25,0.6) node [style=xjp]{$\gamma_{j+1}'^1$};
\path (0+9.25,1.9) node [style=xjp]{$\gamma_{j+1}^2$};
\path (0+9.25,0.6) node [style=xjp]{$\gamma_{j+1}'^2$};
\end{tikzpicture}
\label{p1 deformation}
\end{align}
\end{widetext}

Consider the interpolation of Hamiltonians as following
\begin{align}
H(\theta)=&\cos\theta(i\gamma_j^2\gamma_{j+1}^1+i\gamma_j^2\gamma_{j+1}^1)\nonumber\\
&+\sin\theta(i\gamma_j^2\gamma_j'^2-i\gamma_{j+1}^1\gamma_{j+1}'^1),
\end{align}
when $\theta=0$, the ground state is the one on the left of Eq. (\ref{p1 deformation}); when $\theta=\pi/2$, the ground state is the one in the middle of Eq. (\ref{p1 deformation}). The energy eigenvalue of $H(\theta)$ is actually independent from $\theta$, as
\begin{align}
E=\pm2\sqrt{1-s},
\label{deformation spectrum}
\end{align}
where $s=\pm1$ is the eigenvalue of $\gamma_j^2\gamma_{j+1}^1\gamma_j'^2\gamma_{j+1}'^1$. By this deformation, the inter-site entanglements are fully disentangled, and two Majorana chains are deformed to an atomic insulator.

Similarly, we further consider another intra-site deformation with the following Hamiltonian
\begin{align}
H(\phi)=&\cos\phi(i\gamma_j^2\gamma_j'^2-i\gamma_{j}^1\gamma_{j}'^1)\nonumber\\
&+\sin\phi(i\gamma_j^1\gamma_j^2+i\gamma_j'^1\gamma_j'^2),
\end{align}
when $\phi=0$, the ground state is the one in the middle of Eq. (\ref{p1 deformation}); when $\phi=\pi/2$, the ground state is the one on the right of Eq. (\ref{p1 deformation}). We finally arrive at an atomic insulator with even fermion parity on each site, hence there is no extension between 1D and 0D block-states. Equivalently, the 1D block-state decorations do not extend the 0D block-state decorations, and the ultimate classification with accurate group structure is:
\begin{align}
\mathcal{G}_{p1}^0=\mathcal{G}_{p1}^{1/2}=E_{p1}^{\mathrm{0D}}\times E_{p1}^{\mathrm{1D}}=\mathbb{Z}_2^3
\end{align}
here the symbol ``$\times_{\omega_2}$'' means that 1D and 0D block-states $E_{p1}^{\mathrm{1D}}$ and $E_{p1}^{\mathrm{0D}}$ have no nontrivial extension.
Then we consider the systems with $U^f(1)$ charge conservation. We note that in the main text, we have demonstrated that 1D block-state decoration does not contribute any nontrivial crystalline topological phase because of the absence of nontrivial 1D root phase, and the construction and classification are equivalent for systems with spinless fermions and spin-1/2 fermions. For an arbitrary 0D block, different 0D block-states are characterized by different irreducible representations of the symmetry group as:
\begin{align}
\mathcal{H}^1[U(1),U(1)]=\mathbb{Z}
\end{align}
here $\mathbb{Z}$ represents the complex fermion, and there is no trivialization. Hence the classification for the systems with $U^f(1)$ charge conservation is:
\begin{align}
\mathcal{G}_{p1}^{U(1)}=\mathbb{Z}
\end{align}

\subsection{$pm$}
\begin{figure}
\centering
\includegraphics[width=0.45\textwidth]{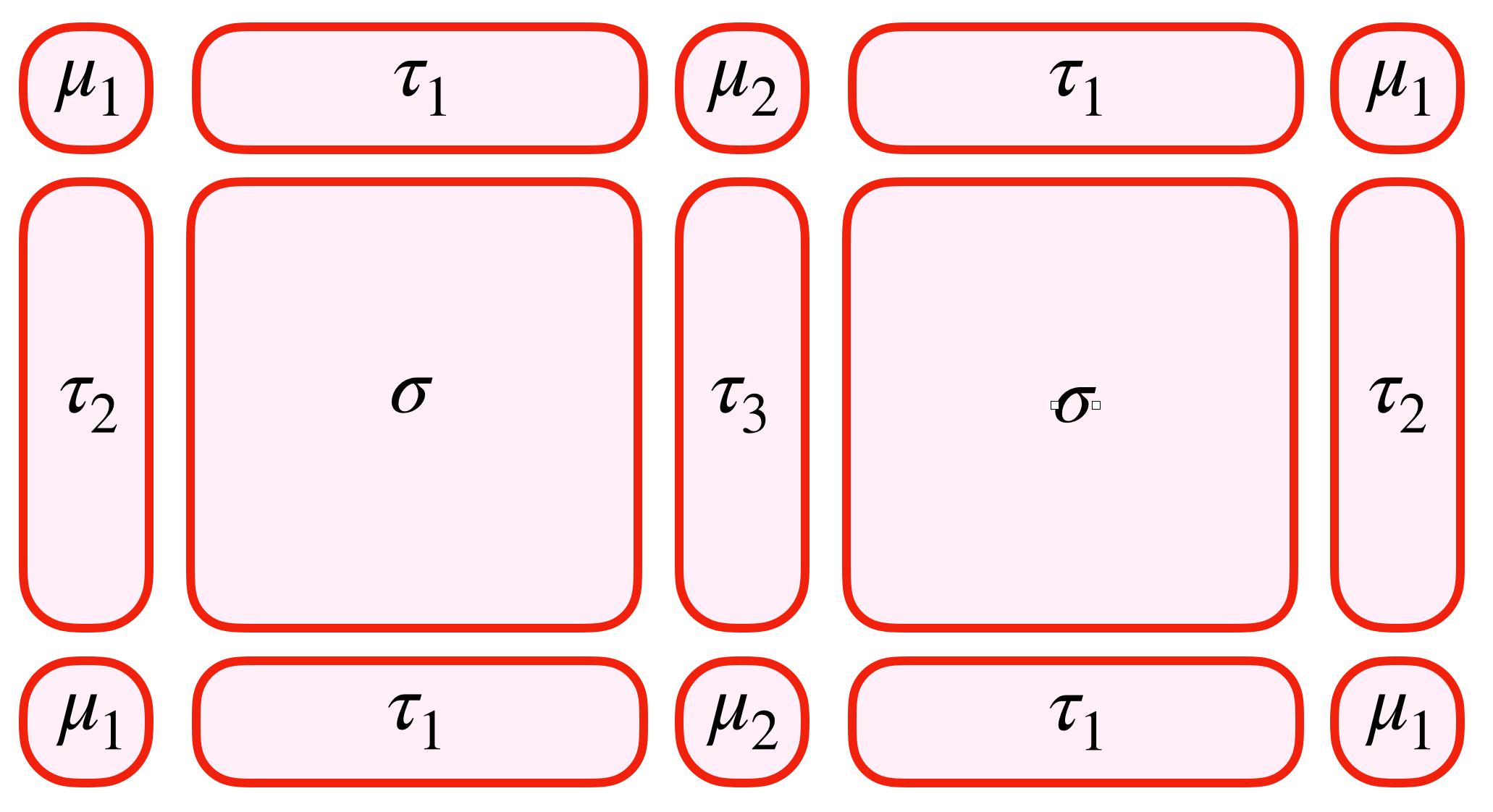}
\caption{\#3 wallpaper group $pm$ and its cell decomposition.}
\label{pm}
\end{figure}

For 2D blocks and 1D blocks labeled by $\tau_1$, there is no on-site symmetry, and for 1D blocks labeled by $\tau_2$/$\tau_3$, the on-site symmetry is $\mathbb{Z}_2$ because the reflection symmetry acts internally, and all 0D blocks have on-site $\mathbb{Z}_2$ symmetry via the reflection symmetry acts internally, see Fig. \ref{pm}. We discuss systems with spinless and spin-1/2 fermions separately.

\subsubsection{Spinless fermions}
First, we consider the 0D block-state decorations, the classification data can be characterized by different 1D irreducible representations of the symmetry group $\mathbb{Z}_2^f\times\mathbb{Z}_2$:
\begin{align}
\mathcal{H}^1\left[\mathbb{Z}_2^f\times\mathbb{Z}_2,U(1)\right]=\mathbb{Z}_2^2
\end{align}
One $\mathbb{Z}_2$ is from the complex fermion, and the other is from the reflection eigenvalue $-1$. So all obstruction-free 0D block-states form the following group:
\begin{align}
\{\mathrm{OFBS}\}_{pm,0}^{\mathrm{0D}}=\mathbb{Z}_2^4
\end{align}

Then consider the 1D block-state decorations: For 1D blocks labeled by $\tau_1$, there is no on-site symmetry, so the unique possible 1D block-state is Majorana chain; for 1D blocks labeled by $\tau_2$ and $\tau_3$, the total symmetry group is $\mathbb{Z}_2^f\times\mathbb{Z}_2$, and the possible 1D block-states are Majorana chain and 1D FSPT states (which can be realized by 2 Majorana chains with different eigenvalues of $\mathbb{Z}_2$ on-site symmetry). So all 1D block-states form the following group:
\begin{align}
\{\mathrm{BS}\}_{pm,0}^{\mathrm{1D}}=\mathbb{Z}_2^5
\end{align}
Then we discuss about the decorations of these two root phases separately.

\paragraph{Majorana chain decoration}First, we consider the Majorana chain decoration on $\tau_1$, which leaves 2 dangling Majorana modes at each 0D block labeled by $\mu_2$ who can be glued by an entanglement pair $i\gamma_1\gamma_2$. Nevertheless, this entanglement pair breaks the reflection symmetry, so the no-open-edge condition is violated.

Subsequently we consider the Majorana chain decoration on $\tau_2$/$\tau_3$, which leaves 2 dangling Majorana modes at $\mu_1/\mu_2$ and can be glued by an entanglement pair without breaking any symmetry. Hence the no-open-edge condition is preserved.

\paragraph{1D FSPT state decoration}The 1D FSPT state decoration can only be decorated on $\tau_2$ and $\tau_3$ due to the on-site symmetry, and this decoration leaves 4 dangling Majorana modes at $\mu_1/\mu_2$ with the following symmetry properties:
\begin{align}
\mathbb{Z}_2:~\left\{
\begin{aligned}
&\gamma_j^1\mapsto\gamma_j^1\\
&\gamma_j^2\mapsto-\gamma_j^2
\end{aligned}
\right.,~~j=A,B.
\label{Z2 symmetry property}
\end{align}
Consider the local fermion parity $P_f=-\gamma_A^1\gamma_A^2\gamma_B^1\gamma_B^2$ which is invariant under $\mathbb{Z}_2$ symmetry, so these 4 dangling Majorana modes can be gapped out by some proper interactions in a symmetric way. Equivalently, the no-open-edge condition is satisfied. Therefore, the obstruction-free 1D block-state decorations form the group:
\begin{align}
\{\mathrm{OFBS}\}_{pm,0}^{\mathrm{1D}}=\mathbb{Z}_2^4
\end{align}

With all obstruction-free block-states, subsequently we discuss about all possible trivializations. Decorate a Majorana chain with anti-PBC on each 2D block and enlarge all ``Majorana bubbles'', near each 1D block labeled by $\tau_1$, ``Majorana bubble'' construction can be deformed to double Majorana chains which can be trivialized because there is no on-site symmetry on $\tau_1$ and the classification of 1D invertible topological phases (i.e., Majorana chain) is $\mathbb{Z}_2$; near each 1D block labeled by $\tau_2/\tau_3$, ``Majorana bubble'' construction can also be deformed to double Majorana chains, nevertheless, these double Majorana chains cannot be trivialized because there is an on-site $\mathbb{Z}_2$ symmetry on each $\tau_2/\tau_3$ by internal action of reflection symmetry, and this $\mathbb{Z}_2$ action exchanges these two Majorana chains, and this is exactly the definition of the nontrivial 1D FSPT phase protected by on-site $\mathbb{Z}_2$ symmetry. Furthermore, 2D ``Majorana bubble'' construction has no effect on the 0D block-state because the reflection symmetry is not compatible with the Majorana chain surrounding each 0D block.

Furthermore, consider the 1D bubble equivalence on 1D blocks $\tau_1$: on each 1D block labeled by $\tau_1$, we decorate a 1D bubble onto it. Here both yellow and red dots represent the complex fermions. Near each 0D block labeled by $\mu_2$, there is 2 complex fermions which form an atomic insulator:
\begin{align}
|\psi\rangle_{pm}^{\mu_2}=c_1^\dag c_2^\dag|0\rangle
\end{align}
with reflection symmetry property as ($\bs{M}_{\tau_3}$ represents the reflection operation with the axis coincide with the 1D block labeled by $\tau_3$):
\begin{align}
\bs{M}_{\tau_3}|\psi\rangle_{pm}^{\mu_2}=c_2^\dag c_1^\dag|0\rangle=-|\psi\rangle_{pm}^{\mu_2}
\end{align}
Hence the reflection eigenvalue $-1$ can be trivialized by the atomic insulator $|\psi\rangle_{pm}^{\mu_2}$. Similar for $\mu_1$, and we can conclude that reflection eigenvalues at 0D blocks labeled by $\mu_1$ and $\mu_2$ are not independent. 

With all possible trivializations, we are ready to study the trivial states. Start from the original 0D trivial state (nothing is decorated on arbitrary blocks):
\[
[(+,+),(+,+)]
\]
If we take 1D bubble equivalences on $\tau_1$ by $l_1$ times, above trivial state will be deformed to a new 0D block-state labeled by:
\begin{align}
[(+,(-1)^{l_1}),(+,(-1)^{l_1})]
\label{pm spinless trivial state}
\end{align}
According to the definition of bubble equivalence, all these states should be trivial. It is easy to see that there are only one independent quantities ($l_1$) in Eq. (\ref{pm spinless trivial state}). Together with the 2D bubble equivalence, all these trivial states form the following group:
\begin{align}
\{\mathrm{TBS}\}_{pm,0}&=\{\mathrm{TBS}\}_{pm,0}^{\mathrm{1D}}\times \{\mathrm{TBS}\}_{pm,0}^{\mathrm{0D}}\nonumber\\
&=\mathbb{Z}_2\times\mathbb{Z}_2=\mathbb{Z}_2^2
\end{align}
here $\{\mathrm{TBS}\}_{pm,0}^{\mathrm{1D}}$ represents the group of trivial states with non-vacuum 1D blocks (i.e., 1D FSPT phase decorations on $\tau_2$ and $\tau_3$ simultaneously), and $\{\mathrm{TBS}\}_{pm,0}^{\mathrm{0D}}$ represents the group of trivial states with non-vacuum 0D blocks.

Therefore, all obstruction and trivialization free 0D/1D block-states are classified as:
\begin{align}
\begin{aligned}
&E_{pm,0}^{\mathrm{0D}}=\{\mathrm{OFBS}\}_{pm,0}^{\mathrm{0D}}/\{\mathrm{TBS}\}_{pm,0}^{\mathrm{0D}}=\mathbb{Z}_2^3\\
&E_{pm,0}^{\mathrm{1D}}=\{\mathrm{OFBS}\}_{pm,0}^{\mathrm{1D}}/\{\mathrm{TBS}\}_{pm,0}^{\mathrm{1D}}=\mathbb{Z}_2^3
\end{aligned}
\end{align}
and all independent nontrivial block-states are labeled by the group elements of the following quotient group:
\begin{align}
E_{pm,0}=E_{pm,0}^{\mathrm{0D}}\times E_{pm,0}^{\mathrm{1D}}=\mathbb{Z}_2^8/\mathbb{Z}_2^2=\mathbb{Z}_2^6
\end{align}
here three $\mathbb{Z}_2$'s are from the nontrivial 0D block-states, and other three $\mathbb{Z}_2$'s are from the nontrivial 1D block-states. 

Similar with $p4m$ case, there is no stacking between 1D and 0D block-states, and the ultimate classification with accurate group structure is:
\begin{align}
\mathcal{G}_{pm}^0=\mathbb{Z}_2^6
\end{align}

\subsubsection{Spin-1/2 fermions}
First, we consider the 0D block-state decorations. The total symmetry of each 0D block labeled by $\mu_1$ and $\mu_2$ is the nontrivial $\mathbb{Z}_2^f$ extension of the on-sity symmetry $\mathbb{Z}_2$: $\mathbb{Z}_4^f$, at which different 0D block-states can be characterized by different 1D irreducible representations of the corresponding symmetry group:
\begin{align}
\mathcal{H}^1\left[\mathbb{Z}_4^f,U(1)\right]=\mathbb{Z}_4
\end{align}
and there is no trivialization on the classification data. Hence the classification attributed to 0D block-state decorations is:
\begin{align}
E_{pm,1/2}^{\mathrm{0D}}=\mathbb{Z}_4^2
\end{align}

Then consider the 1D block-state decorations: for 1D blocks labeled by $\tau_1$, there is no on-site symmetry, hence the unique possible 1D block-state is Majorana chain; for 1D blocks labeled by $\tau_2$ and $\tau_3$, the total symmetry group is $\mathbb{Z}_4^f$, and there is no nontrivial block state due to the trivial classification of the corresponding 1D FSPT phases. Subsequently we consider the Majorana chain decoration on $\tau_1$ who leaves 2 dangling Majorana modes at each 0D block labeled by $\mu_2$ who can be glued by an entanglement pair $i\gamma_1\gamma_2$ and preserve the reflection symmetry. Hence the no-open-edge condition is fulfilled and the classification attributed to 1D block-state decorations is:
\begin{align}
E_{pm,1/2}^{\mathrm{1D}}=\mathbb{Z}_2
\end{align}
Similar with $p2$ case, there is a stacking between 1D and 0D block-states which leads to the classification data $\mathbb{Z}_4\times\mathbb{Z}_2$ to $\mathbb{Z}_8$, and the ultimate classification with accurate group structure is:
\begin{align}
\mathcal{G}_{pm}^{1/2}=E_{pm,1/2}^{\mathrm{0D}}\times_{\omega_2}E_{pm,1/2}^{\mathrm{1D}}=\mathbb{Z}_4\times\mathbb{Z}_8
\end{align}

\subsubsection{With $U^f(1)$ charge conservation}
Then we consider the systems with $U^f(1)$ charge conservation. For an arbitrary 0D block, different 0D block-states are characterized by different irreducible representations of the symmetry group as:
\begin{align}
\mathcal{H}^1[U(1)\times\mathbb{Z}_2,U(1)]=\mathbb{Z}\times\mathbb{Z}_2
\end{align}
Here $\mathbb{Z}$ represents the complex fermion, and $\mathbb{Z}_2$ represents different reflection eigenvalues on each 0D block. Then we consider possible trivializations. For systems with spinless fermions, consider the 1D bubble equivalence on 1D blocks labeled by $\tau_1$ [cf. 1D bubble, here yellow and red dots represent particle and hole, respectively, and they can be trivialized if we shrink them to a point]: Near each 0D block labeled by $\mu_1$, there are 2 fermionic particles $(p_1^\dag,p_2^\dag)$ that form an atomic insulator:
\begin{align}
|\phi\rangle_{\mathrm{0D}}^{\mu_1}=p_1^\dag p_2^\dag|0\rangle
\end{align}
with reflection property as ($\bs{M}_{\tau_2}$ represents the reflection operation with the axis coincide with the 1D block labeled by $\tau_2$):
\begin{align}
\bs{M}_{\tau_2}|\phi\rangle_{\mathrm{0D}}^{\mu_1}=p_2^\dag p_1^\dag|0\rangle=-|\phi\rangle_{\mathrm{0D}}^{\mu_1}
\end{align}
i.e., the reflection eigenvalue $-1$ at $\mu_1$ can be trivialized by atomic insulator $|\phi\rangle_{\mathrm{0D}}^{\mu_1}$. Near each 0D block labeled by $\mu_2$, there are 2 fermionic holes $(h_1^\dag,h_2^\dag)$ that form another atomic insulator:
\begin{align}
|\phi\rangle_{\mathrm{0D}}^{\mu_2}=h_1^\dag h_2^\dag|0\rangle
\end{align}
with reflection property as:
\begin{align}
\bs{M}_{\tau_3}|\phi\rangle_{\mathrm{0D}}^{\mu_2}=h_2^\dag h_1^\dag|0\rangle=-|\phi\rangle_{\mathrm{0D}}^{\mu_2}
\end{align}
i.e., the reflection eigenvalue $-1$ at $\mu_2$ can be trivialized by atomic insulator $|\phi\rangle_{\mathrm{0D}}^{\mu_2}$. Thus the 1D bubble construction on $\tau_{1}$ can change the reflection eigenvalues of $\mu_1$ and $\mu_2$ simultaneously. Equivalently, the reflection eigenvalues of 0D blocks $\mu_1$ and $\mu_2$ are not independent. Furthermore, above particle-hole construction on 1D blocks labeled by $\tau_1$ will add two complex fermions with $U^f(1)$ charge $+1$ (particles) at each 0D block labeled by $\mu_1$ and two complex fermions with $U^f(1)$ charge $-1$ (holes) at each 0D block labeled by $\mu_2$, hence the $U^f(1)$ charges at $\mu_1$ and $\mu_2$ are not independent. 

With the help of above discussions, we consider the 0D blocks $\mu_1$ and $\mu_2$ with 1D block $\tau_1$ connecting them. The 0D block-state decorated on $\mu_1$ can be characterized by $(a,\pm)$, where $a\in\mathbb{Z}$ represents the $U^f(1)$ charges at $\mu_1$ and $\pm$ represents the eigenvalues of reflection symmetry. Hence all candidates of 0D block-states on $\mu_1$ and $\mu_2$ can be labeled by elements of the group $(\mathbb{Z}\times\mathbb{Z}_2)^2$. Then we consider the trivial state labeled by $[(0,+),(0,+)]$, take aforementioned 1D bubble construction on $\tau_1$ one time will lead to a new 0D block-state labeled by $[(2,-),(-2,-)]$ which is also trivial. By parity of reasoning, all trivial 0D block-states will form the following group:
\begin{align}
\left\{\big[(2l,(-1)^l),(-2l,(-1)^l)\big]\Big|l\in\mathbb{Z}\right\}=2\mathbb{Z}
\end{align}
And they can reduce the classification of 0D block-states on $\mu_1$ and $\mu_2$ from $\mathbb{Z}^2\times\mathbb{Z}_2^2$ to $\mathbb{Z}\times\mathbb{Z}_2^3$. Then the classification of crystalline topological phases protected by $pm$ symmetry for 2D systems with spinless fermions is (where the subscript $0$ represents the spinless fermions):
\begin{align}
\mathcal{G}_{pm,0}^{U(1)}=\mathbb{Z}\times\mathbb{Z}_4\times\mathbb{Z}_2
\end{align}

Subsequently we investigate systems with spin-1/2 fermions. Repeatedly consider the atomic insulators $|\phi\rangle_{\mathrm{0D}}^{\mu_1}$ and $|\phi\rangle_{\mathrm{0D}}^{\mu_2}$, with the following reflection properties:
\begin{align}
\left\{
\begin{aligned}
&\bs{M}_{\tau_2}|\phi\rangle_{\mathrm{0D}}^{\mu_1}=-p_2^\dag p_1^\dag|0\rangle=|\phi\rangle_{\mathrm{0D}}^{\mu_1}\\
&\bs{M}_{\tau_3}|\phi\rangle_{\mathrm{0D}}^{\mu_2}=-h_2^\dag h_1^\dag|0\rangle=|\phi\rangle_{\mathrm{0D}}^{\mu_2}
\end{aligned}
\right.
\end{align}
Hence for systems with spin-1/2 fermions, reflection symmetry eigenvalues on different 0D blocks are independent. We repeatedly consider the 0D blocks $\mu_1$ and $\mu_2$ with 1D block $\mu_1$ connecting them. Take aforementioned 1D bubble construction on $\tau_1$ for the trivial state $[(0,+),(0,+)]$ will lead to a new 0D block-state labeled by $[(2,+),(-2,+)]$ which is also trivial. By parity of reasoning, all trivial 0D block-states will form the following group:
\begin{align}
\left\{\big[(2l,+),(-2l,+)\big]\Big|l\in\mathbb{Z}\right\}=2\mathbb{Z}
\end{align}
And they can reduce the classification of 0D block-states from $\mathbb{Z}^2\times\mathbb{Z}_2^2$ to $\mathbb{Z}\times\mathbb{Z}_2^3$. Finally the classification of crystalline topological phases protected by $pm$ symmetry for 2D systems with spin-1/2 fermions is (where the subscript $1/2$ represents the spin-1/2 fermions):
\begin{align}
\mathcal{G}_{pm,1/2}^{U(1)}=\mathbb{Z}\times\mathbb{Z}_4\times\mathbb{Z}_2
\end{align}
All results are summarized in the main text.

\subsection{$pg$}
\begin{figure}
\centering
\includegraphics[width=0.46\textwidth]{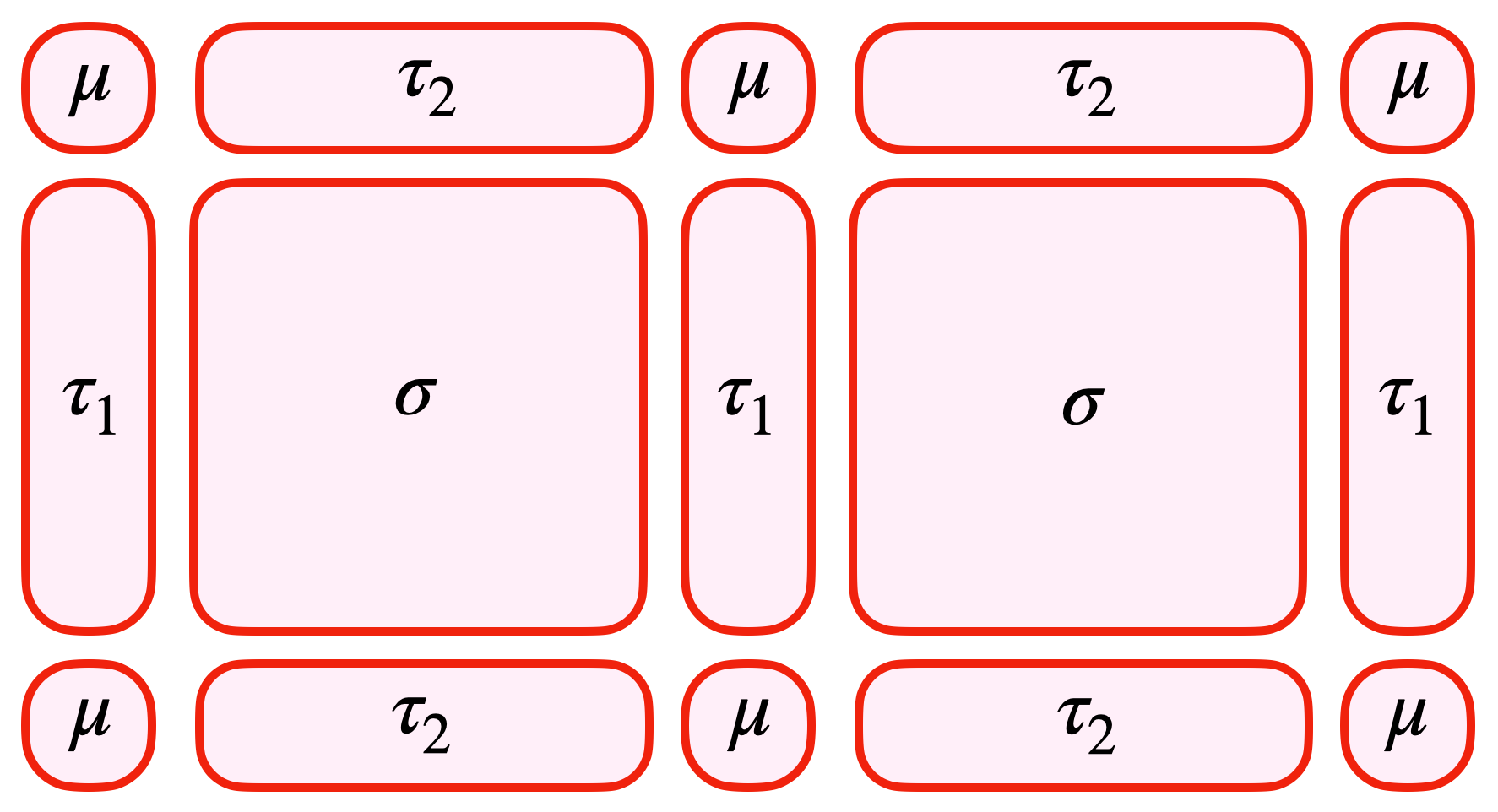}
\caption{\#4 wallpaper group $pg$ and its cell decomposition.}
\label{pg}
\end{figure}

There is no on-site symmetry on blocks with arbitrary dimensions, because the reflection operation is accomplished with a translation operation toward a ``glide reflection'' operation, see Fig. \ref{pg}. The spin of fermions is irrelevant for this case, hence we only need to discuss about the systems with spinless fermions.

\subsubsection{Spinless and spin-1/2 fermions}
First, we consider the 0D block-state decorations. The only relevant on-site symmetry is the local fermion parity $\mathbb{Z}_2^f$. Consider the 2D bubble equivalence: we decorate a Majorana chain with anti-PBC on each 2D block which can be trivialized if it shrinks to a point. By some proper local unitary transformations, this assembly of bubbles can be deformed to an assembly of Majorana chains with anti-PBC surrounding each of 0D block. Nevertheless, the fermion parities of 0D blocks cannot be changed by 2D bubble equivalence because the fermion parity of Majorana chain with anti-PBC is even. As a consequence, the classification contributed by complex fermion decorations on 0D blocks is $G_{\mathrm{0D}}^C=\mathbb{Z}_2$, and subsequently, the classification via 0D block-state decorations is:
\begin{align}
E_{pg}^{\mathrm{0D}}=\mathbb{Z}_2
\end{align}

Then consider the 1D block-state decorations: the unique possible 1D block-state is Majorana chain due to the absence of the on-site symmetry. Majorana chain decoration on $\tau_1/\tau_2$ leaves 2 dangling Majorana modes on each 0D block $\mu$ which can be gapped out by an entanglement pair, hence the no-open-edge condition is satisfied. Thus the classification attributed to 1D block-state decoration is:
\begin{align}
E_{pg}^{\mathrm{1D}}=\mathbb{Z}_2^2
\end{align}
Similar to the $p1$ case, there is no extension between 1D block-state and 0D block-state, and the ultimate classification with accurate group structure is:
\begin{align}
\mathcal{G}_{pg}^0=\mathcal{G}_{pg}^{1/2}=E_{pg}^{\mathrm{0D}}\times E_{pg}^{\mathrm{1D}}=\mathbb{Z}_2^3
\end{align}

\subsubsection{With $U^f(1)$ charge conservation}
Then we consider the systems with $U^f(1)$ charge conservation. For an arbitrary 0D block, different 0D block-states are characterized by different irreducible representations of the symmetry group as:
\begin{align}
\mathcal{H}^1[U(1),U(1)]=\mathbb{Z}
\end{align}
Here $\mathbb{Z}$ represents the complex fermion. There is no trivialization, hence the classification of the systems with $U^f(1)$ charge conservation is:
\begin{align}
\mathcal{G}_{pg,0}^{U(1)}=\mathcal{G}_{pg,1/2}^{U(1)}=\mathbb{Z}
\end{align}

\subsection{$cm$}
\begin{figure}
\centering
\includegraphics[width=0.4\textwidth]{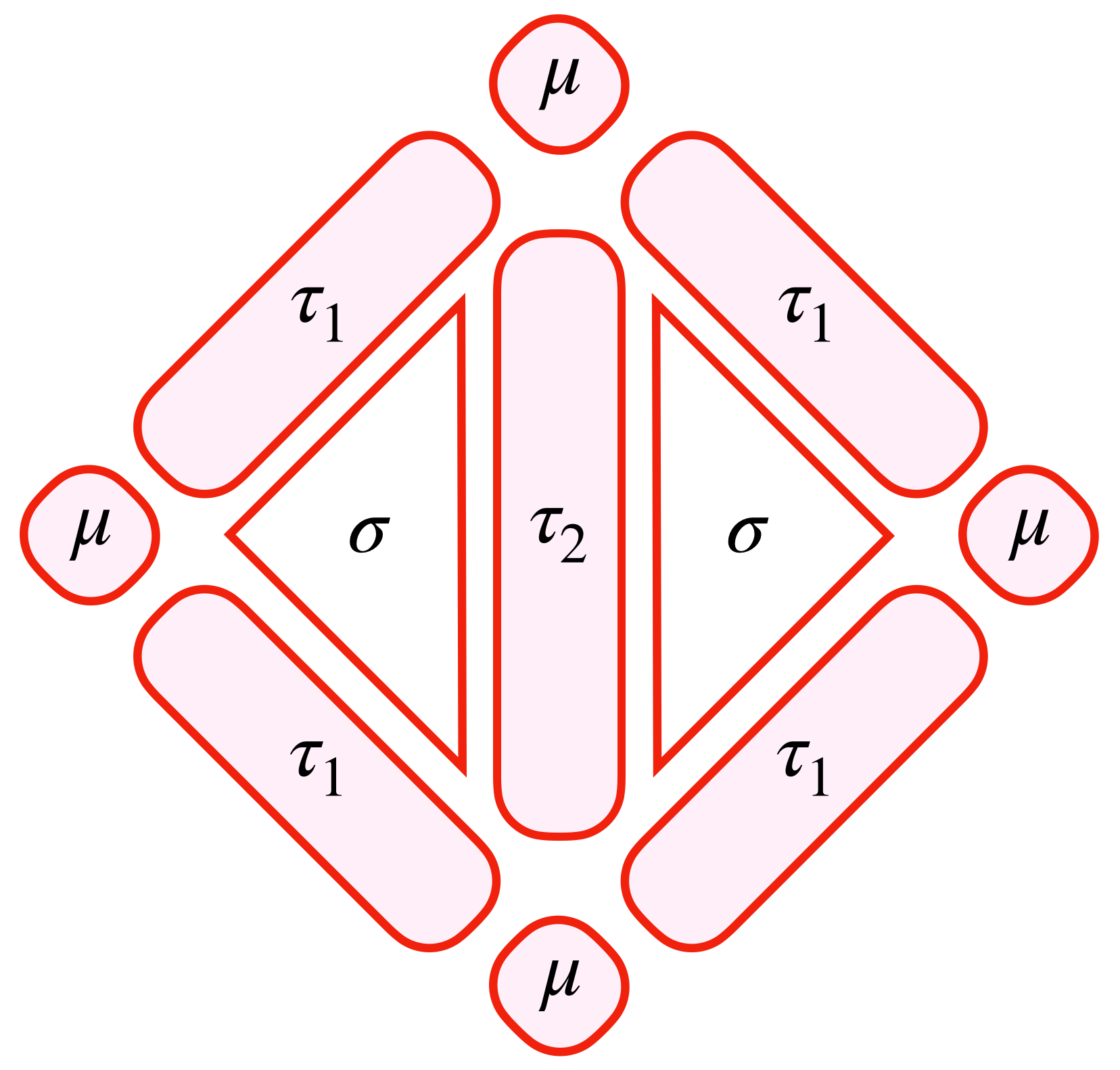}
\caption{\#5 wallpaper group $cm$ and its cell decomposition.}
\label{cm}
\end{figure}

There is no on-site symmetry on 2D block $\sigma$ and 1D blocks $\tau_1$; and there is an on-site $\mathbb{Z}_2$ symmetry on 1D block $\tau_2$ and 0D block $\mu$ because the reflection symmetry action on the corresponding blocks are identical with an on-site $\mathbb{Z}_2$ symmetry, see Fig. \ref{cm}. 

\subsubsection{Spinless fermions}
First, we consider the 0D block-state decorations. The classification data can be characterized by different 1D irreducible representations of the symmetry group $\mathbb{Z}_2^f\times\mathbb{Z}_2$:
\begin{align}
\mathcal{H}^1\left[\mathbb{Z}_2^f\times\mathbb{Z}_2,U(1)\right]=\mathbb{Z}_2^2
\end{align}
One $\mathbb{Z}_2$ represents the complex fermions, and the other represents the reflection eigenvalue $-1$. So all obstruction-free 0D block-states form the group:
\begin{align}
\{\mathrm{OFBS}\}_{cm,0}^{\mathrm{0D}}=\mathbb{Z}_2^2
\end{align}
Subsequently consider the 1D block-state decorations, and the decorations on $\tau_1$ and $\tau_2$ are different: the unique possible block-state on $\tau_1$ is Majorana chain due to the absence of the on-site symmetry, but for $\tau_2$, the total on-site symmetry is $\mathbb{Z}_2^f\times\mathbb{Z}_2$, hence the possible 1D block-states are Majorana chain and 1D FSPT state.

\paragraph{Decoration on $\tau_1$}
The Majorana chain decoration on $\tau_1$ leaves 4 dangling Majorana modes at every 0D block $\mu$; consider the local fermion parity and its reflection symmetry property:
\[
P_f=-\prod\limits_{j=1}^4\gamma_j,~~\bs{M}:~P_f\mapsto P_f
\]
and these 4 Majorana modes can be gapped out by some proper interactions. Equivalently, the no-open-edge condition is satisfied. More precisely, consider the following Hamiltonian containing two entanglement pairs of these four Majorana modes:
\begin{align}
H_\mu=-i\gamma_1\gamma_3-i\gamma_2\gamma_4
\end{align}
It is easy to verify that $H_\mu$ is invariant under all symmetries. 

\paragraph{Decoration on $\tau_2$}
First, we consider the Majorana chain decoration on $\tau_2$ which leaves 2 dangling Majorana modes at each 0D block $\mu$, with the same $\mathbb{Z}_2$-eigenvalue because of the translational symmetry. So these 2 Majorana modes can be symmetrically gapped out by an entanglement pair, and the no-open-edge condition is satisfied. Subsequently we consider the 1D FSPT state decoration on $\tau_2$, which leaves 4 dangling Majorana modes on each $\mu$, with the $\mathbb{Z}_2$ symmetry properties as Eq. (\ref{Z2 symmetry property}). Thus the local fermion parity satisfies the symmetry, and these 4 Majorana modes can be gapped by interaction in a symmetric way. Therefore, the no-open-edge condition is satisfied. Thus all obstruction-free 1D block-states form the group:
\begin{align}
\{\mathrm{OFBS}\}_{cm,0}^{\mathrm{1D}}=\mathbb{Z}_2^3
\end{align}
and different group elements can be labeled by:
\begin{align}
[\pm,\pm;m_1;m_2,m_2']
\end{align}
here $\pm$'s are fermion parity and the reflection eigenvalues at each $\mu$, $m_1$ and $m_2$ represent the number of decorated Majorana chains on the 1D block $\tau_1$ and $\tau_2$, and $m_2'$ represents the number of decorated 1D FSPT phases on the 1D block $\tau_2$.

With all obstruction-free block-states, subsequently we discuss about all possible trivializations. Decorate a Majorana chain with anti-PBC on each 2D block and enlarge all ``Majorana bubble'' construction can be deformed to double Majorana chains at each $\tau_1$ that can be trivialized because there is no on-site symmetry and the classification of 1D invertible topological phases (i.e., Majorana chain) is $\mathbb{Z}_2$; near each 1D block labeled by $\tau_2$, ``Majorana bubble'' construction can also be deformed to double Majorana chains. Nevertheless, these double Majorana chains cannot be trivialized because there is an on-site $\mathbb{Z}_2$ symmetry on each $\tau_2$ by reflection symmetry acting internally, and this $\mathbb{Z}_2$ action exchanges these two Majorana chains, and this is exactly the definition of the nontrivial 1D FSPT phase protected by on-site $\mathbb{Z}_2$ symmetry. Furthermore, similar with the $p4m$ case, 2D ``Majorana bubble'' construction has no effect on arbitrary 0D blocks.

Subsequently we consider the 1D bubble equivalence on 1D blocks $\tau_1$ [cf. 1D bubble, here both yellow and red dots represent the complex fermions]: near each 0D block, there are 4 complex fermions which can form an atomic insulator:
\begin{align}
|\psi\rangle_{cm}=c_1^\dag c_2^\dag c_3^\dag c_4^\dag|0\rangle
\end{align}
with reflection property as ($\bs{M}_{\tau_2}$ represents the reflection operation with the axis coincide with the 1D block labeled by $\tau_2$):
\begin{align}
\bs{M}_{\tau_2}|\psi\rangle_{cm}=c_2^\dag c_1^\dag c_4^\dag c_3^\dag|0\rangle=|\psi\rangle_{cm}
\end{align}
Hence there is no trivialization from 1D bubble equivalence, and all trivial states form the following group:
\begin{align}
\{\mathrm{TBS}\}_{cm,0}=\{\mathrm{TBS}\}_{cm,0}^{\mathrm{1D}}=\mathbb{Z}_2
\end{align}
Therefore, all obstruction and trivialization free 0D/1D block-states are:
\begin{align}
\begin{aligned}
&E_{cm,0}^{\mathrm{0D}}=\{\mathrm{OFBS}\}_{cm,0}^{\mathrm{0D}}=\mathbb{Z}_2^2\\
&E_{cm,0}^{\mathrm{1D}}=\{\mathrm{OFBS}\}_{cm,0}^{\mathrm{1D}}/\{\mathrm{TBS}\}_{cm,0}^{\mathrm{1D}}=\mathbb{Z}_2^2
\end{aligned}
\end{align}
and all independent nontrivial block-states are labeled by the group elements of the following quotient group:
\begin{align}
\mathcal{G}_{cm}^0=E_{cm,0}^{\mathrm{0D}}\times E_{cm,0}^{\mathrm{1D}}=\mathbb{Z}_2^5/\mathbb{Z}_2=\mathbb{Z}_2^4
\label{cm spinless classification data}
\end{align}
Here two $\mathbb{Z}_2$'s are from the nontrivial 0D block-states, and other two $\mathbb{Z}_2$'s are from the nontrivial 1D block-states. 

Similar with $p4m$ case, there is no stacking between 1D and 0D block-states, hence the group structure of the classification data (\ref{cm spinless classification data}) has already been accurate.

\subsubsection{Spin-1/2 fermions}
First, we consider the 0D block-state decorations. The total symmetry of each 0D block labeled by $\mu_1$ and $\mu_2$ is the nontrivial $\mathbb{Z}_2^f$ extension of the on-sity symmetry $\mathbb{Z}_2$: $\mathbb{Z}_4^f$, at which different 0D block-states can be characterized by different 1D irreducible representations of the corresponding symmetry group:
\begin{align}
\mathcal{H}^1\left[\mathbb{Z}_4^f,U(1)\right]=\mathbb{Z}_4
\end{align}
and there is no obstruction and trivialization on the classification data. Hence the classification attributed to 0D block-state decorations is:
\begin{align}
E_{cm,1/2}=\mathbb{Z}_4
\end{align}
Subsequently we consider the 1D block-state decorations. The unique possible block-state for 1D blocks labeled by $\tau_1$ is Majorana chain because of the absence of on-site symmetry. On $\tau_2$, the total symmetry group is $\mathrm{Z}_4^f$ (i.e., nontrivial $\mathbb{Z}_2^f$ extension of $\mathbb{Z}_2$), hence there is no candidate block-state for 1D blocks labeled by $\tau_2$ because of the trivial classification. So the only possible 1D block-state is Majorana chain decoration on $\tau_1$. The Majorana chain decoration on $\tau_1$ leaves 4 dangling Majorana modes at each 0D block which can be gapped out in a symmetric way by some proper interactions because the local fermion parity keeps invariant under arbitrary symmetry actions. Equivalently the no-open-edge condition is satisfied, and the classification attributed to 1D block-state decorations is:
\begin{align}
E_{cm,1/2}^{\mathrm{1D}}=\mathbb{Z}_2
\end{align}
Similar with the $pgg$ case, there is no stacking between 1D and 0D block-states, and the ultimate classification with accurate group structure is:
\begin{align}
\mathcal{G}_{cm}^{1/2}=\mathbb{Z}_4\times\mathbb{Z}_2
\end{align}

\subsubsection{With $U^f(1)$ charge conservation}
Then we consider the systems with $U^f(1)$ charge conservation. For an arbitrary 0D block, different 0D block-states are characterized by different irreducible representations of the symmetry group as:
\begin{align}
\mathcal{H}^1[U(1)\times\mathbb{Z}_2,U(1)]=\mathbb{Z}\times\mathbb{Z}_2
\end{align}
Here $\mathbb{Z}$ represents the complex fermion, and $\mathbb{Z}_2$ represents the eigenvalues of reflection symmetry operation. There is no trivialization, hence the classification of the systems with $U^f(1)$ charge conseration is:
\begin{align}
\mathcal{G}_{cm,0}^{U(1)}=\mathcal{G}_{cm,1/2}^{U(1)}=\mathbb{Z}\times\mathbb{Z}_2
\end{align}

\subsection{$pmm$}
\begin{figure}
\centering
\includegraphics[width=0.45\textwidth]{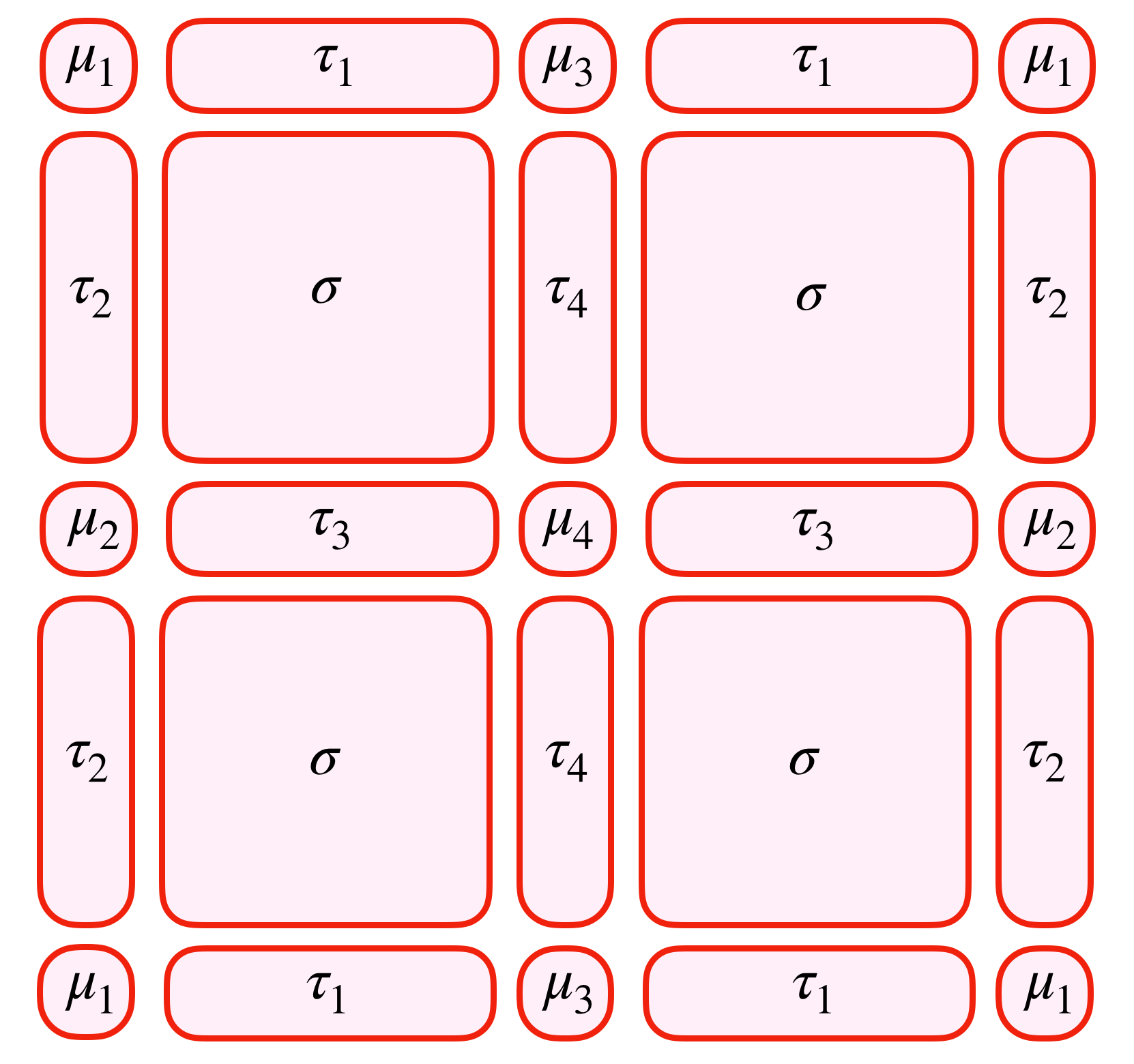}
\caption{\#6 wallpaper group $pmm$ and its cell decomposition.}
\label{pmm}
\end{figure}

The corresponding point group for this case is 2-fold dihedral group $D_2$. For 2D blocks $\sigma$, there is no on-site symmetry; for 1D blocks $\tau_j$ ($j=1,2,3,4$); for 0D blocks $\mu_k$ ($k=1,2,3,4$), the on-site symmetry is $\mathbb{Z}_2\rtimes\mathbb{Z}_2$ via the $D_2$ symmetry acting internally, see Fig. \ref{pmm}.

\subsubsection{Spinless fermions}
First, we investigate the 0D block-state decorations. For an arbitrary 0D block, the classification data can be characterized by different 1D irreducible representations of the full symmetry group $\mathbb{Z}_2\rtimes\mathbb{Z}_2$:
\begin{align}
\mathcal{H}^1\left[\mathbb{Z}_2^f\times(\mathbb{Z}_2\rtimes\mathbb{Z}_2),U(1)\right]=\mathbb{Z}_2^3
\label{pmm classification data}
\end{align}
here three $\mathbb{Z}_2$ in the classification data have different physical meanings: the first $\mathbb{Z}_2$ represents the complex fermion, the second $\mathbb{Z}_2$ represents the rotation eigenvalue $-1$, and the third $\mathbb{Z}_2$ represents the reflection eigenvalue $-1$. So the 0D block-states at $\mu_j$ ($j=1,2,3,4$) can be labeled by $(\pm,\pm,\pm)$, here these three $\pm$'s represent the fermion parity, 2-fold rotation and reflection symmetry eigenvalues (alternatively, the last two $\pm$'s can also represent the eigenvalues of two independent reflection operations because even-fold dihedral group can also be generated by two independent reflections). According to this notation, the obstruction-free 0D block-states form the following group:
\begin{align}
\{\mathrm{OFBS}\}_{pmm}^{\mathrm{0D}}=\mathbb{Z}_2^{12}
\end{align}
and the group elements can be labeled by (three brackets represent the block-states at $\mu_1$, $\mu_2$ and $\mu_3$):
\[
[(\pm,\pm,\pm),(\pm,\pm,\pm),(\pm,\pm,\pm),(\pm,\pm,\pm)]
\]
Subsequently we investigate the 1D block-state decoration. For all 1D blocks, the total symmetry group is $\mathbb{Z}_2^f\times\mathbb{Z}_2$, and the candidate 1D block-state is Majorana chain and 1D FSPT state. So all 1D block-states form a group:
\begin{align}
\{\mathrm{BS}\}_{pmm,0}^{\mathrm{1D}}=\mathbb{Z}_2^8
\end{align}
Then we discuss about the decorations of these two root phases separately.

\paragraph{Majorana chain decoration}Consider Majorana chain decorations on 1D blocks labeled by $\tau_1$ as an example, which leaves 2 dangling Majorana modes at each 0D block labeled by $\mu_1/\mu_3$. Near each $\mu_1$, two dangling Majorana modes have the following rotational symmetry properties:
\begin{align}
\bs{R}_{\mu_1}:~\gamma_1\leftrightarrow\gamma_2
\end{align}
Then consider the local fermion parity and its rotational symmetry property:
\begin{align}
P_f=i\gamma_1\gamma_2,~\bs{R}_{\mu_1}:~P_f\mapsto-P_f
\end{align}
Thus these two dangling Majorana modes form a projective representation of the symmetry group $pmm\times\mathbb{Z}_2^f$, and a non-degenerate ground state is forbidden. Thus Majorana chain decoration on 1D blocks $\tau_1$ is obstructed because of the violation of the no-open-edge condition. Similar arguments can be held on all other 1D blocks, and we can conclude that the Majorana chain decorations on arbitrary 0D blocks are \textit{obstructed}.

\paragraph{1D FSPT state decoration}Then we consider about the 1D FSPT state decorations. As an example, 1D FSPT state decoration on $\tau_1$ leaves four dangling Majorana modes at each $\mu_1$ and $\mu_3$. Similar with arguments about the $p4m$ case, these four Majorana modes form a projective representation of the $D_2$ symmetry group at each 0D block $\mu_1$ and $\mu_3$, and the non-degenerate ground state is forbidden as a consequence, and similar for all other 1D blocks. Therefore, the 1D FSPT state decorations solely on $\tau_j$ ($j=1,2,3,4$) is \textit{obstructed} because of the violation of the no-open-edge condition.

There is one exception: If we decorate a 1D FSPT phase on each 1D block (including $\tau_j,j=1,2,3,4$), the dangling Majorana modes at each 0D block can be gapped out in a symmetric way. For this case, there are two nontrivial projective representations of $D_2$ symmetry group at each 0D block that can be deformed to a linear representation, because there is only one nontrivial projective representation of the $D_2$ symmetry group (acting internally at each 0D block, which is identical with $\mathbb{Z}_2\rtimes\mathbb{Z}_2$ on-site symmetry group). This fact can be checked by the following 2-cohomology:
\begin{align}
\mathcal{H}^2\left[\mathbb{Z}_2\rtimes\mathbb{Z}_2,U(1)\right]=\mathbb{Z}_2
\end{align}
As a consequence, the eight dangling Majorana modes at each 0D block due to the decoration can be gapped out in a symmetric way, and the corresponding 1D block-state is \textit{obstruction-free}. Thus all obstruction-free 1D block-states form the following group:
\begin{align}
\{\mathrm{OFBS}\}_{pmm,0}^{\mathrm{1D}}=\mathbb{Z}_2
\end{align}
and the group elements can be labeled by $m_1=m_2=m_3=m_4$. Here $m_j=0,1$ ($j=1,2,3,4$) represents the number of decorated 1D FSPT states on $\tau_j$, respectively. According to aforementioned discussions, a necessary condition of an obstruction-free block-state is $m_1=m_2=m_3=m_4$.

With all obstruction-free block-states, subsequently we discuss about all possible trivializations. First, we consider about the 2D bubble equivalence: as we discussed in the main text, both types of ``Majorana bubble'' constructions are allowed because all 0D blocks are the centers of $D_2$ point group symmetry, including ``Majorana bubbles'' with both PBC and anti-PBC. Similar with the $p4m$ case, both types of ``Majorana bubbles'' can be deformed to double Majorana chains at each nearby 1D block, and this is exactly the definition of the nontrivial 1D FSPT phase protected by on-site $\mathbb{Z}_2$ symmetry (by reflection symmetry acting internally). As a consequence, 1D FSPT state decorations on all 1D blocks can be deformed to a trivial state via 2D ``Majorana'' bubble equivalences. Furthermore, repeatedly similar with the $p4m$ case, both types of ``Majorana bubble'' constructions have no effect on 0D blocks, and effects of both type of ``Majorana bubble'' constructions are equivalent, so take one of them into account is enough. 

Subsequently we consider the 1D bubble equivalences. 1D bubble equivalence on 1D blocks $\tau_1$ [cf. 1D bubble, here both yellow and red dots represent the complex fermions]: near the 0D block labeled by $\mu_3$, there are 2 complex fermions which form an atomic insulator:
\begin{align}
|\psi\rangle_{pmm}^{\mu_3}=c_1^\dag c_2^\dag|0\rangle
\end{align}
with rotation property as ($\bs{R}_{\mu_3}$ represents the rotation operation centred at the 0D block labeled by $\mu_3$):
\begin{align}
\bs{R}_{\mu_3}|\psi\rangle_{pmm}^{\mu_3}=c_2^\dag c_1^\dag|0\rangle=-|\psi\rangle_{pmm}^{\mu_3}
\end{align}
Hence the rotation eigenvalue $-1$ at $\mu_3$ can be trivialized by the atomic insulator $|\psi\rangle_{pmm}^{\mu_3}$, and the rotation eigenvalues of 0D blocks $\mu_1$ and $\mu_3$ are not independent. Similar discussion can be held on 1D blocks $\tau_2$, $\tau_3$ and $\tau_4$, therefore rotation eigenvalues on all 0D blocks $\mu_j$ ($j=1,2,3,4$) are not independent.

Then we repeatedly consider the atomic insulator $|\psi\rangle_{pmm}^{\mu_3}$, with reflection property as ($\bs{M}_{\tau_4}$ represents the reflection operation with the axis coincide with the 1D block labeled by $\tau_4$):
\begin{align}
\bs{M}_{\tau_4}|\psi\rangle_{pmm}^{\mu_3}=c_2^\dag c_1^\dag|0\rangle=-|\psi\rangle_{pmm}^{\mu_3}
\end{align}
Hence the reflection eigenvalue $-1$ at $\mu_3$ can be trivialized by the atomic insulator $|\psi\rangle_{pmm}^{\mu_3}$, and the reflection eigenvalues of 0D blocks $\mu_1$ and $\mu_3$ are not independent. Similar discussion can be held on 1D blocks $\tau_2$, $\tau_3$ and $\tau_4$, therefore reflection eigenvalues on all 0D blocks $\mu_j$ ($j=1,2,3,4$) are not independent. 

As we mentioned before, $D_2$ symmetry can also be generated by two independent reflection operations, we summarize the effects of 1D bubble constructions in terms of the changes on reflection eigenvalues as following:
\begin{enumerate}[1.]
\item 1D bubble construction on $\tau_1$: simultaneously change the eigenvalue of $\bs{M}_{\tau_2}$ at $\mu_1$ and $\bs{M}_{\tau_4}$ at $\mu_3$;
\item 1D bubble construction on $\tau_2$: simultaneously change the eigenvalue of $\bs{M}_{\tau_1}$ at $\mu_1$ and $\bs{M}_{\tau_3}$ at $\mu_2$;
\item 1D bubble construction on $\tau_3$: simultaneously change the eigenvalues of $\bs{M}_{\tau_2}$ at $\mu_2$ and $\bs{M}_{\tau_4}$ at $\mu_4$;
\item 1D bubble construction on $\tau_4$: simultaneously change the eigenvalues of $\bs{M}_{\tau_1}$ at $\mu_3$ and $\bs{M}_{\tau_3}$ at $\mu_4$;
\end{enumerate}

With all possible bubble constructions, we are ready to investigate the trivial states. Start from the original trivial state (nothing is decorated on arbitrary lower-dimensional blocks):
\[
[(+,+,+),(+,+,+),(+,+,+),(+,+,+)]
\]
if we take 1D bubble constructions on $\tau_j$ by $l_j$ times ($j=1,2,3,4$), above trivial state will be deformed to a new block-state labeled by:
\begin{align}
&\left[(+,(-1)^{l_2},(-1)^{l_1}),(+,(-1)^{l_3},(-1)^{l_2}),\right.\nonumber\\
&\left.(+,(-1)^{l_1},(-1)^{l_4}),(+,(-1)^{l_4},(-1)^{l_3})\right]
\label{pmm spinless trivial state}
\end{align}
According to the definition of bubble equivalence, all these states should be trivial. It is easy to see that there are only four independent quantities in Eq. (\ref{pmm spinless trivial state}): $l_i,i=1,2,3,4$. Together with the 2D bubble equivalence, all trivial states form the group:
\begin{align}
\{\mathrm{TBS}\}_{pmm,0}&=\{\mathrm{TBS}\}_{pmm,0}^{\mathrm{1D}}\times\{\mathrm{TBS}\}_{pmm,0}^{\mathrm{0D}}\nonumber\\
&=\mathbb{Z}_2\times\mathbb{Z}_2^4=\mathbb{Z}_2^5
\end{align}
here $\{\mathrm{TBS}\}_{p6m,0}^{\mathrm{1D}}$ represents the group of trivial states with non-vacuum 1D blocks (i.e., 1D FSPT phase decorations on all 1D blocks simultaneously), and $\{\mathrm{TBS}\}_{p6m,0}^{\mathrm{0D}}$ represents the group of trivial states with non-vacuum 0D blocks.

Therefore, all obstruction and trivialization free 0D/1D block-states are classified by:
\begin{align}
\begin{aligned}
&E_{pmm,0}^{\mathrm{0D}}=\{\mathrm{OFBS}\}_{pmm,0}^{\mathrm{0D}}/\{\mathrm{TBS}\}_{pmm,0}^{\mathrm{0D}}=\mathbb{Z}_2^8\\
&E_{pmm,0}^{\mathrm{1D}}=\{\mathrm{OFBS}\}_{pmm,0}^{\mathrm{1D}}/\{\mathrm{TBS}\}_{pmm,0}^{\mathrm{1D}}=\mathbb{Z}_1
\end{aligned}
\end{align}
and all independent nontrivial block-states are labeled by the group elements of the following quotient group:
\begin{align}
\mathcal{G}_{pmm}^0=E_{pmm,0}^{\mathrm{0D}}=\mathbb{Z}_2^8
\end{align}
here all $\mathbb{Z}_2$'s are from the nontrivial 0D block-states. It is obvious that there is no nontrivial group extension because of the absence of nontrivial 1D block-state, and the group structure of the classification data $E_{pmm,0}$ has already been accurate.

\subsubsection{Spin-1/2 fermions}
Now we turn to discuss systems with spin-1/2 fermions. Consider the 0D block-state decoration, similar with the $p4m$ case, the classification data can be characterized by different 1D irreducible representations of the full symmetry group $\mathbb{Z}_2^f\times_{\omega_2}(\mathbb{Z}_2\rtimes\mathbb{Z}_2)$:
\begin{align}
\mathcal{H}^1\left[\mathbb{Z}_2^f\times_{\omega_2}(\mathbb{Z}_2\rtimes\mathbb{Z}_2),U(1)\right]=\mathbb{Z}_2^2
\end{align}

 Hence nontrivial 2D FSPT states attributed to 0D block-state decoration are classified as:
\begin{align}
E_{pmm,1/2}^{\mathrm{0D}}=\mathbb{Z}_2^8
\end{align}
Then we consider the 1D block-state decorations. The total on-site symmetry on each 1D block is $\mathbb{Z}_4^f$ (nontrivial $\mathbb{Z}_2^f$ extension of on-site symmetry $\mathbb{Z}_2$, the mathematical meaning of spin-1/2 fermions). Hence in this situation, 1D block-state decorations do not contribute any 2D FSPT state due to the trivial classification of the corresponding block-state:
\begin{align}
E_{pmm,1/2}^{\mathrm{1D}}=\mathbb{Z}_1
\end{align}
Therefore, it is obvious that there is no stacking because of the trivial contribution of 1D block-state decoration, and the ultimate classification with accurate group structure of crystalline topological phases protected by $pmm$ symmetry in 2D interacting fermionic systems with spin-1/2 fermions is:
\begin{align}
\mathcal{G}_{pmm}^{1/2}=\mathbb{Z}_2^8.
\end{align}

\subsubsection{With $U^f(1)$ charge conservation}
For an arbitrary 0D block $\mu_j, j=1,2,3,4$, different 0D block-states are characterized by different irreducible representations of the symmetry group as:
\begin{align}
\mathcal{H}^1[U(1)\times(\mathbb{Z}_2\rtimes\mathbb{Z}_2),U(1)]=\mathbb{Z}\times\mathbb{Z}_2^2
\end{align}
Here $\mathbb{Z}$ represents the complex fermion, two $\mathbb{Z}_2$'s represent the eigenvalues $-1$ of two reflection generators of $D_2$. We should further consider possible trivialization: for systems with spinless fermions, consider the 1D bubble equivalence on 1D blocks labeled by $\tau_1$: we decorate a 1D bubble on each $\tau_1$, here yellow and red dots represent the particle and hole, respectively, and they can be trivialized if we shrink them to a point. Near each 0D block labeled by $\mu_1$, there are two particles forming an atomic insulator:
\begin{align}
|\phi\rangle_{pmm}^{\mu_1}=p_1^\dag p_2^\dag|0\rangle
\end{align}
Similar to the crystalline TSC, $|\phi\rangle_{pmm}^{\mu_1}$ changes the eigenvalue of $\bs{M}_{\tau_2}$ at each $\mu_1$. Near $\mu_3$, there are two holes that form another atomic insulator:
\begin{align}
|\phi\rangle_{pmm}^{\mu_3}=h_1^\dag h_2^\dag|0\rangle
\end{align}
$|\phi\rangle_{pmm}^{\mu_3}$ changes the eigenvalue of $\bs{M}_{\tau_4}$ at each $\mu_3$. Similar 1D bubble construction can be held for all 1D blocks $\tau_j, j=1,2,3,4$, which leads to the nonindependence of rotation and reflection eigenvalues of all 0D blocks $\mu_j$.

Subsequently we consider the complex fermion sector. Repeatedly consider the aforementioned 1D bubble construction on $\tau_1$: it adds two complex fermions with $U^f(1)$ charge $+1$ (particles) at each 0D block $\mu_1$ two complex fermions with $U^f(1)$ charge $-1$ (holes) at each 0D block $\mu_3$, hence the $U^f(1)$ charges at $\mu_1$ and $\mu_3$ are not independent. Similar arguments can be held for all 1D blocks $\tau_j, j=1,2,3,4$, which leads to the nonindependence of the $U^f(1)$ charges at all 0D blocks $\mu_j$. 

With the help of above discussions, we consider the 0D block-state decorations. The 0D block-state decorated on $\mu_1$ can be labeled by $(m_1,\pm,\pm)$, where $m_1\in\mathbb{Z}$ represents $U^f(1)$ charges on $\mu_1$ and two $\pm$'s represent the eigenvalues of 2-fold rotational symmetry and reflection symmetry, respectively. Then we consider the trivial state labeled by:
\[
[(0,+,+),(0,+,+),(0,+,+),(0,+,+)]
\]
Take aforementioned 1D bubble constructions on $\tau_j$ by $l_j\in\mathbb{Z}$ times, and it will lead to a new 0D block-state labeled by:
\begin{align}
&\left[\left(2(l_1+l_2),(-1)^{l_2},(-1)^{l_1}\right)\right.,\nonumber\\
&\left(2(-l_2+l_3),(-1)^{l_3},(-1)^{l_3}\right),\nonumber\\
&\left(2(-l_1+l_4),(-1)^{l_1},(-1)^{l_4}\right),\nonumber\\
&\left.\left(-2(l_3+l_4),(-1)^{l_4},(-1)^{l_3}\right)\right]
\label{pmm trivial state}
\end{align}
And this state should be trivial. First, we consider the sector of complex fermion decoration. Alternatively, all 0D block-states can be viewed as vectors of a 12-dimensional vector space $V$, where the complex fermion components are $\mathbb{Z}$-valued and all other components are $\mathbb{Z}_2$-valued. Then all trivial 0D block-states with the form as Eq. (\ref{pmm trivial state}) can be viewed as a vector subspace $V'$ of $V$. It is easy to see that there are only four independent quantities in Eq. (\ref{pmm trivial state}): $l_j, j=1,2,3,4$. So the dimensionality of the vector subspace $V'$ should be 4. For the complex fermion sector, we have the following relation:
\begin{align}
(l_1+l_2)+(-l_2+l_3)+(-l_1+l_4)=(l_3+l_4)
\end{align}
i.e., there are only three independent quantities which serves a $(2\mathbb{Z})^3$ trivialization. The remaining one degree of freedom of the vector subspace $V'$ should be attributed to the eigenvalues of point group symmetry actions, and serves a $\mathbb{Z}_2$ trivialization. Therefore, the ultimate classification of crystalline topological phases protected by $pmm$ symmetry for 2D systems with spinless fermions is:
\begin{align}
\mathcal{G}_{pmm,0}^{U(1)}&=\mathbb{Z}^4\times\mathbb{Z}_2^8/\left[(2\mathbb{Z})^3\times\mathbb{Z}_2\right]\nonumber\\
&=\mathbb{Z}\times\mathbb{Z}_4^3\times\mathbb{Z}_2^{4}
\end{align}

For systems with spin-1/2 fermions, the rotation and reflection properties of $|\phi\rangle_{pmm}^{\mu_1}$ and $|\phi\rangle_{pmm}^{\mu_3}$ at each 0D blocks $\mu_1$ and $\mu_3$ are changed by $-1$, respectively, which leads to no trivialization. Furthermore, like the $p4m$ case, the classification data of the corresponding 0D block-states can be characterized by different 1D irreducible representations of the full symmetry group:
\begin{align}
\mathcal{H}^1\left[U^f(1)\rtimes_{\omega_2}(\mathbb{Z}_2\rtimes\mathbb{Z}_2)\right]=2\mathbb{Z}\times\mathbb{Z}_2^2
\end{align}
we should notice that for systems with spin-1/2 fermions, we can only decorate even $U^f(1)$ charges on each 0D block. Now we repeatedly consider the complex fermion decorations on 0D blocks and 1D bubble constructions (similar to the spinless fermions), the ultimate classification of crystalline topological phases protected by $pmm$ symmetry for 2D systems with spin-1/2 fermions is:
\begin{align}
\mathcal{G}_{pmm,1/2}^{U(1)}=(2\mathbb{Z})^4\times\mathbb{Z}_2^8/(2\mathbb{Z})^3=2\mathbb{Z}\times\mathbb{Z}_2^{8}
\end{align}

\subsection{$pmg$}
\begin{figure}
\centering
\includegraphics[width=0.45\textwidth]{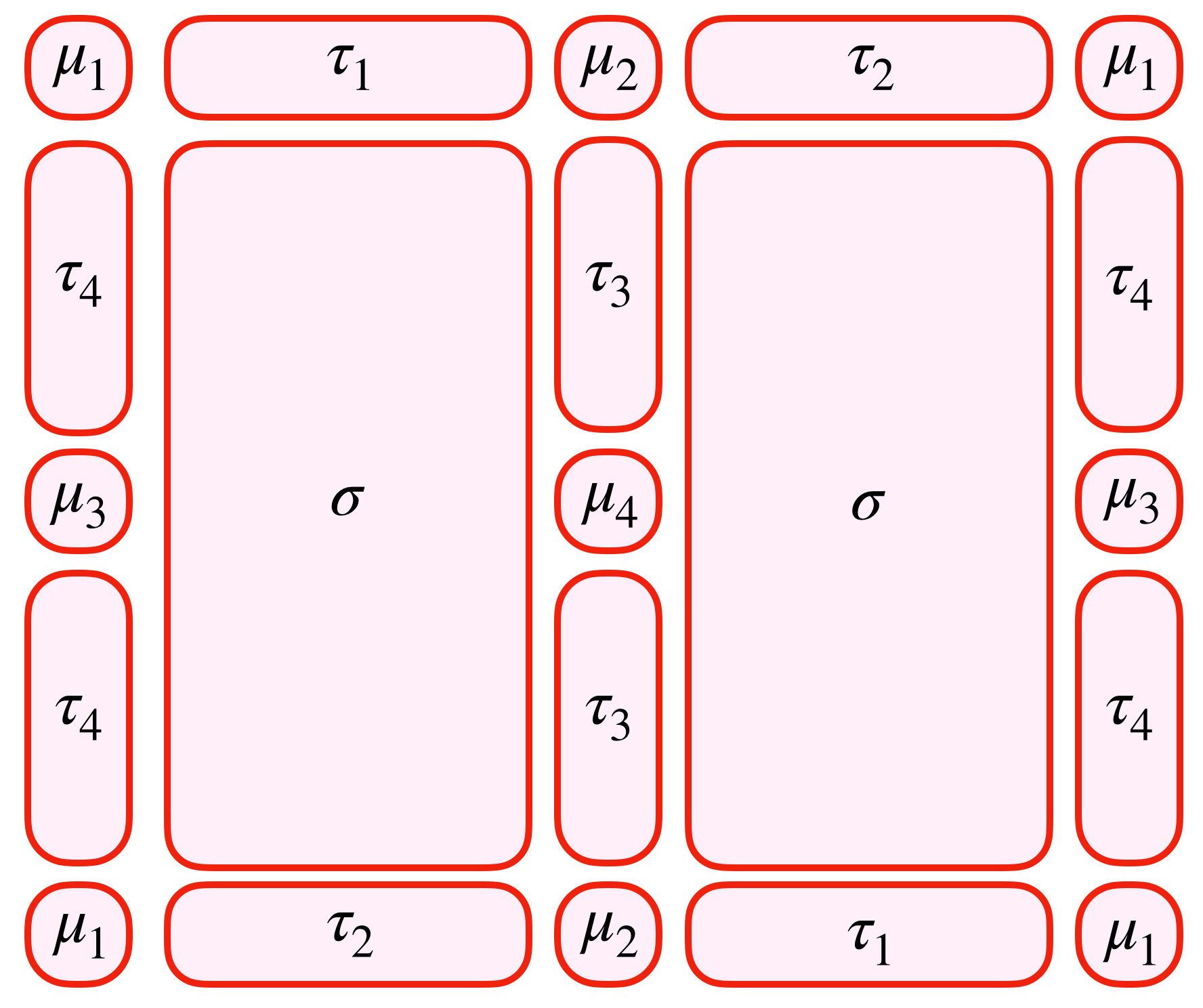}
\caption{\#6 wallpaper group $pmg$ and its cell decomposition.}
\label{pmg}
\end{figure}

The corresponding point group of $pmg$ is dihedral group $D_2$. For 2D blocks $\sigma$, 1D blocks labeled by $\tau_3$ and $\tau_4$, there is no on-site symmetry; for 1D blocks labeled by $\tau_1$ and $\tau_2$, 0D blocks labeled by $\mu_1$ and $\mu_2$, the on-site symmetry is $\mathbb{Z}_2$ via the reflection symmetry acting internally; for 0D blocks labeled by $\mu_3$ and $\mu_4$, the on-site symmetry is $\mathbb{Z}_2$ via the 2-fold rotational symmetry acting internally. The cell decomposition is shown in Fig. \ref{pmg}. We discuss systems with spinless and spin-1/2 fermions separately. 

\subsubsection{Spinless fermions}
For systems with spinless fermions, we consider the 0D block-state decorations: for 0D blocks labeled by $\mu_j$ ($j=1,2,3,4$), the total symmetry group of each is $\mathbb{Z}_2^f\times\mathbb{Z}_2$, and the classification data can be characterized by different 1D irreducible representations of the symmetry group:
\begin{align}
\mathcal{H}^1\left[\mathbb{Z}_2^f\times\mathbb{Z}_2,U(1)\right]=\mathbb{Z}_2^2
\end{align}
one $\mathbb{Z}_2$ is from the fermion parity, and the other is from the rotation or reflection eigenvalue $-1$. So at each 0D block, the block-state can be labeled by $(\pm,\pm)$, here there two $\pm$'s represent the fermion parity and rotation/reflection eigenvalue, respectively. According to this notation, the obstruction-free 0D block-states form the following group:
\begin{align}
\{\mathrm{OFBS}\}_{pmg,0}^{\mathrm{0D}}=\mathbb{Z}_2^8
\end{align}
and different group elements can be labeled by (four brackets represent the block-states at $\mu_j, j=1,2,3,4$):
\[
[(\pm,\pm),(\pm,\pm),(\pm,\pm),(\pm,\pm)]
\]

Subsequently we investigate the 1D block-state decoration. For 1D blocks labeled by $\tau_1$ and $\tau_2$, the total symmetry group is $\mathbb{Z}_2^f\times\mathbb{Z}_2$, hence the candidate 1D block-states are Majorana chain and 1D FSPT state (double Majorana chains); for 1D blocks labeled by $\tau_3$ and $\tau_4$, the only possible 1D block-state is Majorana chain because of the absence of on-site symmetry. So all 1D block-states form a group:
\begin{align}
\{\mathrm{BS}\}_{pmg,0}^{\mathrm{1D}}=\mathbb{Z}_2^6
\end{align}
Then we discuss the decorations on $\tau_1/\tau_2$ and $\tau_3/\tau_4$ separately due to the different on-site symmetry on the corresponding 1D blocks.

\paragraph{Decorations on $\tau_3$ and $\tau_4$}Majorana chain decorations on 1D blocks $\tau_3$ will leave two dangling Majorana modes at each 0D block labeled by $\mu_2/\mu_4$. Near each 0D block $\mu_4$, the nearby two Majorana modes violate the no-open-edge condition: the fermion parity $P_f=i\gamma_1\gamma_2$ is not invariant under the 2-fold rotation $\bs{R}_{\mu_4}$:
\begin{align}
\bs{R}_{\mu_4}:~i\gamma_1\gamma_2\mapsto-i\gamma_1\gamma_2
\end{align}
Similar for 1D blocks $\tau_4$. Therefore, there is no obstruction-free block-state on 1D blocks $\tau_3$ and $\tau_4$. 

\paragraph{Decorations on $\tau_1$ and $\tau_2$}
From Fig. \ref{pmg} we can see that different 1D blocks $\tau_1/\tau_2$ are not connected, hence if we want to get an obstruction-free block-state, we should consider $\tau_1$ and $\tau_2$ together and decorate identical block-states on them. First, we consider the Majorana chain decoration, it leaves two dangling Majorana modes at each 0D block labeled by $\mu_1$ and $\mu_2$. Near each $\mu_1$, these two Majorana modes can be gapped out by an entanglement pair $i\gamma
_1\gamma_2$ without breaking any symmetry, so do $\mu_2$. So Majorana chain decoration satisfy the no-open-edge condition, and contribute an obstruction-free block-state. Subsequently we consider the 1D FSPT state decoration which leaves four dangling Majorana modes at each 0D block labeled by $\mu_1$ and $\mu_2$. Similar with the $pm$ case, these four Majorana modes can be gapped out in a symmetric way, and the no-open-edge condition is fulfilled. Overall, the obstruction-free 1D block-states form the following group:
\begin{align}
\{\mathrm{OFBS}\}_{pmg,0}^{\mathrm{1D}}=\mathbb{Z}_2^2
\end{align}
and the group elements can be labeled by $m_1=m_2$ and $m_1'=m_2'$. Here $m_1/m_2$ represents the number of decorated Majorana chains on $\tau_1/\tau_2$, and $m_1'/m_2'$ represents the number of decorated 1D FSPT states on $\tau_1/\tau_2$. According to aforementioned discussions, a necessary condition of an obstruction-free block-state is $m_1=m_2$ and $m_1'=m_2'$. Different obstruction-free block-states can be labeled by:
\begin{align}
[(\pm,\pm),(\pm,\pm),(\pm,\pm),(\pm,\pm);m_1=m_2;m_1'=m_2']
\end{align}
here the first four brackets represent the 0D block-states at $\mu_j$ ($j=1,2,3,4$), and the last two quantities represent the number of Majorana chains and 1D FSPT states at $\tau_1/\tau_2$. 

With all obstruction-free block-states, subsequently we discuss about all possible trivializations. First, we consider about the 2D bubble equivalences: as we discussed in the main text, only type-\2 (i.e., ``Majorana bubbles'' with anti-PBC) 2D bubble equivalence is valid because there is no 0D block as the center of even-fold dihedral group symmetry. Enlarge all “Majorana bubble” construction can be deformed to double Majorana chains at each $\tau_3$ and $\tau_4$ that can be trivialized because there is no on-site symmetry and the classification of 1D invertible topological phases (i.e., Majorana chain) is $\mathbb{Z}_2$; near each 1D block labeled by $\tau_1$ and $\tau_2$, ``Majorana bubble'' construction can also be deformed to double Majorana chains. Nevertheless, these double Majorana chains cannot be trivialized because there is an on-site $\mathbb{Z}_2$ symmetry on each $\tau_1/\tau_2$ by reflection symmetry acting internally, and this $\mathbb{Z}_2$ action exchanges these two Majorana chains, and this is exactly the definition of the nontrivial 1D FSPT phase protected by on-site $\mathbb{Z}_2$ symmetry. Furthermore, similar with the $p4m$ case, there is no effect on 0D blocks labeled by $\mu_1$ and $\mu_2$ by taking 2D ``Majorana'' bubble equivalence, because the alternative Majorana chain surrounding each $\mu_1/\mu_2$ is not compatible with the reflection operations; nevertheless, similar with the $p2$ case, 2D ``Majorana bubble'' construction changes the fermion parity of each 0D block labeled by $\mu_3/\mu_4$ because there is no reflection operation on 0D block $\mu_3/\mu_4$, and the alternative Majorana chain surrounding each $\mu_3/\mu_4$ is compatible with all other symmetry operations.

Subsequently we consider the 1D bubble equivalences. For instance, we decorate a 1D bubble on each 1D block labeled by $\tau_4$ (here both yellow and red dots represent the complex fermions). Near each 0D block $\mu_3$, there are two complex fermions forming the following atomic insulator:
\begin{align}
|\psi\rangle_{pmg}^{\mu_3}=c_1^\dag c_2^\dag|0\rangle
\end{align}
with rotation property as:
\begin{align}
\bs{R}_{\mu_3}|\psi\rangle_{pmg}^{\mu_3}=c_2^\dag c_1^\dag|0\rangle=-|\psi\rangle_{pmg}^{\mu_3}
\end{align}
i.e., the rotation eigenvalue $-1$ at each 0D block $\mu_3$ can be trivialized by the atomic insulator $|\psi\rangle_{pmg}^{\mu_3}$. Near $\mu_1$, there are another two complex fermions forming another atomic insulator:
\begin{align}
|\psi\rangle_{pmg}^{\mu_1}=c_1'^\dag c_2'^\dag|0\rangle
\end{align}
with reflection symmetry as:
\begin{align}
\bs{M}_{\tau_1}|\psi\rangle_{pmg}^{\mu_1}=c_2'^\dag c_1'^\dag|0\rangle=-|\psi\rangle_{pmg}^{\mu_1}
\end{align}
i.e., the reflection eigenvalue $-1$ at $\mu_1$ can be trivialized by the atomic insulator $|\psi\rangle_{pmg}^{\mu_1}$. Therefore, 1D bubble construction on $\tau_4$ leads to the nonindependence of rotation eigenvalues of $\mu_3$ and reflection eigenvalues of $\mu_1$. Similar 1D bubble construction can be held on $\tau_3$ and leads to the nonindependence of rotation eigenvalues of $\mu_4$ and reflection eigenvalues of $\mu_2$. Then we consider the identical 1D bubble construction on $\tau_1$ and $\tau_2$: it will change the fermion parities of 0D blocks $\mu_1$ and $\mu_2$ simultaneously by adding a complex fermion on each of them. 

There is another type of 1D bubble construction (we denote above type of 1D bubble construction by ``type-\1'', this type of 1D bubble construction by ``type-\2''): we decorate a 1D bubble on each 1D block labeled by $\tau_1$ (here both yellow and red dots represent the 0D FSPT modes characterized by eigenvalue $-1$ of reflection symmetry acting internally). This 1D bubble construction changes the reflection eigenvalues of 0D blocks $\mu_1$ and $\mu_2$ simultaneously. 

With all possible 2D and 1D bubble constructions, we are ready to study the trivial block-states. Start from the original trivial state (nothing is decorated on arbitrary blocks):
\[
[(+,+),(+,+),(+,+),(+,+)]
\]
If we take 2D bubble construction $l_0$ times, take 1D bubble constructions (complex fermions) on $\tau_3/\tau_4$ by $l_3/l_4$ times, and take type-\1 1D bubble constructions (0D $\mathbb{Z}_2$ FSPT modes) on $\tau_1/\tau_2$ by $l_1/l_2$ times and type-\2 1D bubble constructions on $\tau_1/\tau_2$ by $l_1'/l_2'$ times, above trivial state will be deformed to a new 0D block-state labeled by:
\begin{align}
&\left[((-1)^{l_1'+l_2'},(-1)^{l_1+l_2+l_4}),\right.\nonumber\\
&((-1)^{l_1'+l_2'},(-1)^{l_1+l_2+l_3})\nonumber\\
&\left.((-1)^{l_0},(-1)^{l_3}),((-1)^{l_0},(-1)^{l_4})\right]
\label{pmg spinless trivial state}
\end{align}
According to the definition of bubble equivalence, all these 0D block-states should be trivial. Alternatively, all 0D block-states can be viewed as vectors of an 8-dimensional $\mathbb{Z}_2$-valued vector space, and all trivial 0D block-states with the form as Eq. (\ref{pmg spinless trivial state}) can be viewed as a vector of the subspace of aforementioned vector space. The dimensionality of this subspace is 5, hence all trivial block-states form the following group:
\begin{align}
\{\mathrm{TBS}\}_{pmg,0}&=\{\mathrm{TBS}\}_{pmg,0}^{\mathrm{1D}}\times\{\mathrm{TBS}\}_{pmg,0}^{\mathrm{0D}}\nonumber\\
&=\mathbb{Z}_2\times\mathbb{Z}_2^4=\mathbb{Z}_2^5
\end{align}
here $\{\mathrm{TBS}\}_{pmg,0}^{\mathrm{1D}}$ represents the group of trivial states with non-vacuum 1D blocks (i.e., 1D FSPT phase decorations on $\tau_1$ and $\tau_2$ simultaneously), and $\{\mathrm{TBS}\}_{pmg,0}^{\mathrm{0D}}$ represents the group of trivial states with non-vacuum 0D blocks.

Therefore, all obstruction and trivialization free 0D/1D block-states are classified as:
\begin{align}
\begin{aligned}
&E_{pmg,0}^{\mathrm{0D}}=\{\mathrm{OFBS}\}_{pmg,0}^{\mathrm{0D}}/\{\mathrm{TBS}\}_{pmg,0}^{\mathrm{0D}}=\mathbb{Z}_2^4\\
&E_{pmg,0}^{\mathrm{1D}}=\{\mathrm{OFBS}\}_{pmg,0}^{\mathrm{1D}}/\{\mathrm{TBS}\}_{pmg,0}^{\mathrm{1D}}=\mathbb{Z}_2
\end{aligned}
\end{align}
and all independent nontrivial block-states are labeled by the group elements of the following quotient group:
\begin{align}
\mathcal{G}_{cmm}^0=E_{pmg,0}^{\mathrm{0D}}\times E_{pmg,0}^{\mathrm{1D}}=\mathbb{Z}_2^5
\end{align}
here one $\mathbb{Z}_2$ is from the Majorana chain decorations on 1D blocks $\tau_1$ and $\tau_2$ simultaneously, and all other $\mathbb{Z}_2$'s are from the nontrivial 0D block-states. Similar with the $pm$ case, there is no nontrivial group extension because of the absence of nontrivial 1D block-state, and the group structure of $E_{cmm,0}$ has already been accurate.

\subsubsection{Spin-1/2 fermions}
Now we turn to discuss about the systems with spin-1/2 fermions. For arbitrary 0D blocks, the on-site symmetry should be $\mathbb{Z}_4^f$ by nontrivial $\mathbb{Z}_2^f$ (fermion parity) extension of on-site $\mathbb{Z}_2^f$ symmetry. And different 0D block-states at a certain $\mu_j$ ($j=1,2,3,4$) can be characterized by different 1D irreducible representations of $\mathbb{Z}_4^f$ that can be labeled by:
\begin{align}
\mathcal{H}^1\left[\mathbb{Z}_4^f,U(1)\right]=\mathbb{Z}_4=\{1,i,-1,-i\}
\end{align}
So at each 0D block, the block-state can be labeled by an index $\nu\in\{1,i,-1,-i\}$. According to this notation, the obstruction-free 0D block-states form the following group:
\begin{align}
\{\mathrm{OFBS}\}_{pmg,1/2}^{\mathrm{0D}}=\mathbb{Z}_4^4
\end{align}
and the group elements can be labeled by (four indices represent the block-states at $\mu_j,j=1,2,3,4$):
\[
\nu_1,\nu_2,\nu_3,\nu_4
\]

Then we consider possible trivializations. For example, we decorate a 1D bubble on each 1D block labeled by $\tau_1$, here the yellow and red dots represent the 0D FSPT mode characterized by $i\in\mathbb{Z}_4$ and $-i\in\mathbb{Z}_4$, respectively, and they can be trivialized by shrinking them to a point. According to this bubble construction, the eigenvalue of $\mathbb{Z}_4^f$ at each 0D block labeled by $\mu_1/\mu_2$ is changed by $i/-i$. 1D bubble construction on $\tau_2$ is similar. And for 1D blocks $\tau_3$ and $\tau_4$, the unique possible 1D bubble construction is ``complex fermion'' bubble (i.e., both yellow and red dots in 1D bubble represent the complex fermions). Near each 0D block $\mu_3$, it leaves 2 dangling complex fermions forming the following atomic insulator:
\begin{align}
|\phi\rangle_{pmg}^{\mu_3}=a_1^\dag a_2^\dag|0\rangle
\end{align}
with following rotational symmetry property:
\begin{align}
\bs{R}_{\mu_3}|\phi\rangle_{pmg}^{\mu_3}=-a_2^\dag a_1^\dag|0\rangle=|\phi\rangle_{pmg}^{\mu_3}
\end{align}
i.e., 1D bubble constructions on $\tau_3$ and $\tau_4$ lead to no trivialization. 

With all possible bubble constructions, we are ready to study the trivial states. Start from the original trivial state (nothing is decorated on arbitrary 0D block):
\[
[1,1,1,1]
\]
if we take 1D bubble construction on $\tau_1$ and $\tau_2$ by $l_1$ and $l_2$ times, above trivial state will be deformed to a new 0D block-state labeled by:
\begin{align}
[i^{l_1+l_2},(-i)^{l_1+l_2},1,1]
\label{pmg spin-1/2 trivial state}
\end{align}
According to the definition of bubble equivalence, all these states should be trivial. It is easy to see that there is only one independent quantity in the state (\ref{pmg spin-1/2 trivial state}), hence all these trivial states form the following group:
\begin{align}
\{\mathrm{TBS}\}_{pmg,1/2}^{\mathrm{0D}}=\mathbb{Z}_4
\end{align}
Therefore, all independent nontrivial 0D block-states are labeled by different group elements of the following quotient group:
\begin{align}
E_{pmg,1/2}^{\mathrm{0D}}=\{\mathrm{OFBS}\}_{pmg,1/2}^{\mathrm{0D}}/\{\mathrm{TBS}\}_{pmg,1/2}^{\mathrm{0D}}=\mathbb{Z}_4^3
\end{align}

Then we consider the 1D block-state decorations. There is no nontrivial candidate block-state on 1D blocks labeled by $\tau_1$ and $\tau_2$ because of the trivial classification of 1D FSPT phases with $\mathbb{Z}_4^f$ symmetry, and the unique possible block-state for 1D blocks labeled by $\tau_3$ and $\tau_4$ is Majorana chain because of the absence of on-site symmetry. So as an example, we consider the Majorana chain decoration on $\tau_3$: it leaves two dangling Majorana modes at each 0D block $\mu_2/\mu_4$, and can be gapped out by an entanglement pair in a symmetric way: near $\mu_2$, the entanglement pair $i\gamma_1\gamma_2$ has the following reflection symmetry property:
\begin{align}
\bs{M}_{\tau_1}:~i\gamma_1\gamma_2\mapsto-i\gamma_2\gamma_1= i\gamma_1\gamma_2
\end{align}
similar for another two Majorana modes near $\mu_4$. Hence the Majorana chain decorations on 1D blocks $\tau_3$ and $\tau_4$ are obstruction free, and forming the group containing all obstruction-free 1D block-states:
\begin{align}
\{\mathrm{OFBS}\}_{pmg,1/2}^{\mathrm{1D}}=\mathbb{Z}_2^2
\end{align}
And it is obvious that there is no trivialization (i.e., $\{\mathrm{TBS}\}_{pmg,1/2}^{\mathrm{1D}}=\mathbb{Z}_1$). As a consequence, the classification attributed to 1D block-state decorations is:
\begin{align}
E_{pmg,1/2}^{\mathrm{1D}}=\{\mathrm{OFBS}\}_{pmg,1/2}^{\mathrm{1D}}/\{\mathrm{TBS}\}_{pmg,1/2}^{\mathrm{1D}}=\mathbb{Z}_2^2
\end{align}

With all classification data, we consider the group structure of the corresponding classification. Similar with $p2$ case, the Majorana chain decorations on $\tau_3$ and $\tau_4$ have nontrivial extension with 0D block-state decorations on $\mu_4$ and $\mu_3$, and the ultimate classification with accurate group structure is:
\begin{align}
\mathcal{G}_{pmg}^{1/2}=E_{pmg,1/2}^{\mathrm{1D}}\times_{\omega_2}E_{pmg,1/2}^{\mathrm{0D}}=\mathbb{Z}_4\times\mathbb{Z}_8^2
\end{align}
here the symbol $\times_{\omega_2}$ means that independent nontrivial 1D and 0D block-states $E_{pmg,1/2}^{\mathrm{1D}}$ and $E_{pmg,1/2}^{\mathrm{0D}}$ have nontrivial extension, and described by an nontrivial factor system of the following short exact sequence in mathematical language:
\begin{align}
0\rightarrow E_{pmg,1/2}^{\mathrm{1D}}\rightarrow \mathcal{G}_{pmg}^{1/2}\rightarrow E_{pmg,1/2}^{\mathrm{1D}}\rightarrow0
\end{align}

\subsubsection{With $U^f(1)$ charge conservation}
Then we consider the systems with $U^f(1)$ charge conservation. For a 0D block labeled by $\mu_1$ or $\mu_2$, different 0D block-states are characterized by different irreducible representations of symmetry group as:
\begin{align}
\mathcal{H}^1[U(1)\times\mathbb{Z}_2,U(1)]=\mathbb{Z}\times\mathbb{Z}_2
\end{align}
Here $\mathbb{Z}$ represents the complex fermion, and $\mathbb{Z}_2$ represents the eigenvalues of reflection symmetry operation. For a 0D block labeled by $\mu_3$ and $\mu_4$, different 0D block-states are also characterized by different irreducible representations of symmetry group as:
\begin{align}
\mathcal{H}^1[U(1)\times\mathbb{Z}_2,U(1)]=\mathbb{Z}\times\mathbb{Z}_2
\end{align}
Here $\mathbb{Z}$ represents the complex fermion, and $\mathbb{Z}_2$ represents the eigenvalues of rotational symmetry operation. We should further investigate the possible trivializations. For systems with spinless fermions, consider the 1D bubble equivalence on 1D blocks labeled by $\tau_3$ [cf. 1D bubble, here yellow and red dots represent particle and hole, respectively, and they can be trivialized if we shrink them to a point]: Near each 0D block labeled by $\mu_2$, there are two fermionic particles that form an atomic insulator:
\begin{align}
|\phi\rangle_{pmg}^{\mu_2}=p_1^\dag p_2^\dag|0\rangle
\end{align}
with the reflection property as:
\begin{align}
\bs{M}_{\tau_1}|\phi\rangle_{pmg}^{\mu_2}=-p_1^\dag p_2^\dag|0\rangle=-|\phi\rangle_{pmg}^{\mu_2}
\end{align}
i.e., the reflection eigenvalue $-1$ can be trivialized by atomic insulator $|\phi\rangle_{pmg}^{\mu_2}$. Near each 0D block labeled by $\mu_4$, there are two fermionic holes that form another atomic insulator:
\begin{align}
|\phi\rangle_{pmg}^{\mu_4}=h_1^\dag h_2^\dag|0\rangle
\end{align}
with the rotation property as:
\begin{align}
\bs{R}_{\mu_4}|\phi\rangle_{pmg}^{\mu_4}=h_2^\dag h_1^\dag|0\rangle=-|\phi\rangle_{pmg}^{\mu_4}
\end{align}
i.e., the rotation eigenvalue $-1$ can be trivialized by atomic insulator $|\phi\rangle_{pmg}^{\mu_4}$. Hence, this 1D bubble construction can change the reflection eigenvalue of $\mu_2$ and rotation eigenvalue of $\mu_4$ simultaneously. Similar for 1D blocks labeled by $\tau_4$, and 1D bubble construction can change the reflection eigenvalue of $\mu_1$ and rotation eigenvalue of $\mu_3$ simultaneously. We further consider the 1D bubble equivalence on 1D blocks labeled by $\tau_1$ [cf. 1D bubble, here both yellow and red dots represent the 0D FSPT mode characterized by eigenvalue $-1$ of reflection symmetry operation, and they can be trivialized if we shrink them to a point]: this bubble construction can change the reflection eigenvalue of 0D blocks $\mu_1$ and $\mu_2$ simultaneously. Summarize all above trivializations, we know that rotation/reflection eigenvalues at 0D blocks $\mu_j,j=1,2,3,4$ are not independent. 

Subsequently consider the complex fermion sector: consider 1D bubble equivalence on 1D blocks $\tau_1$ [cf. 1D bubble, here yellow and red dots represent particle and hole, respectively, and they can be trivialized if we shrink them to a point]: it adds one complex fermion with $U^f(1)$ charge $+1$ (particle) at each 0D block $\mu_1$ and one complex fermion with $U^f(1)$ charge $-1$ (hole) at each 0D block $\mu_2$, hence the $U^f(1)$ charges at $\mu_1$ and $\mu_2$ are not independent. 

With the help of above discussions, we consider the 0D block-state decorations. The 0D block-state decorated on $\mu_1/\mu_3$ can be labeled by $(m_1/m_3,\pm)$, where $m_1/m_3\in\mathbb{Z}$ represents $U^f(1)$ charges on $\mu_1/\mu_3$ and $\pm$ represents the eigenvalues of reflection/2-fold rotational symmetry on $\mu_1/\mu_3$. Then we consider the trivial state labeled by:
\begin{align}
\left[(0,+),(0,+),(0,+),(0,+)\right]
\label{pmg trivial state}
\end{align}
We should note that there are two different types of 1D bubble constructions on $\tau_1$ and $\tau_2$: particle-hole construction and 0D FSPT mode construction. Take 0D FSPT mode/particle-hole construction on $\tau_1$ and $\tau_2$ by $l_{1,j}/l_{2,j}~(j=1,2)$ times, and particle-hole construction on $\tau_3$ and $\tau_4$ by $l_3$ and $l_4$ times, and it will lead to a new 0D block-state labeled by:
\begin{align}
&\left[\left(l_{1,2}+l_{2,2}+2l_4,(-1)^{l_{1,1}+l_{2,1}+l_4}\right)\right.\nonumber\\
&\left(-l_{1,2}-l_{2,2}+2l_3,(-1)^{l_{1,1}+l_{2,1}+l_3}\right)\nonumber\\
&\left.\left(-2l_4,(-1)^{l_4}\right),~\left(-2l_3,(-1)^{l_3}\right)\right]
\label{pmg U(1) spinless trivial state}
\end{align}
And this state should be trivial. Alternatively, all 0D block-states can be viewed as vectors of an 8-dimensional vector space $V$, where the complex fermion components are $\mathbb{Z}$-valued and all other components are $\mathbb{Z}_2$-valued. Then all trivial 0D block-states with the form as Eq. (\ref{pmg U(1) spinless trivial state}) can be viewed as a vector subspace $V'$ of $V$. We note that $l_{2,1}/l_{2,2}$ should appears together with $l_{1,1}/l_{1,2}$, hence they are not independent. As a consequence, there are only four independent quantities in Eq. (\ref{pmg U(1) spinless trivial state}): $l_{1,1}$, $l_{1,2}$, $l_3$ and $l_4$. So the dimensionality of the vector subspace $V'$ should be 4. For the complex fermion sector, we have the following relation:
\begin{align}
&l_{2,2}+l_4-l_{2,2}+2l_3-2l_4=2l_3
\end{align}
i.e., there are only three independent quantities which serves a $\mathbb{Z}\times(2\mathbb{Z})^2$ trivialization. The remaining one degree of freedom of the vector subspace $V'$ should be attributed to the eigenvalues of point group symmetry actions, and serves a $\mathbb{Z}_2$ trivialization. Therefore, all trivial states with the form as shown in Eq. (\ref{pmg U(1) spinless trivial state}) compose the following group:
\begin{align}
\{\mathrm{TBS}\}_{pmg,0}^{U(1)}=\mathbb{Z}\times(2\mathbb{Z})^2\times\mathbb{Z}_2
\end{align}

Finally, the ultimate classification of crystalline topological phases protected by $pmg$ symmetry for 2D systems with spinless fermions is:
\begin{align}
\mathcal{G}_{pmg,0}^{U(1)}&=\mathbb{Z}^4\times\mathbb{Z}_2^4/\left[\mathbb{Z}\times(2\mathbb{Z})^2\times\mathbb{Z}_2\right]\nonumber\\
&=\mathbb{Z}\times\mathbb{Z}_4^2\times\mathbb{Z}_2
\end{align}

For systems with spin-1/2 fermions, the reflection property of $|\phi\rangle_{pmg}^{\mu_2}$ and rotation property of $|\phi\rangle_{pmg}^{\mu_4}$ are changed by an additional $-1$, which leads to no trivialization. 1D bubble equivalence on 1D blocks $\tau_1$ is identical with spinless fermions which leads to the nonindependence of reflection eigenvalues of 0D blocks $\mu_1$ and $\mu_2$. And it is easy to verify that the complex fermion decorations for spinless and spin-1/2 fermions are identical. Repeatedly take aforementioned 1D bubble construction on the trivial state (\ref{pmg trivial state}), it will lead to the following new 0D block-state:
\begin{align}
&\left[\left(l_{1,2}+l_{2,2}+2l_4,(-1)^{l_{1,1}+l_{2,1}}\right)\right.\nonumber\\
&\left(-l_{1,2}-l_{2,2}+2l_3,(-1)^{l_{1,1}+l_{2,1}}\right)\nonumber\\
&\left.\left(-2l_4,+\right),~\left(-2l_3,+\right)\right]\nonumber
\end{align}
It is straightforward to see that the complex fermion sector is identical with the spinless case, and the 1D bubble construction can only serve a $\mathbb{Z}_2$ trivialization on rotation/reflection eigenvalues because there are only one independent integer $l_{1,1}+l_{2,1}$. Therefore, the ultimate classification of crystalline topological phases protected by $pmg$ symmetry for 2D systems with spin-1/2 fermions is:
\begin{align}
\mathcal{G}_{pmg,1/2}^{U(1)}&=\mathbb{Z}^4\times\mathbb{Z}_2^4/\mathbb{Z}\times(2\mathbb{Z})^2\times\mathbb{Z}_2\nonumber\\
&=\mathbb{Z}\times\mathbb{Z}_4^2\times\mathbb{Z}_2
\end{align}

\subsection{$p4$}
\begin{figure}
\centering
\includegraphics[width=0.46\textwidth]{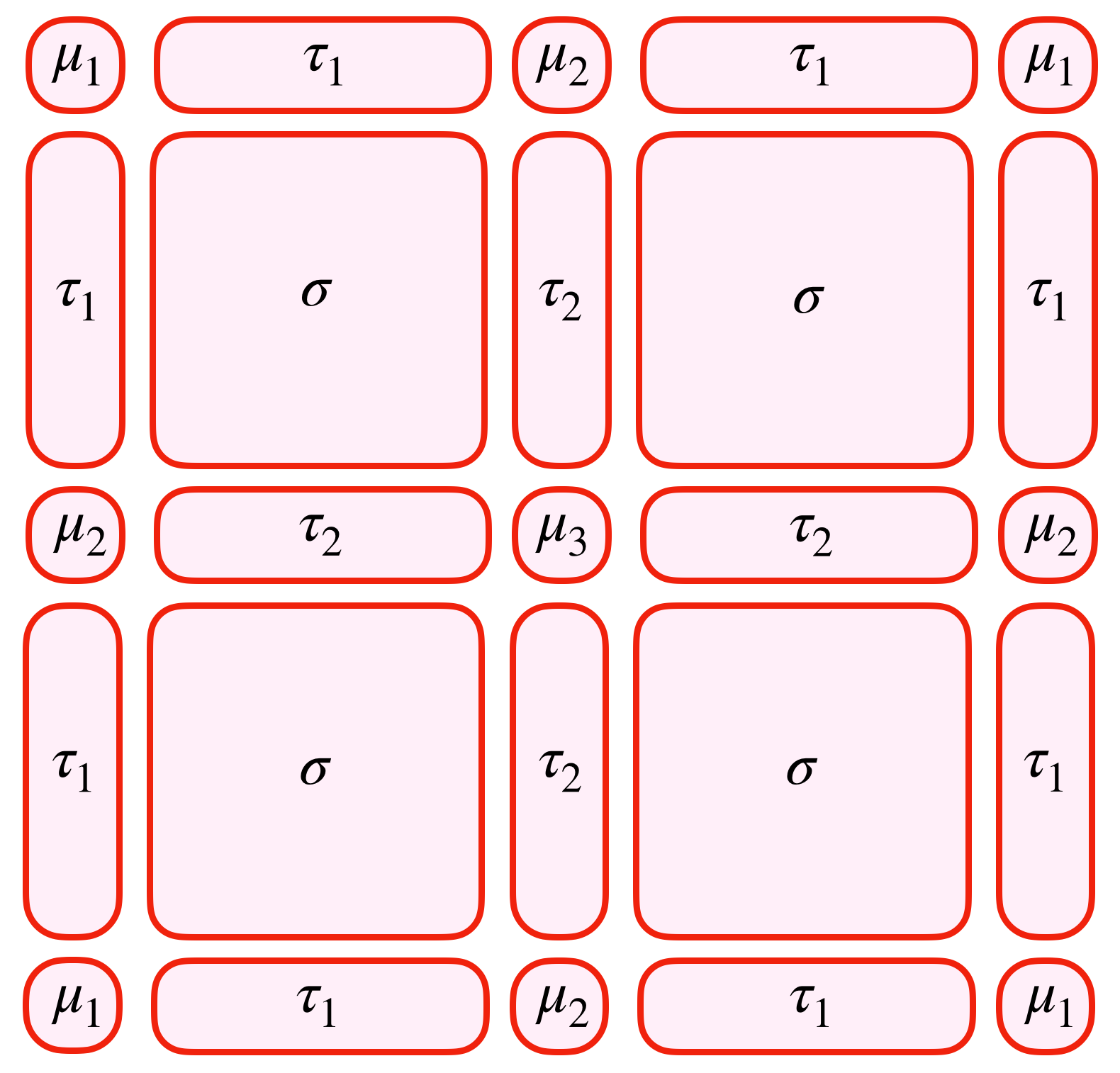}
\caption{\#10 wallpaper group $p4$ and its cell decomposition.}
\label{p4}
\end{figure}

The corresponding point group of this case is 4-fold rotational group $C_4$. For 2D blocks $\sigma$, 1D blocks $\tau_1$ and $\tau_2$, there is no on-site symmetry; for 0D blocks $\mu_1$ and $\mu_3$, the on-site symmetry group is $\mathbb{Z}_4$ via the 4-fold rotational symmetry group $C_4$ acting internally, and for 0D blocks $\mu_2$, the on-site symmetry is $\mathbb{Z}_2$ via the 2-fold rotational symmetry group $C_2\subset C_4$ acting internally, see Fig. \ref{p4}.

\subsubsection{Spinless fermions}
First, we consider the 0D block-state decorations. For 0D blocks labeled by $\mu_1$ and $\mu_3$, the total symmetry group of each of them is $\mathbb{Z}_2^f\times\mathbb{Z}_4$, and the classification data can be characterized by different 1D irreducible representations of the symmetry group $\mathbb{Z}_2^f\times\mathbb{Z}_4$:
\begin{align}
\mathcal{H}^1\left[\mathbb{Z}_2^f\times\mathbb{Z}_4,U(1)\right]=\mathbb{Z}_2\times\mathbb{Z}_4
\end{align}
Here $\mathbb{Z}_2$ is from the fermion parity, and $\mathbb{Z}_4$ is from the rotation eigenvalues. So the corresponding 0D block-state can be labeled by $(\nu_1/\nu_3,\pm)$, here $\nu_1/\nu_3\in\{1,i,-1,-i\}$ represents the rotation eigenvalue, and $\pm$ represents the fermion parity. For 0D blocks labeled by $\mu_2$, the total symmetry group of each of them is $\mathbb{Z}_2^f\times\mathbb{Z}_2$, and the classification data can be characterized by different 1D irreducible representations of the symmetry group $\mathbb{Z}_2^f\times\mathbb{Z}_2$:
\begin{align}
\mathcal{H}^1\left[\mathbb{Z}_2^f\times\mathbb{Z}_2,U(1)\right]=\mathbb{Z}_2^2
\end{align}
One $\mathbb{Z}_2$ is from the fermion parity, and the other is from the rotation eigenvalue $-1$, and at each 0D block, the block-state can be labeled by $(\pm,\pm)$, here these two $\pm$'s represent the fermion parity and rotation eigenvalue, respectively. According to this notation, the obstruction-free 0D block-states form the following group:
\begin{align}
\{\mathrm{OFBS}\}_{p4,0}^{\mathrm{0D}}=\mathbb{Z}_2^4\times\mathbb{Z}_4^2
\end{align}
and the group elements can be labeled by (three brackets represent the block-states at $\mu_j,j=1,2,3$):
\[
[(\nu_1,\pm),(\pm,\pm),(\nu_3,\pm)]
\]

Subsequently we investigate the 1D block-state decoration. The unique possible 1D block-state is Majorana chain because of the absence of the on-site symmetry. On 1D blocks labeled by $\tau_1$, the 1D block-state decoration leaves 4 dangling Majorana modes on each $\mu_1$, with the following rotational symmetry properties:
\begin{align}
\bs{R}_{\mu_1}:~\gamma_j\mapsto\gamma_{j+1},~~j=1,2,3,4.
\end{align}
where $\bs{R}_{\mu_1}$ is the generator of $C_4$ group: 4-fold rotation operation centred at each 0D block labeled by $\mu_1$. and all subscripts are taken with modulo 4. Consider the local fermion parity and its symmetry property:
\begin{align}
P_f=-\prod\limits_{j=1}^4\gamma_j,~~\bs{R}_{\mu_1}:~P_f\mapsto-P_f
\end{align}
Hence these 4 Majorana modes cannot be gapped out in a symmetric way, and the no-open-edge condition is violated. Similar for $\tau_2$, and there is no nontrivial 1D block-state:
\begin{align}
\{\mathrm{OFBS}\}_{p4,0}^{\mathrm{1D}}=\mathbb{Z}_1
\end{align}

With all obstruction-free block-states, subsequently we discuss about all possible trivializations. First, we consider about the 2D bubble equivalences: as we discussed in the main text, only type-\2 (i.e., ``Majorana bubbles'' with anti-PBC) 2D bubble equivalence is valid because there is no 0D block as the center of even-fold dihedral group. Enlarge all “Majorana bubble” construction can be deformed to double Majorana chains at arbitrary 0D block that can be trivialized because there is no on-site symmetry and the classification of 1D invertible topological phases (i.e., Majorana chain) is $\mathbb{Z}_2$. Furthermore, similar with the $p2$ case, 2D ``Majorana bubble'' construction changes the fermion parities of all 0D blocks simultaneously.

Subsequently we consider the 1D bubble equivalences. We study the role of rotational symmetry. Consider the 1D bubble equivalence on 1D blocks labeled by $\tau_1$ [cf. 1D bubble, here both yellow and red dots represent the complex fermions]: near $\mu_1$, there are 4 complex fermions which form an atomic insulator:
\begin{align}
|\psi\rangle_{p4}^{\mu_1}=c_1^\dag c_2^\dag c_3^\dag c_4^\dag |0\rangle
\end{align}
with rotation property as ($\bs{R}_{\mu_1}$ represents the 4-fold rotation operation centred at the 0D block labeled by $\mu_1$):
\begin{align}
\bs{R}_{\mu_1}|\psi\rangle_{p4}^{\mu_1}=c_2^\dag c_3^\dag c_4^\dag c_1^\dag|0\rangle=-|\psi\rangle_{p4}^{\mu_1}
\end{align}
i.e., the rotation eigenvalue $-1$ can be trivialized by the atomic insulator $|\psi\rangle_{p4}^{\mu_1}$. Near $\mu_2$, there are 2 complex fermions which form another atomic insulator:
\begin{align}
|\psi\rangle_{p4}^{\mu_2}=c_1'^\dag c_2'^\dag|0\rangle
\end{align}
with rotation property as ($\bs{R}_{\mu_2}$ represents the 2-fold rotation operation centred at the 0D block labeled by $\mu_2$):
\begin{align}
\bs{R}_{\mu_2}|\psi\rangle_{p4}^{\mu_2}=c_2'^\dag c_1'^\dag|0\rangle=-|\psi\rangle_{p4}^{\mu_2}
\end{align}
i.e., the rotation eigenvalue $-1$ can be trivialized by the atomic insulator $|\psi\rangle_{p4}^{\mu_2}$. Hence the rotation eigenvalues of $\mu_1$ and $\mu_2$ are not independent. 1D bubble equivalence on $\tau_2$ is similar, and therefore the rotation eigenvalues at $\mu_j$, $j=1,2,3$ are not independent. 

With all possible 2D and 1D bubble constructions, we are ready to study the trivial block-states. Start from the original trivial state (nothing is decorated on arbitrary blocks):
\begin{align}
[(1,+),(+,+),(1,+)]
\end{align}
If we take 2D bubble construction $l_0$ times, take 1D bubble equivalences on $\tau_1$ and $\tau_2$ by $l_1$ and $l_2$ times, above trivial state will be deformed to a new 0D block-state labeled by:
\begin{align}
&[((-1)^{l_1},(-1)^{l_0}),((-1)^{l_1+l_2},(-1)^{l_0}),\nonumber\\
&((-1)^{l_2},(-1)^{l_0})]
\label{p4 spinless trivial state}
\end{align}
According to the definition of bubble equivalence, all these 0D block-states should be trivial. It is straightforward to check that there are only three independent quantities ($l_j$, $j=0,1,2$) in Eq. (\ref{p4 spinless trivial state}), hence all trivial block-states form the following group:
\begin{align}
\{\mathrm{TBS}\}_{p4,0}=\{\mathrm{TBS}\}_{p4,0}^{\mathrm{0D}}=\mathbb{Z}_2^3
\end{align}
Therefore, all obstruction and trivialization free 0D/1D block-states are classified as:
\begin{align}
\begin{aligned}
&E_{p4,0}^{\mathrm{0D}}=\{\mathrm{OFBS}\}_{p4,0}^{\mathrm{0D}}/\{\mathrm{TBS}\}_{p4,0}^{\mathrm{0D}}=\mathbb{Z}_2\times\mathbb{Z}_4^2\\
&E_{p4,0}^{\mathrm{0D}}=\{\mathrm{OFBS}\}_{p4,0}^{\mathrm{0D}}/\{\mathrm{TBS}\}_{p4,0}^{\mathrm{0D}}=\mathbb{Z}_1
\end{aligned}
\end{align}
and all independent nontrivial block-states are labeled by the group elements of the following quotient group:
\begin{align}
\mathcal{G}_{p4}^0=E_{p4,0}^{\mathrm{0D}}=\mathbb{Z}_4\times\mathbb{Z}_2^3
\end{align}
We should notice that the group structure should be $\mathbb{Z}_4\times\mathbb{Z}_2^3$ rather than $\mathbb{Z}_2^5$, because rotation eigenvalue $i$ and $-i$ at each 0D block labeled by $\mu_1$ should be a nontrivial 0D block-state. There is no stacking between 1D and 0D block-states because there is no nontrivial 1D block-state, and the group structure of the classification data $E_{p4,0}$ has already been accurate. 

\subsubsection{Spin-1/2 fermions}
First, we investigate the 0D block-state decorations. For 0D blocks labeled by $\mu_1$ and $\mu_3$, the total symmetry group is $\mathbb{Z}_8^f$: nontrivial $\mathbb{Z}_2^f$ extension of the on-site symmetry $\mathbb{Z}_4$. All different 0D block-states can be characterized by different 1D irreducible representations of the corresponding symmetry group:
\begin{align}
\mathcal{H}^1\left[\mathbb{Z}_8^f,U(1)\right]=\mathbb{Z}_8
\end{align}
And there is no more trivialization. For the 0D blocks labeled by $\mu_2$, the total symmetry group is $\mathbb{Z}_4^f$: nontrivial $\mathbb{Z}_2^f$ extension of the on-site symmetry $\mathbb{Z}_2$. All different 0D block-states can be characterized by different 1D irreducible representations of the corresponding symmetry group:
\begin{align}
\mathcal{H}^1\left[\mathbb{Z}_4^f,U(1)\right]=\mathbb{Z}_4
\end{align}
And there is no trivialization. As a consequence, the classification attributed to 0D block-state decorations is:
\begin{align}
E_{p4,1/2}^{\mathrm{0D}}=\{\mathrm{OFBS}\}_{p4,1/2}^{\mathrm{0D}}=\mathbb{Z}_8^2\times\mathbb{Z}_4
\end{align}

Subsequently we investigate the 1D block-state decoration. The unique possible block-state is Majorana chain because of the absence of on-site symmetry. Majorana chain decoration on $\tau_1/\tau_3$ leaves 4 dangling Majorana modes at each 0D block $\mu_1/\mu_3$, and 2 dangling Majorana modes at $\mu_2$. The 4 Majorana modes at $\mu_1/\mu_3$ have the following rotational symmetry properties:
\begin{align}
\bs{R}_{\mu_1/\mu_3}:~(\gamma_1,\gamma_2,\gamma_3,\gamma_4)\mapsto(\gamma_2,\gamma_3,\gamma_4,-\gamma_1)
\end{align}
Where $\bs{R}_{\mu_1/\mu_3}$ is the generator of $C_4$ group: 4-fold rotation operation centred at each 0D block labeled by $\mu_1/\mu_3$. Then consider the local fermion parity and its symmetry property:
\begin{align}
P_f=-\prod\limits_{j=1}^4\gamma_j,~~\bs{R}_{\mu_1/\mu_3}:~P_f\mapsto P_f
\end{align}
Therefore, these 4 Majorana modes can be gapped out by some proper interactions in a symmetric way. Next the 2 Majorana modes at $\mu_2$ can be gapped out by an entanglement pair which respect all symmetry operations, identical with $p2$ case. Therefore the no-open-edge condition is satisfied, and finally the classification attributed to 1D block-state decorations is:
\begin{align}
E_{p4,1/2}^{\mathrm{1D}}=\{\mathrm{OFBS}\}_{p4,1/2}^{\mathrm{1D}}=\mathbb{Z}_2^2
\end{align}

With full classification data, we investigate the possible stacking between 1D and 0D blocks. If we decorate two Majorana chains on each 1D block labeled by $\tau_1$, it can be smoothly deformed to two copies of 0D state that has been shown in the main text ($p2$ case with spin-1/2 fermions) at each 0D block labeled by $\mu_1$ and one at each 0D block labeled by $\mu_2$. Near each $\mu_1$, the deformed 0D block-state has the following rotational property \cite{Srotation}:
\begin{align}
\bs{R}_{\mu_1}|\phi\rangle_{\mathrm{0D}}^{\mu_1}=e^{-i\pi/2}
\end{align}
Hence if a 0D block-state with eigenvalue $e^{i\pi q/4}$ under 4-fold rotation is attached to each 1D block-state (single Majorana chain decoration) near each 0D block labeled by $\mu_1$, the rotation eigenvalue $r$ of the obtained 0D block-state becomes:
\begin{align}
r=e^{-i\pi+i\pi q/2}
\end{align}
And $r=0$ if $q=2$. Therefore, there is an appropriate 1D block-state which itself form a $\mathbb{Z}_2$ structure under stacking, and there is no stacking between 1D and 0D block-states as a consequence. Near each $\mu_2$, similar with the $p2$ case, there is a stacking between 1D and 0D block-states. Therefore, the ultimate classification with accurate group structure is:
\begin{align}
\mathcal{G}_{p4}^{1/2}=E_{p4,1/2}^{\mathrm{1D}}\times_{\omega_2}E_{p4,1/2}^{\mathrm{0D}}=\mathbb{Z}_2\times\mathbb{Z}_8^3
\end{align}
here the symbol ``$\times_{\omega_2}$'' means that 1D and 0D block-states $E_{p4,1/2}^{\mathrm{1D}}$ and $E_{p4,1/2}^{\mathrm{0D}}$ have nontrivial extension, and described by a nontrivial factor system of the following short exact sequence in mathematical language:
\begin{align}
0\rightarrow E_{p4,1/2}^{\mathrm{1D}}\rightarrow G_{p4}^{1/2}\rightarrow E_{p4,1/2}^{\mathrm{0D}}\rightarrow0
\end{align}

\subsubsection{With $U^f(1)$ charge conservation}
Then we consider the systems with $U^f(1)$ charge conservation. For each 0D block labeled by $\mu_1$ and $\mu_3$, different 0D block-states are characterized by different irreducible representations of symmetry group as:
\begin{align}
\mathcal{H}^1[U(1)\times\mathbb{Z}_4,U(1)]=\mathbb{Z}\times\mathbb{Z}_4
\end{align}
Here $\mathbb{Z}$ represents the complex fermion and $\mathbb{Z}_4$ represents the eigenvalues of 4-fold rotational symmetry operation. For each 0D block labeled by $\mu_2$, different 0D block-states are also characterized by different irreducible representations of symmetry group as:
\begin{align}
\mathcal{H}^1[U(1)\times\mathbb{Z}_2,U(1)]=\mathbb{Z}\times\mathbb{Z}_2
\end{align}
Here $\mathbb{Z}$ represents the complex fermion and $\mathbb{Z}_2$ represents the eigenvalues of 2-fold rotational symmetry operation. We should further consider possible trivializations. For systems with spinless fermions, consider the 1D bubble equivalence on $\tau_1$: we decorate a 1D bubble on each 1D block labeled by $\tau_1$, here yellow and red dots represent the fermionic particle and hole, respectively, and can be trivialized if we shrink them to a point. Near each 0D block labeled by $\mu_1$, there are four particles that form an atomic insulator:
\begin{align}
|\xi\rangle_{p4}^{\mu_1}=p_1^\dag p_2^\dag p_3^\dag p_4^\dag|0\rangle
\end{align}
with rotation property as:
\begin{align}
\bs{R}_{\mu_1}|\xi\rangle_{p4}^{\mu_1}=p_2^\dag p_3^\dag p_4^\dag p_1^\dag|0\rangle=-|\xi\rangle_{p4}^{\mu_1}
\end{align}
i.e., rotation eigenvalue $-1$ can be trivialized by the atomic insulator $|\xi\rangle_{p4}^{\mu_1}$ at each 0D block labeled by $\mu_1$. Near each 0D block labeled by $\mu_2$, there are two holes that form another atomic insulator:
\begin{align}
|\xi\rangle_{p4}^{\mu_2}=h_1^\dag h_2^\dag|0\rangle
\end{align}
with rotation property as:
\begin{align}
\bs{R}_{\mu_2}|\xi\rangle_{p4}^{\mu_2}=h_2^\dag h_1^\dag|0\rangle=-|\xi\rangle_{p4}^{\mu_2}
\end{align}
i.e., rotation eigenvalue $-1$ can be trivialized by the atomic insulator $|\xi\rangle_{p4}^{\mu_2}$ at each 0D block labeled by $\mu_2$. Thus the 1D bubble construction on $\tau_1$ can change the rotation eigenvalues of $\mu_1$ and $\mu_2$ simultaneously. Similar for 1D bubble equivalence on 1D blocks $\tau_2$ which can change the rotation eigenvalues of 0D blocks $\mu_2$ and $\mu_3$ simultaneously. Therefore, rotation eigenvalues of 0D blocks $\mu_j, j=1,2,3$ are not independent. 

Subsequently we consider the complex fermion sector. First of all, as shown in Fig. \ref{p4}, we should identify that there is only one 0D block labeled by $\mu_1/\mu_3$ per unit cell, but there are two 0D blocks labeled by $\mu_2$ per unit cell. Then consider 1D bubble equivalence on $\tau_2$ [cf. 1D bubble, here yellow and red dots represent particle and hole, respectively, and they can be trivialized if we shrink them to a point]: it adds four complex fermions with $U^f(1)$ charge $+1$ (particles) at each 0D block $\mu_3$ and two complex fermions with $U^f(1)$ charge $-1$ (holes) at each 0D block $\mu_2$, hence the $U^f(1)$ charges at $\mu_2$ and $\mu_3$ are not independent. Similar arguments can be held for the complex fermion decorations on 0D blocks $\mu_1$ and $\mu_2$. 

With the help of above discussions, we consider the 0D block-state decorations. The 0D block-state decorated on $\mu_j$ ($j=1,2,3$) can be labeled by $(m_j,\phi_j)$, where $m_j\in\mathbb{Z}$ represents $U^f(1)$ charges on $\mu_j$ and $\phi_j$ represent the eigenvalues of rotational symmetry on $\mu_j$. Start from the following trivial state:
\begin{align}
[(0,1),(0,1),(0,1)]
\label{p4 trivial state}
\end{align}
Take aforementioned 1D bubble construction on $\tau_j$ by $l_j$ times ($j=1,2$), and it will lead to the following 0D block-state:
\begin{align}
&\left[\left(4l_1,(-1)^{l_1}\right),\left(-2l_1+2l_2,(-1)^{l_1+l_2}\right)\right.\nonumber\\
&\left.\left(-4l_2,(-1)^{l_2}\right)\right]
\label{p4 U(1) spinless trivial state}
\end{align}
And this state should be trivial. Alternatively, all 0D block-states can be viewed as vectors of an 6-dimensional vector space $V$, where the complex fermion components are $\mathbb{Z}$-valued, two components attributed to $C_4$ centers $\mu_1$ and $\mu_3$ are $\mathbb{Z}_4$-valued, and one component attributed to $C_2$ center $\mu_2$ is $\mathbb{Z}_2$-valued. Then all trivial 0D block-states with the form as Eq. (\ref{p4 U(1) spinless trivial state}) can be viewed as a vector subspace $V'$ of $V$. It is easy to see that there are only two independent quantities in Eq. (\ref{p4 U(1) spinless trivial state}): $l_1$ and $l_2$. So the dimensionality of the vector subspace $V'$ should be 4. For the complex fermion sector,we have the following relation:
\begin{align}
-4l_1-2(-2l_1+2l_2)=-l_2
\end{align}
i.e., there are only two independent quantities which serves a $2\mathbb{Z}\times4\mathbb{Z}$ trivialization. Hence all trivial 0D block-states form the group:
\begin{align} 
\{\mathrm{TBS}\}_{p4,0}^{U(1)}=2\mathbb{Z}\times4\mathbb{Z}
\end{align}

Therefore, the ultimate classification of crystalline topological phases protected by $p4$ symmetry for 2D systems with spinless fermions is:
\begin{align}
\mathcal{G}_{p4,0}^{U(1)}&=\mathbb{Z}^3\times\mathbb{Z}_4^2\times\mathbb{Z}_2/(2\mathbb{Z}\times4\mathbb{Z})\nonumber\\
&=\mathbb{Z}\times\mathbb{Z}_8\times\mathbb{Z}_4^2\times\mathbb{Z}_2
\end{align}

For systems with spin-1/2 fermions, the rotation properties of $|\xi\rangle_{p4}^{\mu_1}$ and $|\xi\rangle_{p4}^{\mu_2}$ are changed by an additional $-1$, which leads to no trivialization. It is easy to verify that the complex fermion decorations for spinless and spin-1/2 fermions are identical for $p4$ symmetric case. For this case, Take aforementioned 1D bubble construction on $\tau_j$ by $l_j$ times ($j=1,2$) on trivial state (\ref{p4 trivial state}) will lead to another state:
\begin{align}
[(4l_1,1),(-2l_1+2l_2,1),(-4l_2,1)]
\end{align}
and the trivialization is almost identical with spinless case. Therefore, the ultimate classification of crystalline topological phases protected by $p4$ symmetry for 2D systems with spin-1/2 fermions is:
\begin{align}
\mathcal{G}_{p4,1/2}^{U(1)}&=\mathbb{Z}^3\times\mathbb{Z}_4^2\times\mathbb{Z}_2/(2\mathbb{Z}\times4\mathbb{Z})\nonumber\\
&=\mathbb{Z}\times\mathbb{Z}_8\times\mathbb{Z}_4^2\times\mathbb{Z}_2
\end{align}

\subsection{$p4g$}
\begin{figure}
\centering
\includegraphics[width=0.46\textwidth]{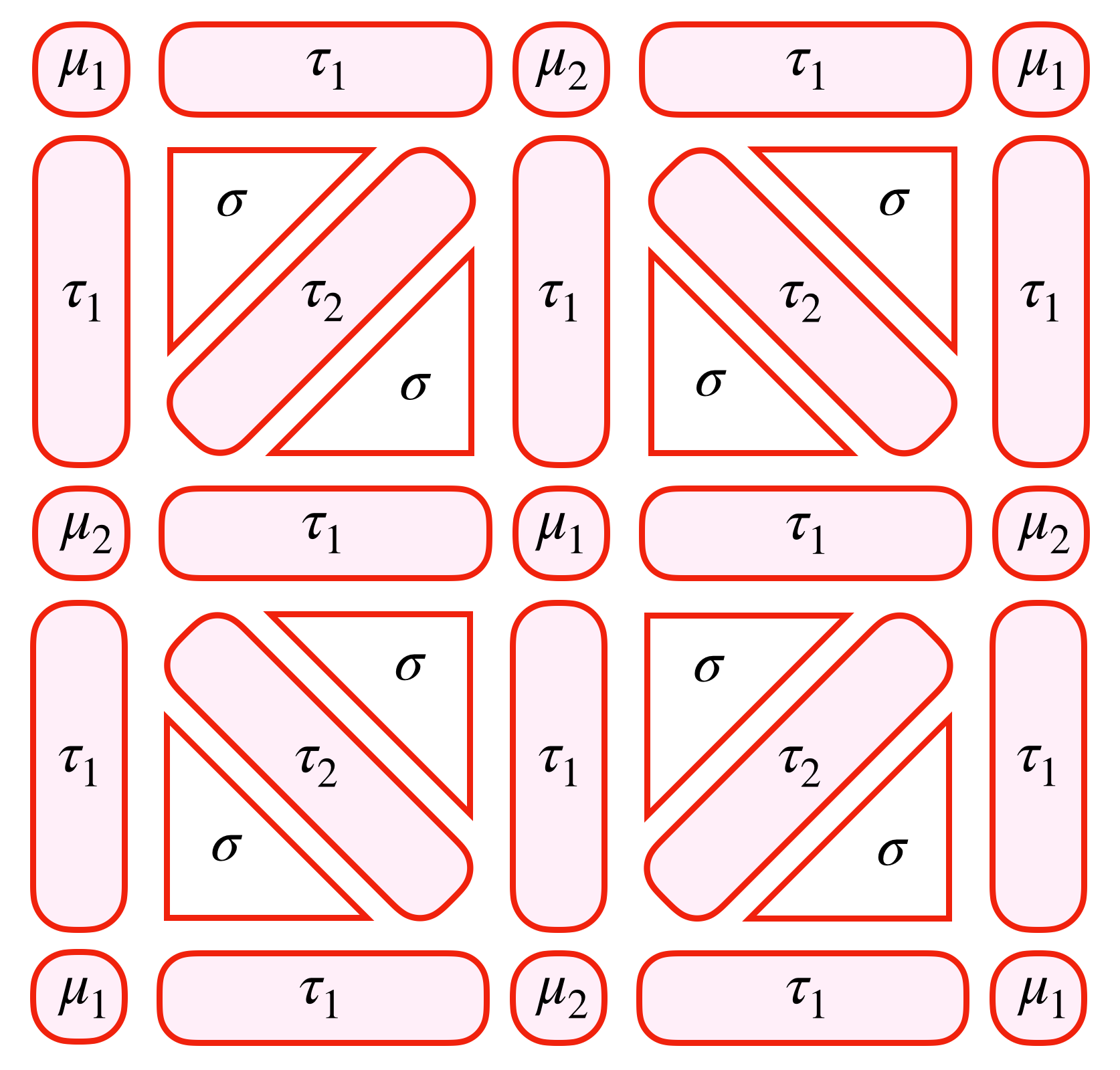}
\caption{\#12 wallpaper group $p4g$ and its cell decomposition.}
\label{p4g}
\end{figure}

The corresponding point group of this case is 4-fold dihedral group. For 2D blocks $\sigma$ and 1D blocks $\tau_1$, there is no on-site symmetry; for 1D blocks $\tau_2$, the on-site symmetry is $\mathbb{Z}_2$ which is attributed to the reflection symmetry acting internally; for 0D blocks $\mu_1$, the on-site symmetry is $\mathbb{Z}_4$ via the symmetry $C_4$ acting internally, and for 0D blocks $\mu_2$, the on-site symmetry is $\mathbb{Z}_2\rtimes\mathbb{Z}_2$ via the symmetry $D_2\in D_4$ acting internally, see Fig. \ref{p4g}. 

\subsubsection{Spinless fermions}
First, we consider the 0D block-state decoration. The different 0D block states at $\mu_1$ can be characterized by different 1-dimensional irreducible representations of the symmetry group:
\begin{align}
\mathcal{H}^1\left[\mathbb{Z}_2^f\times\mathbb{Z}_4,U(1)\right]=\mathbb{Z}_2\times\mathbb{Z}_4
\end{align}
Here $\mathbb{Z}_2$ represents the complex fermion, and $\mathbb{Z}_4$ represents the rotation eigenvalues, and different block-states on each $\mu_1$ can be labeled by $(\pm,\nu_1)$, where $\nu_1\in\{1,i,-1,-i\}$ represents the eigenvalue of 4-fold rotational symmetry, and $\pm$ represents the fermion parity. The classification data of 0D block states at $\mu_2$ can be calculated by general group super-cohomology:
\begin{align}
\mathcal{H}^0(\mathbb{Z}_2\rtimes\mathbb{Z}_2,\mathbb{Z}_2)\times\mathcal{H}^1\left[\mathbb{Z}_2\rtimes\mathbb{Z}_2,U(1)\right]=\mathbb{Z}_2\times\mathbb{Z}_2^2
\end{align}
Three $\mathbb{Z}_2$ have different physical meanings: the first $\mathbb{Z}_2$ represents the complex fermion, the second $\mathbb{Z}_2$ represents the rotation eigenvalue $-1$, and the third $\mathbb{Z}_2$ represents the reflection eigenvalue $-1$. Different block-states on each $\mu_2$ can be labeled by $(\pm,\pm,\pm)$, where these three $\pm$'s represent the fermion parity, the rotation eigenvalue $-1$ and the reflection eigenvalue $-1$, respectively. 

According to this notation, the obstruction-free 0D block-states form the following group:
\begin{align}
\{\mathrm{OFBS}\}_{p4g,0}^{\mathrm{0D}}=\mathbb{Z}_4\times\mathbb{Z}_2^4
\end{align}
and the group elements can be labeled by (two brackets represent the block-states at $\mu_1$ and $\mu_2$):
\begin{align}
[(\nu_1,\pm),(\pm,\pm,\pm)]
\end{align}

Subsequently we investigate the 1D block-state decoration. For $\tau_1$, the unique candidate 1D block-state is Majorana chain due to the absence of on-site symmetry; for $\tau_2$, the total on-site symmetry group is $\mathbb{Z}_2^f\times\mathbb{Z}_2$, hence the candidate 1D block-states are Majorana chain and 1D FSPT state. So all 1D block-states form a group:
\begin{align}
\{\mathrm{BS}\}_{p4g,0}^{\mathrm{1D}}=\mathbb{Z}_2^3
\end{align}
Then we discuss about the decorations of these two root phases separately.

\paragraph{Majorana chain decoration}Majorana chain decoration on 1D blocks labeled by $\tau_1$ leaves 4 dangling Majorana modes at each 0D block labeled by $\mu_1$ and $\mu_2$. At $\mu_1$, these 4 Majorana modes have the following rotational properties:
\begin{align}
\bs{R}_{\mu_1}:~\gamma_j\mapsto\gamma_{j+1},~j=1,2,3,4.
\end{align}
Where $\bs{R}_{\mu_1}$ represents the 4-fold rotation operation centred at each 0D block labeled by $\mu_1$, and all subscripts are taken with modulo 4. Consider the local fermion parity with its symmetry property under rotation:
\begin{align}
P_f=-\prod\limits_{j=1}^4\gamma_j,~~\bs{R}_{\mu_1}:~P_f\mapsto-P_f
\end{align}
Therefore, these 4 Majorana modes cannot be gapped out in a symmetric way. Equivalently, the no-open-edge condition cannot be satisfied.

Majorana chain decoration on 1D blocks labeled by $\tau_2$ leaves 4 dangling Majorana modes at each 0D block labeled by $\mu_2$, with the following rotation and reflection symmetry properties:
\begin{align}
\left.
\begin{aligned}
\bs{R}_{\mu_2}:~&(\gamma_1,\gamma_2,\gamma_3,\gamma_4)\mapsto(\gamma_3,\gamma_4,\gamma_1,\gamma_2)\\
\bs{M}_{\tau_2}:~&(\gamma_1,\gamma_2,\gamma_3,\gamma_4)\mapsto(\gamma_1,\gamma_4,\gamma_3,\gamma_2)
\end{aligned}
\right.
\end{align}
where $\bs{R}_{\mu_2}$ is the generator of $C_4$ group: 2-fold rotation operation centred at each 0D block labeled by $\mu_2$, and $\bs{M}_{\tau_2}$ represents the reflection operation with the axis coincide with the 1D blocks labeled by $\tau_2$. Then consider the local fermion parity and the corresponding rotation and reflection symmetry properties:
\begin{align}
P_f=-\prod\limits_{j=1}^4\gamma_j,~~\left\{
\begin{aligned}
&\bs{R}_{\mu_2}:~P_f\mapsto P_f\\
&\bs{M}_{\tau_2}:~P_f\mapsto-P_f
\end{aligned}
\right.
\end{align}
Hence these 4 Majorana modes cannot be gapped out in a symmetric way, equivalently the no-open-edge condition cannot be satisfied.
\paragraph{1D FSPT state decoration}1D FSPT state decoration on 1D blocks labeled by $\tau_2$ leaves 8 dangling Majorana modes at each 0D block labeled by $\mu_2$, with the following rotation and reflection symmetry properties:
\begin{align}
\left.
\begin{aligned}
\bs{R}_{\mu_2}:~&\gamma_j\mapsto\gamma_{j+2},~\gamma_j’\mapsto\gamma_{j+2}'\\
\bs{M}_{\tau_2}:~&\gamma_j\mapsto-\gamma_{6-j},~\gamma_j'\mapsto\gamma_{6-j}'
\end{aligned}
\right.,~j=1,2,3,4.
\end{align}
Where all subscripts are taken with modulo 4. Then consider the local fermion parity with the corresponding rotation and reflection symmetry properties:
\begin{align}
P_f=\prod\limits_{j=1}^4\gamma_j\gamma_j',~~\bs{R}_{\mu_2},\bs{M}_{\tau_2}:~P_f\mapsto P_f
\end{align}
Thus these 8 Majorana modes can be gapped out by some proper interactions (just like the $p4m$ case) in a symmetric way. Equivalently, the no-open-edge condition is satisfied, and the obstruction-free 1D block-states form the following group:
\begin{align}
\{\mathrm{OFBS}\}_{p4g,0}^{\mathrm{1D}}=\mathbb{Z}_2
\end{align}
and the group elements can be labeled by $n_2=0,1$ that represents the number of decorated 1D FSPT states on $\tau_2$. Different obstruction-free block-states can be labeled by:
\begin{align}
[(\pm,\nu_1),(\pm,\pm,\pm);m_2]
\end{align}
here the first two brackets represent the 0D block-states at $\mu_1$ and $\mu_2$, and the last quantity represents the 1D block-states at $\tau_2$.

With all obstruction-free block-states, subsequently we discuss about all possible trivializations. First, we consider about the 2D bubble equivalence: as we discussed in the main text, both types of ``Majorana bubble'' constructions are allowed because 0D blocks labeled by $\mu_2$ are the centers of $D_2$ point group symmetry, including ``Majorana bubbles'' with both PBC and anti-PBC. Similar with the $p4m$ case, both types of ``Majorana bubbles'' can be deformed to double Majorana chains at each nearby 1D block, but the effects of them are distinct: near each 1D block labeled by $\tau_1$, these double Majorana chains can be trivialized because there is no on-site symmetry on $\tau_1$ and the classification of 1D invertible topological phases (i.e., Majorana chain) is $\mathbb{Z}_2$; near each 1D block labeled by $\tau_2$, these double Majorana chains cannot be trivialized because there is an on-site $\mathbb{Z}_2$ symmetry on each $\tau_2$ by internal action of reflection symmetry, and this $\mathbb{Z}_2$ action exchanges these two Majorana chains, and this is exactly the definition of the nontrivial 1D FSPT phase protected by on-site $\mathbb{Z}_2$ symmetry. Furthermore, similar with the $p4m$ case, there is no effect on 0D blocks labeled by $\mu_2$ by taking 2D ``Majorana'' bubble equivalence, because the alternative Majorana chain surrounding each $\mu_2$ is not compatible with the reflection operations; nevertheless, similar with the $p2$ case, 2D ``Majorana bubble'' construction changes the fermion parity of each 0D block labeled by $\mu_1$ because there is no reflection operation on 0D block $\mu_1$, and the alternative Majorana chain surrounding each $\mu_1$ is compatible with all other symmetry operations. We notice that the ``Majorana bubble'' constructions with both PBC and anti-PBC are equivalent, so take one of them into account is enough.

Subsequently we consider the 1D bubble equivalences. Consider the 1D bubble equivalence on 1D blocks labeled by $\tau_1$ [cf. 1D bubble, here both yellow and red dots represent the complex fermions]: Near each 0D block labeled by $\mu_1$, there are 4 complex fermions which form an atomic insulator:
\begin{align}
|\psi\rangle_{p4g}^{\mu_1}=c_1^\dag c_2^\dag c_3^\dag c_4^\dag|0\rangle
\end{align}
with rotation property as ($\bs{R}_{\mu_1}$ represents the 4-fold rotation operation centred at the 0D block labeled by $\mu_1$):
\begin{align}
\bs{R}_{\mu_1}|\psi\rangle_{p4g}^{\mu_1}=c_2^\dag c_3^\dag c_4^\dag c_1^\dag|0\rangle=-|\psi\rangle_{p4g}^{\mu_1}
\end{align}
i.e., $|\psi\rangle_{p4g}^{\mu_1}$ can trivialized the rotation eigenvalue $-1$ at each 0D block labeled by $\mu_1$; near each 0D block labeled by $\mu_2$, there are 4 complex fermions which form another atomic insulator:
\begin{align}
|\psi\rangle_{p4g}^{\mu_2}=c_1'^\dag c_2'^\dag c_3'^\dag c_4'^\dag|0\rangle
\end{align}
with rotation property as ($\bs{R}_{\mu_2}$ represents the 2-fold rotation operation centred at the 0D block labeled by $\mu_2$):
\begin{align}
\bs{R}_{\mu_2}|\psi\rangle_{p4g}^{\mu_2}=c_3'^\dag c_4'^\dag c_1'^\dag c_2'^\dag|0\rangle=|\psi\rangle_{p4g}^{\mu_2}
\end{align}
i.e., there is no trivialization on rotation eigenvalues at each 0D block labeled by $\mu_2$. Therefore, the 1D bubble construction on $\tau_1$ solely trivializes the rotation eigenvalue $-1$ at each 0D block labeled by $\mu_1$. Then consider the 1D bubble equivalence on 1D blocks labeled by $\tau_2$ [cf. 1D bubble, here both yellow and red dots represent the complex fermions]: near each 0D block labeled by $\mu_2$, there are 4 complex fermions which further forms an atomic insulator:
\begin{align}
|\phi\rangle_{p4g}^{\mu_2}=a_1^\dag a_2^\dag a_3^\dag a_4^\dag|0\rangle
\end{align}
with the rotation property as ($\bs{R}_{\mu_2}$ represents the 2-fold rotation operation centred at 0D block labeled by $\mu_2$):
\begin{align}
\bs{R}_{\mu_2}|\phi\rangle_{p4g}^{\mu_2}=a_3^\dag a_4^\dag a_1^\dag a_2^\dag|0\rangle=|\phi\rangle_{p4g}^{\mu_2}
\end{align}
i.e., there is no trivialization at each 0D block labeled by $\mu_2$. 

And the role of reflection symmetry should also be investigated. Reflection symmetry solely acts on 0D blocks labeled by $\mu_2$ internally. Near each 0D block labeled by $\mu_2$, again we consider the atomic insulator $|\phi\rangle_{p4g}^{\mu_2}$, with the reflection property as ($\bs{M}_{\tau_2}$ represents the reflection operation with the axis coincide with the 1D blocks labeled by $\tau_2$):
\begin{align}
\bs{M}_{\tau_2}|\phi\rangle_{p4g}^{\mu_2}=a_1^\dag a_4^\dag a_3^\dag a_2^\dag|0\rangle=-|\phi\rangle_{p4g}^{\mu_2}
\end{align}
i.e., the reflection eigenvalue $-1$ at $\mu_2$ can be trivialized by the atomic insulator $|\phi\rangle_{p4g}^{\mu_2}$. 

With all possible 2D and 1D bubble constructions, we are ready to study the trivial block-states. Start from the original trivial state (nothing is decorated on arbitrary blocks):
\begin{align}
[(+,1),(+,+,+)]
\end{align}
If we take 2D bubble construction $l_0$ times, take 1D bubble equivalences on $\tau_1$ and $\tau_2$ by $l_1$ and $l_2$ times, above trivial state will be deformed to a new 0D block-state labeled by:
\begin{align}
[((-1)^{l_0},(-1)^{l_1}),(+,+,(-1)^{l_2})]
\label{p4g spinless trivial state}
\end{align}
According to the definition of bubble equivalence, all these 0D block-states should be trivial. It is straightforward to check that there are only three independent quantities ($l_j$, $j=0,1,2$) in Eq. (\ref{p4g spinless trivial state}), hence all trivial block-states form the following group:
\begin{align}
\{\mathrm{TBS}\}_{p4g,0}=\mathbb{Z}_2^3
\end{align}
Therefore, all obstruction and trivialization free 0D/1D block-states are classified as:
\begin{align}
\begin{aligned}
&E_{p4g,0}^{\mathrm{0D}}=\{\mathrm{OFBS}\}_{p4g,0}^{\mathrm{0D}}/\{\mathrm{TBS}\}_{p4g,0}^{\mathrm{0D}}=\mathbb{Z}_2^4\\
&E_{p4g,0}^{\mathrm{1D}}=\{\mathrm{OFBS}\}_{p4g,0}^{\mathrm{1D}}/\{\mathrm{TBS}\}_{p4g,0}^{\mathrm{1D}}=\mathbb{Z}_1
\end{aligned}
\end{align}
and all independent nontrivial block-states are labeled by the group elements of the following quotient group:
\begin{align}
\mathcal{G}_{p4g}^0=E_{p4g,0}^{\mathrm{0D}}=\mathbb{Z}_2^4
\end{align}
and $\mathbb{Z}_2$'s are from nontrivial 0D block-states, so it is obvious that there is no stacking between different block-states, and the group structure of $\mathcal{G}_{p4g}^0$ has already been accurate.

\subsubsection{Spin-1/2 fermions}
First, we consider the 0D block-state decorations. For 0D blocks labeled by $\mu_1$, the total symmetry group at each of them is $\mathbb{Z}_8^f$: nontrivial $\mathbb{Z}_2^f$ extension of the on-site symmetry $\mathbb{Z}_4$, and the different block-states on each of them can be characterized by different 1D irreducible representations of the corresponding symmetry group:
\begin{align}
\mathcal{H}^1\left[\mathbb{Z}_8^f,U(1)\right]=\mathbb{Z}_8
\end{align}
For 0D blocks labeled by $\mu_2$, the classification data can be calculated by general group super-cohomology:
\begin{align}
\mathcal{H}^0(\mathbb{Z}_2\rtimes\mathbb{Z}_2,\mathbb{Z}_2)\times\mathcal{H}^1\left[\mathbb{Z}_2\rtimes\mathbb{Z}_2,U(1)\right]=\mathbb{Z}_2\times\mathbb{Z}_2^2
\end{align}
and the first $\mathbb{Z}_2$ who represents the odd fermion parity is obstructed \cite{Sgeneral2}. As a consequence, the classification attributed to 0D block-state decorations is:
\begin{align}
E_{p4g,1/2}^{\mathrm{0D}}=\mathbb{Z}_8\times\mathbb{Z}_2^2
\end{align}

Subsequently we investigate the 1D block-state decoration. For $\tau_1$, the unique candidate 1D block-state is Majorana chain due to the absence of on-site symmetry; for $\tau_2$, the total symmetry group is $\mathbb{Z}_4^f$, hence there is no nontrivial 1D block-state due to the trivial classification for the corresponding FSPT phases. Majorana chain decoration on $\tau_1$ leaves 4 dangling Majorana modes at each 0D block labeled by $\mu_1$ and $\mu_2$. At each of them, these 4 Majorana modes have the following rotational symmetry properties:
\begin{align}
\bs{R}_{\mu_1}:~(\gamma_1,\gamma_2,\gamma_3,\gamma_4)\mapsto(\gamma_2,\gamma_3,\gamma_4,-\gamma_1)
\end{align}
Then consider the local fermion parity at each 0D block and its symmetry property:
\begin{align}
P_f=-\prod\limits_{j=1}^4\gamma_j,~~\bs{R}_{\mu_1}:~P_f\mapsto P_f
\end{align}
Thus these 4 Majorana modes can be gapped out by some proper interactions in a symmetric way; at each $\mu_1$, the corresponding 2 Majorana modes which can be gapped out by an entanglement pair in a symmetric way. Hence the no-open-edge condition is satisfied. Hence for the system with spin-1/2 fermions, the classification of 2D FSPT phases protected by $p4g$ symmetry attributed to the 1D block-state decoration is:
\begin{align}
E_{p4g,1/2}^{\mathrm{1D}}=\mathbb{Z}_2
\end{align}
Similar with the $p4$ case, there is no stacking between 1D and 0D block-states, and the ultimate classification with accurate group structure is:
\begin{align}
\mathcal{G}_{p4g}^{1/2}=\mathbb{Z}_8\times\mathbb{Z}_2^3
\end{align}

\subsubsection{With $U^f(1)$ charge conservation}
Then we consider the systems with $U^f(1)$ charge conservation. For each 0D block labeled by $\mu_1$, different 0D block-states are characterized by different irreducible representations of symmetry group as:
\begin{align}
\mathcal{H}^1[U(1)\times\mathbb{Z}_4,U(1)]=\mathbb{Z}\times\mathbb{Z}_4
\end{align}
Here $\mathbb{Z}$ represents the complex fermion and $\mathbb{Z}_4$ represents the eigenvalues of 4-fold rotational symmetry operation. For each 0D block labeled by $\mu_2$, different 0D block-states are also characterized by different irreducible representations of symmetry group as:
\begin{align}
\mathcal{H}^1[U(1)\times(\mathbb{Z}_2\rtimes\mathbb{Z}_2),U(1)]=\mathbb{Z}\times\mathbb{Z}_2^2
\end{align}
Here $\mathbb{Z}$ represents the complex fermion, and two $\mathbb{Z}_2$'s represent the rotation eigenvalue $-1$ and the second $\mathbb{Z}_2$ represents the eigenvalue $-1$ of reflection symmetry operation. We should further consider possible trivializations. For systems with spinless fermions, consider the 1D bubble equivalence on $\tau_1$: we decorate a 1D bubble on each 1D block labeled by $\tau_1$, here yellow and red dots represent the fermionic particle and hole, respectively, and can be trivialized if we shrink them to a point. Near each 0D block labeled by $\mu_1$, there are four particles that can form an atomic insulator:
\begin{align}
|\xi\rangle_{p4g}^{\mu_1}=p_1^\dag p_2^\dag p_3^\dag p_4^\dag|0\rangle
\end{align}
with rotation property as:
\begin{align}
\bs{R}_{\mu_1}|\xi\rangle_{p4g}^{\mu_1}=p_2^\dag p_3^\dag p_4^\dag p_1^\dag|0\rangle=-|\xi\rangle_{p4g}^{\mu_1}
\end{align}
i.e., rotation eigenvalue $-1$ at each 0D block labeled by $\mu_1$ can be trivialized by the atomic insulator $|\xi\rangle_{p4g}^{\mu_1}$. Near $\mu_2$, there are four holes that form another atomic insulator:
\begin{align}
|\xi\rangle_{p4g}^{\mu_2}=h_1^\dag h_2^\dag h_3^\dag h_4^\dag|0\rangle
\end{align}
with rotation property as:
\begin{align}
\bs{R}_{\mu_2}|\xi\rangle_{p4g}^{\mu_2}=h_3^\dag h_4^\dag h_1^\dag h_2^\dag|0\rangle=|\xi\rangle_{p4g}^{\mu_2}
\end{align}
i.e., there is no trivialization on rotation eigenvalues at each 0D block labeled by $\mu_2$. Therefore, the 1D bubble construction on $\tau_1$ solely trivializes the rotation eigenvalue $-1$ at each 0D block labeled by $\mu_1$. Then consider the 1D bubble equivalence on $\tau_2$ [cf. 1D bubble, here yellow and red dots represent the fermionic particle and hole, respectively, and can be trivialized if we shrink them to a point]: near each 0D block labeled by $\mu_2$, there are two particles and two holes that form an atomic insulator:
\begin{align}
|\eta\rangle_{p4g}^{\mu_2}=p_1'^\dag p_2'^\dag h_1'^\dag h_2'^\dag|0\rangle
\end{align}
with rotation and reflection property as:
\begin{align}
\begin{aligned}
&\bs{R}_{\mu_2}|\eta\rangle_{p4g}^{\mu_2}=p_2'^\dag p_1'^\dag h_2'^\dag h_1'^\dag=|\eta\rangle_{p4g}^{\mu_2}\\
&\bs{M}_{\tau_2}|\eta\rangle_{p4g}^{\mu_2}=p_1'^\dag p_2'^\dag h_2'^\dag h_1'^\dag=-|\eta\rangle_{p4g}^{\mu_2}
\end{aligned}
\end{align}
i.e., the reflection eigenvalue $-1$ at $\mu_2$ can be trivialized by the atomic insulator $|\eta\rangle_{p4g}^{\mu_2}$. 

Subsequently we consider the complex fermion sector: consider 1D bubble equivalence on 1D blocks $\tau_1$ [cf. 1D bubble, here yellow and red dots represent particle and hole, respectively, and they can be trivialized if we shrink them to a point]: it adds four complex fermions with $U^f(1)$ charge $+1$ (particles) at each 0D block $\mu_1$ and four complex fermions with $U^f(1)$ charge $-1$ (holes) at each 0D block $\mu_2$, hence the $U^f(1)$ charges at $\mu_1$ and $\mu_2$ are not independent. 

With the help of above discussions, we consider the 0D block-state decorations. The 0D block-state decorated on $\mu_1$ can be labeled by $(m_1,\phi_1)$, where $m_1\in\mathbb{Z}$ represents $U^f(1)$ charges on $\mu_1$, $\phi_1$ represents the eigenvalues of 4-fold rotational symmetry on $\mu_1$; the 0D block-state decorated on $\mu_2$ can be labeled by $(m_2,\pm,\pm)$, where $m_2\in\mathbb{Z}$ represents $U^f(1)$ charges on $\mu_2$, two $\pm$'s represent the eigenvalues of 2-fold rotational symmetry and reflection symmetry on $\mu_2$, respectively. Start from the following trivial state:
\begin{align}
[(0,1),(0,+,+)]
\label{p4g trivial state}
\end{align}
Take aforementioned 1D bubble construction on $\tau_j$ by $l_j$ times ($j=1,2$), and it will lead to the following 0D block-state:
\begin{align}
\left[\left(4l_1,(-1)^{l_1}\right),\left(-4l_1,+,(-1)^{l_2}\right)\right]
\end{align}
And this state should be trivial. First, we consider the sector of complex fermion decoration: there are only one independent quantity $4l_1$, hence the 1D bubble construction serves a $4\mathbb{Z}$. Furthermore, once we fix a $l_1$, the rotation eigenvalue $(-1)^{l_1}$ is determined, so there is no trivialization in rotation sector. Subsequently the 1D bubble construction can serve a $\mathbb{Z}_2$ in the reflection sector via the phase factor $(-1)^{l_1+l_2}$. Therefore, the ultimate classification of crystalline topological phases protected by $p4g$ symmetry for 2D systems with spinless fermions is:
\begin{align}
\mathcal{G}_{p4g,0}^{U(1)}&=\mathbb{Z}^2\times\mathbb{Z}_4\times\mathbb{Z}_2^2/\left(4\mathbb{Z}\times\mathbb{Z}_2\right)\nonumber\\
&=\mathbb{Z}\times\mathbb{Z}_8\times\mathbb{Z}_2^2
\end{align}

For systems with spin-1/2 fermions, the rotation property of $|\xi\rangle_{p4g}^{\mu_1}$ at each 0D block $\mu_1$ and reflection property of $|\eta\rangle_{p4g}^{\mu_2}$ at each 0D block $\mu_2$ are changed by an additional $-1$, which leads to no trivialization. Furthermore, like the $p4m$ case, the classification data of the 0D block-states of 0D blocks labeled by $\mu_2$ can be characterized by different 1D irreducible representations of the full symmetry group:
\begin{align}
\mathcal{H}^1\left[U^f(1)\rtimes_{\omega_2}(\mathbb{Z}_2\rtimes\mathbb{Z}_2),U(1)\right]=\mathbb{Z}\times\mathbb{Z}_2^2
\end{align}
we should notice that for systems with spin-1/2 fermions, we can only decorate even $U^f(1)$ charges on each 0D block labeled by $\mu_2$. Repeatedly take aforementioned 1D bubble construction on $\tau_1$ by $l_1$ times on trivial state (\ref{p4g trivial state}) will lead to another state:
\begin{align}
\left[\left(4l_1,1\right),\left(-4l_1,+,+\right)\right]
\end{align}
Hence the 1D bubble construction serves a $4\mathbb{Z}$ in complex fermion sector. Therefore, the ultimate classification of crystalline topological phases protected by $p4g$ symmetry for 2D systems with spin-1/2 fermions is:
\begin{align}
\mathcal{G}_{p4g,1/2}^{U(1)}&=\mathbb{Z}\times2\mathbb{Z}\times\mathbb{Z}_4\times\mathbb{Z}_2^2/4\mathbb{Z}\nonumber\\
&=\mathbb{Z}\times\mathbb{Z}_4^2\times\mathbb{Z}_2
\end{align}

\subsection{$p3$}
\begin{figure}
\centering
\includegraphics[width=0.4\textwidth]{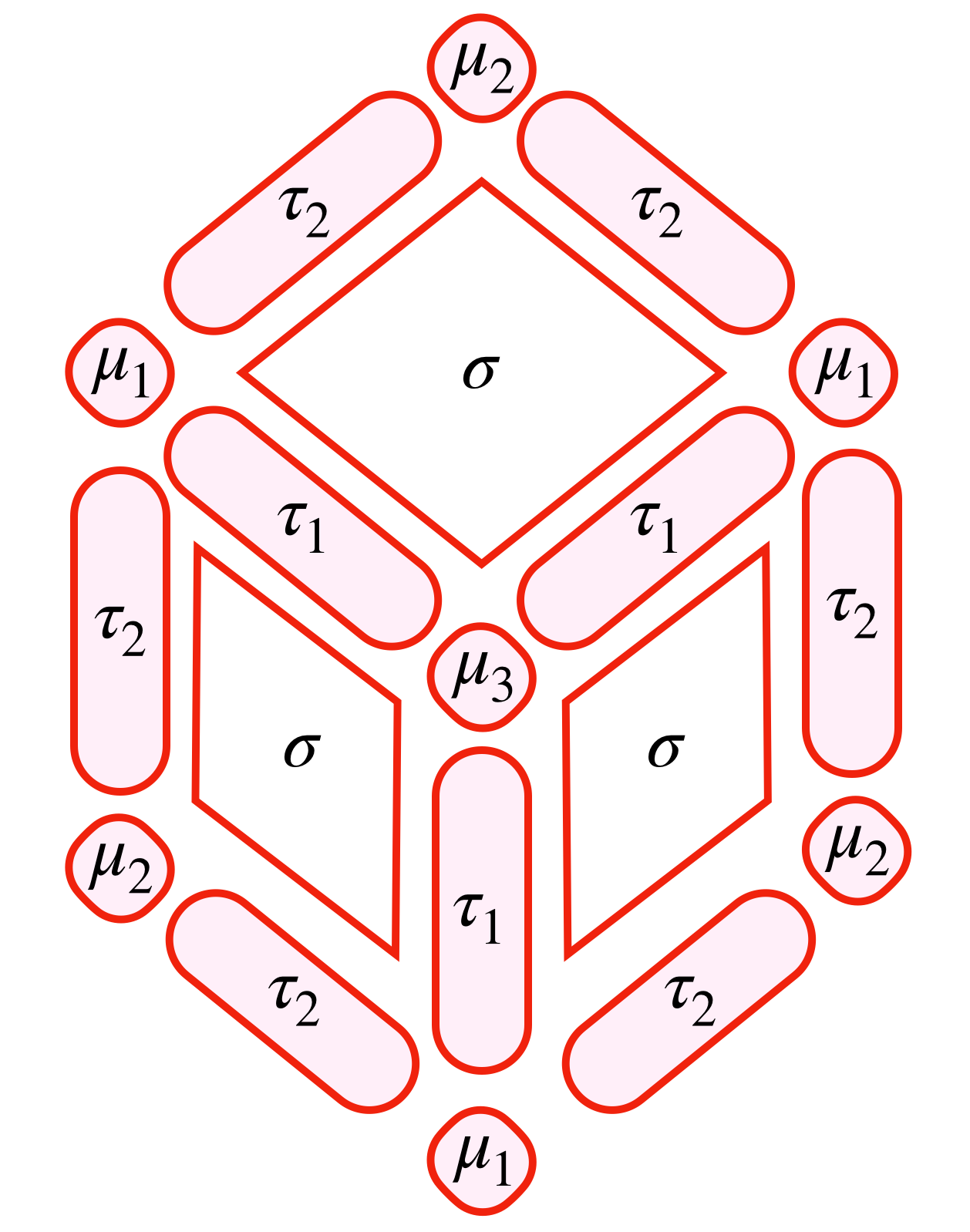}
\caption{\#13 wallpaper group $p3$ and its cell decomposition.}
\label{p3}
\end{figure}

The corresponding point group of this case (by quotient out the translations) is 3-fold rotational symmetry group $C_3$. For 2D blocks $\sigma$ and 1D blocks $\tau_1$ and $\tau_2$, there is no on-site symmetry; for 0D blocks $\mu_j,~j=1,2,3$, the on-site symmetry group is $\mathbb{Z}_3$ due to the 3-fold rotational symmetry acting internally, see Fig. \ref{p3}.

\subsubsection{Spinless fermions}
First, we investigate the 0D block-state decoration. Different 0D block-states at $\mu_j$ ($j=1,2,3$) can be characterized by different 1D irreducible representations of the corresponding on-site symmetry group by space group operations acting internally:
\begin{align}
\mathcal{H}^1\left[\mathbb{Z}_2^f\times\mathbb{Z}_3,U(1)\right]=\mathbb{Z}_2\times\mathbb{Z}_3
\end{align}
Here $\mathbb{Z}_2$ represents the fermion parity, and $\mathbb{Z}_3$ represents the 3-fold rotation eigenvalues, and different 0D block-states on each 0D block can be labeled by $(\pm,\nu_j)$, where $\pm$ represents the fermion parity, and $\nu_j\in\{1,e^{2\pi i/3},e^{4\pi i/3}\}$ represents the eigenvalue of 3-fold rotational symmetry. According to this notation, all obstruction-free 0D block-states form the following group:
\begin{align}
\{\mathrm{OFBS}\}_{p3,0}^{\mathrm{0D}}=\mathbb{Z}_2^3\times\mathbb{Z}_3^3
\end{align}
And the group elements can be labeled by (three brackets represent the block-states at $\mu_j,j=1,2,3$):
\begin{align}
[(\pm,\nu_1),(\pm,\nu_2),(\pm,\nu_3)]
\end{align}

Subsequently we investigate the 1D block-state decoration. The unique candidate 1D block-state for each 1D block is Majorana chain due to the absence of the on-site symmetry, so all 1D block-states form a group:
\begin{align}
\{\mathrm{BS}\}_{p3,0}^{\mathrm{1D}}=\mathbb{Z}_2^3
\end{align}
Majorana chain decoration on $\tau_1/\tau_2$ leaves 3 dangling Majorana modes at each corresponding 0D block. It is well-known that odd number of Majorana modes cannot be gapped out by entanglement pair/ interactions, hence the no-open-edge condition cannot be satisfied, and the classification attributed to 1D block-state decorations is trivial:
\begin{align}
E_{p3,0}^{\mathrm{1D}}=\{\mathrm{OFBS}\}_{p3,0}^{\mathrm{1D}}=\mathbb{Z}_1
\end{align}

With all obstruction-free block-states, subsequently we discuss about all possible trivializations. First, we consider about the 2D bubble equivalences: as we discussed in the main text, only type-\2 (i.e., ``Majorana bubbles'' with anti-PBC) 2D bubble equivalence is valid because there is no 0D block as the center of even-fold dihedral group. And from Ref. \cite{Srotation} we know that there is no effects on both 1D and 0D blocks, so 2D bubble equivalence contributes no trivialization. 

Then we consider the 1D bubble equivalences. Consider the 1D bubble equivalence on 1D blocks labeled by $\tau_1$ [cf. 1D bubble, here both yellow and red dots represent the complex fermions]: Near each 0D block labeled by $\mu_1$ or $\mu_3$, there are 3 complex fermions which form an atomic insulator:
\begin{align}
|\psi\rangle_{p3}^{\mu_3}=c_1^\dag c_2^\dag c_3^\dag|0\rangle
\end{align}
and it is obvious that the fermion parity of $|\psi\rangle_{p3}^{\mu_3}$ is odd, hence $|\psi\rangle_{p3}^{\mu_3}$ can change the fermion parity at each 0D block labeled by $\mu_3$. Similar for each 0D block labeled by $\mu_1$, and the fermion parities of 0D blocks labeled by $\mu_1$ and $\mu_3$ can be changed by 1D bubble construction on $\tau_1$ simultaneously. Similar arguments can also be held on 1D block labeled by $\tau_2$, and we summarize the effects of all 1D bubble equivalences:
\begin{enumerate}[1.]
\item 1D bubble equivalence on $\tau_1$: change the fermion parities of 0D blocks $\mu_1$ and $\mu_3$ simultaneously;
\item 1D bubble equivalence on $\tau_2$: change the fermion parities of 0D blocks $\mu_1$ and $\mu_2$ simultaneously.
\end{enumerate}

With all possible bubble equivalences, we are ready to study the trivial block-states. Start from the original trivial state (nothing is decorated on arbitrary blocks):
\[
[(+,1),(+,1),(+,1)]
\]
If we take 1D bubble equivalences on $\tau_1$ and $\tau_2$ by $l_1$ and $l_2$ times, above trivial state will be deformed to a new 0D block-state labeled by:
\begin{align}
[((-1)^{l_1+l_2},1),((-1)^{l_2},1),((-1)^{l_1},1)]
\label{p3 spinless trivial state}
\end{align}
According to the definition of bubble equivalence, all these 0D block-states should be trivial. It is straightforward to check that there are only two independent quantities ($l_1$ and $l_2$) in Eq. (\ref{p3 spinless trivial state}), hence all trivial block-states form the following group:
\begin{align}
\{\mathrm{TBS}\}_{p3,0}=\mathbb{Z}_2^2
\end{align}
Therefore, all obstruction and trivialization free 0D/1D block-states are classified as:
\begin{align}
\begin{aligned}
&E_{p3,0}^{\mathrm{0D}}=\{\mathrm{OFBS}\}_{p3,0}^{\mathrm{0D}}/\{\mathrm{TBS}\}_{p3,0}^{\mathrm{0D}}=\mathbb{Z}_2\times\mathbb{Z}_3^3\\
&E_{p3,0}^{\mathrm{1D}}=\{\mathrm{OFBS}\}_{p3,0}^{\mathrm{1D}}/\{\mathrm{TBS}\}_{p3,0}^{\mathrm{1D}}=\mathbb{Z}_1
\end{aligned}
\end{align}
and all independent nontrivial block-states are labeled by the group elements of the following quotient group:
\begin{align}
\mathcal{G}_{p3}^0=E_{p3,0}^{\mathrm{0D}}=\mathbb{Z}_2\times\mathbb{Z}_3^3
\end{align}

It is obvious that there is no stacking between 1D and 0D block-states because of the absence of nontrivial 1D root phase. 

\subsubsection{Spin-1/2 fermions}
First, we investigate the 0D block-state decorations. For each 0D block labeled by $\mu_j,~j=1,2,3$, the total symmetry group is $\mathbb{Z}_6^f$: nontrivial $\mathbb{Z}_2^f$ extension of on-site symmetry $\mathbb{Z}_3$, and the different 0D block-states can be characterized by different 1D irreducible representations of the corresponding symmetry group:
\begin{align}
\mathcal{H}^1\left[\mathbb{Z}_6^f,U(1)\right]=\mathbb{Z}_6
\end{align}
Then we investigate the possible trivializations. Consider the 1D bubble equivalence on 1D blocks labeled by $\tau_1$: on each 1D block labeled by $\tau_1$, we decorate a 1D bubble onto it. Here both yellow and red dots represent the complex fermions (Note: the 0D FSPT mode with eigenvalue $-1$ of the symmetry group $\mathbb{Z}_6^f$ is just the complex fermion because $\mathbb{Z}_6^f$ is the nontrivial $\mathbb{Z}_2^f$ extension of the on-site symmetry $\mathbb{Z}_3$). Near each 0D block labeled by $\mu_1$, there are 3 complex fermions which form an atomic insulator with odd fermion parity (equivalently, eigenvalue $-1$ of the symmetry group $\mathbb{Z}_6^f$ on each 0D block labeled by $\mu_1$):
\begin{align}
|\phi\rangle_{p3}^{\mu_1}=a_1^\dag a_2^\dag a_3^\dag|0\rangle
\end{align}
Hence the eigenvalue $-1$ of the symmetry group $\mathbb{Z}_6^f$ at each 0D block labeled by $\mu_1$ can be trivialized by the atomic insulator $|\phi\rangle_{p3}^{\mu_1}$. Near each 0D block labeled by $\mu_3$, there is 3 complex fermions which form an atomic insulator with odd fermion parity (equivalently, eigenvalue $-1$ of the symmetry group $\mathbb{Z}_6^f$ on each 0D block labeled by $\mu_3$):
\begin{align}
|\phi\rangle_{p3}^{\mu_2}=a_1'^\dag a_2'^\dag a_3'^\dag|0\rangle
\end{align}
Hence the eigenvalue $-1$ of the symmetry group $\mathbb{Z}_6^f$ at each 0D block labeled by $\mu_3$ can be trivialized by the atomic insulator $|\phi\rangle_{p3}^{\mu_3}$. Equivalently, the eigenvalue $-1$ of the symmetry group $\mathbb{Z}_6^f$ at 0D block labeled by $\mu_1$ and $\mu_3$ are not independent. Similarly, the 1D bubble equivalence on 1D blocks labeled by $\tau_2$ leads to that the eigenvalue $-1$ of the symmetry group $\mathbb{Z}_6^f$ at 0D block labeled by $\mu_1$ and $\mu_2$ are not independent. Therefore, the eigenvalue $-1$ of the symmetry group $\mathbb{Z}_6^f$ at all 0D blocks are not independent, and the classification attributed to the 0D block-state decorations is:
\begin{align}
E_{p3,1/2}^{\mathrm{0D}}=\mathbb{Z}_6\times\mathbb{Z}_3^2
\end{align}

Subsequently we investigate the 1D block-state decoration, which is similar with the case of spinless fermions: the no-open-edge condition cannot be satisfied, and the classification attributed to 1D block-state decorations is:
\begin{align}
E_{p3,1/2}^{\mathrm{1D}}=\mathbb{Z}_1
\end{align}
It is obvious that there is no stacking between 1D and 0D block-states because of the absence of nontrivial 1D root phase. Therefore, the ultimate classification with accurate group structure is:
\begin{align}
\mathcal{G}_{p3}^{1/2}=\mathbb{Z}_6\times\mathbb{Z}_3^2
\end{align}

\subsubsection{With $U^f(1)$ charge conservation}
Then we consider the systems with $U^f(1)$ charge conservation. For an arbitrary 0D block, different 0D block-states are characterized by different irreducible representations of the symmetry group as:
\begin{align}
\mathcal{H}^1[U(1)\times\mathbb{Z}_3,U(1)]=\mathbb{Z}\times\mathbb{Z}_3
\end{align}
Here $\mathbb{Z}$ represents the complex fermion and $\mathbb{Z}_3$ represents the eigenvalues of 3-fold rotational symmetry operation. For systems with spinless fermions, consider 1D bubble equivalence on 1D blocks $\tau_1$ [cf. 1D bubble, here yellow and red dots represent particle and hole, respectively, and they can be trivialized if we shrink them to a point]: it adds three complex fermions with $U^f(1)$ charge $+1$ (particles) at each 0D block $\mu_1$ and three complex fermions with $U^f(1)$ charge $-1$ (holes) at each 0D block $\mu_3$, hence the $U^f(1)$ charges at $\mu_1$ and $\mu_2$ are not independent. 

With the help of above discussions, we consider the 0D block-state decorations. The 0D block-state at $\mu_j$ ($j=1,2,3$) can be labeled by $(m_j,\phi_j)$, where $m_j\in\mathbb{Z}$ labels the $U^f(1)$ charges on $\mu_j$, $\phi_j$ labels the eigenvalues of 3-fold rotation at $\mu_j$. Start from the following trivial state:
\begin{align}
[(0,1),(0,1),(0,1)]
\end{align}
Take aforementioned 1D bubble constructions on $\tau_j$ by $l_j$ times ($j=1,2,3$), above trivial state will be deformed to the following 0D block-state:
\begin{align}
\left[\left(3l_1,1\right),\left(3l_2,1\right),\left(-3l_1-3l_2,1\right)\right]
\end{align}
And this state should be trivial. There are only two independent indices ($l_1$ and $l_2$). Therefore, the ultimate classification of crystalline topological phases protected by $p3$ symmetry for 2D systems with spinless fermions is:
\begin{align}
\mathcal{G}_{p3,0}^{U(1)}=\mathbb{Z}^3\times\mathbb{Z}_3^3/\left(3\mathbb{Z}\right)^2=\mathbb{Z}\times\mathbb{Z}_3^5
\end{align}

It is easy to verify that for $p3$ symmetry, there is no difference between systems with spinless and spin-1/2 fermions, hence the the ultimate classification of crystalline topological phases protected by $p3$ symmetry for 2D systems with spin-1/2 fermions is:
\begin{align}
\mathcal{G}_{p3,1/2}^{U(1)}=\mathbb{Z}^3\times\mathbb{Z}_3^3/\left(3\mathbb{Z}\right)^2=\mathbb{Z}\times\mathbb{Z}_3^5
\end{align}

\begin{figure}
\centering
\includegraphics[width=0.46\textwidth]{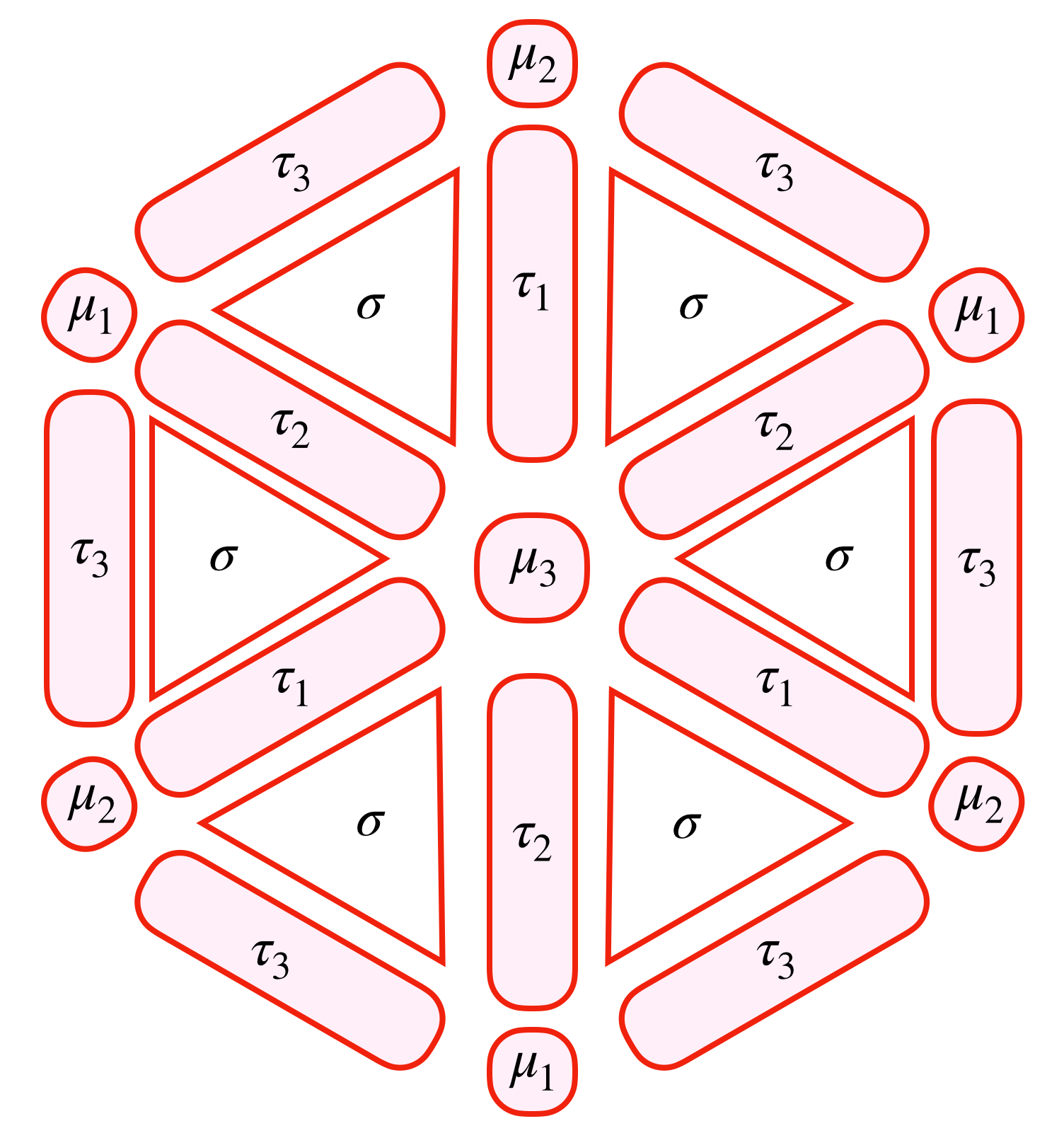}
\caption{\#14 wallpaper group $p3m1$ and its cell decomposition.}
\label{p3m1}
\end{figure}

\subsection{$p3m1$}
The corresponding point group of this case is 3-fold dihedral symmetry group $D_3$. For 2D blocks $\sigma$, there is no on-site symmetry; for 1D blocks $\tau_j,~j=1,2,3$, the on-site symmetry group is $\mathbb{Z}_2$ caused by reflection symmetry acting internally on the corresponding 1D blocks; for 0D blocks $\mu_j,~j=1,2,3$, the on-site symmetry group is $\mathbb{Z}_3\rtimes\mathbb{Z}_2$ which is attributed to the $D_3$ group acting internally, see Fig. \ref{p3m1}.

\subsubsection{Spinless fermions}
First, we consider the 0D block-state decoration. For each 0D block, the classification data can be determined by calculating different 1D irreducible representations of full symmetry group:
\begin{align}
\mathcal{H}^1\left[U^f(1)\times(\mathbb{Z}_3\rtimes\mathbb{Z}_2),U(1)\right]=\mathbb{Z}_2^2
\label{D3 classification data}
\end{align}
Here the two $\mathbb{Z}_2$ have different physical meanings: the first $\mathbb{Z}_2$ represents the complex fermion, and the second $\mathbb{Z}_2$ represents the reflection eigenvalue $-1$ (i.e., the rotational symmetry plays no role in the 0D block-state decorations), and different 0D block-states on each 0D block can be labeled by $(\pm,\pm)$, where these two $\pm$'s represent the fermion parity and eigenvalues of reflection operation, respectively. According to this notation, all obstruction-free 0D block-states form the following group:
\begin{align}
\{\mathrm{OFBS}\}_{p3m1,0}^{\mathrm{0D}}=\mathbb{Z}_2^6
\end{align}
And the group elements can be labeled by (three brackets represent the block-states at $\mu_j,j=1,2,3$):
\begin{align}
[(\pm,\pm),(\pm,\pm),(\pm,\pm)]
\end{align}

Subsequently we consider the 1D block-state decoration. For each 1D block, the total symmetry is $\mathbb{Z}_2^f\times\mathbb{Z}_2$, hence the candidate 1D block-states are Majorana chains and 1D FSPT state, and all 1D block-states form a group:
\begin{align}
\{\mathrm{BS}\}_{p3m1,0}^{\mathrm{1D}}=\mathbb{Z}_2^6
\end{align}

\paragraph{Majorana chain decoration}Majorana chain decoration on $\mu_j,~j=1,2,3$ will leaves 3 dangling Majorana modes at each corresponding 0D block, and it is well-known that odd number of Majorana modes cannot be gapped out by entanglement pairs/interactions, thus the no-open-edge condition is violated. Nevertheless, there is one exception: if we consider all 1D blocks $\tau_1,\tau_2$ and $\tau_3$ together and decorate a Majorana chain on each of them, it will leaves 6 dangling Majorana modes at each 0D block (we discuss the 0D block labeled by $\mu_1$ as an example), with the following rotation and reflection symmetry properties:
\begin{align}
\left.
\begin{aligned}
\bs{R}_{\mu_1}:~&\gamma_j\mapsto\gamma_{j+1},~\gamma_j'\mapsto\gamma_{j+1}'\\
\bs{M}_{\tau_3}:~&\gamma_j\mapsto\gamma_{5-j},~\gamma_j'\mapsto\gamma_{5-j}'
\end{aligned}
\right.
\end{align}

Where $\bs{R}_{\mu_1}$ represents the 3-fold rotation operation centred at each 0D block labeled by $\mu_1$ and $\bs{M}_{\tau_3}$ represents the reflection operation with the axis coincide with the 1D blocks labeled by $\tau_3$. All subscripts are taken with modulo 4. Consider the Hamiltonian who includes 3 entanglement pairs as:
\begin{align}
\mathcal{H}=i\sum\limits_{j=1}^3\gamma_j\gamma_j'
\end{align}
Which is invariant under arbitrary symmetry operations. So the no-open-edge condition is satisfied for this exception. 
\paragraph{1D FSPT state decoration}1D FSPT state decoration on $\tau_j,~j=1,2,3$ will leaves 6 dangling Majorana modes at each corresponding 0D block, with the following rotation and reflection symmetry properties:
\begin{align}
\left.
\begin{aligned}
\bs{R}_{\mu_1}:~&\gamma_j\mapsto\gamma_{j+1},~\gamma_j'\mapsto\gamma_{j+1}'\\
\bs{M}_{\tau_3}:~&\gamma_j\mapsto-\gamma_{5-j},~\gamma_j'\mapsto\gamma_{5-j}'
\end{aligned}
\right.
\end{align}
Then consider the local fermion parity at each corresponding 0D block with its symmetry properties:
\begin{align}
P_f=i\prod\limits_{j=1}^3\gamma_j\gamma_j',~~
\left\{
\begin{aligned}
&\bs{R}_{\mu_1}:~P_f\mapsto P_f\\
&\bs{M}_{\tau_3}:~P_f\mapsto-P_f
\end{aligned}
\right.
\end{align}
Thus these 6 Majorana modes cannot be gapped out in a symmetric way, and the no-open-edge condition is violated. Nevertheless, there is one exception: if we consider all 1D blocks together and decorate a 1D FSPT state on each of them, there is 12 dangling Majorana modes $\gamma_j,\gamma_j',~j=1,...,6$ as the edge modes of the decorated Majorana chains, with the following rotation and reflection symmetry properties:
\begin{align}
\left.
\begin{aligned}
\bs{R}_{\mu_1}:~&\gamma_j\mapsto\gamma_{j+2},~\gamma_j'\mapsto\gamma_{j+2}'\\
\bs{M}_{\tau_3}:~&\gamma_j\mapsto-\gamma_{8-j},~\gamma_j'\mapsto\gamma_{8-j}'
\end{aligned}
\right.
\end{align}
where all subscripts are taken with modulo 6. Then consider the local fermion parity and its symmetry properties:
\begin{align}
P_f=i\prod\limits_{j=1}^6\gamma_j\gamma_j',~~\bs{R}_{\mu_1},\bs{M}_{\tau_3}:~P_f\mapsto P_f
\end{align}
Therefore these 12 Majorana modes can be gapped out by some proper interactions in a symmetric way, and the no-open-edge is satisfied. All obstruction-free 1D block-states form the following group:
\begin{align}
\{\mathrm{OFBS}\}_{p3m1,0}^{\mathrm{1D}}=\mathbb{Z}_2^2
\end{align}
and the group elements can be labeled by $m_1=m_2=m_3$ and $m_1'=m_2'=m_3'$ that represents the number of decorated Majorana chains/1D FSPT states on 1D block labeled by $\tau_j$ ($j=1,2,3$). Different obstruction free block-states can be labeled by:
\begin{align}
[&(\pm,\pm),(\pm,\pm),(\pm,\pm);m_1=m_2=m_3;\nonumber\\
&m_1'=m_2'=m_3']
\end{align}
here the first three brackets represent the 0D block-states at $\mu_j$, and the last two quantities represent the number of Majorana chain/1D FSPT states at $\tau_j$ ($j=1,2,3$). 

With all obstruction-free block-states, we discuss about all possible trivializations. First, we consider about the 2D bubble equivalence: as we discussed in the main text, only type-\2 (i.e., ``Majorana bubbles'' with anti-PBC) 2D bubble equivalence is valid because there is no 0D block as the center of even-fold dihedral group symmetry. Similar with the $p4m$ case, ``Majorana bubbles'' can be deformed to double Majorana chains at each nearby 1D block, and this is exactly the definition of the nontrivial 1D FSPT phase protected by on-site $\mathbb{Z}_2$ symmetry (by reflection symmetry acting internally). As a consequence, 1D FSPT state decorations on all 1D blocks can be deformed to a trivial state via 2D ``Majorana'' bubble equivalences. Furthermore, repeatedly similar with the $p4m$ case, both types of ``Majorana bubble'' constructions have no effect on 0D blocks. 

Subsequently we consider the 1D bubble equivalences. Consider the 1D bubble equivalence on 1D blocks labeled by $\tau_2$ [cf. 1D bubble, here both yellow and red dots represent the complex fermions]: Near each 0D block labeled by $\mu_3$, there are 3 complex fermions which form an atomic insulator with odd fermion parity:
\begin{align}
|\psi\rangle_{p3m1}^{\mu_3}=c_1^\dag c_2^\dag c_3^\dag|0\rangle
\end{align}
i.e., fermion parity at each $\mu_3$ can be changed by the atomic insulator $|\psi\rangle_{p3m1}^{\mu_3}$. Similar for the 0D blocks labeled by $\mu_1$, we can conclude that the fermion parities of 0D blocks labeled by $\mu_1$ and $\mu_3$ are not independent. Hence if we further consider the 1D bubble equivalences on 1D blocks labeled by $\tau_1$ and $\tau_3$, we can obtain that the fermion parities of 0D blocks labeled by $\mu_j$, $j=1,2,3$ are not independent. Furthermore, this 1D bubble construction can change the reflection eigenvalues on 0D block: for instance, consider the reflection property of the atomic insulator $|\psi\rangle_{p3m1}^{\mu_3}$ as:
\begin{align}
\bs{M}_{\tau_1}|\psi\rangle_{p3m1}^{\mu_3}=c_1^\dag c_3^\dag c_2^\dag|0\rangle=-|\psi\rangle_{p3m1}^{\mu_3}
\end{align}
i.e., the atomic insulator $|\psi\rangle_{p3m1}^{\mu_3}$ changes the reflection eigenvalue of the 0D block $\mu_3$, similar for all other 0D blocks. 

Then we study another possible 1D bubble equivalence. Here we consider an alternative 1D bubble equivalence on 1D blocks labeled by $\tau_2$ [cf. 1D bubble, here both yellow and red dots represents the 0D mode with reflection eigenvalue $-1$; we call the aforementioned 1D bubble equivalence ``type-\1'' and this 1D bubble equivalence ``type-\2'']: near each 0D block labeled by $\mu_3$, there are three 0D modes characterized by reflection eigenvalue $-1$ which can form an another type of ``atomic insulator'' (here $d_j^\dag$ is the creation operator of the corresponding 0D mode):
\begin{align}
|\phi\rangle_{p3m1}^{\mu_3}=d_1^\dag d_2^\dag d_3^\dag|0\rangle
\end{align}
who can change the reflection eigenvalue of each 0D block labeled by $\mu_3$: each 0D mode carries an eigenvalue $-1$ of reflection symmetry, hence $|\phi\rangle_{p3m1}^{\mu_3}$ carries an eigenvalue $-1$ of reflection symmetry. Similar for the 0D blocks labeled by $\mu_1$, hence the reflection eigenvalues at $\mu_1$ and $\mu_3$ are not independent. If we consider the similar bubble construction on 1D blocks labeled by $\tau_1$ and $\tau_3$, we obtain that the reflection eigenvalues at $\mu_j$, $j=1,2,3$ are not independent. We summarize the effects of all 1D bubble equivalences:
\begin{enumerate}[1.]
\item Type-\1 1D bubble construction on $\tau_1$: change the fermion parities and reflection eigenvalues of 0D blocks $\mu_2$ and $\mu_3$ simultaneously;
\item Type-\1 1D bubble construction on $\tau_2$: change the fermion parities and reflection eigenvalues of 0D blocks $\mu_1$ and $\mu_3$ simultaneously;
\item Type-\1 1D bubble construction on $\tau_3$: change the fermion parities and reflection eigenvalues of 0D blocks $\mu_1$ and $\mu_2$ simultaneously;
\item Type-\2 1D bubble construction on $\tau_1$: change the reflection eigenvalues of 0D blocks $\mu_2$ and $\mu_3$ simultaneously;
\item Type-\2 1D bubble construction on $\tau_2$: change the reflection eigenvalues of 0D blocks $\mu_1$ and $\mu_3$ simultaneously;
\item Type-\2 1D bubble construction on $\tau_3$: change the reflection eigenvalues of 0D blocks $\mu_1$ and $\mu_2$ simultaneously.
\end{enumerate}

With all possible bubble equivalences, we are ready to study the trivial block-states. Start from the original trivial state (nothing is decorated on arbitrary blocks):
\[
[(+,+),(+,+),(+,+)]
\]
If we take 2D ``Majorana'' bubble equivalences $l_0$ times, take type-\1 1D bubble equivalences on $\tau_j$ by $l_j$ times, and type-\2 1D bubble equivalences on $\tau_j$ by $l_j'$ times ($j=1,2,3$), above trivial state will be deformed to a new 0D block-state labeled by:
\begin{align}
&\left[\left((-1)^{l_2+l_3},(-1)^{l_2+l_3+l_2'+l_3'}\right),\right.\nonumber\\
&\left((-1)^{l_1+l_3},(-1)^{l_1+l_3+l_1'+l_3'}\right),\nonumber\\
&\left.\left((-1)^{l_1+l_2},(-1)^{l_1+l_2+l_1'+l_2'}\right)\right]
\label{p3m1 spinless trivial state}
\end{align}
According to the definition of bubble equivalence, all these 0D block-states should be trivial. It is straightforward to check that there are only four independent quantities in Eq. (\ref{p3m1 spinless trivial state}). Together with the 2D bubble equivalence, all trivial block-states form the following group:
\begin{align}
\{\mathrm{TBS}\}_{p3m1,0}&=\{\mathrm{TBS}\}_{p3m1,0}^{\mathrm{1D}}\times\{\mathrm{TBS}\}_{p3m1,0}^{\mathrm{0D}}\nonumber\\
&=\mathbb{Z}_2\times\mathbb{Z}_2^4=\mathbb{Z}_2^5
\end{align}
here $\{\mathrm{TBS}\}_{p3m1,0}^{\mathrm{1D}}$ represents the group of trivial states with non-vacuum 1D blocks, and $\{\mathrm{TBS}\}_{p3m1,0}^{\mathrm{0D}}$ represents the group of trivial states with non-vacuum 0D blocks.

Therefore, all obstruction and trivialization free 0D/1D block-states are classified as:
\begin{align}
\begin{aligned}
&E_{p3m1,0}^{\mathrm{0D}}=\{\mathrm{OFBS}\}_{p3m1,0}^{\mathrm{0D}}/\{\mathrm{TBS}\}_{p3m1,0}^{\mathrm{0D}}=\mathbb{Z}_2\\
&E_{p3m1,0}^{\mathrm{0D}}=\{\mathrm{OFBS}\}_{p3m1,0}^{\mathrm{0D}}/\{\mathrm{TBS}\}_{p3m1,0}^{\mathrm{0D}}=\mathbb{Z}_2^2
\end{aligned}
\end{align}
and all independent nontrivial block-states are labeled by different group elements of the following quotient group:
\begin{align}
\mathcal{G}_{p3m1}^0=E_{p3m1,0}^{\mathrm{0D}}\times E_{p3m1,0}^{\mathrm{1D}}=\mathbb{Z}_2^3
\end{align}
here one $\mathbb{Z}_2$ is from the Majorana chain decorations on all 1D blocks simultaneously, and all other $\mathbb{Z}_2$'s are from the nontrivial 0D block-states. Similar with the $pm$ case, there is no stacking between 1D and 0D block-states, and the group structure of the classification data $\mathcal{G}_{p3m1}^0$ has already been accurate.

\subsubsection{Spin-1/2 fermions}
First, we investigate the 0D block-state decorations. Similar with the spinless fermions case, the classification data for all 0D blocks can be characterized by different 1D irreducible representations of the full symmetry group:
\begin{align}
\mathcal{H}^1\left[U^f(1)\rtimes_{\omega_2}(\mathbb{Z}_3\rtimes\mathbb{Z}_2),U(1)\right]=\mathbb{Z}_4
\end{align}
Equivalently, we can label different 0D block-states by the group elements of the 4-fold cyclic group:
\begin{align}
\mathbb{Z}_4=\left\{1,i,-1,-i\right\}
\label{Z4A}
\end{align}
Then we investigate the possible trivialization. Consider the 1D bubble equivalence on 1D blocks labeled by $\tau_1$: on each $\tau_1$, the total on-site symmetry is $\mathbb{Z}_4^f$: nontrivial $\mathbb{Z}_2^f$ extension of the on-site symmetry $\mathbb{Z}_2$. Next we decorate a 1D bubble onto each of them, here the yellow/red dots represents the 0D FSPT mode protected by $\mathbb{Z}_4^f$ symmetry which is labeled by $i/-i\in\mathbb{Z}_4$, cf. Eq. (\ref{Z4A}), and they can be trivialized if they shrink to a point. Near each 0D block labeled by $\mu_3$, there are three 0D FSPT modes labeled by $i\in\mathbb{Z}_4$ and they can change the label of 0D block-state decorated at each 0D block $\mu_3$ by $-i\in\mathbb{Z}_4$. Meanwhile, near each 0D block labeled by $\mu_2$, there are three 0D FSPT modes labeled by $-i\in\mathbb{Z}_4$ and they can change the label of 0D block-state decorated at each 0D block $\mu_2$ by $i\in\mathbb{Z}_4$. Therefore, the block-state decorations on 0D blocks labeled by $\mu_2$ and $\mu_3$ are not independent. Similar for the 1D bubble equivalence on 1D blocks labeled by $\tau_2$, we can conclude that the block-state decorations on 0D blocks labeled by $\mu_j$, $j=1,2,3$ are not independent. As a consequence, the classification attributed to the 0D block-state decorations is:
\begin{align}
E_{p3m1,1/2}^{\mathrm{0D}}=\mathbb{Z}_4
\end{align}

Subsequently we investigate the 1D block-state decoration. For all 1D blocks, the total symmetry on each of them is $\mathbb{Z}_4^f$, hence there is no nontrivial 1D block-state because of the trivial classification of the corresponding 1D FSPT phases, and the classification attributed to 1D block-state decorations is trivial:
\begin{align}
E_{p3m1,1/2}^{\mathrm{1D}}=\mathbb{Z}_1
\end{align}
It is obvious that there is no stacking between 1D and 0D block-states because of the absence of nontrivial 1D root phase. Therefore, the ultimate classification with accurate group structure is:
\begin{align}
\mathcal{G}_{p3m1}^{1/2}=\mathbb{Z}_4
\end{align}

\subsubsection{With $U^f(1)$ charge conservation}
Then we consider the systems with $U^f(1)$ charge conservation. For an arbitrary 0D block, different 0D block-states are characterized by different irreducible representations of the symmetry group as:
\begin{align}
\mathcal{H}^1[U(1)\times(\mathbb{Z}_3\rtimes\mathbb{Z}_2),U(1)]=\mathbb{Z}\times\mathbb{Z}_2
\end{align}
Here $\mathbb{Z}$ represents the complex fermion and $\mathbb{Z}_2$ represents the eigenvalues of reflection symmetry operation. We need to further consider the possible trivializations. For systems with spinless fermions, consider the 1D bubble equivalence: on each 1D block labeled by $\tau_1$, we decorate a 1D bubble onto it. Here both yellow and red dots represent the 0D SPT modes characterizing eigenvalue $-1$ of reflection symmetry operation, and they can be trivialized when they shrink to a point. According to this bubble construction, the reflection eigenvalues at 0D blocks $\mu_2$ and $\mu_3$ can be changed simultaneously, hence the reflection eigenvalues at $\mu_2$ and $\mu_3$ are not independent. Similarly, the 1D bubble constructions on other 1D blocks will lead to the fact that the reflection eigenvalues of all 0D blocks are not independent. We call this type of 1D bubble equivalence ``type-\2'' bubble equivalence.

Subsequently we consider the complex fermion sector: consider 1D bubble equivalence on 1D blocks $\tau_1$ [cf. 1D bubble, here yellow and red dots represent particle and hole, respectively, and they can be trivialized if we shrink them to a point]: it adds three complex fermions with $U^f(1)$ charge $+1$ at each 0D block $\mu_2$ and three complex fermions with $U^f(1)$ charge $-1$ at each 0D block $\mu_3$, hence the $U^f(1)$ charges at $\mu_2$ and $\mu_3$ are not independent. Furthermore, three particles near $\mu_2$ forms an atomic insulator:
\begin{align}
|\phi\rangle_{p3m1}^{\mu_2}=p_1^\dag p_2^\dag p_3^\dag|0\rangle
\end{align}
and three holes near $\mu_3$ forms another atomic insulator:
\begin{align}
|\phi\rangle_{p3m1}^{\mu_3}=h_1^\dag h_2^\dag h_3^\dag|0\rangle
\end{align}
Similar to the crystalline TSC, $|\phi\rangle_{p3m1}^{\mu_2}$ changes the reflection eigenvalues at each $\mu_2$, and $|\phi\rangle_{p3m1}^{\mu_3}$ changes the reflection eigenvalues at each $\mu_3$. We call this type of 1D bubble equivalence ``type-\1'' bubble equivalence.

With the help of above discussions, we consider the 0D block-state decorations. The 0D block-state at $\mu_j$ ($j=1,2,3$) can be labeled by $(m_j,\pm)$, where $m_j\in\mathbb{Z}$ labels the $U^f(1)$ charges on $\mu_j$, $\pm$ labels the eigenvalues of reflection at $\mu_j$. Start from the following trivial state:
\begin{align}
[(0,+),(0,+),(0,+)]
\end{align}
Take aforementioned type-\1/type-\2 1D bubble constructions on $\tau_j$ by $l_j/l_j'$ times ($j=1,2,3$), above trivial state will be deformed to the following 0D block-state:
\begin{align}
&\left[\left(3l_2+3l_3,(-1)^{l_2+l_3+l_2'l_3'}\right),\left(3l_1-3l_3,(-1)^{l_1+l_3+l_1'+l_3'}\right),\right.\nonumber\\
&\left.\left(-3l_1-3l_2,(-1)^{l_1+l_2+l_1'+l_2'}\right)\right]
\end{align}
And this state should be trivial. There are four independent indices in above trivial state, therefore, the ultimate classification of crystalline topological phases protected by $p3m1$ symmetry for 2D systems with spinless fermions is:
\begin{align}
\mathcal{G}_{p3m1,0}^{U(1)}&=\mathbb{Z}^3\times\mathbb{Z}_2^3/\left[(3\mathbb{Z})^2\times\mathbb{Z}_2^2\right]\nonumber\\
&=\mathbb{Z}\times\mathbb{Z}_3^2\times\mathbb{Z}_2
\end{align}

It is easy to verify that for $p3m1$ symmetry, there is no difference between systems with spinless and spin-1/2 fermions, hence the the ultimate classification of crystalline topological phases protected by $p3m1$ symmetry for 2D systems with spin-1/2 fermions is:
\begin{align}
\mathcal{G}_{p3m1,1/2}^{U(1)}&=\mathbb{Z}^3\times\mathbb{Z}_2^3/\left[(3\mathbb{Z})^2\times\mathbb{Z}_2^2\right]\nonumber\\
&=\mathbb{Z}\times\mathbb{Z}_3^2\times\mathbb{Z}_2
\end{align}

\subsection{$p31m$}
\begin{figure}
\centering
\includegraphics[width=0.46\textwidth]{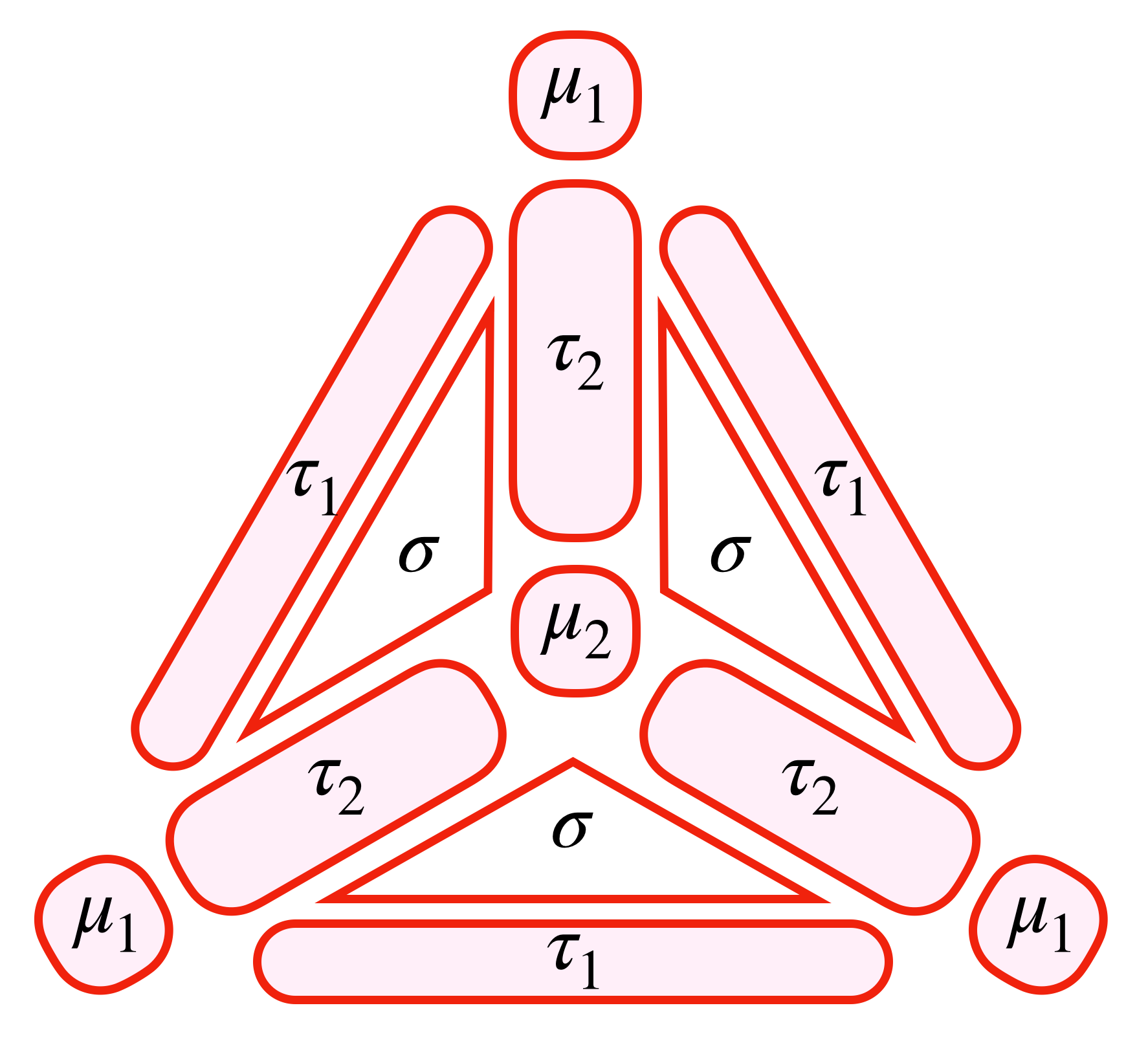}
\caption{\#15 wallpaper group $p31m$ and its cell decomposition.}
\label{p31m}
\end{figure}

The corresponding point group of this case is 3-fold dihedral group $D_3$ by quotient out the translations. For 2D blocks $\sigma$ and 1D blocks $\tau_2$, there is no on-site symmetry; for 1D blocks $\tau_1$, the on-site symmetry is $\mathbb{Z}_2$ attributed to the reflection symmetry acting internally; for 0D blocks $\mu_1$, the on-site symmetry is $\mathbb{Z}_3\rtimes\mathbb{Z}_2$ attributed to the $D_3$ symmetry acting internally; for 0D blocks $\mu_2$, the on-site symmetry is $\mathbb{Z}_3$ attributed to the $C_3$ symmetry acting internally, see Fig. \ref{p31m}.

\subsubsection{Spinless fermions}
First, we investigate the 0D block-state decoration. The classification data of each 0D block labeled by $\mu_1$ can be characterized by different 1D irreducible representations of the full symmetry group:
\begin{align}
\mathcal{H}^1\left[U^f(1)\times(\mathbb{Z}_3\rtimes\mathbb{Z}_2),U(1)\right]=\mathbb{Z}_2^2
\end{align}
Here the two $\mathbb{Z}_2$ have different physical meanings: the first $\mathbb{Z}_2$ represents the complex fermion, and the second $\mathbb{Z}_2$ represents the reflection eigenvalue $-1$, and different 0D block-states on each 0D block $\mu_1$ can be labeled by $(\pm,\pm)$, where these two $\pm$'s represent the fermion parity and eigenvalues of reflection operation, respectively. 

Different 0D block-states on each 0D block labeled by $\mu_2$ can be characterized by different 1-dimensional irreducible representations of the symmetry group:
\begin{align}
\mathcal{H}^1\left[\mathbb{Z}_2^f\times\mathbb{Z}_3,U(1)\right]=\mathbb{Z}_2\times\mathbb{Z}_3
\end{align}
Here $\mathbb{Z}_2$ represents the complex fermion, and $\mathbb{Z}_3$ represents the rotation eigenvalues, and different 0D block-states on each 0D block $\mu_2$ can be labeled by $(\pm,\nu_2)$, where $\pm$ represents the fermion parity, and $\nu_2\in\{1,e^{2\pi i/3},e^{4\pi i/3}\}$ represents the eigenvalue of 3-fold rotational symmetry. According to this notation, all obstruction-free 0D block-states form the following group:
\begin{align}
\{\mathrm{OFBS}\}_{p31m,0}^{\mathrm{0D}}=\mathbb{Z}_2^3\times\mathbb{Z}_3
\end{align}
and different group elements can be labeled by (two brackets represent the block-states at $\mu_1$ and $\mu_2$):
\begin{align}
[(\pm,\pm),(\pm,\nu_2)]
\end{align}

Subsequently we investigate the 1D block-state decoration. For 1D blocks labeled by $\tau_1$, the total symmetry group is $\mathbb{Z}_2^f\times\mathbb{Z}_2$, hence the candidate 1D block-states are Majorana chain and 1D FSPT state who can be realized by double Majorana chains with different $\mathbb{Z}_2$ eigenvalues. For 1D blocks labeled by $\tau_2$, the unique candidate block-state is Majorana chain due to the absence of on-site symmetry. So all 1D block-states form a group:
\begin{align}
\{\mathrm{BS}\}_{p31m,0}^{\mathrm{1D}}=\mathbb{Z}_2^3
\end{align}
We discuss about these two root phases separately.

\paragraph{Majorana chain decoration}Majorana chain decoration on $\tau_1$ leaves 6 dangling Majorana modes at each 0D block labeled by $\mu_1$, with the following rotation and reflection symmetry properties:
\begin{align}
\bs{R}_{\mu_1}:~\gamma_j\mapsto\gamma_{j+2},~~\bs{M}_{\tau_1}:~\gamma_{j}\mapsto\gamma_{8-j}
\end{align}
where $\bs{R}_{\mu_1}$ represents the 3-fold rotation operation centred at each 0D block labeled by $\mu_1$, and $\bs{M}_{\tau_1}$ represents the reflection operation with the axis coincide with the 1D blocks labeled by $\tau_1$. All subscripts are taken with modulo 6. Then consider the local fermion parity with its symmetry properties:
\begin{align}
P_f=i\prod\limits_{j=1}^6\gamma_j,~~\bs{R}_{\mu_1},\bs{M}_{\tau_1}:~P_f\mapsto P_f
\end{align}
Hence these 6 Majorana modes can be gapped out by 3 entanglement pairs in a symmetric way, as shown in the following Hamiltonian who respects arbitrary symmetry actions:
\begin{align}
\mathcal{H}=i\sum\limits_{j=1}^3\gamma_j\gamma_{j+3}
\end{align}
Therefore the no-open-edge condition is satisfied. And the Majorana chain decoration on $\tau_2$ leaves 3 dangling Majorana modes at each 0D block labeled by $\mu_2$. It is well-known that odd number of Majorana modes cannot be gapped out, hence the no-open-edge condition is violated.

\paragraph{1D FSPT state decoration}1D FSPT state decoration on $\tau_1$ leaves 12 dangling Majorana modes at each 0D block labeled by $\mu_1$, with the following rotation and reflection symmetry properties (all subscripts are taken with modulo 6):
\begin{align}
\left.
\begin{aligned}
\bs{R}_{\mu_1}:~&\gamma_j\mapsto\gamma_{j+2},~\gamma_j'\mapsto\gamma_{j+2}'\\
\bs{M}_{\tau_1}:~&\gamma_j\mapsto-\gamma_{8-j},~\gamma_j'\mapsto\gamma_{8-j}'
\end{aligned}
\right.
\end{align}
Then consider the local fermion parity and its rotation and reflection symmetry properties:
\begin{align}
P_f=-\prod\limits_{j=1}^6\gamma_j\gamma_j',~~\bs{R}_{\mu_1},\bs{M}_{\tau_1}:~P_f\mapsto P_f
\end{align}
hence these 12 Majorana modes can be gapped out by some proper interactions in a symmetric way, and the no-open-edge condition is satisfied. As a consequence, all obstruction-free 1D block-states form the following group:
\begin{align}
\{\mathrm{OFBS}\}_{p31m,0}^{\mathrm{1D}}=\mathbb{Z}_2^2
\end{align}
and the group elements can be labeled by $m_1$ and $m_1'$ that represent the number of decorated Majorana chains/1D FSPT states on 1D block labeled by $\tau_1$. Different obstruction free block-states can be labeled by:
\begin{align}
[(\pm,\pm),(\pm,\nu_2);m_1;m_1']
\end{align}
here the first two brackets represent the 0D block-states at $\mu_1$ and $\mu_2$, and the last two quantities represent the number of Majorana chains/1D FSPT states at $\tau_1$.

With all obstruction-free block-states, we discuss above all possible trivializations. First, we consider about the 2D bubble equivalence: as we discussed in the main text, only type-\2 (i.e., ``Majorana bubbles'' with anti-PBC) 2D bubble equivalence is valid because there is no 0D block as the center of even-fold dihedral group symmetry. Similar with the $pmg$ case, ``Majorana bubbles'' can be deformed to double Majorana chains at each nearby 1D block, but the effects of them are distinct: near each 1D block labeled by $\tau_2$, these double Majorana chains can be trivialized because there is no on-site symmetry on $\tau_1$ and the classification of 1D invertible topological phases (i.e., Majorana chain) is $\mathbb{Z}_2$; near each 1D block labeled by $\tau_1$, these double Majorana chains cannot be trivialized because there is an on-site $\mathbb{Z}_2$ symmetry on each $\tau_2$ by internal action of reflection symmetry, and this $\mathbb{Z}_2$ action exchanges these two Majorana chains, and this is exactly the definition of the nontrivial 1D FSPT phase protected by on-site $\mathbb{Z}_2$ symmetry. Furthermore, similar with the $p3m1$ case, there is no effect on 0D blocks labeled by $\mu_2$ by taking 2D ``Majorana'' bubble equivalence. 

Subsequently we consider the 1D bubble equivalence on 1D blocks labeled by $\tau_2$ [cf. 1D bubble, here both yellow and red dots represent the complex fermions]: Near each 0D block labeled by $\mu_1$, there are 6 complex fermions which can form an atomic insulator with even fermion parity:
\begin{align}
|\psi\rangle_{p31m}^{\mu_1}=\prod\limits_{j=1}^6 c_j^\dag|0\rangle
\end{align}
So it cannot change the fermion parity at each 0D block labeled by $\mu_1$. Near each 0D block labeled by $\mu_2$, there are 3 complex fermions which form another atomic insulator:
\begin{align}
|\psi\rangle_{p31m}^{\mu_2}=c_1'^\dag c_2'^\dag c_3'^\dag|0\rangle
\end{align}
and it can change the fermion parity of each 0D block labeled by $\mu_2$. 

Then we study the role of the rotational symmetry. We only need to consider the 0D blocks labeled by $\mu_2$ as aforementioned. We consider the atomic insulator $|\psi\rangle_{p31m}^{\mu_2}$ repeatedly, with the rotation property as ($\bs{R}_{\mu_2}$ represents the 3-fold rotation operation centred at the 0D block labeled by $\mu_2$):
\begin{align}
\bs{R}_{\mu_2}|\psi\rangle_{p31m}^{\mu_2}=c_2'^\dag c_3'^\dag c_1'^\dag|0\rangle=|\psi\rangle_{p31m}^{\mu_2}
\end{align}
i.e., there is no trivialization. 

And the role of reflection symmetry should also be investigated. Reflection symmetry solely acts on 0D blocks labeled by $\mu_1$ internally. We repeatedly consider the atomic insulator $|\psi\rangle_{p31m}^{\mu_1}$, with the reflection property as ($\bs{M}_{\tau_1}$ represents the reflection operation with the axis coincide with the 1D blocks labeled by $\tau_1$):
\begin{align}
\bs{M}_{\tau_1}|\psi\rangle_{p31m}^{\mu_1}=c_6^\dag c_5^\dag c_4^\dag c_3^\dag c_2^\dag c_1^\dag|0\rangle=-|\psi\rangle_{p31m}^{\mu_1}
\end{align}
i.e., the atomic insulator $|\psi\rangle_{p31m}^{\mu_1}$ can trivialize the reflection eigenvalue $-1$. 

With all possible bubble equivalences, we are ready to study the trivial block-states. Start from the original trivial state (nothing is decorated on arbitrary blocks):
\[
[(+,+),(+,1);0;0]
\]
If we take 2D ``Majorana'' bubble equivalences $l_0$ times and 1D bubble equivalences on $\tau_2$ by $l_2$ times, above trivial state will be deformed to a new 0D block-state labeled by:
\begin{align}
\left[(+,(-1)^{l_2}),((-1)^{l_2},1);0;l_0\right]
\label{p31m spinless trivial state}
\end{align}
According to the definition of bubble equivalence, all these 0D block-states should be trivial. It is straightforward to check that there are only two independent quantities in Eq. (\ref{p31m spinless trivial state}): $l_0$ and $l_2$. Hence all trivial block-states form the following group:
\begin{align}
\{\mathrm{TBS}\}_{p31m,0}&=\{\mathrm{TBS}\}_{p31m,0}^{\mathrm{1D}}\times\{\mathrm{TBS}\}_{p31m,0}^{\mathrm{0D}}\nonumber\\
&=\mathbb{Z}_2\times\mathbb{Z}_2
\end{align}
here $\{\mathrm{TBS}\}_{p31m,0}^{\mathrm{1D}}$ represents the group of trivial states with non-vacuum 1D blocks, and $\{\mathrm{TBS}\}_{p31m,0}^{\mathrm{0D}}$ represents the group of trivial states with non-vacuum 0D blocks.

Therefore, all obstruction and trivialization free 0D/1D block-states are classified as:
\begin{align}
\begin{aligned}
&E_{p31m,0}^{\mathrm{0D}}=\{\mathrm{OFBS}\}_{p31m,0}^{\mathrm{0D}}/\{\mathrm{TBS}\}_{p31m,0}^{\mathrm{0D}}=\mathbb{Z}_2^2\times\mathbb{Z}_3\\
&E_{p31m,0}^{\mathrm{1D}}=\{\mathrm{OFBS}\}_{p31m,0}^{\mathrm{1D}}/\{\mathrm{TBS}\}_{p31m,0}^{\mathrm{1D}}=\mathbb{Z}_2
\end{aligned}
\end{align}
and all independent nontrivial block-states are labeled by different group elements of the following quotient group:
\begin{align}
\mathcal{G}_{p31m}^0=E_{p31m,0}^{\mathrm{0D}}\times E_{p31m,0}^{\mathrm{1D}}=\mathbb{Z}_2^3\times\mathbb{Z}_3
\end{align}
here one $\mathbb{Z}_2$ is from the Majorana chain decorations on 1D blocks labeled by $\tau_1$, and $\mathbb{Z}_2^2\times\mathbb{Z}_3$ is from the nontrivial 0D block-states. Similar with the $p3m1$ case, there is no stacking between 1D and 0D block-states, and the group structure of the classification data $E_{p31m,0}$ has already been accurate.

\subsubsection{Spin-1/2 fermions}
First, we investigate the 0D block-state decorations. For each 0D block labeled by $\mu_1$, the classification data can be characterized by different 1D irreducible representations of the full symmetry group:
\begin{align}
\mathcal{H}^1\left[U^f(1)\rtimes_{\omega_2}(\mathbb{Z}_3\rtimes\mathbb{Z}_2),U(1)\right]=\mathbb{Z}_2^2
\end{align}
For each 0D block labeled by $\mu_2$, the total on-site symmetry group is $\mathbb{Z}_6^f$: nontrivial $\mathbb{Z}_2^f$ extension of the on-site symmetry group $\mathbb{Z}_3$, whereas different 0D block-states can be characterized by different 1D irreducible representations of the corresponding symmetry group:
\begin{align}
\mathcal{H}^1\left[\mathbb{Z}_6^f,U(1)\right]=\mathbb{Z}_6
\end{align}
Then we investigate the possible trivialization. Consider the 1D bubble equivalence on 1D blocks labeled by $\tau_2$: we decorate a 1D bubble onto each of them, here both yellow and red dots represent the complex fermions. Near each 0D block labeled by $\mu_1$, there are 6 complex fermions which can form an atomic insulator with even fermion parity who cannot lead to any trivialization:
\begin{align}
|\phi\rangle_{p31m}^{\mu_1}=\prod\limits_{j=1}^6a_j^\dag|0\rangle
\end{align}
Near each 0D block labeled by $\mu_2$, there are 3 complex fermions which can form another atomic insulator with odd fermion parity:
\begin{align}
|\phi\rangle_{p31m}^{\mu_2}=a_1^\dag a_2^\dag a_3^\dag|0\rangle
\end{align}
On each 0D block labeled by $\mu_2$, the eigenvalue $-1$ of the symmetry group $\mathbb{Z}_6^f$ is exactly the complex fermion because the symmetry group $\mathbb{Z}_6^f$ is the on-site symmetry $\mathbb{Z}_3$ extended by fermion parity nontrivially. Therefore, the eigenvalue $-1$ of the symmetry group $\mathbb{Z}_6^f$ can be trivialized by the atomic insulator $|\phi\rangle_{p31m}^{\mu_2}$. As a consequence, the classification attributed to 0D block-state decorations is:
\begin{align}
E_{p31m,1/2}^{\mathrm{0D}}=\mathbb{Z}_4\times\mathbb{Z}_3
\end{align}
and the index $\mathbb{Z}_3$ has the bosonic root phase.

Subsequently we investigate the 1D block-state decoration. For 1D blocks labeled by $\tau_1$, the total symmetry group is $\mathbb{Z}_4^f$, so there is no nontrivial block-state due to the trivial classification of the corresponding 1D FSPT phases; for 1D blocks labeled by $\tau_2$, the unique candidate block-state is Majorana chain due to the absence of the on-site symmetry. Majorana chain decoration on $\tau_2$ leaves 3 dangling Majorana modes at each 0D block labeled by $\mu_2$. It is well-known that the odd number of Majorana modes cannot be gapped out in a symmetric way, and the no-open-edge condition is violated. As a consequence, the classification attributed to 1D block-state decorations is:
\begin{align}
E_{p31m,1/2}^{\mathrm{1D}}=\mathbb{Z}_1
\end{align}
It is obvious that there is no stacking between 1D and 0D block-states because of the absence of nontrivial 1D root phase. Therefore, the ultimate classification with accurate group structure is:
\begin{align}
\mathcal{G}_{p31m}^{1/2}=\mathbb{Z}_4\times\mathbb{Z}_3
\end{align}

\subsubsection{With $U^f(1)$ charge conservation}
Then we consider the systems with $U^f(1)$ charge conservation. For each 0D block labeled by $\tau_1$, different 0D block-states are characterized by different irreducible representations of the symmetry group as:
\begin{align}
\mathcal{H}^1[U(1)\times(\mathbb{Z}_3\rtimes\mathbb{Z}_2)]=\mathbb{Z}\times\mathbb{Z}_2
\end{align}
Here $\mathbb{Z}$ represents the complex fermion and $\mathbb{Z}_2$ represents the eigenvalues of reflection symmetry operation. For each 0D block labeled by $\tau_2$, different 0D block-states are also characterized by different irreducible representations of the symmetry group as:
\begin{align}
\mathcal{H}^1[U(1)\times\mathbb{Z}_3,U(1)]=\mathbb{Z}\times\mathbb{Z}_3
\end{align}
Here $\mathbb{Z}$ represents the complex fermion and $\mathbb{Z}_3$ represents the eigenvalues of 3-fold rotational symmetry operation. We should further consider possible trivializations. For systems with spinless fermions, consider the 1D bubble equivalence: we decorate a 1D bubble on each 1D block labeled by $\tau_2$, here yellow and red dots represent the fermionic particle and hole, respectively, and can be trivialized if we shrink them to a point. Near each 0D block labeled by $\mu_1$, there are six particles that can form an atomic insulator:
\begin{align}
|\xi\rangle_{p31m}^{\mu_1}=\prod\limits_{j=1}^6p_j^\dag|0\rangle
\end{align}
with reflection property as:
\begin{align}
\bs{M}_{\tau_1}|\xi\rangle_{p31m}^{\mu_1}=p_6^\dag p_5^\dag p_4^\dag p_3^\dag p_2^\dag p_1^\dag|0\rangle=-|\xi\rangle_{p31m}^{\mu_1}
\end{align}
i.e., the reflection eigenvalue $-1$ at each 0D block labeled by $\mu_1$ can be trivialized by the atomic insulator $|\xi\rangle_{p31m}^{\mu_1}$. 

Subsequently we consider the complex fermion sector. First of all, as shown in Fig. \ref{p31m}, we should identify that there is only one 0D block labeled by $\mu_1$ per unit cell, but there are two 0D blocks labeled by $\mu_2$ per unit cell. Repeatedly consider 1D bubble equivalence on 1D blocks $\tau_2$ [cf. 1D bubble, here yellow and red dots represent particle and hole, respectively, and they can be trivialized if we shrink them to a point]: it adds six complex fermions with $U^f(1)$ charge $+1$ at each 0D block $\mu_1$ and three complex fermions with $U^f(1)$ charge $-1$ at each 0D block $\mu_2$, hence the $U^f(1)$ charges at $\mu_1$ and $\mu_2$ are not independent. 

With the help of above discussions, we consider the 0D block-state decorations. The 0D block-state at $\mu_1$ can be labeled by $(m_1,\pm)$, where $m_1\in\mathbb{Z}$ labels the $U^f(1)$ charges on $\mu_1$, $\pm$ labels the eigenvalues of reflection at $\mu_1$; the 0D block-state at $\mu_2$ can be labeled by $(m_2,\phi_2)$, where $m_2\in\mathbb{Z}$ labels the $U^f(1)$ charges on $\mu_2$, $\phi_2$ labels the eigenvalues of 3-fold rotation at $\mu_2$. Start from the following trivial state:
\begin{align}
[(0,+),(0,1)]
\end{align}
Take aforementioned 1D bubble constructions on $\tau_2$ by $l_2$ times, above trivial state will be deformed to the following 0D block-state:
\begin{align}
\left[\left(6l_2,(-1)^{l_2}\right),\left(-3l_2,1\right)\right]
\end{align}
And this state should be trivial. There are only one independent index ($l_2$). Therefore, the ultimate classification of crystalline topological phases protected by $p31m$ symmetry for 2D systems with spinless fermions is:
\begin{align}
\mathcal{G}_{p31m,0}^{U(1)}&=\mathbb{Z}^2\times\mathbb{Z}_3\times\mathbb{Z}_2/(3\mathbb{Z}\times\mathbb{Z}_2)\nonumber\\
&=\mathbb{Z}\times\mathbb{Z}_3\times\mathbb{Z}_6
\end{align}

For systems with spin-1/2 fermions, the reflection property of $|\xi\rangle_{p31m}^{\mu_1}$ near the 0D block labeled by $\mu_1$ are changed by an additional $-1$, which leads to no trivialization. It is easy to verify that the complex fermion decorations for spinless and spin-1/2 fermions are identical. Therefore, the ultimate classification of crystalline topological phases protected by $p4g$ symmetry for 2D systems with spin-1/2 fermions is:
\begin{align}
\mathcal{G}_{p31m,1/2}^{U(1)}&=\mathbb{Z}^2\times\mathbb{Z}_3\times\mathbb{Z}_2/3\mathbb{Z}\nonumber\\
&=\mathbb{Z}\times\mathbb{Z}_3\times\mathbb{Z}_6
\end{align}

\subsection{$p6$}
\begin{figure}
\centering
\includegraphics[width=0.46\textwidth]{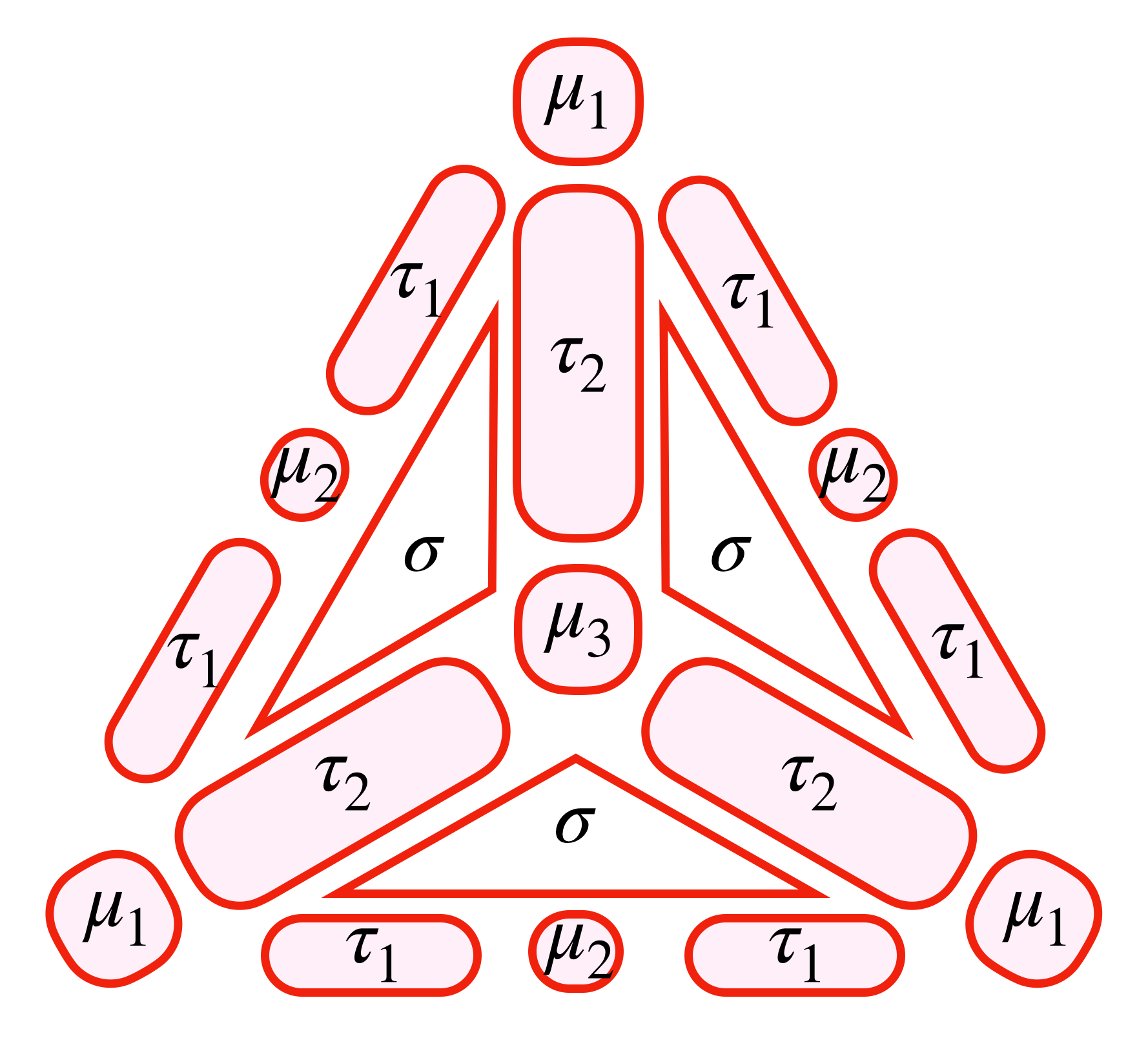}
\caption{\#16 wallpaper group $p6$ and its cell decomposition.}
\label{p6}
\end{figure}

The corresponding point group of this case is 6-fold rotation group $C_6$ by quotient out the translations. For 2D blocks $\sigma$, 1D blocks $\tau_1$ and $\tau_2$, there is no on-site symmetry; for 0D blocks $\mu_1$, the on-site symmetry group is $\mathbb{Z}_6$ attributed to the 6-fold rotational symmetry acting internally; for 0D blocks $\mu_2$, the on-site symmetry group is $\mathbb{Z}_2$ attributed to the 2-fold rotational symmetry acting internally; for 0D blocks $\mu_3$, the on-site symmetry group is $\mathbb{Z}_3$ attributed to the 3-fold rotational symmetry acting internally, see Fig. \ref{p6}.

\subsubsection{Spinless fermions}
First, we consider the 0D block-state decorations. The different 0D block states at $\mu_1$ can be characterized by different 1-dimensional irreducible representations of the symmetry group:
\begin{align}
\mathcal{H}^1\left[\mathbb{Z}_2^f\times\mathbb{Z}_6,U(1)\right]=\mathbb{Z}_2\times\mathbb{Z}_6
\end{align}
Here $\mathbb{Z}_2$ represents the complex fermion, and $\mathbb{Z}_6$ represents the 6-fold rotation eigenvalues, and different 0D block-states on each $\mu_1$ can be labeled by $(\pm,\nu_1)$, where $\pm$ represents the fermion parity, and $\nu_1\in\{e^{i\pi\cdot j/3}|j=0,1,2,3,4,5\}$ represents the eigenvalues of 6-fold rotation operation.

The different 0D block-states at $\mu_2$ can be characterized by different 1-dimensional irreducible representations of the symmetry group:
\begin{align}
\mathcal{H}^1\left[\mathbb{Z}_2^f\times\mathbb{Z}_2,U(1)\right]=\mathbb{Z}_2\times\mathbb{Z}_2
\end{align}
Here the first $\mathbb{Z}_2$ represents the complex fermion, and the second $\mathbb{Z}_2$ represents the 2-fold rotation eigenvalues, and different 0D block-states on each $\mu_1$ can be labeled by $(\pm,\pm)$, where two $\pm$'s represent the fermion parity and 2-fold rotation eigenvalue.

The different 0D block-states at $\mu_3$ can be characterized by different 1-dimensional irreducible representations of the symmetry group:
\begin{align}
\mathcal{H}^1\left[\mathbb{Z}_2^f\times\mathbb{Z}_3,U(1)\right]=\mathbb{Z}_2\times\mathbb{Z}_3
\end{align}
Here the first $\mathbb{Z}_2$ represents the complex fermion, and the second $\mathbb{Z}_3$ represents the 3-fold rotation eigenvalues, and different 0D block-states on each $\mu_1$ can be labeled by $(\pm,\nu_3)$, where $\pm$ represents the fermion parity, and $\nu_3\in\{1,e^{2\pi i/3},e^{4\pi i/3}\}$ represents the eigenvalues of 3-fold rotation operation. According to this notation, all obstruction-free 0D block-states form the following group:
\begin{align}
\{\mathrm{OFBS}\}_{p6,0}^{\mathrm{0D}}=\mathbb{Z}_6\times\mathbb{Z}_3\times\mathbb{Z}_2^4
\end{align}

Subsequently we consider the 1D block-state decoration. For arbitrary 1D block, the unique candidate block-state is Majorana chain because of the absence of on-site symmetry. So all 1D block-states form a group:
\begin{align}
\{\mathrm{BS}\}_{p6,0}^{\mathrm{1D}}=\mathbb{Z}_2^3
\end{align}
Majorana chain decoration on $\tau_1$ leaves 6 dangling Majorana modes at each $\mu_1$ and 2 dangling Majorana modes at each $\mu_2$. Consider the Majorana modes as the edge modes of the decorated Majorana chains at each $\mu_1$, with the following rotational symmetry properties (all subscripts are taken with modulo 6):
\begin{align}
\bs{R}_{\mu_1}:~\gamma_j\mapsto\gamma_{j+1},~~j=1,...,6.
\end{align}
Then consider the local fermion parity and its rotational symmetry property:
\begin{align}
P_f=i\prod\limits_{j=1}^6\gamma_j,~~\bs{R}_{\mu_1}:~P_f\mapsto-P_f
\end{align}
Thus these 6 dangling Majorana modes cannot be gapped out in a symmetric way, and the no-open-edge condition cannot be satisfied.

Majorana chain decoration on $\tau_2$ leaves 6 dangling Majorana modes at each $\mu_1$ and 3 dangling Majorana modes at each $\mu_3$. Consider the Majorana modes as the edge modes of the decorated Majorana chains at each $\mu_3$, it is well-known that the odd number of Majorana modes cannot be gapped out, hence the no-open-edge condition cannot be satisfied. Finally, the classification attributed to the 1D block-state decorations is trivial:
\begin{align}
E_{p6,0}^{\mathrm{1D}}=\{\mathrm{OFBS}\}_{p6,0}^{\mathrm{1D}}=\mathbb{Z}_1
\end{align}

With all possible block-states, we discuss all possible trivializations. First, we consider about the 2D bubble equivalence: as we discussed in the main text, only type-\2 (i.e., ``Majorana bubbles'' with anti-PBC) 2D bubble equivalence is valid because there is no 0D block as the center of even-fold dihedral group. Enlarge all “Majorana bubble” construction can be deformed to double Majorana chains at arbitrary 0D block that can be trivialized because there is no on-site symmetry and the classification of 1D invertible topological phases (i.e., Majorana chain) is $\mathbb{Z}_2$. Furthermore, similar with the $p2$ case, 2D ``Majorana bubble'' construction changes the fermion parities of 0D blocks labeled by $\mu_1$ and $\mu_2$ simultaneously.

Then consider the 1D bubble equivalence on 1D blocks labeled by $\tau_2$ [cf. 1D bubble, here both yellow and red dots represent the complex fermions]: Near each 0D block labeled by $\mu_1$, there are 6 complex fermions which form an atomic insulator with even fermion parity:
\begin{align}
|\psi\rangle_{p6}^{\mu_1}=\prod\limits_{j=1}^6c_j^\dag|0\rangle
\end{align}
Hence it cannot change the fermion parity at each 0D block labeled by $\mu_1$; Near each 0D block labeled by $\mu_3$, there are 3 complex fermions which form another atomic insulator with odd fermion parity:
\begin{align}
|\psi\rangle_{p6}^{\mu_3}=c_1'^\dag c_2'^\dag c_3'^\dag|0\rangle
\end{align}
Hence it can change the fermion parity at $\mu_3$. 

Then we study the role of the rotational symmetry. We consider the atomic insulator $|\psi\rangle_{p6}^{\mu_1}$ and $|\psi\rangle_{p6}^{\mu_3}$ repeatedly. Near each 0D block labeled by $\mu_1$, the atomic insulator $\psi\rangle_{p6}^{\mu_1}$ has the following rotation property ($\bs{R}_{\mu_1}$ represents the 6-fold rotation operation centred at the 0D block labeled by $\mu_1$):
\begin{align}
\bs{R}_{\mu_1}|\psi\rangle_{p6}^{\mu_1}=c_2^\dag c_3^\dag c_4^\dag c_5^\dag c_6^\dag c_1^\dag|0\rangle=-|\psi\rangle_{p6}^{\mu_1}
\end{align}
i.e., the rotation eigenvalue $-1$ can be trivialized by the atomic insulator $|\psi\rangle_{p6}^{\mu_1}$; Near each 0D block labeled by $\mu_3$, the atomic insulator $|\psi\rangle_{p6}^{\mu_3}$ has the following rotation property ($\bs{R}_{\mu_3}$ represents the 3-fold rotation operation centred at the 0D block labeled by $\mu_3$):
\begin{align}
\bs{R}_{\mu_3}|\psi\rangle_{p6}^{\mu_3}=c_2'^\dag c_3'^\dag c_1'^\dag|0\rangle=|\psi\rangle_{p6}^{\mu_3}
\end{align}
i.e., no trivialization. We should further consider the 1D bubble equivalence on 1D blocks labeled by $\tau_1$ [cf. 1D bubble, here both yellow and red dots represent the complex fermions]: Near each 0D block labeled by $\mu_1$, there are 6 complex fermions that form an atomic insulator:
\begin{align}
|\phi\rangle_{p6}^{\mu_1}=\prod\limits_{j=1}^6a_j^\dag|0\rangle
\end{align}
with rotation property as:
\begin{align}
\bs{R}_{\mu_1}|\phi\rangle_{p6}^{\mu_1}=a_2^\dag a_3^\dag a_4^\dag a_5^\dag a_6^\dag a_1^\dag|0\rangle=-|\phi\rangle_{p6}^{\mu_1}
\end{align}
i.e., the rotation eigenvalue $-1$ can be trivialized by the atomic insulator $|\phi\rangle_{p6}^{\mu_1}$. Near each 0D block labeled by $\mu_2$, there are 2 complex fermions that form another atomic insulator:
\begin{align}
|\phi\rangle_{p6}^{\mu_2}=a_1'^\dag a_2'^\dag|0\rangle
\end{align}
with rotation property as:
\begin{align}
\bs{R}_{\mu_2}|\phi\rangle_{p6}^{\mu_2}=a_2'^\dag a_1'^\dag|0\rangle=-|\phi\rangle_{p6}^{\mu_2}
\end{align}
i.e., the rotation eigenvalue $-1$ can be trivialized by the atomic insulator $|\phi\rangle_{p6}^{\mu_2}$. Hence we can conclude that the rotation eigenvalues of $\mu_1$ and $\mu_2$ are not independent. 

With all possible 2D and 1D bubble equivalences, we are ready to study the trivial block-states. Start from the original trivial state (nothing is decorated on arbitrary blocks):
\[
[(+,1),(+,+),(+,1)]
\]
If we take 2D bubble construction $l_0$ times, take 1D bubble equivalences on $\tau_1$ and $\tau_2$ by $l_1$ and $l_2$ times, above trivial state will be deformed to a new block-state labeled by:
\begin{align}
&\left[\left((-1)^{l_0},(-1)^{l_1+l_2}\right),\left((-1)^{l_0},(-1)^{l_1}\right),\right.\nonumber\\
&\left.\left((-1)^{l_2},1\right)\right]
\label{p6 spinless trivial state}
\end{align}
According to the definition of bubble equivalence, all these 0D block-states should be trivial. It is straightforward to see that there are only three independent quantities in Eq. (\ref{p6 spinless trivial state}), hence all trivial block-states form the following group:
\begin{align}
\{\mathrm{TBS}\}_{p6,0}=\{\mathrm{TBS}\}_{p6,0}^{\mathrm{0D}}=\mathbb{Z}_2^3
\end{align}
Therefore, all obstruction and trivialization free 0D/1D block-states are classified as:
\begin{align}
\begin{aligned}
&E_{p6,0}^{\mathrm{0D}}=\{\mathrm{OFBS}\}_{p6,0}^{\mathrm{0D}}/\{\mathrm{TBS}\}_{p6,0}^{\mathrm{0D}}=\mathbb{Z}_2^2\times\mathbb{Z}_3^2\\
&E_{p6,0}^{\mathrm{1D}}=\{\mathrm{OFBS}\}_{p6,0}^{\mathrm{1D}}/\{\mathrm{TBS}\}_{p6,0}^{\mathrm{1D}}=\mathbb{Z}_1
\end{aligned}
\end{align}
and the ultimate classification with accurate group structure is:
\begin{align}
\mathcal{G}_{p6}^0=E_{p6,0}^{\mathrm{0D}}=\mathbb{Z}_2^2\times\mathbb{Z}_3^2
\end{align}
We should note that the group structure should be $\mathbb{Z}_2^2\times\mathbb{Z}_3^2$ rather than $\mathbb{Z}_2\times\mathbb{Z}_3\times\mathbb{Z}_6$ because the eigenvalue $-1$ of the 6-fold rotational symmetry at each 0D block labeled by $\mu_1$ is trivialized by 1D bubble equivalence. It is obvious that there is no stacking between 1D and 0D block-states because there is no nontrivial 1D block-state, and the group structure of the classification data $E_{p6,0}$ has already been accurate.

\subsubsection{Spin-1/2 fermions}
First, we investigate the 0D block-state decoration. For each 0D block labeled by $\mu_1$, the total symmetry group is $\mathbb{Z}_{12}^f$: on-site $\mathbb{Z}_6$ symmetry with nontrivial extension of fermion parity, and the different block-states on each of them can be characterized by different 1D irreducible representations of the corresponding symmetry group:
\begin{align}
\mathcal{H}^1\left[\mathbb{Z}_{12}^f,U(1)\right]=\mathbb{Z}_{12}
\end{align}
For each 0D block labeled by $\mu_2$, the total symmetry group is $\mathbb{Z}_{4}^f$: on-site $\mathbb{Z}_2$ symmetry with nontrivial extension of fermion parity, and the different block-states on each of them can be characterized by different 1D irreducible representations of the corresponding symmetry group:
\begin{align}
\mathcal{H}^1\left[\mathbb{Z}_{4}^f,U(1)\right]=\mathbb{Z}_{4}
\end{align}
For each 0D block labeled by $\mu_3$, the total symmetry group is $\mathbb{Z}_6^f$: on-site $\mathbb{Z}_3$ symmetry with nontrivial extension of fermion parity, and the different block-states on each of them can be characterized by different 1D irreducible representations of the corresponding symmetry group:
\begin{align}
\mathcal{H}^1\left[\mathbb{Z}_{6}^f,U(1)\right]=\mathbb{Z}_{6}
\end{align}
Then we consider the possible trivializations. Consider the 1D bubble equivalence on the 1D blocks labeled by $\tau_2$: we decorate a 1D bubble onto each of them, whereas both yellow and red dots represent the complex fermions. Near each 0D block labeled by $\mu_1$, there are 6 complex fermions which can form an atomic insulator with even fermion parity who cannot leads to any trivialization:
\begin{align}
|\phi\rangle_{p6}^{\mu_1}=\prod\limits_{j=1}^6a_j^\dag|0\rangle
\end{align}
Near each 0D block labeled by $\mu_3$, there are 3 complex fermions which can form another atomic insulator with odd fermion parity:
\begin{align}
|\phi\rangle_{p6}^{\mu_3}=a_1^\dag a_2^\dag a_3^\dag|0\rangle
\end{align}
On each 0D block labeled by $\mu_3$, the eigenvalue $-1$ of the symmetry group $\mathbb{Z}_6^f$ is exactly the fermion parity. 
Therefore, the eigenvalue $-1$ of the symmetry group $\mathbb{Z}_6^f$ can be trivialized by the atomic insulator $|\phi\rangle_{p6}^{\mu_3}$. As a consequence, the classification attributed to 0D block-state decorations is:
\begin{align}
E_{p6,1/2}^{\mathrm{0D}}=\mathbb{Z}_{12}\times\mathbb{Z}_4\times\mathbb{Z}_6/\mathbb{Z}_2=\mathbb{Z}_{12}\times\mathbb{Z}_4\times\mathbb{Z}_3
\end{align}

Subsequently we consider the 1D block-state decoration. For arbitrary 1D block, the unique candidate 1D block-state is Majorana chain due to the absence of on-site symmetry. Majorana chain decoration on $\tau_1$ leaves 6 dangling Majorana modes at each $\mu_1$ and 2 dangling Majorana modes at each $\mu_2$. Consider the Majorana modes as the edge modes of the decorated Majorana chains at $\mu_1$, with the following rotational symmetry properties (here $\bs{R}_{\mu_1}$ represents the 6-fold rotation operation centred at each 0D block labeled by $\mu_1$):
\begin{align}
\bs{R}_{\mu_1}:~\gamma_j\mapsto\gamma_{j+1},~j=1,...,5;~\gamma_6\mapsto-\gamma_1.
\end{align}
Consider the Hamiltonian including 3 entanglement pairs which is invariant under the rotational symmetry:
\begin{align}
\mathcal{H}=i\sum\limits_{j=1}^3\gamma_{j}\gamma_{j+3}
\end{align}
and these Majorana chains are gapped out in a symmetric way. For Majorana modes at each $\mu_2$, identical with the $p2$ case, these 2 Majorana modes can be gapped out by an entanglement pair who respects all symmetry. Therefore, the no-open-edge condition is satisfied.

Majorana chain decoration on $\tau_2$ leaves 6 dangling Majorana modes at each $\mu_1$ and 3 dangling Majorana modes at each $\mu_3$. It is well-known that odd number of Majorana modes cannot be gapped out, hence the no-open-edge condition cannot be satisfied, and finally the classification attributed to 1D block-state decorations is:
\begin{align}
E_{p6,1/2}^{\mathrm{1D}}=\mathbb{Z}_2
\end{align}

With full data of classification, we investigate the possible stacking between 1D and 0D block-states. If we decorate two Majorana chains on each 1D block labeled by $\tau_1$, it can be deformed to an assembly of 0D block-states at 0D blocks labeled by $\mu_1$ and $\mu_2$. Similar with the $pmm$ case, 1D block-states extend 0D block-states at 0D blocks $\mu_1$ and $\mu_2$ simultaneously. Therefore, the ultimate classification with accurate group structure is:
\begin{align}
\mathcal{G}_{p6}^{1/2}=\mathbb{Z}_{12}\times\mathbb{Z}_8\times\mathbb{Z}_3
\end{align}

\subsubsection{With $U^f(1)$ charge conservation}
Then we consider the systems with $U^f(1)$ charge conservation. For each 0D block labeled by $\mu_1$, different 0D block-states are characterized by different irreducible representations of the symmetry group as:
\begin{align}
\mathcal{H}^1[U(1)\times\mathbb{Z}_6,U(1)]=\mathbb{Z}\times\mathbb{Z}_6
\end{align}
Here $\mathbb{Z}$ represents the complex fermion and $\mathbb{Z}_6$ represents the eigenvalues of 6-fold rotational symmetry operation. For each 0D block labeled by $\mu_1$, different 0D block-states are also characterized by different irreducible representations of the symmetry group as:
\begin{align}
\mathcal{H}^1[U(1)\times\mathbb{Z}_2,U(1)]=\mathbb{Z}\times\mathbb{Z}_2
\end{align}
Here $\mathbb{Z}$ represents the complex fermion and $\mathbb{Z}_2$ represents the eigenvalues of 2-fold rotational symmetry operation. For each 0D block labeled by $\mu_1$, different 0D block-states are still characterized by different irreducible representations of the symmetry group as:
\begin{align}
\mathcal{H}^1[U(1)\times\mathbb{Z}_3,U(1)]=\mathbb{Z}\times\mathbb{Z}_3
\end{align}
Here $\mathbb{Z}$ represents the complex fermion and $\mathbb{Z}_3$ represents the eigenvalues of 3-fold rotational symmetry operation. We should further consider possible trivializations. For systems with spinless fermions, consider the 1D bubble equivalence: we decorate a 1D bubble on each 1D block labeled by $\tau_1$, here yellow and red dots represent the fermionic particle and hole, respectively, and can be trivialized if we shrink them to a point. Near each 0D block labeled by $\mu_1$, there are 6 particles that can form an atomic insulator:
\begin{align}
|\xi\rangle_{p6}^{\mu_1}=\prod\limits_{j=1}^6p_j^\dag|0\rangle
\end{align}
with rotation property as:
\begin{align}
\bs{R}_{\mu_1}|\xi\rangle_{p6}^{\mu_1}=p_2^\dag p_3^\dag p_4^\dag p_5^\dag p_6^\dag p_1^\dag|0\rangle=-|\xi\rangle_{p6}^{\mu_1}
\end{align}
i.e., the rotation eigenvalue $-1$ at 0D blocks $\mu_1$ are trivialized by the atomic insulator $|\xi\rangle_{p6}^{\mu_1}$. Near $\mu_2$, there are two holes that can form another atomic insulator:
\begin{align}
|\xi\rangle_{p6}^{\mu_2}=h_1^\dag h_2^\dag|0\rangle
\end{align}
with rotation property as:
\begin{align}
\bs{R}_{\mu_2}|\xi\rangle_{p6}^{\mu_2}=h_2^\dag h_1^\dag|0\rangle=-|\xi\rangle_{p6}^{\mu_2}
\end{align}
i.e., the rotation eigenvalue $-1$ at 0D blocks $\mu_2$ are trivialized by the atomic insulator $|\xi\rangle_{p6}^{\mu_2}$. Equivalently, the rotation eigenvalues of $\mu_1$ and $\mu_2$ are not independent. Moreover, consider the 1D bubble equivalence on $\tau_2$: we decorate a 1D bubble on each of them, here yellow and red dots represent the fermionic particle and hole, respectively, and can be trivialized if we shrink them to a point. Near each 0D block labeled by $\mu_1$, there are 6 particles that can form an atomic insulator:
\begin{align}
|\eta\rangle_{p6}^{\mu_1}=\prod\limits_{j=1}^6p_j'^\dag|0\rangle
\end{align}
with rotation property as:
\begin{align}
\bs{R}_{\mu_1}|\eta\rangle_{p6}^{\mu_1}=p_2'^\dag p_3'^\dag p_4'^\dag p_5'^\dag p_6'^\dag p_1'^\dag|0\rangle=-|\eta\rangle_{p6}^{\mu_1}
\end{align}
i.e., the rotation eigenvalue $-1$ at 0D blocks $\mu_1$ are trivialized by the atomic insulator $|\eta\rangle_{p6}^{\mu_1}$. Near $\mu_3$, there are three holes that can form another atomic insulator:
\begin{align}
|\eta\rangle_{p6}^{\mu_3}=h_1'^\dag h_2'^\dag h_3'^\dag|0\rangle
\end{align}
with rotation property as:
\begin{align}
\bs{R}_{\mu_3}|\eta\rangle_{p6}^{\mu_3}=h_2'^\dag h_3'^\dag h_1'^\dag|0\rangle=|\eta\rangle_{p6}^{\mu_3}
\end{align}
so there is no trivialization near $\mu_3$. 

Subsequently we consider the complex fermion sector. First of all, as shown in Fig. \ref{p6}, we should identify that there is one 0D block labeled by $\mu_1$ per unit cell, two 0D block labeled by $\mu_3$ per unit cell and three 0D block labeled by $\mu_2$ per unit cell. Repeatedly consider 1D bubble equivalence on 1D blocks $\tau_1$ [cf. 1D bubble, here yellow and red dots represent particle and hole, respectively, and they can be trivialized if we shrink them to a point]: it add six complex fermions with $U^f(1)$ charge $+1$ at each 0D block $\mu_1$ and two complex fermions with $U^f(1)$ charge $-1$ at each 0D block $\mu_2$, hence the $U^f(1)$ charges at $\mu_1$ and $\mu_2$ are not independent. 

Then consider 1D bubble equivalence on 1D blocks $\tau_2$: it adds six complex fermions with $U^f(1)$ charge $+1$ at each 0D block $\mu_1$ and three complex fermions with $U^f(1)$ charge $-1$ at each 0D block $\mu_3$, hence the $U^f(1)$ charges at $\mu_1$ and $\mu_3$ are not independent. 

With the help of above 1D bubble constructions, we consider the 0D block-state decorations. The 0D block-state at $\mu_j$ ($j=1,2,3$) can be labeled by $(m_j,\phi_j)$, where $m_j\in\mathbb{Z}$ labels the $U^f(1)$ charges on $\mu_j$, $\phi_j$ labels the eigenvalues of rotations at $\mu_j$ (6-fold for $\mu_1$, 2-fold for $\mu_2$ and 3-fold for $\mu_3$). Start from the following trivial state:
\begin{align}
[(0,1),(0,1),(0,1)]
\end{align}
Take aforementioned 1D bubble construction on $\tau_j$ by $l_j$ times ($j=1,2$), above trivial state will be deformed to the following 0D block-state:
\begin{align}
\left[\left(6l_1+6l_2,(-1)^{l_1+l_2}\right),\left(-2l_1,(-1)^{l_1}\right),\left(-3l_2,1\right)\right]
\end{align}
And this state should be trivial. There are two independent indices $l_1$ and $l_2$, therefore, the ultimate classification of crystalline topological phases protected by $p6$ symmetry for 2D systems with spinless fermions is:
\begin{align}
\mathcal{G}_{p6,0}^{U(1)}&=\mathbb{Z}^2\times\mathbb{Z}_6\times\mathbb{Z}_3\times\mathbb{Z}_2/(2\mathbb{Z}\times3\mathbb{Z})\nonumber\\
&=\mathbb{Z}\times\mathbb{Z}_{12}\times\mathbb{Z}_6\times\mathbb{Z}_3
\end{align}

For systems with spin-1/2 fermions, the rotation properties of $|\xi\rangle_{p6}^{\mu_1}$ and $|\eta\rangle_{p6}^{\mu_1}$ at 0D block $\mu_1$, $|\xi\rangle_{p6}^{\mu_2}$ at 0D block $\mu_2$ are changed by an additional $-1$, which leads to no trivialization. Furthermore, it is easy to verify that the complex fermion decorations for spinless and spin-1/2 fermions are identical, by the same 1D bubble equivalence. Finally the ultimate classification of crystalline topological phases protected by $p6$ symmetry for 2D systems with spin-1/2 fermions is:
\begin{align}
\mathcal{G}_{p6,1/2}^{U(1)}&=\mathbb{Z}^2\times\mathbb{Z}_6\times\mathbb{Z}_3\times\mathbb{Z}_2/(2\mathbb{Z}\times 3\mathbb{Z})\nonumber\\
&=\mathbb{Z}\times\mathbb{Z}_{12}\times\mathbb{Z}_6\times\mathbb{Z}_3
\end{align}

\subsection{Remarks}
In this section, together with other five examples in the main text, we systematically constructed and classified all possible cases of 2D interacting fermionic crystalline TSC and TI by explicit real-space construction (see Sec. {\color{red}\2} in the main text), for both spinless and spin-1/2 fermions. All results of classification, together with accurate group structures (i.e., possible stacking between block-states with different dimensions) are summarized in three tables in the main text. Compare our results with the classifications of FSPT phases protected by the corresponding internal symmetries, we verify the fermionic \textit{crystalline equivalence principle} for all cases.

\providecommand{\noopsort}[1]{}\providecommand{\singleletter}[1]{#1}%

\end{document}